\documentclass{bredele}
\usepackage{physics}
\usepackage{comment}
\usepackage{subfigure}
\usepackage{bbold}
\usepackage{amsmath}
\usepackage[nottoc, notlof, notlot]{tocbibind}
\usepackage[LGR,T1]{fontenc}
\newcommand{\textgreek}[1]{\begingroup\fontencoding{LGR}\selectfont#1\endgroup}
\DeclareUnicodeCharacter{00A0}{}
\title{La Contextualité}

\lesoustitre{Figure de l'\'Etrangeté Quantique}

\discipline{Fondements Quantiques}

\dirdethese{Alexei Grinbaum}
\titredudirdethese{chercheur au LARSIM}
\author{Hippolyte Dourdent}

\date{janvier - août 2017}

\nomdeuniversite{Institut d'Optique Graduate School - CEA Saclay}
\logouniversite{logo_univ} 
\scalelogouniversite{0.5} 
\logolabo{logo_labo}
\scalelogolabo{0.2}

\unite{Laboratoire des Recherches sur les Sciences de la Matière}
\ecoledoc{LARSIM (IRFU)}

\begin{document}

\maketitle

\clearemptydoublepage

\frontmatter
\mainmatter

\clearemptydoublepage
\vspace{-1cm}
\begin{center}\large
{\textbf{La Contextualité, Figure de l'\'Etrangeté Quantique\\} \vspace{0.5cm}}

\normalsize
Hippolyte Dourdent\footnote{(e-mail : hippolyte.dourdent@institutoptique.fr)\newline Supervised by \texttt{Alexei Grinbaum}, \textit{CEA-Saclay/IRFU/LARSIM, 91191 Gif-sur-Yvette, France} }\\ \centering\textit{Institut d'Optique Graduate School, Université Paris Saclay, F91127 Palaiseau, France}\vspace{5mm}

\thispagestyle{empty}

\textbf{Abstract}\\
\end{center}
\footnotesize{The notion of contextuality, which emerges from a theorem established by Simon Kochen and Ernst Specker (1960-1967) and by John Bell (1964-1966), is certainly one of the most fundamental aspects of quantum weirdness. If it is a questioning on scholastic philosophy and a study of contrafactual logic that led Specker to his demonstration with Kochen, it was a criticism of von Neumann's "proof" that led John Bell to the result. A misinterpretation of this famous "proof" will lead them to diametrically opposite conclusions. Over the last decades, remarkable theoretical progresses have been made on the subject in the context of the study of quantum foundations and quantum information. Thus, the graphic generalizations of Cabello-Severini-Winter and Ac\'in-Fritz-Leverrier-Sainz raise the question of the connection between non-locality and contextuality. It is also the case of the sheaf-theoretic approach of Samson Abramsky et al., which also invites us to compare contextuality with the logical structure of certain classical logical paradoxes. Another approach, initiated by Robert Spekkens, generalizes the concept to any type of experimental procedure. This new form of "universal" contextuality has been raised as a criterion of non-classicality, i.e. of weirdness. It notably led to identify the nature of curious quantum paradoxes involving post-selections and weak measurements. In the light of the fiftieth anniversary of the publication of the Kochen-Specker theorem, this report aims to introduce these results little known to the French scientific public, in the context of an investigation on the nature of the weirdness of quantum physics.}

\begin{center} \textbf{Résumé\\}\end{center}
\footnotesize{La notion de contextualité, associée à un théorème établi par Simon Kochen et Ernst Specker (1960-1967) et par John Bell (1964-1966), est certainement l'une des facettes les plus fondamentales de l'étrangeté quantique. Si c'est un questionnement issu de la philosophie scholastique et d'une étude de la logique contrafactuelle qui mena Specker à sa démonstration avec Kochen, c'est une critique de la "preuve" de von Neumann qui conduisit John Bell au résultat. Une mauvaise interprétation de cette fameuse "preuve" les guidera par ailleurs vers des conclusions diamétralement opposées. Au cours des dernières décennies, des avancées théoriques remarquables ont été réalisées autour de ce théorème dans le contexte de l'étude des fondements quantiques et de l'information quantique. Les généralisations graphiques de Cabello-Severini-Winter et de Ac\'in-Fritz-Leverrier-Sainz posent ainsi la question du rapport entre non-localité et contextualité. C'est aussi le cas de l'approche faisceau de Samson Abramsky et al., qui nous invite également à comparer la contextualité à la structure logique de certains paradoxes logiques classiques. Une autre approche, initiée par Robert Spekkens, généralise le concept à tout type de procédure expérimentale . Cette nouvelle forme de contextualité, "universelle", a été érigée en critère de non-classicité, i.e. d’étrangeté, et a notamment permis d’identifier la nature de curieux paradoxes quantiques impliquant post-sélections et mesures faibles. À l’aune du cinquantenaire de la publication du théorème de Kochen-Specker, ce mémoire vise ainsi à introduire ces résultats peu connus du public scientifique français, dans le contexte d’un questionnement sur la nature de l’étrangeté de la physique quantique.}

\normalsize

\chapter*{Remerciements}
\thispagestyle{empty}

(\textit{Mars 2017})\\

Je tiens avant tout à sincèrement remercier mon maître de stage, Alexei Grinbaum, pour ses conseils avisés, son honnêteté et sa confiance ; pour avoir répondu à toutes mes questions avec une impressionnante clarté ; et pour m'avoir inculqué une rigueur dans l'écriture et la réflexion. Merci pour m'avoir permis de m'épanouir pendant ces six mois, et m'avoir donné l'opportunité de vivre pleinement ma passion.\\

Je souhaite également exprimer ma gratitude à l'ensemble des membres du LARSIM, pour TOUT :\\

Merci à Etienne Klein pour sa sollicitude ; pour avoir décuplé mon intérêt pour la physique il y a 5 ans ; pour m'avoir indirectement conduit à intégrer l'IOGS et à me pencher sur les questions des fondements quantiques.\\

Merci à Vincent Bontems pour sa jovialité et sa gentillesse, ses "cours magistraux" de politique, de philosophie et de culture générale ; pour Bachelard, Nikopol, les Idées Noires, les Deux du Balcon, Stranger Things et j'en passe !\\

Merci à Vincent Minier pour son soutien et sa bienveillance.\\

Merci à Gilles Cohen-Tannoudji et Roland Lehoucq (SAp) pour les petites dicussions aux restos du CEA, furtives mais particulièrement enrichissantes.\\

Enfin, un très grand merci aux "jeunes" du LARSIM, Alice, JC et Thomas, pour les thés, les cafés, les fou-rires, le covoit', les afterwork... ; pour leur irremplaçable amitié.\\

\textit{Mise à jour, Août 2017\\}

Mes recherches sur la contextualité ont temporellement débordé du cadre du stage, et la liste des remerciements s'en voit rallongée.\\

Je tiens tout d'abord à réitérer mes remerciements aux personnes précédemment mentionnées, ainsi qu'à ma famille.\\

Merci également à Elisabeth et Jonathan, pour m'avoir offert, l'espace de quelques jours, le cadre idéal pour avancer dans l'écriture de ce mémoire.\\

Je suis profondément reconnaissant envers Lídia del Rio et les autres organisateurs de l'école d'été "Solstice of Foundations". Un grand merci également à Ana Belén Sainz, Ravi Kunjwal, David Schmid, Matthew Leifer et Robert Spekkens pour avoir pris le temps de répondre à toutes mes questions.\\

Enfin, merci à Ad\'an Cabello pour avoir partagé avec moi de magnifiques images d'archives, et merci à Chris Fuchs, Antoine Suarez et (une nouvelle fois) à Robert Spekkens pour m'avoir aidé et guidé dans ma quête sur les origines du théorème de Kochen-Specker.

\clearemptydoublepage

\chapter*{Contextualisation}

Qu'est ce qu'un contexte ? Il s'agit, d'après le Larousse, de "l'ensemble des conditions naturelles, sociales, culturelles dans lesquelles se situe un énoncé, un discours"; ou encore de "l'ensemble des circonstances dans lesquelles se produit un événement, se situe une action" ; de "l'ensemble du texte à l'intérieur duquel se situe un élément d'un énoncé et dont il tire sa signification." La définition à choisir dépend du contexte.\\

Ce mémoire est né dans un contexte particulier. Il est, à l'origine, le fruit d'un stage de six mois effectué au LARSIM - le Laboratoire des Recherches sur les Sciences de la Matière dirigé par Etienne Klein au CEA Saclay - et supervisé par Alexei Grinbaum. La première moitié de ce stage consista en une initiation au domaine des fondements quantiques (quantum foundations), qui vise à une meilleure compréhension et une meilleure formulation de  la théorie quantique en étudiant notamment sa structure conceptuelle et mathématique. La seconde moitié fut consacrée à l'étude de la contextualité, qui est l'objet du mémoire. \\

La "mise en contexte" est ce qui nous permet de donner du sens à un acte, à un évènement. Un rire peut par exemple être nerveux, moqueur ou sincère, en fonction du contexte. "Sortie de son contexte", une phrase peut être mal interprétée, et mener à de terribles quiproquos. Notre vie et notre langage sont contextuels. Pourtant, la notion de non-contextualité, i.e. de ce qui est indépendant du contexte, nous est également familière, lorsqu'elle s'applique à nos observations. Que je contemple à un temps $t$ un cratère lunaire avec un télescope, des jumelles, ou mes lunettes, cela n'a aucune différence sur le résultat : le cratère est toujours là, a toujours la même circonférence, la forme, et la même profondeur. Si nos mesures étaient contextuelles, alors il faudrait s'attendre à ce que les caractéristiques de l'objet observé change en fonction du regard que l'on porte sur lui. \\

L'illustration suivante, bien connue, pourrait apparaître comme un contre-exemple évident. Les couleurs des cercles semblent différentes, mais sont pourtant identiques. L'environnement de l'objet influence simplement notre perception de celui-ci. 

\begin{figure}[ht!]
\centering
\includegraphics[width=4cm]{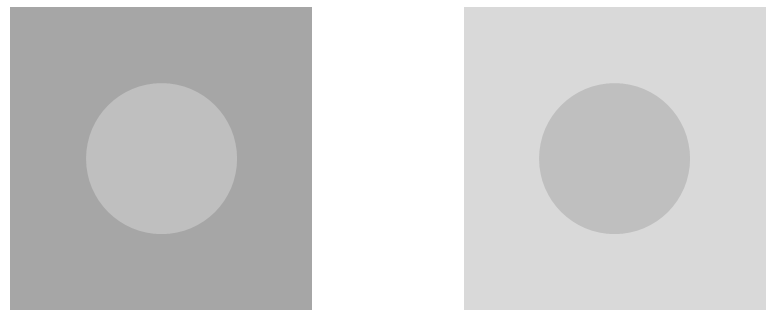}
\caption{Ilusion d'optique : les deux cercles sont de même couleur.}
\end{figure}

Néanmoins, la nature de l'objet n'est pas affectée. Ce n'est pas la réalité de l'objet qui est modifiée, mais notre interprétation personnelle de cette réalité. Pour un appareil de mesure doté d'un détecteur de teintes, l'étrangeté apparente de ce phénomène serait balayée. Le résultat de mesure serait le même, indépendamment du type d'instrument utilisé.\\

Dans notre contexte de vie quotidienne, classique, nos mesures sont donc bien non-contextuelles. Dans celui de la mécanique quantique, il en est autrement. En effet, la nature de la lumière ne change-t-elle pas en fonction de l'appareil de mesure considéré ? Dans un contexte particulier - la détection d'interférences - elle se comporte comme une onde. Dans un autre - la détection en coïncidence - elle se comporte comme un corpuscule. Or ces deux images classiques, onde et corpuscule, sont contradictoires. Tout étudiant en physique a certainement déjà entendu la phrase : "En mécanique quantique, le résultat de mesure dépend du contexte expérimental", où le "contexte" désigne l'appareil de mesure utilisé. Celle-ci est souvent maladroitement associée à l'interprétation philosophique de Niels Bohr, et son principe de \textit{complémentarité}\footnote{Une brève présentation peut être trouvée en annexe.}, bien qu'il n'ait jamais fait usage des termes "\textit{contexte}" ou "\textit{contextualité}". \\

Au cours de mon initiation au domaine des fondements quantiques, j'ai néanmoins découvert que ces notions de contextes et de contextualité étaient précisément définis par les physiciens. Un \textit{contexte d'une mesure} désigne ainsi un ensemble de mesures compatibles auquel appartient celle que l'on effectue. La \textit{contextualité}, quant à elle, est associée à un théorème, dit "théorème de Kochen-Specker". Ce dernier m'était jusqu'alors inconnu, malgré mon profond intérêt pour la mécanique quantique. Plus étonnant encore, j'ai appris qu'il était considéré comme un résultat particulièrement fondamental, d'importance comparable au théorème de Bell. Il y a trois ans, une équipe de chercheurs en information quantique a ainsi montré qu'il pourrait être la clé permettant de construire un ordinateur quantique universel. Aussi, lorsque mon maître de stage m'a proposé d'en faire le sujet d'un mémoire, j'ai accepté sans hésitation.\\

 Bien que scolairement satisfaisant, la première version de mon rapport manquait paradoxalement de "contexte". La définition du statut de la "contextualité" était ambigüe, et l'origine du concept n'était pas exposée. Ainsi, malgré la fin du stage, j'ai décidé de poursuivre mes recherches en "free-lance", en envisageant de réécrire mon travail.\\

Une version 2.0, plus complète, plus rigoureuse, s'apprêtait à voir le jour trois mois plus tard, lorsque je me rendis à ETH Zurich pour suivre une école d'été de quelques jours, le "Solstice of Foundations", organisée par Lidia del Rio. Là-bas, j'eus l'occasion de discuter avec la plupart des auteurs des articles qui constituent la bibliographie du mémoire, de les harceler de questions, et d'assister notamment aux cours de Ana Bel Sainz (sur la contextualité de Kochen-Specker), de Matthew Leifer (sur les théorèmes no-go) et de Robert Spekkens (sur les réalisations expérimentales dans le domaine des fondements quantiques). \\

La version 3.0, ci-présente, résulte du nouveau regard que cette semaine zurichoise m'a permis de porter sur la contextualité.\\

\`A l'aune du cinquantenaire de la publication du théorème de Kochen-Specker, le présent mémoire vise ainsi à introduire ce résultat peu connu du public scientifique français, dans le contexte d'un questionnement sur la nature de l'étrangeté de la physique quantique (chapitre \ref{chapetrange}). Après une introduction historique du théorème originel de Kochen et Specker (KS) (chapitre \ref{chapKS}), différentes approches de généralisation (graphique, combinatoire et logique) seront brièvement présentées. Chacune conduit à un résultat étonnant : la non-localité pourrait être considérée comme un cas particulier de contextualité (chapitre \ref{chapnlcont}). Une autre approche, basée sur le concept de modèles ontologiques de théories opératoires, généralise la notion de contextualité à tout type de procédure expérimentale (chapitre \ref{chapspek}). Cette nouvelle forme "universelle", inspirée d'un principe philosophique de Leibniz, a été érigée en critère de non-classicité, i.e. d'étrangeté. Elle a par exemple permis d'identifier la nature de curieux paradoxes issus d'une double sélection indépendante des états initial et final d'un système quantique (chapitre \ref{chappara}).\\

\tableofcontents

\chapter{Un contexte : l'étrangeté quantique.}
\label{chapetrange}

La physique quantique est étrange. C'est du moins ainsi qu'elle semble perçue par la société, par la presse scientifique populaire, et par la plupart de la communauté scientifique. C'est aussi la raison pour laquelle elle me tient tant à cœur et que je désire l'explorer. J'ai toujours été fasciné par l'étrange, l'absurde, le paradoxal, par ces histoires qui sortent des sentiers battus et qui vous donne le vertige, ces histoires qui vous pousse au bord du précipice du raisonnable et vous conduisent à remettre en question vos rassurantes idées préconçues. \\

La physique vise à décrire le monde et à en dissoudre l'étrangeté, ou, du moins, à se familiariser avec elle. Pourtant, la physique quantique est entourée d'une certaine aura de complexité et de stupeur, qui fait la part belle au mysticisme et aux pseudosciences.\footnote{cf. aux divers exemples de médecine, thérapie, voyance ou encore "paranormal" quantiques répertoriés et dénoncés par Richard Monvoisin dans "\textit{Quantox. Mésusages idéologiques de la mécanique quantique.}" }\\

Des phrases chocs de grands physiciens avouant eux-mêmes avoir été bousculés par la théorie sont ainsi souvent utilisées et sorties de leur contexte pour rassurer l'audience grand public, ou, au contraire, dans le but d'accentuer cet halo de mystère et faire sensation. En voici un florilège (à utiliser avec parcimonie, donc) :\\
\begin{quotation}
"\textit{La mécanique quantique n’a absolument aucun sens.}" Roger Penrose\\
\end{quotation}
\begin{quotation}
“\textit{Je peux dire de manière certaine que personne ne comprend la physique quantique.}” Richard Feynman \\
\end{quotation}
\begin{quotation}
“\textit{Si vous n’êtes pas complètement désorienté par la mécanique quantique, vous ne la comprenez pas.}” John Wheeler\\
\end{quotation}
\begin{quotation}
"\textit{Ceux qui ne sont pas choqués quand ils rencontrent pour la première fois la théorie quantique ne l’ont probablement pas comprise.}" Niels Bohr\\
\end{quotation}
\begin{quotation}
“\textit{Si la théorie quantique est correcte, cela signifie la fin de la physique en tant que science.}” Albert Einstein\\
\end{quotation}
\begin{quotation}
“\textit{Je n’aime pas la mécanique quantique, et je suis désolé d’avoir eu quelque chose à voir avec ça.}” Erwin Schrödinger\\
\end{quotation}
\begin{quotation}
“\textit{La mécanique quantique est magique.}” Daniel Greenberger\\
\end{quotation}

En quoi la mécanique quantique est-elle si étrange ? Après l'argument d'autorité, vient les exemples. Cette fois, ce sont des "phénomènes quantiques", plus fantasques et bizarres les uns que les autres, qui sont invoqués. L'insaisissable nature de la lumière et de la matière, "parfois ondes, parfois corpuscules" ; le chat "mort-et-vivant" de Schrödinger, ou encore les facétieuses "influences cachées" de la non-localité. Qu'ont en commun ces résultats ? Ils ne sont pas \textit{classiques}, i.e. ils ne peuvent pas être expliqués par la mécanique newtonienne, l'électrodynamique de Maxwell ou la thermodynamique. En revanche, la théorie qui les décrit est belle et bien née dans ce contexte classique.\\

\`A la fin du XIXe siècle, la physique parvient à décrire la quasi-totalité des phénomènes de la Nature. La quête du réel semble arriver à son terme. En 1894, Albert Michelson déclare
\begin{quotation}
\textit{[...] qu'il semble probable que la plupart des grands principes fondamentaux aient été fermement établis. [...]  Un éminent physicien a remarqué que les vérités futures de la science physique doivent être recherchées à la sixième place des décimales. }
\end{quotation}

Cependant, six années plus tard, à l'aube du XXe siècle, Lord Kelvin explique, lors d'un cours devant la Royal Instiution, que 
\begin{quotation}
\textit{la beauté et la clarté de la théorie dynamique [...] est à présent obscurcit par deux nuages.}
\end{quotation}

Le premier nuage, la question de l'existence de l'éther luminifère, conduira Einstein à sa théorie de la relativité restreinte en 1905. La dissipation du second nuage, "la doctrine de Maxwell-Boltzmann concernant la partition de l'énergie", sera apportée par Max Planck l'année même du discours de Kelvin, et sera considérée comme l'acte de naissance de la théorie des quanta. \footnote{On peut cependant noter "qu'il n'est pas certain que Planck ait eu conscience d'avoir inventé la quantification des échanges d'énergie entre la matière et la lumière." (cf. annexe de \cite{klein}). Il est peut-être donc plus juste d'attribuer l'acte de naissance de la physique quantique aux travaux d'Einstein de 1905 sur la lumière. } Pendant les trois décennies suivantes, une vingtaine de brillants physiciens et mathématiciens mèneront une véritable révolution conceptuelle qui aboutira à l'élaboration d'une des plus belles construction intellectuelle de notre histoire. L'édifice théorique de la mécanique quantique prendra sa forme finale à l'aube des années 30, lorsque John von Neumann parviendra  à clarifier la réconciliation, ébauchée par Paul Dirac en 1926, de l'approche matricielle de Heisenberg (1925) et de l'approche ondulatoire de Schrödinger (1926), en démontrant rigoureusement leur équivalence au sein du formalisme des espaces de Hilbert. Les "\textit{Fondements mathématiques de la mécanique quantique}", publiés en 1932, fixent ainsi la structure mathématique moderne de la nouvelle mécanique quantique, et sonnent la fin de la (première) révolution. Elle continuera, tout au long de la seconde moitié du XXe siècle, à faire vibrer la science et la société à travers ses extensions théoriques (théorie quantique des champs) et ses retombées technologiques (laser, transistor, circuits intégrés, ordinateurs). \\

Cette première révolution quantique est issue d'un contexte atomiste, bien que le "second nuage" de la physique dissipé par Planck soit un problème de physique statistique. La mécanique quantique fut construite en tant que théorie de physique atomique et, encore aujourd'hui, c'est ainsi qu'elle semble perçue et enseignée par une grande majorité.\\

La plupart du grand public et du public scientifique s'accorde à dire que "\textit{la mécanique quantique décrit le monde de l'infiniment petit}". Le concept d'infini, s'il peut provoquer une forme de vertige métaphysique, transporte aussi l'idée de quelque chose d'intangible, d'inatteignable, d'ineffable. La mécanique quantique a beau être étrange, dire qu'elle décrit "\textit{l'infiniment petit}" peut véhiculer une forme de distanciation rassurante, rendant notre cohabitation plus facile. Nous ne serions jamais amenés à être directement confrontés à cette étrangeté, qui resterait une forme de curiosité de laboratoire. En somme, une étrangeté occulte, qui ne se plie pas à notre regard, entourée d'un halo de mystère. Cette magie quantique ne serait-elle qu'une histoire d'échelle ? Existe-t-il deux mondes - le monde classique macroscopique, dans lequel nous vivons, et le monde quantique microscopique - reliés par un pont secret, caché à l'échelle mésoscopique intermédiaire ? \\

La paroi me semble poreuse. La découverte de "phénomènes" quantiques de plus en plus grand, voir macroscopiques, tels que la supraconductivité, la superfluidité ou encore la superposition de molécules de fullerène ($C_{60}$), en témoigne. Le contexte historique de création de la théorie quantique est certes atomiste, mais le formalisme ne fait pas de distinction entre objets microscopiques et macroscopiques. C'est d'ailleurs la raison de l'ennui de Schrödinger, qui semble déplorer que l'état quantique, ce "catalogue d'espérances", pouvait aussi bien porté sur un atome radioactif que sur un chat. Niels Bohr se méfiait aussi de cette distinction d'échelles, et préconisait de chercher la frontière dans "\textit{l'utilisation des termes du langage ordinaire pour décrire les propriétés d'un objet}". \cite{alexei6} \\

Les concepts classiques ordinaires ne semblent pas s'appliquer aux objets quantiques. Dirac et von Neumann ont fixé la structure mathématique de la théorie, mais qu'en est-il de son \textit{sens} ? Que disent les équations de la mécanique quantique ?  Le contexte atomiste, teinté de classicisme, est obsolète. Le nuage de Kelvin s'est dissipé, mais a pris la forme d'un épais brouillard. On arrive à la conclusion que pour parler de et faire parler la mécanique quantique,  un nouveau contexte est nécessaire. Physiciens et philosophes se transforment alors en chaman, interprétant les vapeurs du formalisme afin d'en extraire un énoncé sur la Nature. Malheureusement, les chamans ne parviennent pas à se mettre d'accord sur une interprétation commune. \\
\begin{quotation}
\textit{La théorie quantique est la théorie la plus utile et la plus puissante que les physiciens aient jamais conçue. Aujourd'hui pourtant, près de 90 ans après sa formulation, les désaccords sur le sens de la théorie sont plus forts que jamais. De nouvelles interprétations apparaissent tous les jours. Jamais aucune ne disparait.} \hspace{1cm}   David Mermin (cité dans \cite{cabello13})
\end{quotation}

Deux grands camps interprétatifs sont généralement distingués : \\
\begin{itemize}
\item les "\textit{réalistes}" ("réalisme naïf", "réalisme intrinsèque", "réalisme entitique"), pour lesquels la mécanique quantique doit, au même titre que toute théorie physique, décrire un monde indépendant de tout observateur et, si ce n'est pas le cas, doit être complétée. On trouve dans cette catégorie l'interprétation de Dirac-von Neumann, la mécanique bohmienne, l'effondrement spontané, les interprétations modales ou encore les mondes multiples d'Everett.\\  
\item les  "\textit{copenhaguiens}" ("réalisme participatif" \cite{fuchs}, parfois appelés à tord "anti-réalisme"), pour lesquels la mécanique quantique décrit l'état de connaissance (ou de croyance, ou l'information) d'un observateur qui "participe" aux réalisations expérimentales étudiées. Les interprétations de Copenhague (e.g. celles de Bohr, de Heisenberg et de Pauli), l'interprétation relationnelle de Carlo Rovelli, l'interprétation informationnelle de Jeff Bub, ou encore le QBisme entrent dans cette classe.\\
\end{itemize}

Au début de ce stage, encore novice en matière de fondements quantiques, je trouvais que ce pluralisme d'interprétations avait quelque chose d'exaltant. Parcourir le catalogue, lire et apprendre les descriptifs d'interprétations plus exotiques les unes que les autres, à la recherche de celle qui me "parlerait le plus", était passionnant. Aujourd'hui, avec plus de recul, je réalise que cette exaltation était certainement d'origine existentielle, guidée par l'envie d'effectuer un choix pour mieux définir son identité, par le plaisir rassurant d'appartenir à une communauté qui partage les mêmes idées que soi. C'est cette dérive, aux accents quasi-religieux, que Chris Fuchs dénonce dans son fameux article : "\textit{Quantum Mechanics as Quantum Information (and only a little more)}"\footnote{Cette critique vise principalement les théories réalistes, qui selon lui, désirent une "réalité solide - i.e., un Dieu unique qu'ils [les réalistes] peuvent pointer du doigt et déclarer : "Là, ce terme est ce qui est réel dans l'univers même lorsqu'aucun physicien n'est présent" - sans vraiment avoir essayé de sortir du royaume platonicien des mathématiques pures pour la trouver." } \cite{fuchs3}. Toutes les interprétations sont extraites de la même source : "\textit{la formulation des axiomes de la théorie quantique issue des manuels standards.}"\\
\begin{center}
 Les "\textit{Tables de la Loi Quantique}"\\
\end{center}
\vspace{3mm}
  1. Pour tout système physique $A$, on associe un espace de Hilbert $\mathcal{H}_A$. \\
  2. Un état d'un système physique est un vecteur unité $\ket{\psi}_A \in \mathcal{H}_A$  \\
  3.Un système isolé evolue selon l'équation de Schrödinger : $i\hbar \frac{\partial \ket{\psi}}{\partial t} = H \ket{\psi}$\\
  4. Les quantités physiques mesurables sont représentées par des opérateurs auto-adjoints. $M= M^\dagger = \sum_j m_j \Pi_j $ \\
  5. Règle de Born : Quand $M$ est mesuré sur un système $A$ dans l'état $\ket{\psi}$, le résultat $m_j$ est obtenu avec la probabilité $p(m_j)=\bra{\psi}\Pi_j\ket{\psi}$ \\
  6. On associe au système $AB$ composé des deux sous-systèmes $A$ (avec l'espace de Hilbert $\mathcal{H}_A$) et $B$ (avec $\mathcal{H}_B$) l'espace des états $\mathcal{H}_{AB}=\mathcal{H}_A \otimes \mathcal{H}_B$\\

\vspace{3mm}

Fuchs invite ainsi, comme Rovelli avant lui\footnote{\begin{quotation}\textit{La mécanique quantique cessera de paraître déroutante seulement lorsque nous serons capable de dériver le formalisme de la théorie à partir d'un ensemble d'assertions physiques simples ("postulats", "principes") sur le monde. Ainsi nous ne devrions pas essayer d'ajouter une interprétation raisonnable au formalisme de la mécanique quantique, mais plutôt de dériver le formalisme à partir d'un ensemble de postulats motivés expérimentalement.}
\cite{rovelli}\end{quotation}}, à chercher des principes physiques à partir desquels ces axiomes pourraient être dérivés, plutôt que de se contenter de les interpréter. La mécanique quantique ne devrait pas être perçue comme une simple théorie constructive, i.e. fondée sur l'hypothèse d'une réalité matérielle concrète, mais devrait être étudiée comme théorie principielle.\\

\begin{quotation}
\textit{La tâche ne consiste pas à donner du sens aux axiomes quantiques en entassant plus de structure, de définitions d'images de science-fiction sur eux, mais à les rejeter en bloc et à recommencer. Nous devrions être implacables en nous demandant : De quels principes physiques profonds pouvons-nous dériver cette structure mathématique raffinée ?}
( cf. chapitre 4 " \textit{Relation to principle theories}" de \cite{alexei2})
\end{quotation}

On peut ainsi comparer la position actuelle des chercheurs vis-à-vis des fondements quantiques à celle d'Einstein avant l'élaboration de la théorie de la relativité restreinte. Les axiomes mathématiques de la relativité étaient présents dès 1895, sous le nom de "\textit{transformations de Lorentz}". Or, malgré leur adéquation empirique, ces transformations restaient un mystère. Le génie d'Einstein est d'être parvenu à les dériver de principes physiques simples :\\

\begin{itemize}
\item la vitesse de la lumière dans le vide est indépendante de la vitesse de sa source ;
\item les lois de la physique devraient être les mêmes dans tout référentiel inertiel.\\
\end{itemize}

Le programme de reconstruction de la théorie quantique vise un objectif similaire : refonder la théorie à partir de nouveaux axiomes physiques simples (cf. \cite{alexei7} pour une étude détaillée de ce programme) ; répondre à la question de John Wheeler, "\textit{Why the quantum ?}", non pas en tentant d'extraire du sens à partir du formalisme mathématique, mais en élaborant une liste d'axiomes qui permette de distinguer la théorie quantique de la théorie classique et de toutes autres théories. En somme, identifier clairement l'étrangeté quantique. \\

Aujourd'hui, le brouillard n'est toujours pas dissipé, mais une nouvelle dynamique est en marche : plusieurs interprétations ont commencé à être "axiomatisées", et des résultats prometteurs ont été obtenus. Lucian Hardy a ainsi identifié un ensemble de postulats propres à la théorie quantique et la théorie classique des probabilités. Afin de distinguer clairement la théorie quantique, il suffit d'ajouter à cette liste tout principe quantique incohérent avec la théorie classique des probabilités \cite{hardy2}. \\ 

Pour comparer les théories classique et non-classiques (quantiques et autres théories alternatives plus exotiques), un cadre formel commun est nécessaire. Il est possible de l'obtenir en formulant l'ensemble des théories en termes purement opératoires. Dans une théorie opératoire, les concepts primitifs sont des procédures de préparation, de transformation, et de mesure, chacune comprise comme une liste d'instructions qu'un expérimentateur doit respecter. La théorie donne un algorithme mathématique qui fixe la distribution de probabilité sur les résultats (sorties) d'une mesure donnée, pour toute les préparations et transformations possibles. Diverses théories opératoires se distinguent alors par le type de statistiques expérimentales qu'elles autorisent.\\

Au sein d'un tel cadre, il est possible de découvrir des principes physiques quantiques en explorant des alternatives à la théorie. \footnote{En anglais : foil theories. Un foil de X est quelque chose qui permet de souligner les caractéristiques différente de X par contraste avec lui.} Ces "mathématiques de science-fiction" ne sont pas des théories concurrentes ; elles ne sont pas "physiques", i.e. pas empiriquement valables. Cependant, si une théorie alternative partage un ensemble de caractéristiques avec la théorie quantique, alors cet ensemble de caractéristiques ne peut pas être un ensemble complet d'axiomes quantiques. \\

Robert Spekkens a ainsi montré qu'un grand nombre de résultats supposés "étranges" et exclusivement quantiques - non-commutativité, cohérence (superposition), effondrement, complémentarité, interférences, téléportation, effets de post-sélections, théorème de non-clonage et bien d'autres - émergeaient de théories statistiques classiques auxquelles a été ajoutée une condition de restriction épistémologique, i.e. une restriction sur la capacité de connaître l'état du système étudié. Ces "théories épistreintes"\footnote{"Epistricted theories" en anglais} sont en revanche incapables de reproduire d'autres "aspects" de la théorie, comme la violation des inégalités de Bell \cite{spekkens7}.\\

Remarquons qu'un grand nombre des pères fondateurs, qu'ils soient réalistes (Einstein et Schrödinger) ou copenhaguiens (Bohr, Heisenberg, Pauli), semblaient considérer la théorie quantique comme une restriction de notre capacité à obtenir une information sur le monde microscopique.\\

Einstein pensait ainsi que la mécanique quantique était une théorie incomplète, et cherchait comment l'information manquante pouvait être complétée grâce à des variables cachées \cite{harrigan}. Wolfgang Pauli écrivit un jour à Markus Fierz : " \textit{La bien-connue 'incomplétude' de la mécanique quantique (Einstein) est certainement d'une manière ou d'une autre, un fait existant, mais ne peut certainement pas être supprimée en revenant à la physique classique des champs. }"\\
 
Enfin, John von Neumann, qui fut le premier à tenter, dès 1932, une reconstruction principielle de la théorie quantique, enquêta sur l'éventualité que la nature statistique de  la théorie soit une manifestation de notre connaissance lacunaire (auquel cas l'existence de variables cachées expliquerait le besoin d'énoncés probabilistes). Cette vision des choses me semble être un héritage du contexte atomiste, qui transporte l'idée que l'on ne peut pas complètement comprendre ce que l'on ne peut pas appréhender (les objets de l'infiniment petit), que la magie, l'étrangeté quantique est de nature \textit{occulte}, issue d'un réel qui se dérobe à la connaissance. \\

Cependant,  depuis les expériences révolutionnaires d'Alain Aspect en 1980-82 confirmant la violation des inégalités de Bell \cite{aspect2} \cite{aspect3}, la physique quantique est entrée dans une nouvelle ère, un nouveau contexte, informationnel. Cette "seconde révolution quantique" \cite{aspect2} a renversé l'idée que nous nous faisions de l'étrangeté quantique : ce qui était vu comme une limite est devenu source de traitements de l'information jusque-là inimaginables. L'occulte atomiste est devenu source de pouvoir informatique. La non-classicité de la mécanique quantique n'est plus vue comme une restriction, mais comme catalyseur de nouvelles possibilités informatiques extraordinaires, de tâches qui ne peuvent pas être réalisées classiquement. Contrairement aux simples tentatives d'associer une saveur philosophique au formalisme, le  programme de reconstruction axiomatique s'ancre parfaitement dans ce nouveau contexte, et lui fait écho : Quels principes physiques de la théorie quantique permettent de tels avantages ? La possibilité d'effectuer un avantage implique-t-elle la possibilité d'autres avantages ? La théorie quantique est-elle la seule théorie à partir de laquelle émerge ces avantages ? Telles sont questions qui animent à présent non seulement les chercheurs, mais aussi les industriels \cite{eco1}\cite{eco2} .\\

Les géants de l'informatique Google et Microsoft ont  récemment fait part de leur souhait d'accélérer le processus de transition des technologies quantiques du laboratoire à l'ingénierie \cite{nat1}\cite{nat2}\cite{nat3}. En 2018, la Commission Européenne lancera un flagship d'un milliard d'euro dans le but d'effectuer cette transition. Parmi ces technologies, l'ordinateur quantique universel incarne incontestablement le Graal : la possibilité de sa réalisation semblant prendre un nouveau tournant, il devient nécessaire d'identifier les ressources qui lui sont nécessaires. De nombreuses propositions ont été formulées, de l'hypothétique "parallélisme quantique" \cite{para} que certains associent à la superposition, à la quantité d'intrication disponible \cite{quint}, ou encore la discorde \cite{dis}.  Malheureusement, aucune d'entre elles ne semble satisfaisante \cite{nop1}\cite{nop2}\cite{nop3}\cite{nop4}.\\

L'année 2017 marque le cinquantenaire du théorème de Simon Kochen et d'Ernst Specker, qui stipule que la théorie quantique ne peut être expliquée par des modèles non-contextuels, i.e. des modèles où l'assignation de valeurs prédéterminées dépend de l'ensemble des observables comesurables considéré, autrement dit du \textit{contexte} de la mesure.\\

Malgré le fait que le théorème de Kochen-Specker figure parmi les résultats fondamentaux de la théorie quantique, les questions qu'il soulève ont longtemps reçues très peu d'attention. En outre, au cours de la dernière décennie, de formidables avancées ont insufflé une nouvelle vie au sujet, en revisitant d'anciens problèmes conceptuels à l'aune de l'approche de reconstruction informationnelle. 
Ainsi, un nombre croissant d'études invite à penser que la contextualité serait la ressource fondamentale derrière l'avantage quantique pour le calcul et le traitement de l'information. Par exemple, il a récemment été démontré qu'elle est à la base d'une approche permettant de construire des dispositifs quantiques résistants au bruit environnant, appelée "distillation d'états magiques"\cite{comput}, qui sont un ingrédient essentiel à la construction d'un ordinateur quantique universel \cite{magic}.\\ 

Surtout, de nombreux travaux suggèrent que l'impossibilité des modèles non-contexutels serait le meilleur critère permettant de capturer l'étrangeté quantique, de saisir en quoi la théorie quantique est si différente de la physique classique. De plus en plus de chercheurs s'engagent activement à approfondir notre compréhension de la distinction entre théories quantiques et classiques par le biais de la contextualité.\\

\chapter{Infuturabilia : le théorème de Kochen-Specker et ses origines.}
\label{chapKS}

La notion de contextualité est associée à un théorème (section \ref{sKS}) indépendamment établi par Simon Kochen et Ernst Specker (1960-1967) \cite{ks} et par John Bell (1964-66) \cite{bell}. Si la démonstration de Bell (section \ref{contbell}) s'appuie sur une critique de la preuve "d'impossibilité des variables cachées" de von Neumann et un corollaire du théorème de Gleason (section \ref{sgleason}), c'est un questionnement issu de la philosophie scholastique et l'étude de la logique contrefactuelle qui mena Ernst Specker \cite{specker} (sections \ref{sinfu} et \ref{sparabole}) au théorème démontré avec Simon Kochen (section \ref{sks}). Or, Bell qualifie son résultat "d'idiot", et Kochen et Specker l'interprètent comme une preuve de l'inexistence des variables cachées, ce qui est incorrect (cf. interprétation de Broglie-Bohm). Une mauvaise compréhension de la preuve de Von Neumann (section \ref{svon}), réhabilitée dans \cite{bub3} et \cite{dieks}, en est certainement la cause (section \ref{scritique}). En plus d'apporter la lumière sur les positions respectives de Bell et Kochen et Specker, cette relecture permet d'identifier la fameuse "preuve d'impossibilité de variables cachées" comme une première démonstration (partielle et implicite) de contextualité.

\section{Enoncé du théorème KS}
\label{sKS}
On peut énoncer le théorème de Bell-Kochen-Specker de la façon suivante
:\\

Considérons deux observables $A$ et $B$, issues d'un espace de Hilbert de dimension quelconque, qui commutent : $[A,B]=0$.
Il existe ainsi une base de décomposition propre de projecteurs ${\Pi_k}$ commune à $A$ et $B$ :
\[ A = \sum_j a_j \Pi_j   \hspace{3cm} B = \sum_j b_j \Pi_j
\]
On aura donc :
\[ AB = \sum_j a_jb_j \Pi_j   \hspace{3cm} A+B = \sum_j(a_j + b_j) \Pi_j
\]
Faisons à présent l'hypothèse dite de "\textit{déterminisme des résultats}" ou "\textit{pré-détermination}". 
Celle-ci consiste à admettre que les valeurs d'une observable définie pour un système existent indépendamment de la mesure. Si la mesure d'une observable n'est pas effectuée, l'observable a tout de même une valeur bien déterminée. En associant une fonction d'affectation de valeur $v(*)$ aux observables (représentées par des opérateurs auto-adjoints) : \[ v(A) = \sum_j a_j v(\Pi_j) \]
le \textit{déterminisme des résultats} (DR) implique que \[v(A)\in\{0,1\}\]
Or, d'après l'hypothèse de \textit{non-contextualité des mesures}(NC), la valeur assignée à un projecteur ne dépend pas du \textit{contexte} de la mesure considérée, i.e. de la base dans laquelle celui-ci est considéré. Dans toutes les décompositions, le même projecteur se verra donc affecter la valeur 1 lorsque A et B commutent. Il en résulte que la fonction de valeurs suit les conditions : 
\[v(AB) = v(A)v(B) \hspace{3cm} v(A+B) = v(A)+v(B) \]
Le théorème de Kochen-Specker énonce que dans un espace de Hilbert de dimension supérieure ou égale à 3, les prédictions de la mécanique quantique (MQ) sont logiquement incompatibles avec ces deux hypothèses. \\
\[DR \wedge NC \wedge MQ \rightarrow Contradiction\]
Ainsi, accepter la mécanique quantique nous contraint à renoncer au déterminisme, à la noncontextualité des mesures, ou aux deux à la fois.\\

Le choix de l'hypothèse à abandonner est dicté par l'interprétation que l'on a de la mécanique quantique. Un réaliste, en général favorable au déterminisme, aura tendance à conclure que si selon le résultat de Bell, la Nature est non-locale, elle est également contextuelle d'après le théorème KS. Un tel parallèle doit être néanmoins employé avec prudence, puisque, comme nous le verrons dans le chapitre \ref{chapspek}, la nature du postulat de pré-détermination n'est pas la même dans les deux théorèmes. La terminologie "\textit{contextualité}" est donc, au même titre que la "\textit{non-localité}", implicitement réaliste. Par abus de langage, ces termes sont souvent employés comme s'ils étaient indépendants de l'interprétation que l'orateur a de la mécanique quantique, ce qui n'est pas le cas. Afin d'éviter que "contextualité" devienne synonyme de "théorème de Kochen-Specker", nous nous efforcerons de désigner ce dernier par l'expression "résultat (ou théorème) KS".\\

Tout comme le théorème de Bell, le théorème KS est un \textit{théorème no-go} : il peut dissuader d'opter pour une interprétation réaliste qui imposerait des contraintes sur la réalité. Un copenhaguien concluera donc plutôt que ces résultats invitent à penser que "Des mesures non-effectuées non pas de résultats." \cite{peres2}, évitant ainsi un choix cornélien entre deux postulats fondamentalement classiques.\\

Historiquement, le théorème a été élaboré dans un contexte réaliste, en réponse au "problème des variables cachées en mécanique quantique"\footnote{les deux articles originaux sont intitulés "On the Problem of hidden Variables in Quantum mechanics", et "The Problem of hidden Variables in Quantum mechanics"}. En 1952, David Bohm remet au goût du jour une théorie à variables cachées émise trente-deux ans auparavant par Louis de Broglie, que ce dernier avait très vite abandonnée du fait de son caractère non-local. Le "problème" est une contradiction évidente entre compatibilité de cette théorie avec les prédictions quantiques et l'existence d'une preuve d'impossibilité de ce type de théories, démontrée par John von Neumann en 1932. Cependant, si Kochen et Specker semblent prendre parti pour von Neumann en expliquant que leur théorème constitue "une preuve de l'inexistence des variables cachées"\footnote{\textit{L'objectif principal de cet article est de donner une preuve de la non-existence des variables cachées. [...] la présente preuve a été comparée à la preuve bien connue de von Neumann de anon-existence des variables cachées. La preuve de von Neumann est essentiellement basée sur la non-existence d'une fonction réelle srl'ensemble des observables de la mécanique quantique qui sont multiplicatives sur des observables commutantes et linéaires. Dans notre preuve, nous montrons que la on-existence d'une fonction réelle qui est à la fois multipicative et linéaire uniquement sur des observabes qui commutent. Ainsi, formellement, notre résultat est plus fort que celui de von Neumann.} \cite{ks}}, Bell utilise au contraire son résultat comme critique de la preuve de von Neumann, et défend l'idée que les variables cachées peuvent exister.\\

Un même résultat, le théorème KS, semble donc être utilisé pour défendre une chose et son contraire. Les deux interprétations originelles de ce résultat sont contradictoires. Un quiproquo au sujet de la nature la démonstration de von Neumann pourrait en être la cause.

\section{L'article "oublié" de Bell}
\label{contbell}

\`A l’été 1964, avant la soumission de son article sur la fameuse inégalité qui portera son nom \cite{bell1}, John S. Bell soumet un autre papier \cite{bell} qui, pour certaines raisons, ne sera publié qu’en 1966.\footnote{Le manuscrit (\textit{On the Problem of Hidden Variables in Quantum Mechanics}) soumis par Bell à l’été 1964 au \textit{Reviews of Modern Physics} lui fut renvoyé avec la demande de fournir plus de détails sur “les mesures impliquées”. Bell s’exécuta et envoya une version augmentée à l’éditeur en janvier 1965 ; version qui fut mal classée et égarée dans les bureaux des \textit{Reviews}. La correspondance avec Bell à ce sujet retarda encore plus la publication en raison du retour de Bell au CERN, ce qu’ignorait l'éditeur. C'est ce qui explique que le premier article sur les variables cachées de Bell ait été publié deux ans plus tard que le second. On remarquera par ailleurs que la preuve qu'il ne puisse y avoir de variables cachées locales du deuxième (et plus célèbre) article ne peut être pleinement appréciée que s'il a déjà été prouvé qu'il puisse exister des variables cachées pour la mécanique quantique. \cite{jammer1} \cite{jammer2}} Dans cet article, Bell, motivé par une forte opposition aux interprétations de type Copenhague qui donnaient selon lui une trop grande place au "vague" concept d’\textit{observation} \cite{bell2}, cherche une faille dans le raisonnement de von Neumann. Il utilise pour cela le résultat suivant\footnote{Il s'agit d'un corollaire du théorème de Gleason, que l'on introduira dans la section (\ref{sgleason})} :\\

si la dimension $d$ d’un espace de Hilbert est supérieure ou égale à 3, alors il existe un ensemble de tests élémentaires tels que l’assignation de valeurs 1 (vrai) ou 0 (faux) à ces tests est impossible si elle respecte les deux règles suivantes : \\ (i) 1 ne peut pas être assigné à deux tests mutuellement exclusifs\footnote{i.e. deux tests représentés par des vecteurs orthogonaux ou des observables non-commutantes} ; \\(ii) 1 doit être assigné à exactement un des $d$ tests mutuellement exclusifs.\\ 

Bell démontre ce corollaire en construisant un ensemble infini explicite de tests élémentaires dans un espace des états à dimension $d=3$ pour lequel une telle assignation est impossible. Il déclare alors avoir mis en évidence une erreur dans la preuve de John von Neumann, qui “\textit{supposait tacitement que la mesure d’une observable doive fournir la même valeur indépendamment de quelles autres mesures peuvent être simultanément faites.}\cite{bell}”\\  Selon lui (\cite{bell2}, pp. 8–9 et 164–6), son théorème montre que les résultats de mesure doivent dépendre non seulement de l’observable mesurée et de l’état caché du système, mais aussi du \textit{dispositif expérimental complet}. Bell voit ainsi la contextualité comme une manifestation, en terme de variables cachées, de "\textit{l'impossibilité de toute séparation précise entre le comportement des objets atomiques et l'interaction avec les instruments de mesure qui servent à définir les conditions sous lesquelles le phénomène apparait}", qui fut préalablement décrite par Bohr en ces termes :
\begin{quotation}
\textit{le résultat d'une observation pourrait ne pas dépendre uniquement de l'état du système (incluant les variables cachées) mais aussi de la disposition complète de l'appareil de mesure.}
\end{quotation}
Cette concession à Bohr est étonnante, lorsqu'on sait que Bell était profondément réaliste. Loin d'être anodine, elle pourrait être due à une tentative de sabotage de son propre théorème. 
Bell était en effet obnubilé par le fait que le résultat KS  puisse être vu comme un théorème no-go, i.e. une preuve d’impossibilité ou du moins de restriction sur les variables cachées. Or, c'est exactement l'interprétation qu'en donneront Kochen et Specker. Il ira ainsi jusqu'à le qualifier "d'idiot" (\cite{bell2} p. 166, cf. \cite{mermin1}). 
\begin{quotation}
\textit{Ce mot} [measurement] \textit{suggère très fortement le constat de quelque propriété préexistante de quelque chose, tout instrument impliqué ne jouant qu’un rôle purement passif. Les expérimentations quantiques ne sont juste pas comme cela, comme nous l’avons appris, en particulier par Bohr. Les résultats doivent être vus comme le produit d’une articulation entre « système » et « appareil de mesure », le dispositif expérimental complet.} […] \textit{je suis convaincu que l’on a aujourd’hui tellement abusé du mot « mesure »} [measurement] \textit{que le champ serait beaucoup plus avancé si l’on interdisait son utilisation, en faveur par exemple du mot « expérience »} [experiment].
\end{quotation}
Cette invocation infortunée de Bohr ressemble à s'y méprendre à une "manœuvre de judo" selon Abner Shimony \cite{shimony} . Bell chercherait à retourner le propre poids de son théorème contre lui.  Il minimisera également sa propre contribution, appelant notamment le théorème "preuve de Gleason-Jauch", et omettant de mentionner son nom dans la liste des chercheurs l'ayant découvert (\cite{bell2}, p. 164)) :
\begin{quotation}
 […] \textit{J.M Jauch m’en a parlé en 1963. Ce n’est pas la totalité du théorème de Gleason qui est nécessaire, mais seulement un corollaire qui se prouve facilement par lui-même. (L’idée fut plus tard redécouverte par Kochen et Specker ; voir aussi Belinfante, Fine et Teller)} […]
\end{quotation}
Bell s’acharne à faire passer son théorème comme quelque chose que Jauch lui aurait dit, puis comme un héritage de Bohr. \\

Curieusement, il mentionnera à nouveau le physicien danois dans sa définition des beables - "\textit{ces éléments d'une théorie qui "doivent être pris au sérieux, correspondant à quelque chose de réel"}", et à partir desquelles toutes observables doivent être construites (\cite{bell2}, p.234) -
plutôt que d'invoquer le critère de réalité de l'article EPR\footnote{"
\textit{Si, sans perturber en aucune façon un système physique nous pouvons prévoir avec certitude la valeur d’une quantité physique, alors il existe un élément de la réalité physique qui correspond à cette quantité.}" \cite{epr}}. Celles-ci doivent ainsi inclure  : "\textit{les dispositifs d'interrupteurs et de boutons sur un équipement expérimental, le courant dans les bobines, et les lectures des instruments}" \cite{bell3} , et qu'elles peuvent toujours être "\textit{décrites en termes classiques}", contrairement aux observables. Il s'agit de la même concession mentionnée ci-dessus, dont l'objectif est de déprécier le résultat du théorème KS.\\ 

Or, ces beables "sont là", même lorsqu’aucune observation n’est effectuée. Interpréter son résultat comme  "\textit{l'inobservabilité de valeurs pré-existantes}" \cite{appleby} implique donc que ces propriétés pré-existantes, les beables, doivent être "cachées".
Le théorème KS pourrait donc suggérer l'inexistence des interprétations à variables "non-cachées", ou "exposées" de la mécanique quantique, mais Bell refuse d'arriver à une telle conclusion.\\ 

D'une part, il interprète cette "inobservabilité" comme une simple conséquence du point de vue de Bohr, issue de l'importance d'intégrer la totalité du dispositif expérimental au sysème étudié. Marcus Appleby démontrera qu'il n'en est rien : ces deux propositions sont généralement indépendante l'une de l'autre. L'erreur de Bell est d'identifier l'inobservabilité issue de son théorème à celle présente dans la théorie de de Broglie-Bohm, qui est, elle, issue du fait que l'interaction de l'appareil de mesure avec le système engendre une valeur qui n'existait pas précédemment. L'inobservabilité de valeurs pré-existantes n'est, dans le cas général, pas due à l'influence de l'appareil de mesure, mais émergerait plutôt de la "discontinuité "pathologique" de l'affectation de ces valeurs" \cite{appleby}.\\

D'autre part, Bell reste réticent à l'idée que les beables soient nécessairement cachées
: \begin{quotation}
\textit{Les individus pragmatiques peuvent très bien se demander pourquoi se préoccuper d'entités cachées qui n'ont aucun effet sur rien ?}
(\cite{bell2}, p.92)
\end{quotation}

Bien sûr, des quantités non pas besoin d'être directement observables pour être physiquements pertinentes. Cependant, les valeurs pré-existantes semblent avoir été introduites "gratuitement" \cite{appleby}, leur réalité prenant un aspect  "métaphysique, tout comme la réalité de l'éther de l'électrodynamique de Maxwell." \\

L'interprétation que Bell fait de la contexutalité, ou, plus exactement, de son théorème, aboutit donc à une impasse. Guidé par son attachement à le déprécier, de crainte qu'il ne soit vu comme un théorème no-go, il concède l'existence d'éléments de réalité "bohriens" \footnote{L'association peut paraître paradoxale, Bohr se situant dans le camp des copenhaguiens (ou partisans d'un réalisme participatif). Néanmoins, il existe des interprétations qui tentent d'aller au-delà de ce clivage. C'est notamment le cas d'une nouvelle approche de la mécanique quantique intitulée "Contextes-Systèmes-Modalités" (CSM) \cite{csm1}, dans laquelle Alexia Auffèves et Philippe Grangier proposent de "replacer la mécanique quantique dans le cadre du réalisme physique" en lui donnant une "nouvelle ontologie", dans laquelle les "éléments de réalité" ne seraient pas simplement attachés à un système, mais prendrait également en compte son contexte expérimental.},  lui qui était pourtant farouchement opposé à l'interprétation de Copenhague. Selon lui, ce résultat n'est pas un "problème sérieux", contrairement à la non-localité. \footnote{Cette attitude explique peut-être  en partie pourquoi la contextualité a été délaissée pendant plusieurs années au profit de la non-localité.} \\

Bell identifierait la contextualité à ce qui est en fait un cas particulier, une forme "contextualité forte", issue du comportement contextuel de la théorie de de Broglie-Bohm, où l'appareil de mesure engendre une valeur non pré-déterminée en interagissant avec le système. Le rapprochement avec la philosophie complémentaire de Niels Bohr (cf. annexe), qui peut être justifié dans ce cas particulier, n'est pas extrapolable au résultat du théorème KS, dans lequel aucune mention de l'utilisation d'appareils de mesure est nécessaire.\\

\`A la différence du physicien irlandais, le problème des variables cachées n'était pas la motivation première d'Ernst Specker (1920-2011), mathématicien zurichois\footnote{Pour une biographie introductive, cf. \cite{connorrob} \cite{specker3}} qui obtint pour la première fois l'ébauche du théorème KS en 1960. 

\section{"Infuturabilités"} 
\label{sinfu}
\textit{Cette section constitue l'esquisse d'une enquête personnelle sur les origines du théorème de Kochen-Specker.\\}

En 1961, Ernst Specker, alors professeur en mathématiques à ETH Zurich depuis six ans, effectue une année sabbatique à l'université de Cornell (Ithaca, New-York).\footnote{Il s'agit de son second séjour à Cornell, le premier ayant eu lieu en 1958.} Lors d'un colloque de mathématiques, il donna un cours au sujet d'un de ses articles récemment publié. Sa présentation attira l'attention de Simon Kochen, mathématicien canadien, et ils collaborèrent pendant plusieurs années sur l'étude de la logique quantique et le problème des variables cachées, entre Ithaca et Zurich. Leurs travaux aboutirent notamment à une reformulation enrichie du résultat de Specker, adaptée au formalisme quantique, qui sera connue sous le nom de théorème de Kochen-Specker.\\

L'article originel de Specker est intitulé "\textit{Die Logik nicht gleichzeitig entscheidbarer Aussagen}", "La logique des propositions non simultanément décidables" \cite{specker}, et fut publié en septembre 1960 dans le revue \textit{Dialectica}.\\

Cette revue trimestrielle, essentiellement consacrée à la diffusion de travaux en philosophie des sciences et de la connaissance, fut fondée en 1947 par Gaston Bachelard, Paul Bernays\footnote{Paul Bernays (1888-1977) est un mathématicien suisse qui contribua de manière significative à la logique mathématique, à la théorie axiomatique des ensembles et à la philosophie des mathématiques. Il fut assistant et un étroit collaborateur de David Hilbert, et il joua un rôle crucial dans la crise des fondements.}  et Ferdinand Gonseth\footnote{Ferdinand Gonseth (1890-1975) est un mathématicien et philosophe suisse. Proche ami de Bachelard, qui déclarait que Gonseth "l'aidait à être meux lui-même", ils développèrent ensemble une "\textit{méthode non-cartésienne}, une méthode de recherche qui n'impose aucun principe avant la connaissance elle-même. Comme la science, cette méthode doit être \textit{ouverte} et être en mesure de changer ses fondements selon l'expérimentation \cite{bontems2}. Pour plus d'informations sur la philosophie de Gonseth, cf. par exemple \cite{gilles1}. }. Entre 1942 et 1945, Specker fut l'élève de ces deux derniers, et il semble leur vouer une admiration particulière.

\begin{quotation}
[Il]\textit{assista aux cours de Gonseth, Hopf et Bernays, les hommes qu'il considère comme ses professeurs.} \cite{specker3}
\end{quotation}

Entre 1945 et 1948, Ernst Specker fut assistant en mathématiques à ETH Zurich. Au cours de cette période, il assista à un séminaire organisé par Gonseth et auquel participa Wolfgang Pauli\footnote{Pauli, qui fut assistant scientifique de Gonseth en 1951, était particulièrement intéressé par les séminaires d'histoire et philosophie des sciences organisés par Gonseth et par Bernays \cite{primas}. Il fit également parti du conseil consultatif de la revue \textit{Dialectica}, et dirigea un numéro spécial sur "L'idée de complémentarité" paru en 1948, auquel ont contribué Einstein, Bohr, Heisenberg, de Broglie, Reichenbach, les Destouches et Gonseth \cite{gilles}.}, sur "les fondements de la théorie quantique", qu'il cite comme l'une des influences principales de son article.

\begin{quotation}
[Le séminaire] \textit{portait essentiellement sur le livre de von Neumann} [probablement Mathematical Foundations of Quantum Mechanics], \textit{mais il y avais également des présentations sur l'article de Birkhoff - von Neumann et les travaux de Destouches. Armand Borel et Res Jost figuraient parmi les participants; les discussions entre mathématiciens et physiciens avaient tendance à être plutôt enflammées et il n'était pas toujours facile pour l'orateur de finir sa présentation}\footnote{Au cours d'une conversation privée enregistrée par Ad\'an Cabello, Specker rapporte qu'à la fin du séminaire, Borel aurait déclaré : "La prochaine fois, je prendrai un pistolet avec moi !"}. \textit{Le théorème de base de l'article} [Die Logik nicht gleichzeitig entscheidbarer Aussagen] \textit{fut prouvé peu de temps après, mais publié bien plus tard. Quand un jour je demandai à Ernst pour quelles raisons il avait reporté la publication si longtemps après, il me parla de certains résultats que William Craig avait déjà prouvé à Zurich et publié bien plus tard. Et il ajouta : Certaines personnes aiment qu'on leur dise de publier.} \cite{specker3}\footnote{L'article autobiographique "A Story of a Friend", qui retrace la vie d'Ernst Specker, est signé Jonas Meon. Il s'agit d'un pseudonyme, composé de l'inversion des mots "on me" (Meon) et du nom d'un prophète qui le fascinait. \cite{suarez}}
\end{quotation}

Specker attendit près de dix ans avant de chercher à publier son travail, "principalement parce qu'il n'imaginait pas qu'un journal accepterait de le [faire]". Cependant, en 1960, Gonseth, qui était favorable à l'idée de Specker selon laquelle la théorie quantique aurait son mot à dire sur la logique, encouragea son ancien étudiant \cite{spekkens8} à publier ses idées dans un numéro spécial de \textit{Dialectica} en l'honneur de son soixante-dixième anniversaire. Specker rend d'ailleurs hommage à son professeur  dans l'épigraphe et l'introduction de l'article, et le taquine même secrètement dans une parabole  (cf. section \ref{sparabole}).

\begin{quotation}
"La logique est d'abord une science naturelle." \textit{Ferdinand Gonseth\\}

\textit{L'épigraphe attaché à ce travail est le sous-titre du chapitre} La physique de l'objet quelconque \textit{du livre} Les mathématiques et la réalité. \textit{Cette physique se révèle être essentiellement une forme de logique propositionnelle classique, par laquelle, d'une part, elle obtient une réalisation typique et, d'autre part, elle est, de manière presque évidente, privée de sa prétention à l'absolutisme, duquel elle est occasionnellement habillée. Les remarques suivantes concordent avec ce point de vue et peuvent être comprises dans ce même sens empirique. \cite{specker}}
\end{quotation}

La thèse de Gonseth \cite{emery}, qui s'opposait farouchement au néo-positivisme du cercle de Vienne, est que la logique suppose l'existence d'objets sur lesquels portent les propositions. Or, les objets de la logique sont obtenus par abstraction. Ils correspondent à des objets quelconque du monde physique, dépourvus de toute spécificité. Ils sont aux objets du monde réel ce que la droite mathématique est au "fil tendu, [au] faîte d'un toit, [à] l'arête d'une règle de dessin...", i.e. une image schématique, conforme à notre expérience de la réalité et à son évolution. Ce "réalisme" n'est cependant pas fermé, au sens où les objets abstraits ne sont pas identifiés aux Idées immuables et éternelles de Platon, mais sont construits à partir de faits réels empiriques. La philosophie gonséthienne est une philosophie \textit{ouverte} et \textit{idoine}, qui après diverses évaluations, s'accorde sur ce qui apparait comme la solution la meilleure possible, sans pour autant renier aux droits fondamentaux de l'erreur et de la révision.\\

Gonseth dégage trois lois empiriques de ce concept d'objets logiques.\\

\hspace{-6mm}1- Tout objet est ou il n'est pas.\\
2- Un objet ne peut être et à la fois ne pas être.\\
3- Tout objet est identique à lui-même.\\

Il voit dans ces trois lois de la physique de l'objet quelconque les formes primitives du principe du tiers-exclu (ou de bivalence\footnote{Pour une étude sur la distinction de ces deux principes, cf. par exemple cours 17 de \cite{bouveresse}}), du principe de contradiction, et du principe d'identité, qui sont les trois piliers de la logique classique, introduits par Aristote\footnote{Contrairement aux principes du tiers-exclu, de bivalence et de contradiction, le principe d'identité n'est pas explicitement posé par Aristote.}.\\

Or, dans le chapitre 9 de son "De Interpretatione", le philosophe grec remarqua que la stabilité de ces piliers semble, dans certains cas, être mise à rude épreuve. Il exposa le problème suivant : \\

Supposons qu'une bataille navale n'aura pas lieue demain. Si cet énoncé est vrai, alors il l'était également hier. Il était aussi vrai le mois dernier, et l'année passée. Cependant, toutes les vérités passées sont nécessairement vraies au présent. Il en résulte qu'il est impossible que la bataille ait lieue demain. 

\begin{quotation}
\textit{Telles sont donc, avec d'autres de même nature, les absurdités où l'on est entraîné si l'on admet que, pour toute affirmation et négation (qu'il s'agisse soit de propositions portant sur les universels et prises universellement, soit de propositions portant sur le singulier), nécessairement l'une des opposées est vraie, et l'autre fausse, et qu'il n'existe aucune indétermination dans le devenir, mais qu'au contraire toutes choses sont et deviennent par l'effet de la nécessité.} \cite{aristote}\\
\end{quotation}

Une résolution célèbre du paradoxe fut apportée par Diodore Cronos (IV-IIIe siècle avant J.C.) un philosophe grec de l'école mégarique, qui considère que la bataille est impossible ou nécessaire. Dans son argument dit "dominateur", \textgreek{ὁ κυριεύων λὸγος},
il expose un conflit entre les trois propositions suivantes :
\begin{itemize}
\item 1 : Toute proposition vraie concernant le passé est nécessaire. 
\item 2 : L'impossible ne suit pas logiquement du possible. 
\item 3 : Est possible ce qui n'est pas actuellement vrai et ne le sera pas. \\
\end{itemize}

Diodore renonce à la dernière proposition. Selon lui, le possible est nécessairement vrai. Ou bien il est impossible que la bataille ait lieue demain et ce ne sera pas le cas, ou bien il est possible qu'elle ait lieue et donc ce sera nécessairement le cas.\\

Contrairement à Diodore, Aristote résout le paradoxe en invoquant une violation exceptionnelle du principe de bivalence à travers l'introduction d'une nouvelle modalité : la \textit{contingence}\footnote{Le principe de bivalence est un principe logique selon lequel toute proposition $P$ ne peut prendre qu'une et une seule des deux valeurs de vérité : vrai ou faux. Il est postulé dans l'argument dominateur de Diodore.}.  Les énoncés "la bataille n'aura pas lieu demain" et "la bataille aura lieue demain" sont exclusifs : il est impossible de leur affecter la même valeur de vérité. Or, si demain, un et un seul de ces énoncés prendra la valeur "vrai", il est néanmoins impossible d'assigner cette valeur aujourd'hui. 

\begin{quotation}
\textit{Que ce qui est soit, quand il est, et que ce qui n'est pas ne soit pas, quand il n'est pas, voilà qui est vraiment nécessaire. Mais cela ne veut pas dire que tout ce qui est doive nécessairement exister, et que tout ce qui n'est pas doive nécessairement ne pas exister ; car ce n'est pas la même chose de dire que tout être, quand il est, est nécessairement, et de dire, d'une manière absolue, qu'il est nécessairement. Il en est de même pour tout ce qui n'est pas. — C'est la même distinction qui s'applique aux propositions contradictoires. Chaque chose, nécessairement, est ou n'est pas, sera ou ne sera pas, et cependant si on envisage séparément ces alternatives, on ne peut pas dire laquelle des deux est nécessaire.} \cite{aristote}\\
\end{quotation}

Selon Aristote, les énoncés sur le futur sont contingents. Ils ne prennent une valeur de vérité que lorsque le potentiel devient actuel.\\

Par définition, toute hypothèse de contingence, i.e. tout hypothèse portant sur l'existence de propriétés non mesurées, ou de faits non actuels, conduit à l'invocation de \textit{contrafactuelles} \cite{svozil1}. La \textit{contrafactualité} (ou contrefactualité) est une notion de logique conditionnelle, qui consiste à étudier des évènements qui ne se sont pas produits, mais qui auraient pu se produire sous certaines conditions. Un énoncé contrafactuel prend la forme suivante : 

\begin{quotation}
\textit{Si l'évènement $A$ avait eu lieu, alors l'évènement $B$ se serait produit.\footnote{Pour une étude de l'usage des contrafactuelles en physique, cf. \cite{vaidman2}.}}\\
\end{quotation}

L'étude du problème des futurs contingents et des contrafactualités a traversé les siècles (pour un rapide tour d'horizon (loin d'être exhaustif), cf. annexe).\\

Au Moyen-Âge, ils sont étudiés par la \textit{scolastique} (du latin schola, ae, "école ", issu lui-même du grec \textgreek{σχολή}, 
 "oisiveté", "loisir consacré à l'étude "), qui tente de concilier la philosophie aristotélicienne et la théologie chrétienne. De la notion de contrafactualité émerge alors des interrogations sur la nature de l'omniscience de Dieu. \\

Dans le traité \textit{Somme théologique} de Thomas d'Aquin (1225-1274), on trouve ainsi une analyse\footnote{Selon Thomas d'Aquin, Dieu a connaissance aussi bien des choses qui sont en acte que de celles qui ne le sont pas, ces dernières étant tout de même \textit{en puissance}. Comme Dieu connait toute chose y compris celles qui sont uniquement en puissance, et que parmi ces choses en puissances se trouve des "contingents futurs pour nous", il s'ensuit que "Dieu connaît les futurs contingents". La science de Dieu n'est cause des choses que si sa volonté s'y adjoint. Il n'est donc pas nécessaire que tout ce que Dieu sait existe, ait existé ou doive un jour exister. En revanche, tout ce qu'il veut ou qu'il permet d'être l'est. Ce qui est dans la science de Dieu, ce n'est pas que ces choses sont, mais qu'elles peuvent être.} des questionnements suivants : 
\begin{quotation}
 \textit{Dieu a-t-il connaissance des choses qui ne sont pas ?
Dieu connaît-il les futurs contingents ?}
\end{quotation}

Ces interrogations sont au cœur de la démarche de Specker, et constitue certainement la motivation première de l'article de 1960. Dans \cite{fuchs4}, Chris Fuchs relate ainsi avoir rencontré Specker à l'occasion d'un séminaire sur les "Informations Primitives et les Lois de la Nature", qui s'est tenu à ETH Zurich en mai 2008. 

\begin{quotation}
\textit{Il m'a expliqué comment le théorème de Kochen-Specker (qu'il a pour la première fois obtenu de lui-même cinq ans avant le papier habituellement cité) émerge d'une question théologique. \`A l'époque, il voulait vraiment savoir si Dieu pouvait savoir ce que le monde aurait été si Hitler n'était jamais né.}\\
\end{quotation}

Ou retrouve également une référence à cette motivation dans l'article de Robert Spekkens en hommage à Specker :

\begin{quotation}
\textit{Ce n'est pas exagérer que de dire que ce théorème - le théorème de Kochen-Specker - est un des faits les plus profonds sur les fondements de la théorie quantique. L'histoire de comment Specker est parvenu à ce résultat est tout à fait merveilleuse. Elle montre que, même dans une ère où "shut up and calculate" est le mantra de nombreux chercheurs, des questions philosophiques profondes peuvent encore conduire à de grands progrès dans notre compréhension du monde. C'est une histoire qui va réchauffer le cœur de toute personne qui croit que la physique devrait être poursuivie dans un style romantique ... [Specker] a été conduit à la question critique : Dieu pourrait-il connaître le résultat qui aurait été obtenu si une mesure quantique différente avait été effectuée à la place de celle qui a actuellement été effectuée, sans entrer en contradiction ? La réponse, qu'il a trouvée, est qu'Il ne peut pas.} \cite{spekkens8}\\
\end{quotation}

Specker se réfère à ce questionnement sous le terme "infuturabilité" ("infuturabilia") dans l'article en question \cite{specker}.

\begin{quotation}
\textit{D'une certaine façon, les spéculations de la scholastique à propos des "infuturabilités" ont aussi leur place ici, i.e. la question de l'extension de l'omniscience de Dieu aux évènements qui auraient eu lieu si quelque chose qui ne s'est pas produite s'était effectivement produite. (cf. e.g. [3], Vol. 3, p.363.)}\\
\end{quotation}

La référence "(cf. e.g. [3], Vol. 3, p.363.)" renvoie à l'ouvrage \textit{Historia de la filosofía española} de Marcial Solana, et au chapitre "Teodicea" du troisième volume \cite{solana}. L'auteur y traite des interrogations scholastiques sur l'adéquation entre omniscience de Dieu et libre-arbitre humain, à travers la philosophie de Pedro da Fonseca (1528-199), théologien jésuite portugais, surnommé "`l'Aristote portugais". Fonseca est le père de la notion de \textit{science moyenne} (sciencia media), qu'il définit comme "une connaissance, antérieure au décret divin de prédestination, que Dieu aurait de l’usage que chacun fera des dons de grâce à lui accordés". Une connaissance qui permet donc de réconcilier omniscience divine et libre-arbitre.\\

\begin{quotation}
\textit{L'objet propre et caractéristique de cette science moyenne est ce que l'on appelle "futuribles", les futurs conditionnels contingents, c'est à dire, ces futurs dont la réalisation dépend d'une condition qui résulte du libre arbitre de la créature.}\\
\end{quotation}

Ce concept de science moyenne (connu en anglais sous le terme de \textit{middle knowledge}) sera embrassé et développé par Luis de Molina (1535-1600), autre théologien jésuite, dans son oeuvre : \textit{De concordia liberi arbitrii cum gratiae donis, divina praescientia, providentia, praedestinatione, et reprobatione ad nonnullos primae partis divi Thomae articulos} (1588), en français : \textit{Accord du libre arbitre avec le don de la grâce, la prescience divine, la providence, la prédestination et la réprobation...}\\

Pour un moliniste, l'omniscience de Dieu est telle qu'avant même la Création, il connait à l'avance tous les actes de toute créature. Nos actes n'exercent aucune influence causale sur la volonté divine, mais ont un effet contrafactuel sur elle : si j'avais décidé de ne pas faire certaines choses, la connaissance divine aurait été modifiée.\\

Cette philosophie fait écho à une autre influence explicite revendiquée par Specker (cf. 29-05-08 Infuturabilia (to E. Specker) dans \cite{fuchs4}: la notion de "double-vérité d'Ibn-Rushd", aussi connu sous le nom d'Averroès (1126-1198, le "Commentateur" du "Philosophe" (Aristote)), qui dépeint une forme de coexistence entre vérités religieuses et scientifiques. Il était ainsi convaincu "qu'il y a quelque chose au delà de la physique et des mathématiques... et de la religion. Mais c'est très difficile à formuler...".\footnote{S'il n'y a pas de référence directe à Averroès dans son autobiographie, on y trouve toute fois une mention du mysticisme musulman : "\textit{Ernst aime les mathématiques et il aime enseigner. Et donc, sans surprise (bien que ce ne soit en aucune façon une conséquence logique) il aimait enseigner les mathématiques. "Enseigner les mathématiques à de bons étudiants, c'est comme raconter des contes de fées à des enfants. Un monde à part est dévoilé. Ceux qui y pénètrent peuvent l'explorer davantage, et même participer à sa construction ; la faculté utilisé est "l'imagination active" ou la "conscience imaginaire". Par conséquent, il ne peut y avoir de distinctions entre "identifier" et "développer", entre "exploration" et "création"". Ces mots d'Ernst sont inspirés d'un texte décrivant le "alam al-mithral", - le monde où les images étaient réelles - du mysticisme islamique.} D'autres influences et références sont certainement à trouver dans ses Sermons \cite{specker4}.}\\

\vspace{-1cm}
\section{La parabole de Specker}
\label{sparabole}
Dans "La logique des propositions qui ne sont pas décidables simultanément" \cite{specker}, Ernst Specker  met en exergue un exemple de difficultés qui émergent des \textit{infuturabilités} sous la forme d’une parabole : un voyant défie les prétendants à la main de sa fille en les soumettant à un jeu de prédiction impossible à gagner. L'intrigue moraliste du conte est un clin d'oeil à son ancien professeur, Ferdinand Gonseth, qui, à l'époque se plaignait à ses collègues de tous les copains que sa famille avait amené à la maison \cite{spekkens8}.
                                                   
\begin{quotation}
\textit{\`A l’école assyrienne des Prophètes de Arba’ilu au temps du Roi Asarhaddon (681-669 avant JC), enseignait un voyant, originaire de Ninive. Il était un représentant distingué de sa faculté et, mis à part l'étude des corps célestes, son intérêt était exclusivement centré sur sa fille. Ses prouesses d’enseignant étaient en effet limitées, le sujet était rude et demandait une connaissance en mathématiques qui était rarement disponible. Mais, s’il ne parvenait pas à intéresser les étudiants, il leur découvrit un grand intérêt pour une toute autre affaire.\\
Sa fille avait à peine atteint l’âge de se marier qu’il croulait sous les requêtes de jeunes diplômés pour l’épouser. Et bien qu’il fût conscient qu’il ne pourrait pas être pour toujours à ses côtés, elle restait tout de même trop jeune pour partir dans les bras de l'un ces prétendants, dont aucun ne semblait dignes de sa main. Afin qu'ils puissent se convaincre de leur indignité, il leur promit qu’elle serait mariée à celui qui résoudrait l’énigme qui leur serait posée.\\
Chaque prétendant fut amené devant une table sur laquelle étaient disposées trois petites boîtes alignées les unes à côtés des autres} [chacune susceptible de contenir, ou non, une gemme]. \textit{On demanda alors aux prétendants de prédire quelles boîtes contenaient une gemme, et quelles boîtes étaient vides.  Mais peu importe le nombre de leurs tentatives, il semblait impossible de mener à bien cette tâche. Après que chaque prétendant ait donné sa prédiction, il était invité par le père à ouvrir la paire de boîtes qu’il avait désigné comme étant toutes les deux vides (dans le cas où le prétendant avait prédit qu’il n’y avait qu’une seule gemme) ou la paire de boîtes qu’il avait désigné comme étant toutes les deux pleines (dans le cas où le prétendant avait prédit qu’il y avait deux gemmes). Mais, dans chaque cas, il se trouva qu’une boîte contenait une gemme et que l’autre n’en contenait pas. De plus, la pierre se trouvait parfois dans la première boîte ouverte, parfois dans la seconde. Comment était-il possible, étant donné trois boîtes, de n’être ni capable d’en choisir deux comme étant vides, ni deux comme étant pleines ?} \\
\textit{La fille serait restée célibataire jusqu'à la mort de son père si, après la prédiction du prétendant} [qu'elle aimait], \textit{elle ouvrit rapidement deux boîtes elle-même, une qu’il avait indiquée comme étant vide, et l’autre qu’il avait prédit pleine.} [Et, miracle !] \textit{La prédiction, pour ces deux boîtes, se trouva être correcte ! Après la faible protestation de son père qui aurait voulu qu’un autre couple de boîtes fut ouvert, elle essaya de regarder ce que contenait la troisième boîte, ce qui s'avéra impossible} [une force obscure empêchant l’ouverture de cette boîte]. \textit{Le père finit par admettre à contrecœur que la prédiction, ne pouvant être falsifiée, était validée.}[La fille et le prétendant se marièrent et menèrent une vie de joie et de bonheur.]
[traduction personnelle, entre crochets : ajouts personnels]
\end{quotation}

\begin{figure}[ht!]
\centering
\includegraphics[width=7cm]{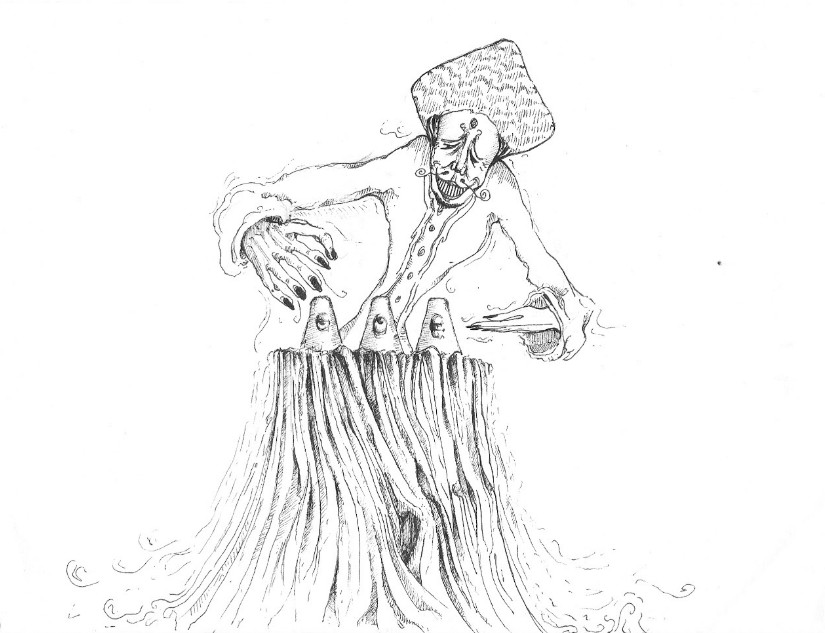}
\caption{Le voyant et ses gobelets magiques, illustration par Louison Dourdent.}
\end{figure}

Le contexte narratif de cette parabole n'est certainement pas anodin. Comme Specker, Asarhaddon eut une santé plutôt fragile. Le roi Assyrien s'inquiétait ainsi souvent du futur, et il donna une place très importante à la cour aux prophètes et aux astrologues. Son nom apparaît dans (au moins) deux textes bibliques : "2 Rois" et le "Livre de Tobit". Il s'agit peut-être d'une simple coïncidence, mais un autre prophète célèbre, chargé par Dieu de délivrer un message aux habitants de Ninive, apparaît également dans ces deux textes : Jonas.\\
\vspace{-1mm}
La passion que vouait Specker pour ce personnage biblique pourrait être intimement liée aux motivations qui l'ont conduit à son théorème. Le livre de Jonas peut en effet être interprété comme un argument moliniste : le peuple et Dieu peuvent tous deux changer d'avis, et sont donc chacun doté d'une forme de libre-arbitre. La prophétie de Jonas peut être vue comme étant conditionnelle, la prédiction du Jugement étant conditionnée par la réponse que donneraient les habitants de Ninive. "Encore quarante jours, et Ninive sera détruite !". Cette prophétie était une communication divine portant sur le futur. Il aurait ainsi pu paraître que lorsque celle-ci fut énoncée, la future destruction de Ninive aurait été nécessaire. Cependant, il n'en est rien, puisqu'il s'agit d'un futur contingent. Il existerait donc une place pour le libre-arbitre humain, malgré l'omniscience divine. Specker aurait pu être inspiré, influencé par ce récit biblique dans sa réflexion sur les "futurs contingents".\\

\`A la suite de la parabole, Specker la reformule mathématiquement \cite{specker} :
Considérons que $A_i$ indique que la ième boîte est pleine ; $A_i^*$ indique que la boîte $i$ est vide.
Pour tous couples de boîtes $(i,j)$ retournées parmis 1,2,3 ; les corrélations de la parabole sont décrites par les propositions suivantes :\\

(a) $A_i \rightarrow A_j^*$ et $A_i^* \rightarrow A_j$  (“dans chaque cas, il se trouva qu’une boîte contenait une gemme et que l’autre n’en contenait pas”)\\
      (b) $A_i \rightarrow A_i$ ; $A_i^*\rightarrow A_i^*$ (la gemme ne se téléporte pas)\\

Ces corrélations impliquent les propositions suivantes :
$A_1\rightarrow A_2^*$ , $A_2^*\rightarrow A_3$ , $A_1\rightarrow A_3^*$ \\

Si l’on retourne les boites 1 et 2 et que l’on découvre que la première est pleine et donc que la seconde est vide ; alors on sait que la boite 3 est vide. Si l’on avait retourné les boites 2 et 3, comme la boite 2 est vide, on aurait trouvé la boite 3 pleine. En revanche, si l’on avait retourné la boite 1 et la boite 3, la boite 1 étant pleine, alors la boite 3 serait vide, en vertu de (a) et (b). la boîte 3 semble donc être vide et pleine. La prise en compte d'énoncés contrafactuels a priori compatibles aboutit à une contradiction. \\

La contradiction est évitée du fait que le système est tel que une mesure conjointe des trois boîtes est impossible. Ainsi, si l’on veut expliquer les mesures (l'ouverture d’une boîte) comme révélant une propriété préexistante (l'existence d'une gemme sous une boîte), alors on doit imaginer que le résultat de la mesure dépende du contexte : que la gemme soit vue ou non dans la première boîte dépend du choix de l'autre boîte. Le défi du voyant ne peut être relevé par les prétendants car en leur demandant de spécifier si une gemme sera trouvée ou non dans chaque boîte, et ce indépendamment de l’autre boîte ouverte, il leur demande d’effectuer une assignation des résultats qui ne dépend justement pas du contexte de mesure. \\

Specker remarque que les difficultés qui émergent d'énoncés “\textit{non décidables simultanément}” se retrouvent dans les propositions issues de la mécanique quantique. \\

Cependant, en mécanique quantique, si chaque paire d'observables est une paire compatible, alors contrairement aux boites du voyant, l'ensemble des trois observables est simultanément compatible. On peut assigner simultanément une valeur à chaque observable du triplet.\\

Très souvent, on considère un ensemble de trois observables $A$,$B$, $C$, telles que\\ $[A,B] = 0$ ; $[A,C] =0$ et $[B,C]\neq 0$. La valeur $v(A)$ obtenue lors d'une mesure de l'observable A dépend du contexte de la mesure, i.e. si $A$ a été mesurée avec $B$ ou avec $C$.\\ Si $[B,C] =0$, la contextualité est toujours présente dans le cas de la parabole de Specker mais disparait pour la mécanique quantique. Dans le cas de mesures projectives, il est impossible de trouver un triplet d'observables quantiques vérifiant cette propriété. 
Specker voyait là un aspect fondamental de la théorie quantique (cf. chapitre \ref{chapnlcont}).
\begin{quotation}
\textit{Si l'on a, disons, ... trois questions, et que l'on peut répondre à chaque paire de questions, alors on peut aussi répondre aux trois questions. Il me semble que ceci est très fondamental.} \cite{specker2}
\end{quotation}  
Néanmoins, l'usage des contrafactuels en mécanique quantique peut sembler problématique. La physique produit des énoncés conditionnels : "Si les conditions $A_i$ sont respectées, alors on observera les résultats $B_i$". 
Par définition, toute mesure est effectuée dans un contexte. Tout énoncé portant sur des résultats qui n'ont pas été réalisés "en fait" conduit donc à des contrafactuelles . En physique classique, celles-ci  sont considérées comme de simples conditionnelles, les valeurs des observables étant définies et toutes les prémisses étant compatibles entre elles. Le choix des observables à mesurer n'a pas d'importance : on suppose que chaque observable possède une valeur définie, indépendante de toute mesure. Il y a un contexte global, constitué de toutes les observables concevables. 
Cependant, si l'on considère, pour un système unique, des conditionnelles avec des prémisses $A_i$ incompatibles, il est nécessaire de différencier factuelles et contrafactuelles.  Une proposition quantique ne satisfait pas une logique à deux valeurs de vérité. Du fait de la non-commutabilité de certaines observables, les contextes de mesures quantiques ne sont pas globaux et la structure de ces contextes ainsi que les probabilités de mesures associées soulèvent des questions loin d'êtres triviales.
\boitesimple{Un contexte correspond à un ensemble complet d'observables co-mesurables auxquelles sont associées des opérateurs commutant entre eux. Tout contexte peut aussi être caractérisé par un simple (mais non unique) opérateur. En logique quantique, les contextes sont représentés par des sous-algèbres booléennes ou des "blocs collés ensemble pour former le treillis hilbertien" \cite{svozil1}.}
Einstein, Podolsky et Rosen \cite{epr} ont par exemple suggéré de mesurer et d'inférer contrafactuellement deux contextes simultanément en considérant des éléments de réalité physique (la position et le moment) qui ne peuvent pas être mesurés simultanément sur le même système quantique. L'usage de contrafactualité n'a d'ailleurs été  critiqué par Bohr, qui semble considérer qu'il est légitime de considérer des alternatives contrafactuelles  \cite{peres}  : 
\begin{quotation}
\textit{notre liberté de manipulation des instruments de mesure est caractéristique de l'idée même d'expérience... Nous sommes complètement libre dans notre choix si nous voulons déterminer l'une ou l'autre de ces quantités... }
\end{quotation}
Comme le remarque Specker, l'étude des contrafactuels en mécanique quantique est "\textit{\textbf{équivalent à la question de savoir si la logique de la mécanique quantique est essentiellement la même que la logique classique}}". \cite{specker}  En se réferrant aux travaux de Birkhoff et von Neumann, Specker obtient un théorème analogue à celui de Gleason, qui servira de piédestal à l'article co-écrit avec Simon B. Kochen.

\section{Le théorème de Gleason}
\label{sgleason}
Dans les années 1930, Garrett Birkhoff et John von Neumann se lancent le défi de refonder la logique en l’adaptant aux résultats contre-intuitifs de la mécanique quantique. La recherche axiomatique de von Neumann finit par le conduire à douter du formalisme même de l'espace de Hilbert\footnote{...qu'il avait pourtant introduit avec David Hilbert en 1927 !}. En 1935, il écrit à Birkhoff : "Je voudrais faire une confession qui peut paraitre immorale. Je ne crois plus de façon absolue en l'espace de Hilbert." \cite{alexei1}
Une année plus tard, ils publient un article intitulé “The logic of quantum mechanics”\cite{birkhoff}, qui marque l’acte de naissance de la logique quantique, et qui commence ainsi : 
\begin{quotation}
\textit{Un des aspects de la théorie quantique qui a suscité le plus d’attention est celui de la nouveauté des notions logiques qu’elle présuppose. Elle affirme ainsi qu’une description mathématique d’un système physique S, même complète, ne permet pas en général de prédire avec certitude le résultat d’une expérience sur S, et l’on ne peut en particulier jamais prédire avec certitude à la fois la position et la quantité de mouvement de S (principe d’incertitude d’Heisenberg). Elle affirme de plus que la plupart des paires d’observations ne peuvent pas être effectuées simultanément sur S (principe de non-commutativité des observations).} [...] \textit{L’objet du présent article est de découvrir quelle structure logique peut-on espérer trouver dans des théories qui, comme la mécanique quantique, ne se conforme pas à la logique classique.} $[traduction\hspace{3mm} personnelle]$
\end{quotation}
Von Neumann et Birkhoff y élaborent une théorie axiomatique où la structure de l’espace de Hilbert serait dérivée d’un ensemble d’axiomes reposant sur des propositions logiques non distributives, appelées “propositions expérimentales” : des lois “\textit{et}” et “\textit{ou}” non classiques et adaptées aux résultats expérimentaux de la mécanique quantique. \\

George Mackey reprend l’idée de von Neumann de “projections en tant que propositions” \cite{mackey} \footnote{ Comme les projections ont seulement deux valeurs propres, 0 et 1, on peut se représenter la proposition associée à une projection comme une réponse “oui” ou “non” à la question correspondante. }, et constitue un nouvel édifice axiomatique. \\

Se pose alors la question de savoir si la règle de Born, qui régit les distributions de probabilité quantique, découle nécessairement des axiomes trouvés par Mackey.  En 1957, Andrew Gleason, qui fut son étudiant, donne une réponse positive à cette question, à condition que l’espace de Hilbert soit de dimension supérieure ou égale à 3. Sa preuve porte aujourd'hui le nom de “théorème de Gleason” \cite{gleason}. \\

Soit $H$ un espace séparable de Hilbert de dimension supérieure ou égale à 3. Soit $Q$ un treillis \footnote{Un treillis est un ensemble ordonné où toute paire d'éléments a une borne supérieure et une borne inférieure} de projections auto-adjointes sur $H$. Une mesure de probabilité sur $Q$ est une application $\mu$ : $Q \rightarrow$ [0,1] telle que $\mu(I)=1$ et, pour toute famille $P_i$ d’éléments de $Q$ disjoints deux à deux,  
\[  \mu (\oplus_i P_i) = \sum _i \mu (P_i)  \]
(Théorème) Pour toute projection P appartenant à Q,  il existe un unique opérateur de classe trace W tel que la mesure de probabilité $\mu(P)$  est :  \[\mu(P) = Tr(WP)\]
Le théorème de Gleason prescrit les descriptions probabilistes permises. Il donne un moyen de passer de la structure des mesures à celle des états.

Appliqué à la mécanique quantique, il stipule que les probabilités quantiques peuvent être dérivées de l'hypothèse que la théorie des probabilités classique est cohérente au sein des contextes. Le théorème peut donc être interprété de la façon suivante :\\

\begin{itemize}
\item l’opérateur de classe trace $W$ correspond à la matrice densité d’un état quantique, toute combinaison convexe d'opérateurs de projections de rang 1 qui représente l'état du système. ;
\item le treillis de projections $Q$ peut être identifié à un ensemble de propositions quantiques sur l’espace de Hilbert, i.e. des observables (questions) fermées sur le système quantique.
\item la mesure de probabilité représente l’assignation d’une valeur (0 ou 1) à chaque projection, i.e. une réponse positive (1 $\rightarrow$ “oui”) ou négative ( 0 $\rightarrow$ “non”)\\
\end{itemize}

 Pour une structure linéaire, chaque probabilité de résultat est un produit scalaire du projecteur de la mesure correspondante et d'un opérateur densité du système. L'hypothèse de la non-contextualité de l'assignation des probabilités est implicite : si les projecteurs représentent des résultats de mesures physiques alors la probabilité associée à un projecteur doit être la même indépendamment de la mesure. \\
 
\section{Le théorème de Kochen-Specker}
\label{sks}
Le théorème de Gleason implique ainsi immédiatement le résultat du théorème de Kochen-Specker. En mécanique classique, la description du système peut être donnée par un point dans l'espace des phases. Ce point est le "vrai" point - les autres sont "faux" - de sorte que le résultat d'une mesure peut être prédit avec certitude. Le théorème de Kochen-Specker interdit une telle description concrète : il n'y a aucun moyen d'affecter des valeurs de vérités à tous les projecteurs de rang 1 de sorte que pour toute mesure il n'y ait qu'un seul vrai résultat. Autrement dit, il n'y a pas d'opérateurs densité qui produisent des distribution de probabilité pour toutes les mesures évaluées uniquement sur 0 et 1. L'ajout de l'hypothèse du déterminisme des valeurs conduit donc à une contradiction. \\

Quelles conclusions "scolastiques"\footnote{L'étude de la science de Dieu, la question de l'étendue de son omniscience peut être ramenée, comme le remarque Specker, à la question philosophique de ce que le scientifique peut dire de la Nature. L'entité de Leibniz et de Laplace ne doit pas nécessairement être interprétée comme divine (cf. annexe chapitre 8).} en déduire ?\footnote{Les réponses à cette question sont inspirées de \cite{svozil2}.}\\

On peut par exemple conclure que l'omniscience ne s'étend pas aux contrafactuelles, et donc abandonner l'idée d'une omniscience "classique" telle que Thomas d'Aquin la concevait, où les observables  en puissance co-existent entre elles. Si la non-calssicité de l'omniscience est admise, on peut alors considérer qu'elle est réduite au contexte de l'observable actuellement mesurée (molinisme), ou bien qu'elle n'est tout simplement pas, et déclarer qu'il est vain de penser que les entités physiques existent sans être mesurées. \\

Si, au contraire, on persiste à considérer que l'omniscience s'étend aux contrafactuelles, il faut alors choisir entre :
\begin{itemize}
\item une vision superdéterministe du monde, renoncer avec Diodore à la proposition 3 de l'argument dominateur, et admettre que le possible est nécessairement vrai. Cette hypothèse a été discutée par John Bell dans les années 1980, sous la forme d'une échappatoire ultime dans le cadre des tests expérimentaux de ses inégalités : 
\begin{quotation}
\textit{Il existe une façon d'échapper à l'inférence de vitesses supraluminique et d'action fantôme à distance. Mais elle implique le déterminisme absolu dans l'univers, l'absence complète de libre-arbitre. Supposez que le monde est super-déterministe, pas seulement avec la nature inanimée qui court en coulisse, mais avec notre comportement, en incluant notre croyance que nous sommes libres de choisir de faire une expérience plutôt que l'autre, absoluement déterminé, incluant la "décision" de l'expérimentateur d'effectuer un ensemble de mesures plutôt qu'un autre, les difficultés disparaissent. Un signal plus rapide que la lumière n'est pas nécessaire pour dire à la particule A quelle mesure a été effectuée sur la particule B, parce que l'univers, particule A inclue, "connait" déjà ce que seront cette mesure, et son résultat.} \cite{davies}
\end{quotation}

Cette approche semble cependant remettre en cause la liberté de l'expérimentaliste, qui constitue une hypothèse fondamentale à la démarche scientifique : \begin{quotation}
\textit{Nous supposons toujours implicitement la liberté de l'expérimentateur... Cette hypothèse fondamentale est essentielle pour faire de la science. Si elle n'était pas vraie, alors, je suppose que poser des questions à la nature dans une expérience n'aurait plus aucun sens, puisque la nature pourrait alors déterminer quelles sont nos questions, et cela pourrait guider nos questions de telle sorte que nous arrivions à une fausse image de la nature.} Anton Zeilinger (\cite{zeilinger}, p.266)
\end{quotation}

\item renoncer à au moins l'une des deux autres propositions de l'argument dominateur : admettre donc qu'une proposition vraie dans le passé n'est pas nécessaire, ou que l'impossible peut émerger du possible.
\item admettre la \textit{contextualité des mesures}, renoncer à l'idée que le résultat de mesure d'une observable est indépendant de l'ensemble des observables co-mesurables considéré. Autrement dit, renoncer à l'idée que le factuel ne dépende pas du contrafactuel. Bien qu'il s'inscrive au final dans une vision superdéterministe, le concept de compossibilité de Leibniz me semble être relativement proche de cette position.\\
\end{itemize}

L'interprétation que donne Kochen et Specker de leur théorème  diffère cependant de ces propositions. Leur tentative de "capturer l'intuition" d'une symétrie logico-mathématique de la mécanique quantique qui serait violée par les théories à variables cachées sous la forme d'un critère formel les auraient conduit à une preuve de l'inexistence de ces variables cachées \cite{appleby}.
Si celles-ci ne sont pas explicitement utilisées dans le théorème, Kochen et Specker établissent qu'une théorie à variables cachées "réussite" doit affecter à chaque observable $A$ une fonction réelle $f(A)$ définie sur l'espace des phases $\Omega$ telle que pour toute fonction de Borel\footnote{Une fonction de Borel est une fonction de classe $\mathcal{C}^\infty$ à variables rélles et valeurs complexes, définie au voisinage de 0, telle que pour toute suite $(a_n)$ de nombres complexes, $\forall n \in \mathcal{N}$, $f^{(n)}(0)=a_n$} \[f_{g(A)} = g \circ f_A \]
Leur résultat démontrant qu'une telle affectation est impossible, par conséquent, les théories à variables cachées n'existent pas. 
Cependant, comme l'a remarqué Appleby, si une telle affectation est en effet impossible pour toute les observables quantique de l'espace, Meyer, Clifton et Kent (\cite{meyer} \cite{kent} \cite{clifton} \cite{clifton2}) ont montré que ces fonctions pouvaient être assignées à toutes les observables d'un sous-ensemble dense\footnote{La densité du sous-ensemble peut se traduire par le fait que tout complémentaire de celui-ci dans l'espace est d'intérieur vide.} de l'espace considéré \cite{appleby}.
Kochen et Specker avaient d'ailleurs connaissance du fait qu'une telle démonstration invaliderait leur argument :
\begin{quotation}
\textit{Ce résultat [Le théorème de Kochen-Specker],  exposé pour la première fois par Specker dans [référence à l'article de 1960 \cite{specker}], peut être obtenu plus simplement par un argument topologique directe ou en appliquant un théorème de Gleason [référence \cite{gleason}].  Cependant, il nous semble important pour la démonstration de la non-existence des variables cachées que nous utilisions une algèbre booléenne partielle petite et finie. Car autrement, une objection raisonnable peut être soulevée dans le fait qu'il n'y a pas de sens physique à supposer qu'il existe un nombre continu de propositions quantiques.}\\
\cite{ks}
\end{quotation}
Par ailleurs, la conclusion de Kochen et Specker entre ouvertement en contradiction avec l'existence de théorie à variables cachées compatibles avec les prédictions de la mécanique quantique, comme la théorie de de Broglie-Bohm. \\

Ainsi, si Bell tente par tous les moyens d'éviter que le théorème KS soit interprété comme une restriction sur l'éventuel existence des variables cachées, Kochen et Specker prennent le contre-pied de son interprétation et affirment avoir démontré l'impossibilité d'une telle existence. Il ne s'agit pas là d'une contradiction contextuelle, qui serait simplement issue de positions philosophiques différentes, mais bien d'un désaccord profond concernant la signification d'un même théorème.\\

Je pense néanmoins qu'une des sources de cette divergence peut être trouvée dans l'article original de von Neumann. Tandis que Bell le critique avec virulence, Kochen et Specker le défendent et indiquent avoir amélioré sa preuve. Tous les trois identifient néanmoins le travail de von Neumann à une tentative de preuve d'impossibilité des variables cachées. Cette interprétation reste encore aujourd'hui profondément ancrée dans les esprits. Deux "réévaluations" de la démonstration de von Neumann, par Jeffrey Bub \cite{bub3} et Dennis Dieks \cite{dieks}, défendent cependant que celle-ci est bien plus subtile et est loin d'être "stupide" (\cite{bell2} p. 166, cf. \cite{mermin1}).\\

Le résultat de von Neumann ne stipule pas que les modèles à variables cachées sont impossibles, mais qu'ils doivent nécessairement "exhiber une forme de contextualité" \cite{dieks}, autrement dit que la structure logique de la mécanique quantique ne peut être classique, annonçant ainsi implicitement le futur résultat du théorème KS.\\

\section{La preuve de Von Neumann}
\label{svon}
Dans son fameux livre \textit{Fondements Mathématiques de la Mécanique Quantique}, publié en 1932, John von Neumann s'intéresse notamment à la structure générale de la théorie quantique, et en particulier à l'émergence des probabilités. Il propose ainsi en introduction d'étudier si les probabilités quantiques sont issues d'une lacune épistémologique, i.e. une limitation de la connaissance de l'observateur. Si c'est le cas, l'existence de "variables cachées"  pourrait expliquer la nature statistique des propositions quantiques. \footnote{Cette section est basée sur l'article de Dieks \cite{dieks}}

Von Neumann aboutit à la conclusion suivante : 

\begin{quotation} \textit{Il se trouve cependant que [l'introduction de variables cachées] sera difficilement satisfaisante - plus précisément, une telle explication et incompatible avec certains postulats qualitatifs à la base de la mécanique quantique.} \cite{vonneum} \end{quotation}

Il ne déclare pas qu'il est impossible d'établir une théorie à variables cachées en accord avec la structure de la théorie quantique, mais que ces théories sont incompatibles avec ce qu'il considère comme un "principe quantique" : l'association d'un opérateur hermitien à chaque propriété physique dans l'espace de Hilbert. Les théories à variables cachées ne respectant pas cet axiome, elles doivent donc avoir une structure logique différente de la mécanique quantique.
\begin{quotation}
\textit{Une introduction de paramètre cachés n'est certainement pas possible sans changer des aspects essentiels de la théorie présente.}
\end{quotation}
La démonstration de von Neumann ne part pas du formalisme quantique mais de l'hypothèse naturelle que les systèmes physiques sont définis par des quantités mesurables. Dans le cas de quantités physiques $\textbf{R}$ et $\textbf{S}$ co-mesurables, on peut inclure toute fonction arbitraire de  $\textbf{R}$ et $\textbf{S}$ dans l'ensemble des quantités physiques. $f(\textbf{R},\textbf{S})$ est définie par l'application de la fonction $f$ aux résultats de mesure des quantités $\textbf{R}$ et $\textbf{S}$, mesurées conjointement.\\

Von Neumann remarque alors que dans le cas d'un système unique, la fonction $f(\textbf{R},\textbf{S})$ n'a pas de sens si $\textbf{R}$ et $\textbf{S}$ ne sont pas co-mesurables. Cependant, il est toujours possible de considérer des ensembles statistiques plutôt que de raisonner sur un système unique. Dans un tel ensemble, on peut mesurer la quantité $\textbf{R}$ dans un certain sous-ensemble, et la quantité $\textbf{S}$ dans un autre sous-ensemble. Von Neumann en déduit qu'il est permis de définir une quantité physique $\textbf{R+S}$, malgré la non co-mesurabilité de $\textbf{R}$ et $\textbf{S}$. Il suffit que cette quantité ne soit pas directement mesurable, et opérationnellement indépendante de $\textbf{R}$ et $\textbf{S}$. La nouvelle quantité introduite satisfait alors la condition d'additivité des espérances dans l'ensemble considéré :
\[ E (\textbf{R+S}) = E (\textbf{R}) + E (\textbf{S}) \]
Von Neumann ayant défini opérationnellement les quantités physiques pour un système, il est alors possible de considérer des sommes de ces quantités, que celles-ci soient co-mesurables ou non. Si les quantités sont co-mesurables, il suffit d'ajouter les valeurs obtenues. Si ce n'est pas le cas, on définit implicitement la quantité $\textbf{R+S}$ pour laquelle l'espérance dans un ensemble statistique arbitraire est égal à la somme des espérances de $\textbf{R}$ et $\textbf{S}$.\\

\`A partir de deux hypothèses,
\begin{itemize}
\item I - il existe une correspondance bijective entre quantités physiques de tout système et opérateurs hermitiens dans l'espace de Hilbert.
\item II - Cette correspondance respecte une relation d'additivité.
\end{itemize}
Von Neumann dérive la relation suivante : pour toute quantité physique arbitraire $\textbf{R}$ dans un ensemble statistique arbitraire du système physique considéré, il existe un opérateur hermitien $R$ tel que
\[ E(R) = Tr(UR)\]
avec $U$ un opérateur hermitien non négatif de trace 1 indépendant de $R$ (U correspond à un opérateur densité du système).
On peut alors tirer de cette expression le corollaire suivant :
Pour tout opérateur densité $U$, il existe toujours un opérateur $R$ tel que :
\[ E(R^2) - E(R)^2 > 0 \]
Ainsi, il n'existe pas d'ensemble pour lequel les quantités physiques n'ont pas de dispersion.
Or l'idée même des variables cachées est qu'il puisse justement exister des ensembles sans dispersion statistique, où les paramètres cachés ont des valeurs fixes.\\

Si l'on ajoute des propriétés physiques considérées comme cachées à la description quantique, on ne peut pas associer d'opérateurs hermitiens aux variables cachées de la théorie à la façon des quantités quantiques standards. Si c'était le cas, ces opérateurs violerait la relation d'additivité.\\

Ainsi, la fameuse "preuve" de von Neumann ne montre pas l'impossibilité des théories à variables cachées, mais plutôt l'impossibilité de dériver les probabilités quantiques d'une théorie à variables cachées où les états sont déterminés (sans dispersion).  Si l'on souhaite tout de même utiliser une théorie à variables cachées, les probabilités quantiques ne peuvent pas refléter la distribution de valeurs prédéterminées,  mais doivent être dérivées autrement. Dans la théorie de Bohm, les probabilités sont ainsi issues d'un processus dynamique qui n'est pas une mesure de valeur prédéterminée. \\

Selon von Neumann, il ne semble pas y avoir de raison de douter de la validité de la structure théorique de la mécanique quantique. A fortiori, il ne pense pas que l'introduction de variables cachées soit nécessaire. De telles variables n'obéiraient en effet pas à la même relation que les observables quantiques. \\

Par ailleurs, des quantités physiques définies implicitement ne correspondraient pas, en général, à des résultats de mesure. La valeur cachée de la quantité $\textbf{R+S}$ sera donnée par la somme des valeurs des quantités $\textbf{R}$ et $\textbf{S}$, quand bien même celles-ci ne sont pas co-mesurables. Si l'on accepte l'adéquation empirique de la mécanique quantique au niveau statistique, cette valeur cachée ne correspondra pas à un résultat de mesure, représenté par une valeur propre d'un opérateur. Ainsi, dans une théorie à variables cachées, les résultats expérimentaux ne reflètent pas toutes les propriétés présentent dans le système : les mesures ne représentent pas le système tel qu'il est. Une théorie à variables cachées en accord avec les prédictions quantiques implique donc une forme de contextualité de mesure, puisqu'elle nécessite qu'une des conditions $I$ ou $II$ soit violée. Le résultat de von Neumann annonce ainsi celui de KS : \textit{\textbf{la structure logique de la mécanique quantique ne peut être classique}}.

\section{La critique de Bell et l'amélioration de Kochen-Specker}
\label{scritique}
Dans une interview pour le magasine Omni (mai 1998, p.88) (source : \cite{bub3}), John Bell déclare :
\begin{quotation}
\textit{Pourtant, la preuve de von Neumann, si vous arrivez à la saisir, s'effondre dans vos mains ! Il n'y a rien. Ce n'est pas seulement imparfait, c'est stupide ! . . . Lorsque vous traduisez [ses hypothèses] en termes de disposition physique, elle sont absurdes. Vous pouvez me citer sur cela: La preuve de von Neumann n'est pas seulement fausse mais idiote !}
\end{quotation}

\`A partir des années 30, il semblerait que la communauté scientifique soit convaincue de la complétude du formalisme quantique, a priori démontrée par von Neumann. Mais, lorsqu'en 1952, David Bohm redécouvre la théorie de l'onde pilote de Louis de Broglie, Bell "\textit{[voit] l'impossible se réaliser}." (\cite{bell2}, p.160) Il décide alors de se pencher sur la preuve de von Neumann, et met à jour ce qui semble être une erreur grossière de raisonnement :
\begin{quotation}
\textit{[Dans la preuve de von Neumann] ce ne [sont] pas les prédictions mesurables objectives de la mécanique quantique qui excluent les variables cachées. [C'est] l'hypothèse arbitraire d'une relation particulière (et impossible) entre les résultats de mesures incompatibles dont l'une ou l'autre pourrait être effectuée dans une occasion donnée, mais dont une seule peut être effectuée en fait. }
\cite{bell}
\end{quotation}
Bell s'étonne d'ailleurs d'avoir été le premier a remarqué cette faute. Ce n'est en fait pas le cas, puisque Grete Hermann avait formulée une critique assez similaire en 1935. Cependant, pour diverses raisons, cette critique  semble être restée dans l'ombre \cite{dieks}. Dans son article \cite{bell}; il résume la preuve de von Neumann ainsi :
\begin{quotation}
\textit{Toute combinaisons réelles de deux opérateurs hermitiens représente une observable, et la même combinaison linéaire des espérances est l'espérance de la combinaison.}
\end{quotation}
Si l'on prend l'exemple d'une mesure spin 1/2 dans les directions $x$ et $y$, la mesure de $\sigma_x + \sigma_y$ serait alors une mesure dans une troisième direction différente, qui requierait un dispositif expérimental différent et indépendant. 
Or, pour von Neumann, les variables cachées ne peuvent pas être représentées de la même façon que les quantités physiques dans le formalisme quantique standard. Si c'était le cas, cela conduirait à la violation de la proposition I ou II. Il traite de quantités physiques indépendamment de la mécanique quantique et de leur représentation mathématique. Il définit la combinaison linéaire de ces quantités de sorte que leur espérance (valeur attendue) satisfasse par définition la même relation linéaire que les quantités. \\

Contrairement à ce que déclare Bell, von Neumann considèrait le fait que certaines quantités ne peuvent pas être co-mesurables et était conscient que cela jette un doute sur le sens des sommes de telles quantités. Cette difficulté est la raison pour laquelle il définit la somme de deux quantités non co-mesurables de façon implicite, sécurisant ainsi l'additivité des espérances. \\

De l'hypothèse d'une association bijective entre quantité physique et opérateur hermitien dans un espace de Hilbert, von Neumann déduit en effet qu'il ne peut y avoir d'ensemble d'états sans dispersion. Cependant, ce que Bell semble ne pas prendre en compte, c'est que von Neumann démontre également que si de tels états existent dans une théorie à variables cachées quelconque, l'ensemble des quantités physiques telles que von Neumann les définit ne peuvent pas être associés à des opérateurs hermitiens dans $H$.\\

Bell interprète incorrectement les prémisses de la preuve de von Neumann comme des contraintes imposées sur les espérances des quantités physiques, via leur représentation par opérateurs dans le formalisme quantique standard.\\

Kochen et Specker semblent faire une erreur d'interprétation similaire à celle de Bell. Mais, contrairement à lui, ils considérent que leur résultat va dans le sens de celui de von Neumann, et qu'ils l'améliorent d'un point de vue formel.

En conclusion de leur article \cite{ks}, Kochen et Specker déclarent :
\begin{quotation}
\textit{[Ces résultats] nous semble présenter une nouvelle caractéristique de la mécanique quantique par son changement radical par rapport à la mécanique classique. Bien sûr, le principe d'incertitude marque déjà un changement radical par rapport à la mécanique classique. Cependant, l'énoncé du principe d'incertitude implique deux observables qui ne sont pas co-mesurables, et pourrait ainsi être réfuté par l'ajout de nouveaux états. C'est le point de vue de ceux qui croient aux variables cachées. [...] L'énoncé [...] que nous avons construit met uniquement en jeu, à chaque étape de sa construction, des observables co-mesurables, et ne pourra ainsi pas être réfuté.}
\end{quotation}
Ils seraient convaincus d'avoir fourni une "preuve de la non-existence des variables cachées", qu'ils comparent à "celle de von Neumann" :
\begin{quotation}
\textit{La preuve de von Neumann est essentiellement basée sur la non-existence d'une fonction à valeurs réelles sur l'ensemble des observables quantique qui est multiplicative sur les observables commutantes, et linéaire. Dans notre preuve nous montrons que la non-existence d'une fonction à valeurs réelles qui est à la fois multiplicative et linéaire uniquement pour des observables commutantes. Ainsi, dans un sens formel, notre résultat est plus solide que celui de von Neumann.}
\cite{ks}
\end{quotation}
Cependant, n'ayant probablement pas eu accès à l'article de Bell paru quelques mois avant le leur, ils se réfèrent à la critique moins connue de Grete Hermann (également discutée dans \cite{dieks}).
\begin{quotation}
\textit{Le critère de von Neumann a été critiqué dans la littérature du fait qu'elle requiert l'additivité [des espérances] même pour des opérateurs qui ne commutent pas [...] (voir par exemple Hermann [9,pp.99-104].)}
\cite{ks}
\end{quotation}

Là où von Neumann utilise la condition : 
\[ E(R+S) = E(R) + E(S) \hbox{ pour tout } R,S \in H  \]
pour laquelle il est (injustement) critiqué, les théorèmes de Gleason et de Kochen-Specker adoptent le critère suivant : 
\[ E(R+S) = E(R) + E(S) \hbox{ avec } R,S \in H \hbox{ tels que } RS = SR \]
  
Comme le remarque Jeffrey Bub, " Le théorème de Gleason permet d'arriver à la même conclusion [que le théorème de von Neumann]. Sa preuve est cependant fondée sur des hypothèses plus faibles pour des espaces de Hilbert de dimension supérieures à deux, et n'est valable que pour des quantités simultanément mesurables."\\

Apprécié à sa juste valeur,  le résultat de von Neumann semble ainsi sous-tendre, ou du moins annoncé, le résultat KS.

\section{Conclusion}
\begin{center}
Qu'appelle-t-on contextualité quantique ? \\
\end{center}

La mécanique quantique n'est pas contextuelle : les probabilités que son formalisme permet de calculer sont indépendantes du contexte de la mesure. Ce qui est "contextuel", ou plutôt ce qui pourrait l'être au sein d'une interprétation réaliste de la mécanique quantique, c'est l'assignation de valeurs aux quantités physiques que l'on mesure, i.e. aux observables, qui dépend de l'ensemble d'observables co-mesurables considéré.  On parle alors de \textit{contextualité de la mesure}.\\

Cette contextualité est \textit{étrange} au sens où elle semble témoigner du comportement non-classique des objets quantiques. Mais cette étrangeté est-elle pour autant inquiétante ? Bouscule-t-elle fondamentalement notre conception du monde ? \\

Bell pensait que non. Il trouve ici un point d'entente avec Bohr, qui en avait déjà l'intuition, lorsqu'il déclarait  que " \textit{le résultat d'une observation pourrait ne pas dépendre uniquement de l'état du système (incluant les variables cachées) mais aussi de la disposition complète de l'appareil de mesure}" \cite{bell}. D'autre part, les variables cachées de théories réalistes "naturelles" (à trajectoires classiques) comme la théorie de de Broglie-Bohm sont contextuelles : le résultat d'une mesure dépend de la façon dont est effectué la mesure. Cependant, la contextualité "à la de Broglie-Bohm" que défendent Bohr et  Bell est parachutée.  Elle est basée sur le concept philosophique bohrien de \textit{complémentarité}, et ne peut être rigoureusement démontrée à partir du formalisme quantique, car elle est valable pour des mesures projectives sur des qubits (espace de Hilbert de dimension 2). Cette "extra-contextualité" a quelque chose d'artificiel, et n'est pas \textit{nécessaire}.  \\

La contextualité quantique peut en effet être déduite d'arguments logiques et géométriques à partir théorèmes des Gleason et de Kochen-Specker,  mais à condition que les mesures soient projectives, que l'espace de Hilbert considéré soit de dimension supérieure ou égale à trois, et d'admettre l'hypothèse du déterminisme des résultats.  La fameuse preuve de von Neumann peut être considérée comme un résultat précurseur de ces théorèmes, mais il ne peut constituer une véritable preuve de contextualité, n'imposant pas la même condition sur la dimension.\\

Il est important de noter qu'aucun des protagonistes - Bohr, von Neumann, Gleason, Bell, Kochen et Specker - n'utilise les termes "contexte" ou "contextualité" dans leurs articles originaux respectifs. L'adjectif “contextualiste” a été introduit pour la première fois par A. Shimony : Experimental test of local hidden variable theories, dans B. d’Espagnat ed. : Foundations of Quantum Mechanics (Academic Press, New York and London, 1971). Il a ensuite été raccourci en “contextuel” par E.G. Beltrametti et C. Cassinelli : The Logic of Quantum Mechanics (SIAM, 1983) pour traiter des implications du théorème de Kochen-Specker. Néanmoins, encore une fois, le théorème KS n'implique pas nécessairement la contextualité de mesure.\\

La théorie quantique est fondée sur des axiomes mathématiques, mais ne donne aucune indication sur le sens physique de ces postulats, laissant libre cours aux interprétations philosophiques. Cette lacune nous invite à rester prudent face aux raccourcis du langage, qui sortent les concepts de leur cadre interprétatif au risque de semer la confusion.\\

Ainsi,  le résultat KS convie (ou contraint) un réaliste qui croit en la détermination des valeurs des quantités physiques, à considérer que l'affectation de ces valeurs dépend du contexte de la mesure. Contrairement au contexte de la théorie de de Broglie-Bohm (et par extension, ce que Bell et Bohr pourraient appeler "contexte"), qui correspond à une description détaillée du dispositif qui mesure une unique observable, le contexte du théorème de Koche-Specker correspond à un ensemble de mesures d'observables qui \textit{auraient pu} être effectuées simultanément à celle actuellement réalisée.  Une interprétation réaliste pourrait conduire à conclure que le théorème de Kochen-Specker s'appuie sur la contrafactualité, et les infuturabilités qui ont inspirées Specker. \\

Les partisans d'un "réalisme participatif" (QBisme, reconstruction informationnelle etc.)  considèreraient plutôt que le théorème conforte l'intuition selon laquelle "les mesures non effectuées n'ont pas de résultats" \cite{peres2}.\footnote{Chris Fuchs fait du résultat de Kochen-Specker un des pivot de son interprétation "QBiste" : "\textit{Je pense que la conclusion la plus raisonnable et la plus simple que l'on puisse tirer des résultats de Kochen-Specker et de la violation des inégalités de Bell est, comme le dit Asher Peres, que "des mesures non-effectuées n'ont pas de résultat." La mesure provoque l'existence de la "valeur de vérité", qui n'existe pas préalablement.}
\cite{fuchs4}}  Il n'existe pas, selon eux, de valeurs pré-déterminés. Il ne s'agit pas ici d'un abandon de la pré-détermination au profit de la noncontextualité, mais d'un refus de choisir : les résultats de Bell et de Kochen-Specker sont des théorèmes no-go, qui invitent à ne plus considérer les mesures comme révélatrices, mais comme créatrices, et ainsi à abandonner toute vision réaliste "naïve" et classique comme les théories à variables cachées.\\

Enfin, on peut remarquer que d'autres réalistes considèrent que leur interprétation n'est pas "concernée" par le théorème KS. C'est semble-t-il le cas des adhérents à l'interprétation d'Everett (ou MWI) qui considèrent que "l'ontologie de la théorie quantique est la fonction d'onde de l'Univers $\Psi(t)$. Rien d'autre. [...] Par conséquent, les relations d'incertitude de Heisenberg, les relations d'incertitude de Robertson, le théorème de Kochen-Specker, l'argument EPR, GHZ, et les inégalités de Bell ne sont pas pertinents pour l'analyse fondamental des propriétés de la Nature.  " \cite{vaidman3}. Dans cette interprétation des mondes multiples, les concepts d'observateur et d'observation disparaissent du formalisme et avec eux, la notion de contrafactualité.\\

Les conclusions que l'on peut tirer du théorème de Kochen-Specker diffèrent ainsi du contexte interprétatif considéré. Il n'en est pas pour autant dénué d'intérêt. Le théorème, basé sur un argument géométrique et fondé sur une réflexion logico-mathématique, semble issu du cœur même du formalisme quantique. Le fait que, contrairement au théorème de Bell, il porte sur un système unique et soit indépendant de l'état du système renforce l'intuition qu'il pourrait être d'autant plus fondamental. \\

Néanmoins, comme le remarque Asher Peres, "\textit{le problème ne peut se résumer à la pure logique. Toute discussion sur la physique doit au final établir un lien avec les faits expérimentaux.}" \cite{peres}  La question de la faisabilité de tests expérimentaux du théorème KS a suscité de nombreux débats au sein de la communauté scientifique. La découverte d'inégalités de non-contextualité  semble avoir clos ces discussions et de nombreux résultats ont été obtenus à ce jour, malgré les divers échappatoires ("loopholes") à prendre en compte par les expérimentateurs. (cf. annexe pour détails)\\

Divers pistes d'études du résultat KS ont été ouvertes, et semblent montrer qu'il pourrait sous-tendre la violation des inégalités de Bell. Cette dernière ne repose-t-elle pas aussi sur un raisonnement contrafactuel ?  Un principe, tapis derrière la parabole de Specker, pourrait permettre de comprendre contextualité et non-localité au sein d'une ossature formelle commune, et être ainsi potentiellement érigé au rang d'axiome de la théorie quantique. \\

\chapter{Figures Impossibles: généralisation du théorème KS }
\label{chapnlcont}
En 1982, Alain Aspect, Philippe Grangier, Gérard Roger et Jean Dalibard parviennent pour la première fois à mettre en évidence de manière convaincante une violation expérimentale des inégalités de Bell, observant ainsi la nature non-locale de la mécanique quantique \cite{aspect2} \cite{aspect3}. En 2015, trois expériences indépendantes ont confirmé ce résultat en fermant les dernières échappatoires susceptibles de sauver la localité \cite{hensen}, \cite{giustina}, \cite{shalm}. Qu'en est-il pour la contextualité ? Il y a encore quelques années, la question de la réalisation d'expériences analogues de test de contextualité était très controversée, et ne semblait qu'un vague idéal à atteindre. La réussite expérimentale et la notoriété du théorème de Bell contraste avec l'impopularité du résultat KS, totalement obscurcit par l'éclat de son "petit frère" \footnote{ Si du point de vue des publications, les articles de Bell et de Kochen-Specker sont postérieurs à l'article de 1964, il n'en reste pas moins que les bourgeons de la contextualité étaient déjà profondément ancrés dans les esprits (cf. chapitre précédent : von Neumann, Bohr, Gleason et l'article de 1960 de Specker).}. Aujourd'hui encore, malgré des avancés conceptuelles particulièrement importantes, la contextualité reste pour la plupart (à l'exception des experts) inconnue, dans l'ombre de la non-localité.\\

Cette "obscurité relative" \cite{mermin1} n'est sans doute pas indissociable du traitement de Bell à l'égard de son propre théorème (cf. chapitre précédent). Par ailleurs, dans son article intitulé "\textit{Hidden variables and the two theorems of John Bell}" publié en 1993, David Mermin a remarqué que les restrictions issues du résultat KS, qui s'appliquent a priori à toutes les théories à  variables cachées, peuvent uniquement être formulées au sein de la structure formelle de la mécanique quantique. 

\begin{quotation} \textit{On ne peut pas décrire le théorème de Bell-KS au grand public, en termes d'une collection d'expériences} de pensées [\textit{gedanken}]\textit{ sous formes de boîtes noires, l'unique rôle de la mécanique quantique étant de fournir des résultats de pensée} [gedanken]\textit{, qui impliquent par eux-mêmes qu'au moins une de ces expériences n'aurait pas pu révéler un résultat préexistant. Le théorème de Bell, cependant, peut être précisément exprimé dans ces termes.}
\end{quotation}

Enfin, une autre raison évoquée par Mermin est qu'à la différence de la clarté de la démonstration du théorème de Bell, le résultat KS implique "\textit{un exercice modérément élaboré en géométrie}". 

\begin{quotation}
\textit{Les physiciens sont simplement moins enclins que les philosophes à souffrir à travers quelques pages d'analyses obscures pour prouver quelque chose dont ils n'ont jamais douté en premier lieu.}
\end{quotation}

Cet argument est cependant devenu obsolète à partir des années 1990, de nouvelles preuves particulièrement simples du résultat KS (i.e. nécessitant moins d'observables que la preuve originale de Kochen et Specker)  ayant été découvertes, notamment par Mermin et Peres \cite{mermin1}. Cette recherche de simplification des preuves s'est poursuivit au cours des deux dernières décennies. Néanmoins, comme le remarque Ana Belén Sainz \cite{sainz3},
\begin{quotation}
\textit{une grande partie des travaux portant sur la contextualité jusqu'à aujourd'hui s'est concentrée sur des exemples spécifiques et des preuves concrètes du théorème de Kochen-Specker, tandis que définitions générales et théorèmes sur la contextualité sont restées rares. Par exemple, il est communément admis que la non-localité est un cas particulier de contextualité, mais qu'est-ce que cela signifie exactement ?}
\end{quotation}

Cette dernière considération ne parait guère triviale. Non-localité et contextualité apparaissent, à premier abord, comme deux notions de non-classicité intrinsèquement différentes, et ces différences nous amènent à nous questionner sur la nature de leur relation.\\

\begin{tabular}{|c|c|c|}
  \hline
  Caractéristiques & Théorème de Bell & Résultat KS  \\
  \hline
  Valable pour un système unique & Non & Oui \\
  \hline
  Valable pour un système multipartite & Oui & Oui \\
  \hline
  Indépendant de la théorie (boites noires) & Oui & Non \\
  \hline
  Indépendant de l'état du système & Non & Oui \\
  \hline
  Valable pour un qubit & Oui & Non\\
  \hline
\end{tabular}
\vspace{3mm}

David Mermin a remarqué, dans son article de 1993, que certaines preuves du théorème KS pouvaient être "converties" en preuve du théorème de Bell. Il expose ainsi un scénario particulier où l'hypothèse de non-contextualité peut être interprétée comme conséquence d'une hypothèse de localité, qui constituerait simplement une hypothèse d'indépendance contextuelle où les contextes seraient éloignés.

\begin{quotation}
\textit{Supposons que l'expérience qui mesure les observables commutantes $A$,$B$,$C$,... utilise des pièces d'équipement éloignées les unes des autres, qui enregistrent séparément les valeurs de $A$,$B$,$C$,... . Et supposons que l'expérience qui mesure $A$ avec les observables commutantes $L$,$M$,..., qui ne commutent pas toutes avec $B$,$C$,..., requiert des changements dans l'appareillage complet qui remplacent seulement les parties qui enregistrent les valeurs de  $B$,$C$,... avec des pièces d'équipement différentes qui enregistrent les valeurs de $L$,$M$,... . Et supposons que tous ces changements d'équipement sont fait loin des pièces inchangées de l'appareil qui enregistre la valeur de $A$. En l'absence d'action à distance, il est peu probable que de tels changements dans la disposition complète de l'appareil n'aient aucun effet sur le résultat de la mesure $A$ sur un système individuel.  Dans ce cas, l'hypothèse problématique de non-contextualité peut être remplacée par une simple hypothèse de localité. \cite{mermin1}}
\end{quotation}

Dans ce cas particulier, une démonstration de non-localité implique donc un argument de contextualité.
En revanche, contrairement aux preuves traditionnelles de contextualité, celle-ci n'est valable que pour un état particulier du système et, à la différence d'une inégalité de Bell, elle ne s'applique qu'à la mécanique quantique.\\

Les différences entre ces deux notions non-classiques sont aussi au coeur de la controverse portant sur la question de la faisabilité et du sens de tests expérimentaux de contextualité. (\cite{appleby}, \cite{appleby2}, \cite{barrett1}, \cite{havlicek}, \cite{cabello5}, \cite{breuer}, \cite{simon1}, \cite{cabello6}) \`A l'aune de ces débats, différentes approches ont été proposées : en s'inspirant d'une inégalité découverte par Klyachko, Can, Binicioglu et Shumovsky (section \ref{kcbs}), Ad\'an Cabello est parvenu à transformer une preuve indépendante du théorème KS établie par Mermin et Peres en inégalité (section \ref{merminperescab}). Ces travaux l'ont conduit à formaliser, en collaboration avec Simone Severini et Andreas Winter, une approche graphique du résultat KS (section \ref{csw}). Celle-ci semblerait permettre d'obtenir les bornes maximales quantiques et générales d'inégalités de corrélations, et de traiter les inégalités de Bell comme cas particuliers d'inégalités de "non-contextualité". En se basant sur cette approche, Antonio Ac\'in, Tobias Fritz, Anthony Leverrier et Ana Belén Sainz (AFLS) ont construit un formalisme permettant d'étudier et de comparer le modèles probabilistes classiques et quantiques (section \ref{safls}). Au sein de cette approche, un scenario de type Bell correspond au produit de Foulis-Randall de scenarii de contextualité. Ces approches graphiques ont conduit à identifier ce qui pourrait être un principe nécessaire pour comprendre les corrélations non-classiques. Enfin, une autre approche, développée par Samson Abramsky et basée sur le langage des catégories (section \ref{sabram}), permet d'envisager de raconter le résultat KS au grand public à l'aide de paradoxes logiques simples, ou encore de les illustrer via des figures impossibles présentes dans certains tableaux de M.C. Escher (section \ref{sraconter}).\\

\section{L'inégalité de Klyachko (ou KCBS)}
\label{kcbs}

La première inégalité de "non-contextualité" fut indépendamment développée en 2000 par Larsson \cite{larsson1} et par Simon, Brukner et Zeilinger \cite{simon2}. Cependant, du fait qu'elle ne repose pas exclusivement sur l'hypothèse de non-contextualité mais également sur des prédictions issues du formalisme quantique, elle n'est pas indépendante de ce dernier, et n'est donc pas satisfaisante.\\

Il faut attendre 2007 pour que quatre chercheurs - Klyachko, Can, Binicioglu et Shumovsky (KCBS) \cite{kcbs} - introduisent la première "\textit{inégalité de non-contextualité-KS}"\footnote{L'appelation "inégalité de non-contextualité" peut porter à confusion. En effet, comme nous le verrons, l'inégalité ne porte pas sur la non-contextualité, mais sur le résultat KS. Une violation de l'inégalité n'est donc pas nécessairement une preuve de contextualité, mais pourrait être interprétée comme une violation du déterminisme des résultats. Même si cette dernière hypothèse est considérée comme admise, il est plus juste d'utiliser l'expression "\textit{inégalité de non-contextualité KS}". Nous y reviendrons ultérieurement (cf. fin du chapitre). }, i.e. satisfaite par toutes les théories à variables cachées non-contextuelles, et violée par la mécanique quantique. Il s'agit de l'inégalité de non-contextualité la plus simple qui n'est pas respectée par un qutrit. Elle est souvent mise en parallèle avec l'inégalité CHSH, qui est l'inégalité de Bell la plus simple violée par un qubit.\\

Considérons un système physique quelconque, auquel on peut poser cinq questions fermées, représentées par des projecteurs $P_i$, avec $i$ compris entre 1 et 5. Toutes questions $P_i$ et $P_{(i+1)mod5}$ sont compatibles, i.e. peuvent être posées conjointement. Elles sont aussi exclusives, et ne peuvent donc pas être toutes les deux vraies.\\

Un contexte est défini comme étant un ensemble de questions mutuellement compatibles. Ainsi $\{P_i , P_{i+1} \}$  et $\{P_{i-1} , P_i \}$ sont deux contextes différents. $\{P_{i-1} , P_{i+1} \}$ en revanche, n’en est pas un.\\

Si l’on suppose que les réponses aux questions sont prédéterminées de façon non-contextuelle, la réponse à $P_i$ sera la même dans les contextes $\{P_i , P_{i+1} \}$ et $\{P_{i-1} , P_i \}$.\\

Les projections peuvent être représentées sous la forme d’un pentagone, dont chaque côté représente une relation de commutabilité et d’exclusivité, et lie ainsi deux projections au sein d'un contexte.\\

\begin{figure}[ht!]
\centering
\includegraphics[width=4cm]{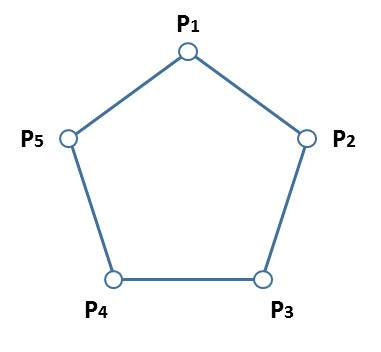}
\caption{Pentagone des projections}
\end{figure}

On pose alors la question suivante : Quel nombre maximal de réponses affirmatives peut-on espérer obtenir lorsque l'on pose les cinq questions au système physique ?\\

Sur le pentagone, représentons une réponse affirmative par un coin orange ; une réponse négative par un coin bleu. Le coloriage, qui correspond à une assignation non-contextuelle de valeurs, doit respecter la règle d'exclusivité suivante : \\
Il est interdit de colorier en orange deux coins reliés dans le pentagone.

\begin{figure}[ht!]
\centering
\includegraphics[width=4cm]{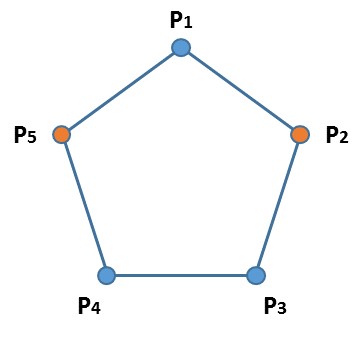}
\caption{Pentagone des projections}
\label{fpent2}
\end{figure}

Il en résulte que l'on ne peut pas donner plus de 2 réponses affirmatives à ces questions (cf. figure \ref{fpent2}).

Ainsi, si l'on pose individuellement chacune des cinq questions à plusieurs systèmes identiquement préparés, et que l'on calcule à partir des résultats obtenus la moyenne des réponses positives, on obtient l'inégalité suivante (en associant 1 à une réponse positive et 0 à une réponse négative) :

\[ \beta : = \sum_{i=1}^5 <P_i> \leq 2  \]

Il s'agit de l'inégalité de Klyachko, aussi appelée inégalité KCBS.
Cette inégalité est violée par la mécanique quantique : $\beta_{MQ}$ = $\sqrt{5} \approx$ 2,236.\\

La violation maximale est obtenue pour l'état $\bra{\psi}$ = (0,0,1) et les questions\\ $P_i = \ket{v_i}\bra{v_i}$ avec\footnote{Pour comprendre la construction de l'inégalité et notamment le choix des projecteurs (observables) permettant de la violer, cf. \cite{kcbs} ou \cite{bub4}} :\\

$\ket{\nu_1} = \frac{1}{\sqrt{2}} \begin{pmatrix} 1 \\ 0 \\  \sqrt{\cos(\pi/5)}  \end{pmatrix} $ \hspace{6mm} 
$\ket{\nu_2} = \frac{1}{\sqrt{3}} \begin{pmatrix} \cos(4\pi/5) \\ \sin(4\pi/5) \\  \sqrt{\cos(\pi/5)}  \end{pmatrix} $ \hspace{6mm} 
$\ket{\nu_3} = \frac{1}{\sqrt{3}} \begin{pmatrix} \cos(2\pi/5) \\ - \sin(2\pi/5) \\  \sqrt{\cos(\pi/5)}  \end{pmatrix} $ 

\vspace{1cm}

$\ket{\nu_4} = \frac{1}{\sqrt{3}}  \begin{pmatrix} \cos(4\pi/5) \\ - \sin(4\pi/5) \\  \sqrt{\cos(\pi/5)}  \end{pmatrix} $ \hspace{6mm}
$\ket{\nu_5} = \frac{1}{\sqrt{3}} \begin{pmatrix} \cos(2\pi/5) \\ \sin(2\pi/5) \\  \sqrt{\cos(\pi/5)}  \end{pmatrix} $

\vspace{1cm}

Une autre version de l'inégalité de Klyachko peut être donnée en utilisant les observables $A_i = 2\ket{v_i}\bra{v_i} - I$ (avec ($i=1,2,...,5$). Ces observables admettent les valeurs propres $\pm 1$.\\

On considère les cinq paires d'observables suivantes, qui correspondent à cinq contextes de mesure : $\{A_1,A_2\}$, $\{A_2,A_3\}$,$\{A_3,A_4\}$,$\{A_4,A_5\}$,$\{A_5,A_1\}$, où chaque observable est mesurable dans deux contextes différents. L'inégalité de Klyachko prédit que dans tout modèle à variables cachées non-contextuelles, les corrélations entre résultats de mesures sont limités par :
\[\alpha:=\langle A_1A_2\rangle + \langle A_2A_3\rangle + \langle A_3A_4\rangle + \langle A_4A_5\rangle + \langle A_5A_1\rangle  \geq -3\]

Pour un qutrit dans l'état fondamental $\ket{\psi}$ :

\[\alpha_{MQ} = 5 - 4\sqrt{5} \simeq -3,944 \]

L'inégalité de Klyachko est donc bien violée par la mécanique quantique, mais dépend de l'état dans lequel se trouve le système, ce qui fut considéré comme un "inconvénient" ("drawback") par ses créateurs. Elle pourrait, de ce fait, être considérée comme une preuve de contextualité KS plus "faible" que celles qui en sont indépendantes. \\

Le \textit{carré de Mermin-Peres}, présenté dans l'article de 1993 de Mermin, et inspiré d'une preuve de Peres \cite{peres3} ( elle-même basée sur la version GHZ (Greenberger-Horne-Zeilinger) du théorème de Bell \cite{ghz}) constitue une telle preuve indépendante de l'état du système.\footnote{La version de Peres utilise un plus petit nombre d'observables, mais s'applique uniquement à un système préparé dans un état particulier. \cite{mermin1}}

\section{Le carré de Mermin-Peres et l'inégalité de Cabello}
\label{merminperescab}

La preuve de Mermin-Peres met en jeu un ensemble de 9 observables $\{A,B,C,a,b,c,\alpha,\beta,\gamma\}$ dans un espace de Hilbert de dimension 4, qui peuvent en pratique être représentées par des matrices de Pauli s'appliquant sur un couple de qubits\footnote{Mermin donne l'exemple d'un système composé de deux spin-1/2.} indépendants. Elles sont rapportées dans le tableau 3x3 suivant :\\
\begin{center}
\begin{tabular}{|c|c|c|}
  \hline
  $A=\sigma_z\otimes \mathbb{1}$ & $B=\mathbb{1}\otimes\sigma_z$ & $C=\sigma_z\otimes\sigma_z$ \\
  \hline
  $a=\mathbb{1}\otimes\sigma_x$ & $b=\sigma_x\otimes \mathbb{1}$ & $c=\sigma_x\otimes\sigma_x$ \\
  \hline
  $\alpha=\sigma_z\otimes\sigma_x$ & $\beta=\sigma_x\otimes\sigma_z$ & $\gamma=\sigma_y\otimes\sigma_y$ \\
  \hline
\end{tabular}
\end{center}

Les observables dans chacune des trois lignes et chacune des trois colonnes sont mutuellement commutantes. Ainsi, on peut considérer que chaque ligne et chaque colonne correspond à un contexte de mesure. Par exemple, l'observable $A$ peut être mesurée dans le contexte "ligne" $\{A,B,C\}$, ou dans le contexte "colonne" $\{A,a,\alpha\}$.\\

D'après l'hypothèse de déterminisme des résultats de mesure, il est possible d'affecter une valeur propre $\pm 1$ à chaque observable. L'hypothèse de non-contextualité nous indique que la valeur assignée à une observable ne dépend pas du fait que celle-ci soit considérée dans son contexte "ligne" ou dans son contexte "colonne". \\

Le produit des trois observables de chaque ligne est égal à la matrice identité $+ \mathbb{1}$ :
$ABC = abc = \alpha\beta\gamma = + \mathbb{1}$. La valeur propre associée à $+ \mathbb{1}$ étant $+1$, il en résulte que dans chaque ligne, le nombre de $-1$ affecté doit être pair, et donc que le nombre total de valeurs $-1$ dans le tableau est pair. \\

Pour les deux premières colonnes, on obtient le même résultat : $Aa\alpha = Bb\beta = + \mathbb{1} $. En revanche, pour la troisième colonne :  $Cc\gamma = - \mathbb{1} $. Le nombre de valeurs $-1$ affectées dans cette colonne est donc impair, et, par extension, le nombre total de valeur $-1$ dans le tableau est impair, ce qui entre en contradiction avec le raisonnement précédent. \\

Ainsi, on retrouve le résultat KS : il faut renoncer à l'hypothèse du déterminisme des résultats ou à l'hypothèse de non-contextualité. \\

La preuve du carré de Mermin-Peres, particulièrement simple, s'adapte parfaitement à un cadre éducatif. Elle peut par exemple être reformulée de la façon suivante : imaginons que chaque observable soit représentée par une carte. Neuf cartes sont positionnées, face cachée, pour former une grille de trois lignes et trois colonnes. Chaque carte peut être noire (valeur $+1$), ou rouge (valeur $-1$), et la couleur de la carte ne dépend pas du fait qu'on la retourne ou non (hypothèse de déterminisme des résultats). Par ailleurs, si l'on s'intéresse à la couleur d'une carte en particulier, et que l'on retourne simultanément les deux autres cartes de la colonne considérée lors de la mesure, on suppose que la couleur obtenue aurait été la même si l'on avait plutôt retourné simultanément les deux cartes de la ligne considérée. (hypothèse de non-contextualité). Essayez à présent de construire une grille telle qu'il existe un nombre pair de cartes rouges dans chaque ligne et dans les deux premières colonnes, et un nombre impair de cartes rouges dans la dernière colonne. Si, classiquement, un telle configuration est impossible, la mécanique quantique et ses "cartes magiques" y parvient. C'est le résultat obtenu par Bell-Kochen-Specker : les mesures quantiques ne peuvent pas être interprétées comme révélant une propriété préexistante et indépendante du contexte de mesure considéré.\\

En 2008, le carré de Mermin-Peres fut converti en "inégalité de non-contextualité KS" par Ad\'an Cabello \cite{cabello14}. Nous représentons la version proposée dans \cite{kleinmann}.\\

Considérons un système unique, neuf observables $\{A,B,C,a,b,c,\alpha,\beta,\gamma\}$, et le corrélateur 
\[\langle\chi\rangle = \langle ABC \rangle + \langle abc \rangle + \langle \alpha\beta\gamma \rangle + \langle Aa\alpha \rangle + \langle Bb\beta \rangle - \langle Cc\gamma \rangle\]

Chaque terme de ce corrélateur correspond à la valeur moyenne du produit des résultats de mesures de trois observables, si celles-ci étaient mesurées simultanément ou dans une même séquence.\footnote{Si, pour chaque espérance, les trois observables sont compatibles, le fait qu'elles soient mesurées simultanément ou dans une même séquence n'a pas d'importance.} Tout modèle à variables cachées (KS-)non-contextuel doit ainsi affecter une valeur fixe $\pm 1$ à chacune des neufs observables, indépendamment de la colonne ou de la ligne utilisée comme contexte de mesure. Afin d'obtenir la borne maximale du corrélateur, on peut imaginer que la valeur $+1$ est affectée à chaque observable. On a alors $\langle ABC \rangle = \langle abc \rangle = \langle \alpha\beta\gamma \rangle = \langle Aa\alpha \rangle = \langle Bb\beta \rangle = \langle Cc\gamma \rangle = + 1$, et obtient :
\[\langle\chi\rangle \leq 4\]

On considère à présent que le système évolue dans un espace de Hilbert de dimension 4, et que les neufs observables en jeu correspondent à celles présentes dans le carré de Mermin-Peres. Comme $ABC = abc = \alpha\beta\gamma = Aa\alpha = Bb\beta = + \mathbb{1}$, on a $\langle ABC \rangle = \langle abc \rangle = \langle \alpha\beta\gamma \rangle = \langle Aa\alpha \rangle = \langle Bb\beta \rangle = +1$. Par ailleurs, $Cc\gamma = -\mathbb{1}$, et donc $\langle Cc\gamma \rangle = -1$. Il en résulte que, pour la mécanique quantique, 

\[\langle\chi\rangle_{MQ} = 6\]

La mécanique quantique viole ainsi l'inégalité indépendamment de l'état dans lequel le système se trouve, à condition que les mesures soient choisies de manière appropriée. \\

Il semblerait qu'il existe aussi bien des inégalités pour les corrélations décrites dans le théorème de Bell que pour celles présentes dans le théorème KS. Néanmoins, les relations conceptuelles entre ces corrélations ne sont toujours pas claires. Pour mieux les comprendre, un cadre commun est nécessaire.\\

D'autre part, de la violation quantique d'inégalités de non-contextualité-KS émerge instantanément une nouvelle interrogation : "Pourquoi ?". Plus précisément, "pourquoi cette valeur ?" ; "pourquoi la mécanique quantique n'est-elle pas plus contextuelle ?". Cette question fait écho aux travaux portant sur la nature de la limite de la non-localité quantique. En effet, si la mécanique quantique viole les inégalités de Bell, il ne s'agit pas de la violation maximale autorisée par le principe de non-communication, i.e. l'impossibilité de communiquer instantanément. Sandu Popescu et Daniel Rohrlich, deux pionniers de ce domaine, ont ainsi montré que la non-communication ne suffisait pas à décrire l'ensemble des corrélations quantiques. Ils ont mis en évidence des corrélations "super-quantiques"  entre deux parties non-communicants (des "boîtes PR"), qui ne nécessitent aucune réalisation quantique \cite{pr}. L'identification d'un principe physique fondamental justifiant l'étrangeté de cette limite offrirait une avancée exceptionnelle au programme de reconstruction axiomatique de la mécanique quantique. Depuis les travaux de Tsirelson, Popescu et Rhorlich, de nombreux candidats ont été proposés, comme par exemple la complexité non-trivial des communications \cite{brassard}, la causalité informationnelle \cite{pawlowski}, ou encore le localité macroscopique \cite{navascues}. Cependant, la plupart des principes présentés sont formulés dans le cadre d'un scenario bipartite (i.e. avec deux joueurs distincts), et leurs généralisations multipartites restent indéfinies. Il a par ailleurs été prouvé qu'il existe des corrélations super-quantiques tripartites indétectables dans un scenario bipartite \cite{ghz2}, confortant ainsi l'idée que la réponse à la question  "pourquoi la mécanique quantique n'est-elle pas plus non-locale ?" reste également ouverte. Et si les corrélations unipartites de la contextualité et les corrélations multipartites de la non-localité étaient guidées par un principe commun ?

\section{L'approche graphique de CSW}
\label{csw}
En 2010, Ad\'an Cabello, Simone Severini et Andreas Winter (CSW) remarquent que l'inégalité de Klyachko (KCBS) appartient à une famille plus large d'inégalités, qui sont chacune associées à un graphe \cite{cabello9}. CSW utilisent alors la \textit{théorie des graphes} pour construire un cadre général permettant d'étudier les corrélations non-locales et contextuelles \cite{cabello10}.\footnote{Nous présentons ici succinctement une partie de leurs résultats, sans nous attarder sur la théorie des graphes, qui mériterait une introduction détaillée, mais trop minutieuse dans le cadre de ce mémoire. Pour une exposition rigoureuse, cf. par exemple \cite{cabello10} ou \cite{desilva} }\\

\subsection{Les inégalités CSW}

Dans l'approche CSW, un scenario expérimental est représenté par un graphe d'exclusivité. Les sommets d'un tel graphe représentent des évènements, i.e. les réponses du système à une question posée. Ceux-ci sont connectés par des segments qui correspondent à des relations d'exclusivité. Autrement dit, deux évènements reliés entre eux sont exclusifs, i.e. ils ne peuvent être tous les deux vrais. Dans le cas du scénario proposé par KCBS, les évènements sont $\{(0|P_i),(1|P_i)\}_i$ avec $ 1\leq i \leq 5$.\\

Les données opératoires issues des réalisations répétées du scenario expérimental sont tabulées par des probabilités associées à chaque sommet : $V=\{p(0|P_i),p(1|P_i)\}_i$. Les inégalités CSW correspondent à des bornes supérieures de combinaisons linéaires de ces probabilités : $\sum_i w_i p_i$ où $p_i$ est la probabilité d'observer le $i$-ème évènement et $w_i \geq 0$ est un coefficient associé. Cette combinaison linéaire est représentée par un graphe pondéré $(G,w)$. L'inégalité KCBS est ainsi donnée par la borne supérieure de  \[\sum_{i=1}^5 w_{(1|P_i)}p(1|P_i) + w_{(0|P_i)}p(0|P_i) \].

CSW définissent un \textit{modèle classique} pour des données opératoires $(G,p_i)$ par un espace $\Lambda$, un évènement $e_i \subset \Lambda$ pour chaque sommet tel que les évènements correspondant à des sommets adjacents sont mutuellement exclusifs et donc des sous-ensembles disjoints de $\Lambda$, et une distribution de probabilités $\mu$ sur $\Lambda$ telle que $p_i$ est la probabilité que $\mu$ soit affectée à $e_i$. \cite{desilva}\\

Une inégalité de non-contextualité CSW peut être obtenue en cherchant la valeur maximale d'une combinaison linéaire $\sum_i w_i p_i$ représentée par le graphe pondéré $(G,w)$, si les données $p_i$ appartiennent à un modèle classique. CSW ont démontré que ce maximum correspond précisément au \textit{nombre d'indépendance pondéré} $\alpha$ du graphe $(G,w)$. \\

Quelques définitions s'imposent :
Un ensemble de sommets indépendants d'un graphe $G$ est un sous-ensemble $I$ de $V$ tel qu'aucun couple de sommets dans $I$ n'est adjacent. Le nombre d'indépendance $\alpha$ d'un graphe $G$ correspond à la taille du plus grand ensemble indépendant dans $G$. Pour un graphe pondéré $(G,w)$, il s'agit de la valeur maximale de la somme $\sum_{i\in I} w_i$ où $I$ est un ensemble indépendant quelconque. \cite{desilva}  On remarque que pour un graphe simple, le nombre d'indépendance peut être facilement trouvé par une méthode de "coloriage" : on colore les sommets correspondant à un évènement vrai en respectant les relations d'exclusivité, et on compte le nombre maximal de sommets coloriés, qui correspond au nombre d'indépendance du graphe. \\

Les \textit{inégalités de non-contextualité CSW} sont donc de la forme 

\[ \sum_i w_i p_i \leq \alpha(G,w)\]

\begin{figure}[ht!]
\centering
\includegraphics[width=4cm]{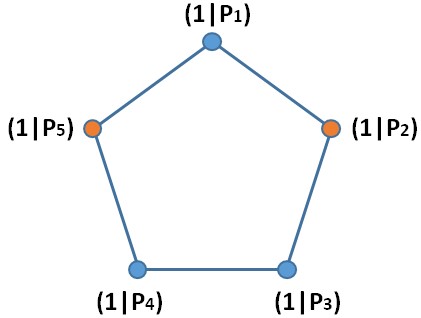}
\caption{Graphe du scenario KCBS}
\label{fpentagone3}
\end{figure}

Pour le scénario KCBS, on suppose que $w_{(1|P_i)} = 1$ et $w_{(0|P_i)}=0$. Le graphe pondéré correspondant est alors le pentagone représenté sur la figure \ref{fpentagone3}.\\

Le nombre d'indépendance du pentagone est $\alpha = 2$. On retrouve alors l'inégalité de non-contextualité KCBS : 

\[ \sum_{i=1}^5 <P_i> \leq 2  \]

Nous avons précédemment démontré que cette inégalité est violée par la mécanique quantique. L'approche graphique de CSW permet non seulement de retrouver cette violation, mais aussi d'obtenir la violation quantique maximale de n'importe quel scénario expérimental décrit dans ce cadre, qui correspond au \textit{nombre theta de Lovasz} $\theta(G,w)$.\\

Avant de présenter cette propriété graphique, il nous faut d'abord définir le concept de représentation orthonormée d'un graphe $G$. Celle-ci consiste à choisir un vecteur unitaire $\ket{\psi} \in \mathbb{R}^d$ et à assigner un vecteur unitaire $\ket{\phi_i} \in \mathbb{R}^d$ à chaque sommet $v_i \in V$ du graphe $G$, de telle sorte que $\bra{\phi_i}\ket{\phi_j}=0$ si deux les sommets $v_i$ et $v_j$ sont adjacents (le choix de la dimension $d$ pouvant être arbitraire). Le nombre theta de Lovasz $\theta(G,w)$ est définit comme étant la valeur maximal de la somme $\sum_{i\in V}w_i |\bra{\phi_i}\ket{\psi}|^2$ sur l'ensemble des représentations orthonormées possibles du graphe $G$.\cite{desilva} \\

CSW définit le concept de \textit{modèle quantique} à partir de cette représentation :
ce modèle est constitué d'un espace de Hilbert $H$, de projecteurs $P_i = \ket{\phi_i}\bra{\phi_i}$  associé à chaque sommet du graphe de sorte que des sommets adjacents correspondent à des projecteurs orthogonaux , et un état pur $\ket{\psi} \in H$ tel que les probabilités $p_i$ soient égales à $\bra{\psi}Pi\ket{\psi}$.
\\
Dans ce cas, la valeur maximum atteignable par $\sum_i w_i p_i$ sur l'ensemble des modèles quantiques est bornée par le nombre theta de Lovasz : 

\[ \sum_i w_i p_i \leq \theta(G,w)\]

Pour le scenario KCBS,
\[ \theta(G,w) = \sqrt{5} > 2 \]

Si le lien entre preuve de contextualité-KS et théorie des graphes n'est pas particulièrement surprenant (les preuves étant de nature géométrique), la généralisation graphique de la borne maximale quantique obtenue par CSW est remarquable. En effet, celle-ci ne se limite pas aux inégalités de non-contextualité, mais s'étend également aux scénarii de type Bell. Il est par exemple possible de retrouver la borne de Tsirelson du scénario CHSH, en étudiant simplement ce dernier par l'approche graphique CSW.\\

\subsection{L'inégalité CHSH par CSW}

Dans le scénario CHSH, deux joueurs, Alice et Bob possèdent chacun une boîte noire. Ils disposent d'entrées ($x$ pour Alice, et $y$ pour Bob, avec $x,y \in \{0,1\}$) qu'ils envoient dans leur boite noire respective, et observent les sorties ($a$ pour Alice, $b$ pour Bob, avec $a,b \in \{0,1\}$). Le jeu est gagné s'ils observent la relation $a\oplus b =xy$. Si les deux joueurs enregistrent leurs entrées et sorties pour chaque réalisation, ils peuvent établir alors une étude statistique et éventuellement obtenir la distribution de probabilités conditionnelles $p(ab|xy)$ (souvent nommées "corrélations"). Toute la physique de ce scénario opératoire est encapsulée dans la probabilité $p(ab|xy)$, qui permet de décrire les éventuelles corrélations observées entre les entrées et sorties d'Alice et Bob. Les deux joueurs sont par ailleurs spatialement séparés : chaque réalisation du scénario a lieu avant que toute information sur l'entrée et la sortie de l'autre partie ne puisse être communiquée. \\

\begin{figure}[ht!]
\centering
\includegraphics[width=5cm]{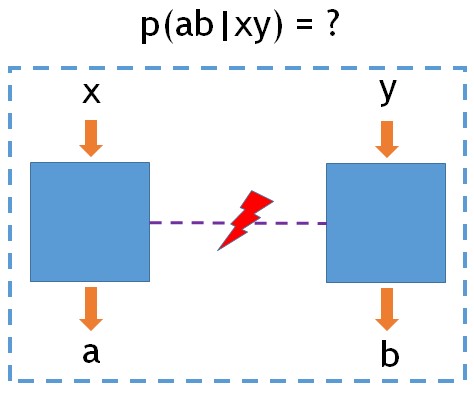}
\caption{Scenario Bipartite}
\end{figure}

Les évènements sont donc décrits par :
\[ \{(ab|xy) : a,b,x,y = 0,1\} \]

L'expression $(a,b,c|x,y,z)$ désigne l'évènement "les résultats $a,b,c$ sont obtenus lorsque les tests respectifs $x,y,z$ sont effectués". Deux évènements $(a,b,c|x,y,z)$ et $(a',b',c'|x',y',z')$ sont exclusifs s'ils ne peuvent pas être simultanément vrais; autrement dit si $x=x'$ et $a\neq a'$, ou si $y=y'$ et $b\neq b'$, ou bien $z=z'$ et $c \neq c'$.\\

Fritz, Sainz, Augusiak, Brask, Chaves, Leverrier et Ac\'in ont introduit le terme d'\textit{orthogonalité locale} pour décrire ce principe d'exclusivité dans le cas de scenarios multipartites \cite{fritz2} \cite{sainz2}. Deux évènements sont dit localement orthogonaux (ou exclusifs) s'ils mettent en jeu des résultats différents pour une même mesure locale par (au moins) un des parties. Par exemple, les évènements $(00|00)$ et $(10|01)$ sont localement orthogonaux, puisque les résultats obtenus par Alice sont différents alors que celle-ci effectue la même mesure. En revanche, les évènements $(00|00)$ et $(01|01)$ ne le sont pas.\\

\begin{figure}[ht!]
\centering
\includegraphics[width=5cm]{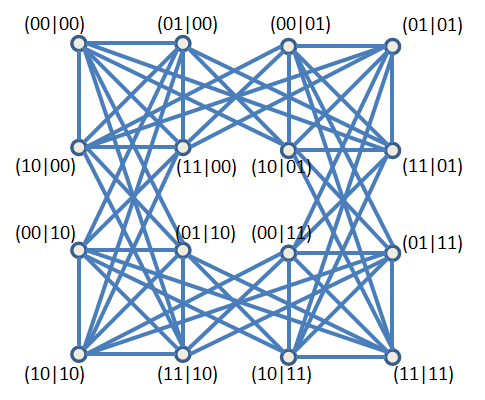}
\caption{Graphe d'orthogonalité d'un scenario de Bell (2,2,2). D'après : \cite{fritz2} }
\label{fgrapheor}
\end{figure}

Le graphe d'exclusivité (ou orthogonalité) d'un scenario de Bell (2,2,2)\footnote{(2,2,2) = 2 joueurs disposant chacun  de deux mesures (entrées) possibles, et pouvant obtenir deux résultats (sorties) possibles.} est représenté en figure \ref{fgrapheor}.

\begin{figure}[ht!]
\centering
\includegraphics[width=5cm]{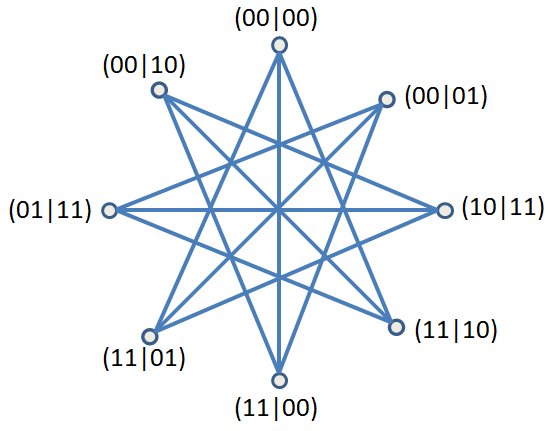}
\caption{Graphe d'orthogonalité du scenario CHSH. D'après : \cite{fritz2} }
\label{fgrapheor2}
\end{figure}

En équipant ce graphe avec des poids \[ w(ab|xy) = \delta_{a\oplus b=xy} \]
on obtient le graphe pondéré associé au scenario CHSH, cf. figure \ref{fgrapheor2}.

Par un simple coloriage, on constate que le nombre d'indépendance du graphe est \[ \alpha(G,w) = 3\]
Or, il s'agit précisément de la limite classique de l'inégalité CHSH :
\[\sum_{\substack{a\oplus b=xy\\a,b,x,y\in \{0,1\}}} p(ab|xy) \leq 3 \]
Le nombre theta de Lovasz correspondant est quant à lui donné par la théorie des graphes :
\[ \theta(G,w) = 2 + \sqrt{2}\]
Il est exactement égal à la borne de Tsirelson \cite{tsirelson}, qui désigne la limite maximale atteignable par des corrélations quantiques dans un jeu CHSH.\\

L'approche graphique de la contextualité-KS semble non seulement permettre d'obtenir des inégalités de non-contextualité-KS et leur violation maximale par la mécanique quantique, mais également des inégalités de Bell, qui ont une structure similaire. Il est ainsi particulièrement tentant de voir les scénarii de Bell comme des cas particuliers de scénarii de contextualité, et d'ériger la contextualité en tant qu'axiome (cf. titre de \cite{cabello9}). Cependant, comme le remarque CSW, pour l'ensemble des scenarii de Bell, le nombre theta de Lovasz ne correspond en général qu'à une borne supérieure, qui n'est pas nécessairement atteinte. C'est par exemple le cas dans un scenario $(2,3,2)$ (aussi désigné par $I_{3322}$ \cite{brunner}), mettant en jeu deux joueurs, disposant chacun de trois entrées possibles, et pouvant chacun enregistrer deux sorties possibles. Pour un tel scenario, on obtient ainsi la borne supérieure $\theta(G,w)=0.25147$, tandis que la borne de type-Tsirelson est de  $0.25087556$ \cite{pal}. Cet écart est notamment dû au fait que contrairement aux scenarii de Bell, l'approche CSW utilise des "probabilités sous-normalisées, qui semblent nécessaire à la dérivation des invariants graphiques". \cite{acin1} \\

\section{L'approche combinatoire d'AFLS}
\label{safls}
\textit{Cette section est basée sur la version 5 de l'article \cite{acin1}, ainsi que sur un cours dispensé par Ana Belén Sainz lors du Solstice of Foundations (Zurich). Rappelons que cette présentation est loin d'être exhaustive : de nombreux résultats et nuances intéressants (comme par exemple la chaine complète d'inclusions des divers modèles probabilistes, des modèles classiques aux modèles les plus généraux, et leurs invariants graphiques associés) n'ont pas été reproduits dans ce mémoire, et nous invitons le lecteur à consulter l'article original pour un aperçu plus complet. \\} 

Une autre approche de généralisation des corrélations quantiques, parue sur \textit{arxiv} en 2012 entre alors en jeu. Cette approche \textit{combinatoire}, développée par Ac\'in, Fritz, Leverrier et Sainz, acronyme AFLS, se base sur la théorie des hypergraphes, utilisée en logique quantique \cite{coecke}, et intègrerait l'approche CSW comme cas particulier. \`A la différence de celle-ci, AFLS conservent la normalisation des probabilités, et montrent que cette normalisation n'entrave pas l'obtention des relations entre invariants graphiques. De fait, plutôt que d'étudier des inégalités de contextualité, AFLS s'intéressent directement aux ensembles de modèles probabilistes (classiques, quantiques et généraux).\\

\subsection{Scenario de Contextualité et Hypergraphe}
Dans l'approche combinatoire, un modèle probabiliste correspond à l'affectation d'une probabilité à chaque résultat de mesure, de sorte que la somme des probabilités d'obtenir les résultats d'une mesure donnée soit égale à 1.\\

Un \textit{scenario de contextualité} tel qu'il est défini dans l'approche AFLS spécifie, pour une collection de mesures données, le nombre de sorties (résultats) de chaque mesure et quelles mesures partagent un (ou plusieurs) résultat(s) identique(s). Mathématiquement, un tel scenario est représenté par un hypergraphe $H$, une notion généralisée de graphe dans laquelle les arêtes relient un nombre quelconque de sommets. Les sommets $V(H)$ d'un hypergraphe de contextualité représentent des évènements, i.e. des résultats de mesures ; tandis que les arêtes $E(H) \subseteq 2^{V(H)}$ correspondent à des mesures complètes, i.e. l'ensemble des résultats possibles d'une mesure ; de sorte que $\bigcup_{e\in E(H)}e = V(H)$. Cette condition stipule simplement que chaque résultat de mesure devrait avoir lieu dans au moins une des mesures considérées. \\

Il est important de noter que la notion de contexte jusqu'alors considérée n'est pas utilisée par AFLS. En effet, dans le théorème KS et l'approche CSW, un contexte de mesure désignait un ensemble d'observables liées entre elles par des relations de compatibilité, i.e. le fait qu'elles puissent être mesurées simultanément. L'hypothèse de compatibilité est abandonnée par AFLS au profit d'une \textit{équivalence opératoire} : deux mesures sont équivalentes si elles partagent la même probabilité pour un résultat commun. La notion de contexte de AFLS est donc plus générale.\\

Prenons l'exemple de trois mesures projectives (quantiques), représentées par trois mesures projectives distinctes, représentées par trois Projective Valued Measures\footnote{Un PVM est un ensemble de projecteurs $\{\Pi_i\}$ tel que $\sum_i \Pi_i = \mathbb{1}$ et $\forall i$, $Pi_i^2 = \Pi_i$.}, $M_1$, $M_2$ et $M_3$, tels que :
\[M_1 = \{\Pi_1, \Pi_2, \Pi_3\} \]
\[M_2 = \{\Pi_3, \Pi_4, \Pi_5\} \]
\[M_3 = \{\Pi_1, \Pi_5, \Pi_6\} \]
avec $\sum_{i=1}^3 \Pi_i = \sum_{i=3}^5 \Pi_i = \sum_{i=1,5,6} \Pi_i = \mathbb{1} $ 
On associe à $M_1$ les sommets $\{v_1,v_2,v_3\}$, les sommets $\{v_3',v_4,v_5\}$ à $M_2$, et les sommets $\{v_1',v_5',v_6\}$ à $M_3$.\\

\begin{figure}[ht!]
\centering
\includegraphics[width=8cm]{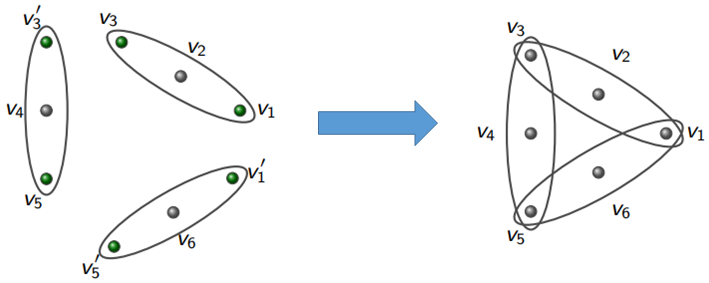}
\caption{Représentation graphique des trois mesures projectives et de l'hypergraphe de contextualité associé. Source : \cite{sainz3} }
\label{fhypgraph}
\end{figure}

La règle de Born nous donne la probabilité d'obtenir le résultat $v_i$ sachant que la mesure projective $M_k$ est effectuée, pour un système dans l'état $\rho$. On obtient notamment les probabilités
\[ p(v_3)=p(v_3')=tr(\Pi_3 \rho) \hspace{2cm} \forall \rho\]
\[ p(v_1)=p(v_1')=tr(\Pi_1 \rho) \hspace{2cm} \forall \rho\]
\[ p(v_5)=p(v_5')=tr(\Pi_5 \rho) \hspace{2cm} \forall \rho\]
De ce fait, il en résulte trois relations d'équivalences opératoires : entre $M_1$ et $M_2$ pour les résultats $v_3$ et $v_3'$ , entre $M_2$ et $M_3$ pour les résultats $v_5$ et $v_5'$, et entre $M_3$ et $M_1$ pour les résultats $v_1$ et $v_1'$. L'hypergraphe de contextualité correspondant est représenté sur la figure \ref{fhypgraph}.

Etant donné un hypergraphe $H(V,E)$, on peut lui associer un modèle probabiliste, i.e. des probabilités conditionnelles de résultats de mesures $p: V\rightarrow [0,1]$, qui respectent les relations d'équivalences opératoires, de sorte que $\forall e\in E$, $\sum_{v\in e}p(v)  = 1$. Pour le graphe ci-dessus, tout modèle probabiliste doit ainsi respecter les conditions
\[ p(v_1)+p(v_2)+p(v_3)=1\]
\[ p(v_3)+p(v_4)+p(v_5)=1\]
\[ p(v_5)+p(v_6)+p(v_1)=1\]

\begin {figure}[!ht]
\centering
\subfigure{\includegraphics [width=4cm]{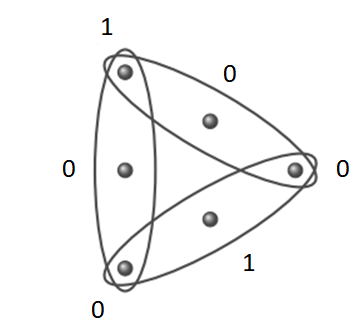}}
\quad
\subfigure{\includegraphics [width=4.5cm]{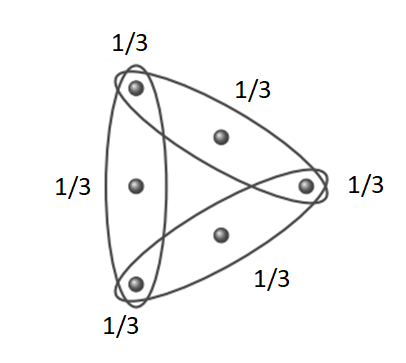}}
\quad
\subfigure{\includegraphics [width=4cm]{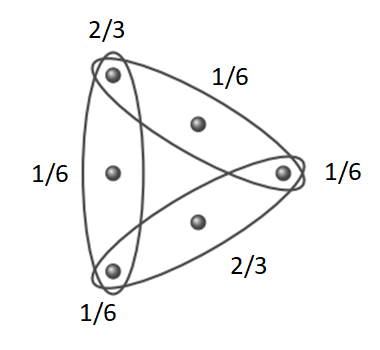}}
\caption {Exemples de modèles probabilistes}
\end {figure}

L'ensemble des modèles probabilistes est noté $\mathcal{G}(H)$.
On peut en distinguer différents types.\\ 

Les \textit{modèles probabilistes classiques} correspondent à des théories à variables cachées $\lambda$ déterministes et non-contextuelles. Ces variables cachées déterminent avec certitude le résultat de mesure qui sera effectivement obtenu. On observe uniquement une moyenne sur l'ensemble de ces variables, pour une préparation du système donné. 
Un modèle probabiliste $p:V\rightarrow [0,1]$ est déterministe si $\forall v\in V(H)$, $p(v) \in \{0,1\}$.
Un modèle probabiliste est classique ($p \in \mathcal{C}(H)$) si et seulement s'il est une combinaison convexe de modèles déterministes, i.e. il existe des poids $q_\lambda \in [0,1]$ indexés par un paramètre $\lambda$ tel que $\sum_\lambda q_\lambda =1$, et des modèles déterministes $p_\lambda \in \{0,1\}$ pour chaque $\lambda$, tels que
\[ p(v) = \sum_\lambda q_\lambda p_\lambda (v) \hspace{2cm \forall v \in V(H)}\]

Les \textit{modèles probabilistes quantiques} sont des modèles probabilistes compatibles avec les lois de la théorie quantique. 
Soit $H$ un scenario de contextualité. Une affectation de probabilités $p: V(H) \rightarrow [0,1]$ est un modèle quantique ($p \in \mathcal{Q}(H)$) s'il existe un espace de Hilbert $\mathcal{H}$, un état quantique $\rho$ et un opérateur de projection associé à chaque sommet de l'hypergraphe $\{P_v : v \in V\} $, représentant des mesures projectives
\[ \sum_{v\in e} P_v = \mathbb{1}_\mathcal{H} \hspace{2cm} \forall e\in E(H)\]
et si l'assignation respecte la loi de Born
\[p(v) = tr(\rho P_v) \hspace{2cm} \forall v \in V(H) \]

Il est intéressant de noté que, dans une telle approche, tout modèle classique est un modèle quantique, i.e. $\mathcal{C}(H)\subseteq\mathcal{Q}(H)$. (cf. \cite{acin1} pour démonstration). Par ailleurs, l'approche AFLS permet de reformuler le théorème KS de façon simple et naturelle :

\boitesimple{\begin{center}\textit{Il existe un scénario de contextualité $H_{KS}$ tel que} \[\mathcal{C}(H_{KS})=\emptyset \hspace{3cm} \mathcal{Q}(H_{KS})\neq\emptyset\] \end{center}}

\begin{figure}[ht!]
\centering
\includegraphics[width=6cm]{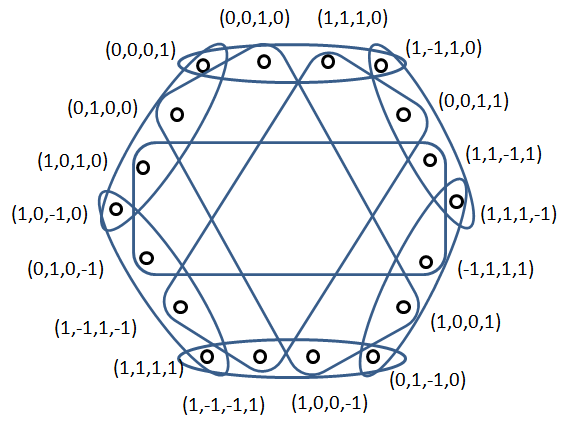}
\caption{Hypergraphe de contextualité de la preuve de Cabello-Esterbaranz-Garc\'{\i}a Alcaine. Source :  \cite{kunjwal3} }
\label{fcab}
\end{figure}

Tout hypergraphe de contextualité pour lequel il n'existe aucun modèle classique mais au moins un modèle quantique constitue une preuve de contextualité-KS. La non-classicité obtenue a la particularité d'être indépendante de l'état du système. Par exemple, Cabello, Estebaranz et Garc\'{\i}a Alcaine ont proposé une preuve du théorème KS indépendante de l'état du système en utilisant uniquement 18 vecteurs unitaires, dans un espace de dimension 4 (alors que la preuve de Mermin-Peres nécessitait quant à elle 24 vecteurs) \cite{cabello12} \cite{cabello14}. Ces 18 vecteurs sont représentés par des sommets sur l'hypergraphe de la figure \ref{fcab}. Chaque arête de l'hypergraphe correspond donc à une mesure à 4 résultats possibles. On constate que chaque sommet apparait dans deux arêtes différentes. Autrement dit, certaines mesures partagent un même résultat. Tout modèle respectant l'hypothèse de \textit{non-contextualité de mesure} doit affecter à un tel résultat partagé une probabilité indépendante de la mesure particulière pour laquelle le résultat est effectivement obtenu.

Or, on constate, par un simple coloriage, qu'il est impossible d'affecter des probabilités déterministes $p(v)\in\{0,1\}$ aux sommets de cet hypergraphe de telle sorte que chaque arête contienne un sommet auquel est assigner la valeur 1 (afin de respecter la règle de normalisation, cf. définition du scenario de contextualité). Il en résulte qu'aucun modèle classique ne peut lui être associé, ce qui constitue une preuve du théorème KS en dimension 4.

\subsection{Scenario de Bell}
L'approche AFSL ne se limite pas uniquement au théorème KS, et permet également d'étudier le résultat du théorème de Bell. Intuitivement, un "scenario de Bell" peut être vu comme le "produit" de plusieurs scénarios de contextualité, chacun étant attribué à un des parties. La question de la nature de la combinaison de deux scenarii de contextualité en un scenario joint n'est pas triviale. Une première solution a été proposée par Foulis et Randall en 1981 \cite{foulis}, dans le cadre de l'étude de la logique quantique.\footnote{En logique quantique, ce que AFLS appellent scénario de contextualité est connu sous le nom d'\textit{espace de test}. \cite{wilce}}\\

Le \textit{produit de Foulis-Randall} de deux scenarii représentés par les hypergraphes $H_A$ et $H_B$ est le scenario $H_A \otimes H_B$ tel que 
\[ V(H_A \otimes H_B) = v(H_A)\times V(H_B) \hspace{2cm} E(H_A \otimes H_B) = E_{A\rightarrow B}\cup E_{A\leftarrow B}\]
où $E_{A\rightarrow B}$ (resp. $E_{A\leftarrow B}$) correspond à un ensemble d'arêtes décrivant les mesures jointes pour lesquelles Alice (resp. Bob) effectue en premier une mesure et communique le résultat à Bob (resp. Alice), qui choisit sa mesure en fonction de ce résultat. De cette façon, les arêtes d'un hypergraphe $H_A\otimes H_B$ sont des éléments de $E_{A\rightarrow B}$, ou de $E_{A\leftarrow B}$, ou de leur intersection. \\

\begin{figure}[ht!]
\centering
\includegraphics[width=4cm]{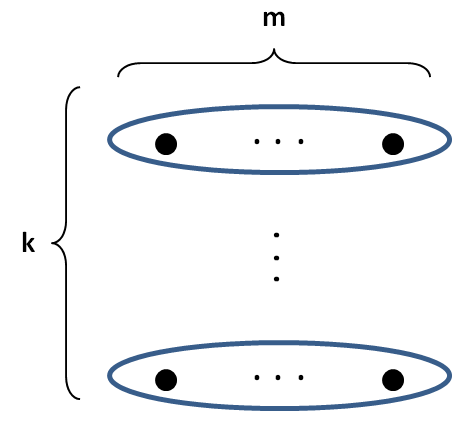}
\caption{Le scenario de contextualité $B_{1,k,m}$, un scénario de Bell à un partie. Source :  \cite{acin1} }
\label{fhy}
\end{figure}

Un \textit{scenario de Bell} $B_{n,k,m}$ est décrit par $n$ parties ayant chacun  accès à $k$ mesures locales, chacune pouvant donner $m$ résultats possibles. Dans le cas d'un scenario à une partie, on obtient un scénario de contextualité $B_{1,k,m}$ décrit par l'hypergraphe de la figure \ref{fhy}.

AFLS définissent alors un scénario de Bell par
\[B_{n,k,m} := B_{1,k,m}\otimes ... \otimes B_{1,k,m} \]

où $\otimes$ désigne le \textit{produit de Foulis-Randall}. \\

\begin{figure}[ht!]
\centering
\includegraphics[width=6cm]{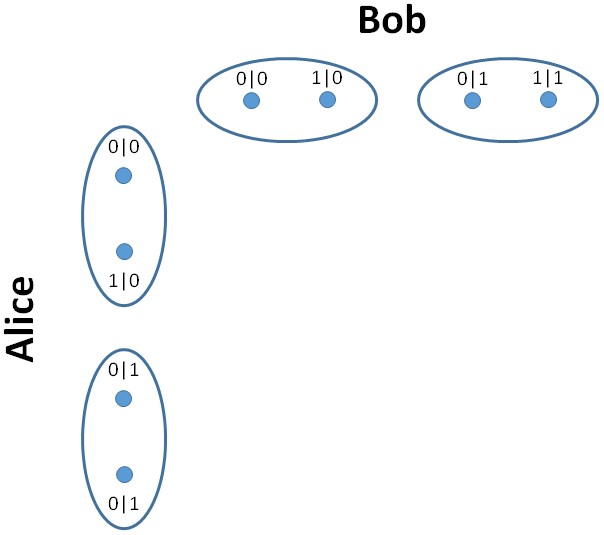}
\caption{Représentation des scenarii de contextualité $B_{1,2,2}$ d'Alice et Bob, d'après \cite{acin1} }
\label{falicebob}
\end{figure}

Prenons l'exemple du scenario CHSH $B_{2,2,2}=B_{1,2,2}\otimes B_{1,2,2}$.
On attribue un scenario de contextualité $B_{1,2,2}$ à chacun des parties, Alice et Bob. Ce scenario est constitué de deux arrêtes, qui correspondent aux deux choix de mesures d'un partie, chacune contenant deux sommets, les deux résultats possibles (0 ou 1) d'une mesure, cf. figure \ref{falicebob}.\\

\begin{figure}[ht!]
\centering
\includegraphics[width=4cm]{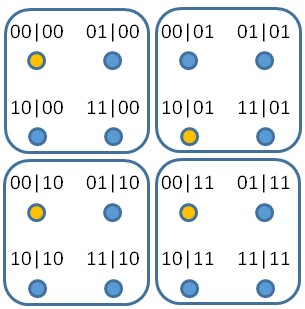}
\caption{Hypergraphe des mesures simultanées d'Alice et Bob $B_{1,2,2}\times B_{1,2,2}$, d'après \cite{acin1} }
\label{messimu}
\end{figure}

Une composition "naïve" des scenarii d'Alice et de Bob, où le scenario de contextualité composite est issu d'un simple produit $H_{A\times B} = H_A \times H_B$, est l'hypergraphe des mesure simultanées, qui répertorie 16 résultats de mesures différents, i.e. 4 résultats pour chacune des quatre mesures simultanées possibles. Dans un hypergraphe composé, les sommets représentent des évènements $(ab|xy)$ avec $x,y,a,b \in \{0,1\}$ (x et y correspondent aux mesures respectives d'Alice et Bob, a et b aux résultats respectifs de ces mesures), et les arêtes dessinent les mesures corrélées, cf. figure \ref{messimu}.\\

Néanmoins, un tel hypergraphe est insuffisant pour rendre compte du résultat de Bell. En effet, supposons qu'on lui attribue le modèle probabiliste suivant : 
\[p : (00|00)\rightarrow 1, \hspace{2cm} (10|01)\rightarrow 1, \hspace{2cm} (00|10)\rightarrow 1, \hspace{2cm} (00|11)\rightarrow 1 \]

Il en résulte que \[ p_A(0|00) = 1, \hspace{2cm} p_A(0|01)=0\]
Chaque fois qu'Alice obtient le résultat $0$ (resp. $1$) lorsqu'elle choisit la mesure "$0$", elle sait avec certitude que Bob a effectué la mesure "0" (resp. "1").
Autrement dit, une telle affectation, et par un extension, ce choix d'hypergraphe, autorise Alice et Bob à communiquer entre eux, ce qui est contraire au principe de non-communication du théorème de Bell.\\

\boitesimple{Le principe de non-communication stipule que les probabilités associées aux résultats de Bob ne devraient pas dépendre des choix de mesure d'Alice et vice-versa. 
Un modèle probabiliste respecte le principe de non-communication  si pour tout partie $i = 1,...,n$, toutes les arêtes $e,e' \in E(H_i)$ et tous les sommets $v_j \in V(H_j)$ pour $j\neq i$, 
\[\sum_{v_i \in e} p(v_1,...,v_n) = \sum_{v_i \in e'} p(v_1,...,v_n) \]}

Pour palier à ce problème, AFLS propose donc de considérer une autre forme de composition : le produit de Foulis-Randall. On introduit alors deux nouveaux hypergraphes : l'hypergraphe $H_{A\rightarrow B}$, qui représente le scenario où les choix de mesure de Bob dépendent des résultats d'Alice ; et l'hypergraphe $H_{B\rightarrow A}$, correspondant au scenario où les choix de mesure d'Alice dépendent des résultats de Bob. \\
L'ensemble $E_{A\rightarrow B}$ (resp. $E_{B\rightarrow A}$)  contient les arrêtes des hypergraphes $H_{A\times B}$ et $H_{A\rightarrow B}$ (resp. $H_{B\rightarrow A}$).\\

\begin{figure}[ht!]
\centering
\includegraphics[width=8cm]{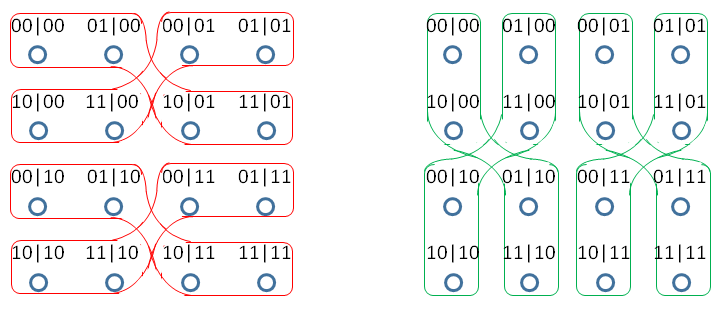}
\caption{Hypergraphe $H_{A\times B}$ (à gauche) et $H_{B\times A}$ (à droite), d'après \cite{acin1} }
\end{figure}

L'hypergraphe résultant (figure), issu du produit de Foulis-Randall des hypergraphes élémentaires d'Alice et Bob, est le scenario CHSH $B_{2,2,2}$.\\

\begin{figure}[ht!]
\centering
\includegraphics[width=4cm]{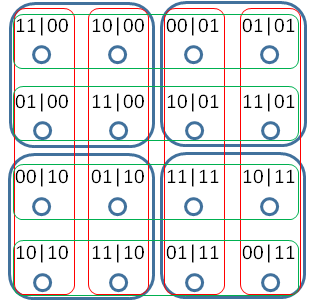}
\caption{Hypergraphe réarrangé du scenario de Bell $B_{2,2,2}=B_{1,2,2}\otimes B_{1,2,2}$, d'après \cite{acin1} }
\end{figure}
Par construction, tout modèle probabiliste  associé respecte nécessairement le principe de non-communication : $\mathcal{G}(B_{n,k,m}) = \mathcal{NS}(k,n,m)$, et coïncident avec ce qui est désigner dans la littérature par "boîte non-communicante" \cite{popescu}. De même, tout modèle classique associé à un scenario de Bell correspond à une boîte non-communicante \textit{locale}. Autrement dit, les résultats obtenus par chaque partie sont indépendant du résultat et du choix de mesure de l'autre partie : pour une variable cachée $\lambda$ donnée, $p(a,b|x,y,\lambda)=p(a|x,\lambda)p(b|y,\lambda)$. On parle de variable cachée locale, ou encore de l'hypothèse de \textit{causalité locale}. \\

Il est possible de démontrer (cf. article original \cite{acin1}) que tout modèle classique peut être représenté par une telle boîte, i.e. $\mathcal{C}(\mathcal{B}_{n,k,m})=\mathcal{C}(n,k,m)$. On peut par exemple essayer d'associer un modèle classique au scenario CHSH, en coloriant des sommets de telle sorte qu'aucun des sommets coloriés n'appartienne à une même arête (afin de toujours respecter la normalisation des probabilités). Par exemple, on associe la probabilité $1$ aux sommets $(11|00), (11|01), (01|10), (01|11)$, et la probabilité $0$ à tous les autres. Le modèle classique obtenu correspond à deux boîtes classiques non-communicantes. D'autre part, on en déduit :
\[p_A(0|1) = 1 \hspace{2cm} p_A(1|0) = 1 \]
\[p_B(1|1) = 1 \hspace{2cm} p_A(1|0) = 1 \]
Il s'agit donc bien d'un modèle local.\\

L'ensemble des modèles quantiques peuvent être associés à un scenario de Bell\\ $\mathcal{Q}(\mathcal{B}_{n,k,m})=\mathcal{Q}(n,k,m)$, et dans chacun de ces modèles, viole l'hypothèse de causalité locale. Autrement dit, tout modèle quantique respectant le principe de non-communication correspond à une théorie à variables cachées non-locales. L'approche AFLS reproduit bien le théorème de Bell.\\

Les corrélations quantiques non-locales mises en évidence par Bell dans son célèbre article de 1964 sont un "cas particulier de contextualité", au sens où un scenario de Bell peut être obtenu à partir du produit de Foulis-Randall de scenarii, locaux, de contextualité.\\

\section{Le principe de Specker}

AFLS ont donc donné une réponse à la question de la nature de la relation entre non-localité et contextualité. Il est tentant d'en déduire que la contextualité est un résultat plus fondamental, puisqu'elle semble se cacher derrière les corrélations quantiques, indépendemment du nombre de parties ou de la distance qui les sépare. Néanmoins, la contextualité ne permettrait pas de singulariser la théorie quantique. De même qu'il existe des scenarii non-locaux n'admettant pas pas de modèles quantiques, il existe des scenarri contextuelles qui ne reproduisent pas les prédictions quantiques. C'est le cas de la parabole du voyant surprotecteur, inventée par Specker dans son article “La logique des propositions qui ne sont pas décidables simultanément” (1960), un jeu à trois boîtes notamment étudié par Liang, Spekkens et Wiseman en 2010 dans un article intitulé “Specker’s parable of the overprotective seer: A road to contextuality, nonlocality and complementarity” \cite{liang}. Cette parabole (cf. chapitre 1) exhibe une forme de contextualité particulière, puisqu'elle ne peut être expliquée par aucun modèle classique ni quantique. \\

\begin{figure}[ht!]
\centering
\includegraphics[width=4cm]{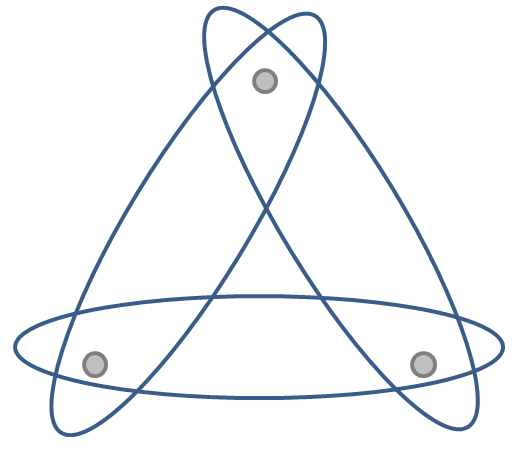}
\caption{Scenario du voyant surprotecteur }
\end{figure}

En effet, pour des mesures projectives en théorie quantique,  deux opérateurs peuvent être mesurés conjointement si seulement s'ils sont diagonalisables dans une base commune. Pour trois opérateurs, si chaque paire d'opérateurs est conjointement mesurable, alors les trois sont conjointement mesurables.\\

Cette propriété était considérée comme "\textit{fondamentale}" par Ernst Specker. \\

Lors d'une conversation privée, filmée par Ad\'an Cabello en juin 2009, il déclare que 
le "théorème fondamental de la mécanique quantique" pourrait être que "si vous disposez de quelques questions et que vous pouvez répondre à n'importe quelle paire d'entre elles [i.r. si les propositions (ou évènements) sont décidables deux à deux], alors vous pouvez aussi répondre à toutes les questions [i.e. les propositions correspondantes sont simultanément (ou conjointement) décidables]". \cite{specker2}\\

Dans son article "Specker's fundamental principle of quantum mechanics" \cite{cabello8}, Cabello relate également une discussion qu'il eut avec Simon Kochen en novembre 2012 :
\begin{quotation}
"\textit{Ernst et moi avont passé de nombreuses heures à discuter du principe (...). La difficulté consiste à essayer de le justifier sur des bases physiques générales, sans déjà supposer le formalisme de l'espace de Hilbert de la mécanique quantique. Nous avons décidé d'incorporer le principe comme axiome dans notre définition des algèbres booléennes partielles. Il apparaît aux pp. 65-66 comme suit: Une algèbre booléenne partielle $C$ est une union d'une famille $F$ d'algèbres booléennes qui est (i) fermée sous l'intersection par paires d'algèbres booléennes, et que (ii) si toute paire d'un ensemble fini $S$ d'éléments de $C$ se trouvent dans une algèbre booléenne commune dans $F$ alors tous les éléments de $S$ se trouvent dans une algèbre booléenne commune dans $F$. (...) Je n'ai jamais trouvé de justification physique générale pour (ii)}

\end{quotation}

Le principe de Specker pourrait être un candidat sérieux dans la quête des axiomes fondamentaux de la mécanique quantique. Issu de l'essence même du formalisme, il serait ainsi à l'origine de l'étrangeté de la non-localité et de la contextualité.\\

CSW ont dérivé des conséquences particulièrement intéressantes au sein de leur formalisme en supposant ce principe \cite{cabello10}. Cette "\textit{exclusivité globale}", comme ils la nomment, fournit notamment une borne supérieure (le nombre theta de Lovasz) aux corrélations quantiques, et permet même parfois de les singulariser (c'est le cas pour l'inégalité KCBS) \cite{cabello15}.\\

AFLS ont aussi étudié le principe de Specker au sein de leur formalisme \cite{acin1}, sous le nom d'\textit{exclusivité cohérente}\footnote{Pour un débat sur l'appelation de ce principe, voir notamment (\cite{henson} et \cite{cabello8})}. \\

Un modèle probabiliste $p \in \mathcal{G}(H)$ satisfait le principe d'\textbf{exclusivité cohérente} si et seulement si \[\sum_{v \in I} p(v) \leq 1 \] pour tout ensemble indépendant $I \subseteq V(NO(H))$. $NO(H)$ désigne le \textit{graphe de non-orthogonalité} d'un hypergraphe $H$, i.e. qui relie tous les sommets qui n'ont pas d'arêtes communes.

\begin{figure}[ht!]
\centering
\includegraphics[width=8cm]{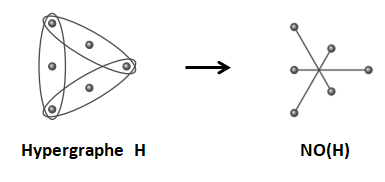}
\caption{Exemple d'hypergraphe et de son graphe de non-orthogonalité}
\end{figure}

La probabilité totale de tout ensemble de résultats exclusifs deux à deux est donc inférieur ou égal à 1.\footnote{Une autre formulation stipule que $p \in \mathcal{G}(H)$ satisfait le principe si et seulement si le nombre d'indépendance du graphe de non-orthogonalité $\alpha(No(H),p)$ est égal à 1.}\\ 

Au sein de leur approche, AFLS remarquent également que l'orthogonalité locale (\cite{sainz} \cite{fritz2}) n'est qu'un cas particulier "d'exclusivité cohérente", appliqué à un scenario de Bell. Ce principe, en expliquant la nature de la magie des corrélations quantiques, semblerait donc être une des clés du mystère de la mécanique quantique. Néanmoins, AFLS ont démontré qu'il ne caractériserait finalement pas les modèles quantiques, puisqu'il existe des modèles non quantiques qui le respectent également, prouvant ainsi qu'il ne constituerait pas une caractéristique exclusivement quantique. Bien qu'il semble nécessaire, le principe de Specker ne serait pas suffisant pour distinguer la théorie quantique des autres modèles probabilistes, d'après ce résultat.  (cf. article original pour démonstration \cite{acin1})\\

L'approche AFLS inclue non seulement les résultats de l'approche graphique de CSW, mais également ceux de l'approche faisceau, créée par Abramsky et Brandenburger \cite{abramsky1}. Cette dernière, formulée dans le langage des catégories, nous semble particulièrement difficile à aborder dans le cadre de ce mémoire. Néanmoins, si l'on fait fit des difficultés formelles, on obtient - malgré le risque de simplification et de perte de résultats importants  - non seulement un outil théorique particulièrement fécond, mais également une image permettant de représenter l'étrangeté du résultat KS.
\newpage
\section{L'approche faisceau d'Abramsky-Brandenburger}
\label{sabram}
Dans un article intitulé \textit{The Sheaf-Theoretic Structure Of Non-Locality and Contextuality} \cite{abramsky1}, publié en 2011, Samson Abramsky et Adam Brandenburger, utilisent le cadre des catégories pour donner une structure mathématique générale au concept de contextualité. Selon eux, 
\begin{quotation}
\textit{La contextualité émerge lorsque une famille de données est localement cohérente, mais globalement incohérente.}
\end{quotation}  La théorie des faisceaux, developpée par Jean Leray, Henri Cartan et Jean-Pierre Serre \cite{houzel}, permet d'étudier le passage de données locales à des données globales. 
Topologiquement, cette approche distingue l'espace des contextes, i.e. l'ensemble des variables comesurables, et l'espace des données (resultats des observations). Ces données sont cohérentes localement, mais ne le sont pas globalement. Pour certaines famille de "sections locales", i.e. pour diverses affectations de valeurs possibles (0 ou 1) pour un contexte donné, il n'existe pas de "section globale", i.e. de collection de sections locales compatibles, définie sur l'ensemble des variables qui permette de réconcilier toutes les données locales. La contextualité est, selon Abramsky et Brandenburger, à l'origine d'une telle \textit{obstruction}.\\

L'approche faisceau utilise une notion logique plus générale de la contextualité, au sens où celle-ci englobe bien la contextualité "standard" du théorème KS (preuve de Cabello, carré de Mermin-Peres), mais aussi celle de la parabole de Specker, ainsi que la non-localité du théorème de Bell, du modèle GHZ, du paradoxe de Hardy ou encore des boîtes PR. \\

\subsection{Hiérarchie logique de contextualité}

Abramsky \textit{et al.} (\cite{abramsky1} \cite{abramsky2} \cite{abramsky3}) distinguent trois types de contextualité : la contextualité "probabiliste" [appelation personnelle], la contextualité logique, et la contextualité forte.\\

Afin d'étudier cette hierarchie à travers différents exemples de scénarii, on introduit le concept de table de probabilités (et de possibilité) suivant :\\

\boitesimple{Pour un jeu d'observables $\mathcal{X} = \{A,B\}$, chaque observable pouvant donner un résultat parmis $\mathcal{O} = \{0,1\}$, on obtient par exemple la ligne suivante :\\
\begin{center}
\begin{tabular}{c|cccc}
  &  00 & 01 & 10 & 11 \\
  \hline
  A B & 1/2 & 0 & 0 & 1/2 \\
\end{tabular}
\end{center}
\vspace{7mm}
Cette ligne signifie que dans le contexte de mesure $\mathcal{C}=\{A,B\}$, les résultats de mesures $AB \rightarrow 00$ et $AB \rightarrow 11$ ont une probabilité d'1/2 d'advenir, tandis que l'obtention des résultats $AB \rightarrow 01$ et $AB \rightarrow 10$ ont une probabilité nulle. Une \textit{table de probabilité} répertorie les distributions de probabilités associées aux résultats d'une mesure $\mathcal{M} = \{\mathcal{C}_i\} $. Une \textit{table de possibilité} est construite à partir d'une table de probabilité, en associant la valeur "1" (pour "possible") à toutes les distributions de probabilités strictement positives, et la valeur "0" (pour "impossible") à toutes les distributions de probabilités nulles.  \\}

La contextualité "la plus faible" dans l'approche faisceau, la \textit{contextualité "probabiliste"}, correspond a des scenarios qui n'exhibent pas de propriétés particulières au niveau "possibiliste". C'est le cas notamment de la non-localité des corrélations du scenario CHSH, dont le tableau de probabilité associé est le suivant :\\

\begin{center}
\begin{tabular}{c|cccc}
  &  00 & 01 & 10 & 11 \\
  \hline
  A B & 1/2 & 0 & 0 & 1/2 \\
  A B' & 3/8 & 1/8 & 1/8 & 3/8 \\
  A' B & 3/8 & 1/8 & 1/8 & 3/8 \\
  A' B' & 1/8 & 3/8 & 3/8 & 1/8 \\
\end{tabular}
\end{center}

La version "possibiliste" du scenario CHSH est donnée par 

\begin{center}
\begin{tabular}{c|cccc}
  &  00 & 01 & 10 & 11 \\
  \hline
  A B & 1 & 0 & 0 & 1 \\
  A B' & 1 & 1 & 1 & 1 \\
  A' B & 1 & 1 & 1 & 1 \\
  A' B' & 1 & 1 & 1 & 1 \\
\end{tabular}
\end{center}

Afin de rendre la structure topologique du scenario apparente, on construit un "faisceau de contextualité" ("\textit{contextuality bundle}"). On associe pour cela un sommet à chaque observable, et l'on relie entre eux les sommets qui représentent des observables compatibles. \`A partir de chacun d'entre eux, on trace une "fibre" des valeurs qui peuvent lui être affecter (0 ou 1). Lorsqu'un résultat joint est possible (entrée 1 dans le tableau de possibilité), on connecte les deux valeurs correpondantes des fibres adjacentes. Par exemple, pour le résultat possible $AB \rightarrow 00$, on relie le point "0" de la fibre de $A$ ou point "0" de la fibre de $B$. \\

\boitesimple{Une section globale correspond à un chemin fermé traversant toutes les fibres exactement une fois, i.e. un chemin univoque.}

\begin{figure}[ht!]
\centering
\includegraphics[width=4cm]{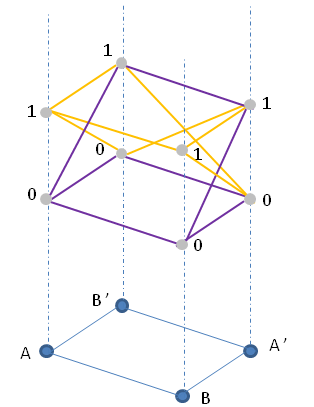}
\caption{Faisceau de contextualité du scenario CHSH}
\label{fcchsh}
\end{figure}

Le faisceau de contextualité du scenario CHSH (figure \ref{fcchsh}) admet des chemins univoques dans le cas "possibiliste" et n'admet donc de contextaulité qu'au niveau probabiliste.\\

Le niveau de contextualité supérieur est appelé \textit{contextualité logique}, et correspond au cas où des assignations globales cohérentes sont possibles, mais où il existe au moins une assignation locale qui ne peut être étendue globalement. Le paradoxe de Hardy - une expérience de pensée créée par Lucian Hardy en 1992-93 (\cite{hardy3} \cite{hardy4}) qui permet de décrire une forme de non-localité quantique sans inégalité (cf. annexe pour une introduction vulgarisée)- est un exemple de scenario exhibant une forme de contextualité logique. \\

\begin{figure}[ht!]
\centering
\includegraphics[width=4cm]{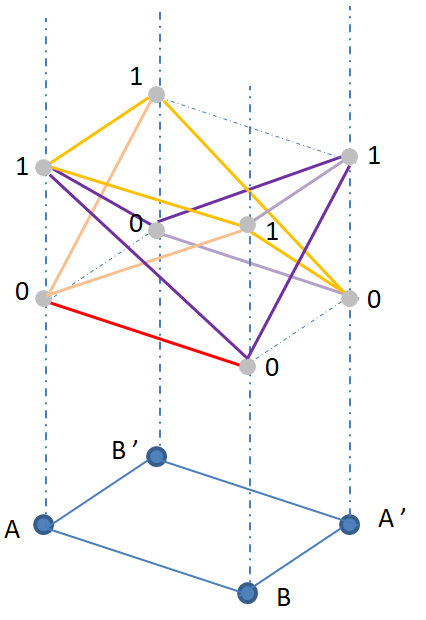}
\caption{Faisceau de contextualité du paradoxe de Hardy}
\label{fcph}
\end{figure}

Le tableau de possibilité et le faisceau de contextualité  correspondant à ce scenario sont représentés sur la figure \ref{fcph}.

\begin{center}
\begin{tabular}{c|cccc}
  &  00 & 01 & 10 & 11 \\
  \hline
  A B & 1 & 1 & 1 & 1 \\
  A B' & 0 & 1 & 1 & 1 \\
  A' B & 0 & 1 & 1 & 1 \\
  A' B' & 1 & 1 & 1 & 0 \\
\end{tabular}
\end{center}

On constate qu'il existe bien des assignations locales globalement cohérentes :\\ $\{A\rightarrow 1 , A'\rightarrow 1, B\rightarrow 0, B'\rightarrow 0 \}$ ;\\ $\{A\rightarrow 0 , A'\rightarrow 0, B\rightarrow 1, B'\rightarrow 1 \}$\\ ou encore $\{A\rightarrow 1 , A'\rightarrow 0, B\rightarrow 1, B'\rightarrow 0 \}$.\\
Cependant, toute assignation locale telle que $\{A\rightarrow 0 , B\rightarrow 0 \}$ ne peut être étendue globalement. Il est imposible de tracer un chemin fermé univoque incluant la connexion colorée en rouge sur le faisceau de contextualité.\\

Le dernier niveau de contextualité est appelé \textit{contextualité forte}, et correspond aux scenarios pour lesquels aucune assignation globale cohérente n'est possible.\\

\begin{figure}[ht!]
\centering
\includegraphics[width=3cm]{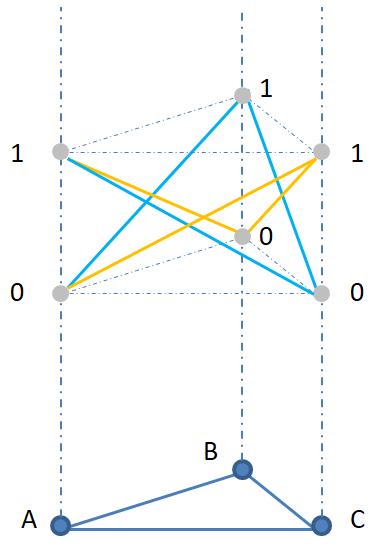}
\caption{Faisceau de contextualité de la parabole du voyant surprotecteur}
\label{fcpvs}
\end{figure}

\begin{figure}[ht!]
\centering
\includegraphics[width=3cm]{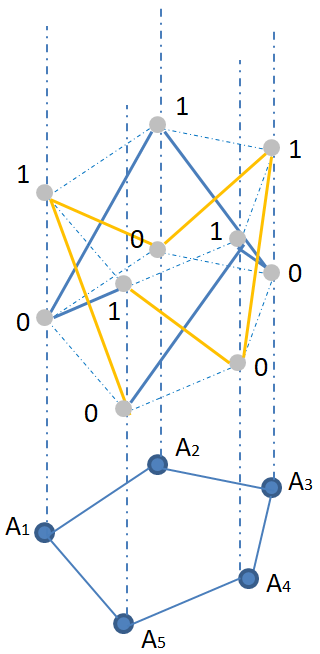}
\caption{Faisceau de contextualité du scénario KCBS}
\label{fcsk}
\end{figure}

\begin{figure}[ht!]
\centering
\includegraphics[width=3cm]{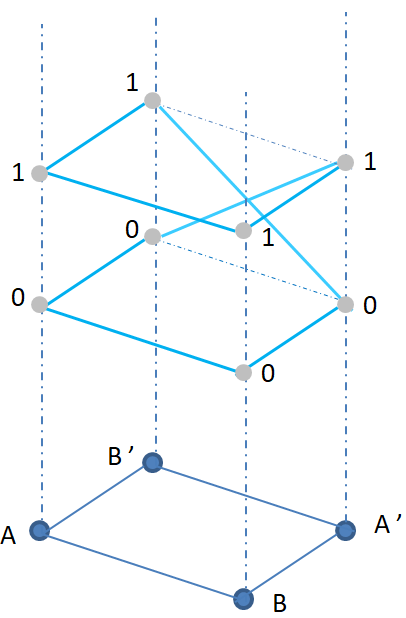}
\caption{Faisceau de contextualité de la boite PR}
\label{fcbpr}
\end{figure}

C'est notamment le cas du carré de Mermin-Peres (cf. section \ref{merminperescab}), du scenario KCBS (cf. tableau et figure \ref{fcsk}), de la parabole du voyant surprotecteur (cf. tableau et figure \ref{fcpvs}), et des boîtes PR (cf. tableau et figure \ref{fcbpr}).\\

\textbf{Parabole du voyant surprotecteur} :
\begin{center}
\begin{tabular}{c|cccc}
  &  00 & 01 & 10 & 11 \\
  \hline
  A B & 0 & 1 & 1 & 0 \\
  B C & 0 & 1 & 1 & 0 \\
  C A & 0 & 1 & 1 & 0 \\
\end{tabular}
\end{center}

\hspace{1cm}

\textbf{Scenario KCBS :}
\begin{center}
\begin{tabular}{c|ccccc}
  &  00 & 01 & 10 & 11 \\
  \hline
  $A_1 A_2$ & 0 & 1 & 1 & 0 \\
  $A_2 A_3$ & 0 & 1 & 1 & 0 \\
  $A_3 A_4$ & 0 & 1 & 1 & 0 \\
  $A_4 A_5$ & 0 & 1 & 1 & 0 \\
  $A_5 A_1$ & 0 & 1 & 1 & 0 \\
\end{tabular}
\end{center}

\textbf{Boites PR }:
\begin{center}
\begin{tabular}{c|cccc}
  &  00 & 01 & 10 & 11 \\
  \hline
  A B & 1 & 0 & 0 & 1 \\
  A B' & 1 & 0 & 0 & 1 \\
  A' B & 1 & 0 & 0 & 1 \\
  A' B' & 0 & 1 & 1 & 0 \\
\end{tabular}
\end{center}

\subsection{Contextualité et paradoxes d'auto-référence}
Un résultat particulièrement intéressant de l'approche faisceau est la connection directe qu'elle établit entre la structure topologique de la contextualité et celle des paradoxes sémantiques classiques, que Abramsky \textit{et al.} \cite{abramsky3} appellent "cycles menteurs", i.e. des \textit{paradoxes d'auto-référence}. 
Un exemple de tel "paradoxe sémantique" est le \textit{paradoxe du menteur}, souvent associé à la figure d'Epiménide de Knossos (cité crétoise), poète et chaman hyperboréen du VIIe siècle avant JC, auquel on attribue l'énoncé 
\begin{center}
Tous les Crétois sont des menteurs.
\end{center}
Un autre  paradoxe antique, dit \textit{paradoxe de Ménon}, exhibe une structure similaire. Celui-ci stipule qu'on ne peut ni chercher quelque chose que l'on connait, ni quelque chose que l'on ne connait pas. On ne peut pas chercher quelque chose que l'on sait ; puisqu'on le sait, il n'y a besoin de le chercher - ni quelque chose que l'on ne sait pas ; puisqu'on ne sait pas ce que c'est. Ce paradoxe fait ainsi en quelque sorte écho à la maxime socratique : "Je sais que je ne sais rien."; ou encore au paradoxe du nihilisme : l'absence de sens semblerait être une sorte de sens.\\ 

Soit $\{S_i\}_{i=1,...,N}$ un ensemble de phrases, chacune d'entre elle, prise séparément, pouvant être vraie ou fausse. Lorsque l'on considère deux phrases $S_i$ et $S_j$ simultanément, on écrit $S_i : S_j$. Les phrases d'un couple peuvent être compatibles, auquel cas le couple est "vrai" ; ou incompatibles : le couple est "faux". \\ 

Un cycle menteur est désigné par la construction  suivante :
\[S_1:S_2 \hspace{5mm} vrai \hspace{1cm} S_2:S_3 \hspace{5mm} vrai \hspace{1cm} ... \hspace{1cm} S_{N-1}:S_N \hspace{5mm} vrai \hspace{1cm} S_N:S_1 \hspace{5mm} faux \]

Par exemple, pour $N=1$, on obtient $S_1 : S_1 faux $, i.e. un mensonge classique. \\

On peut associer à chaque phrase une variables booléenne $x_i \in \{ 0,1\}$, prenant la valeur 0 pour "faux", ou la valeur "1" pour vraie, afin de réécrire le cycle menteur sous la forme d'une équation logique. Si deux phrases sont compatibles, leurs valeurs associées sont égales (si elles sont incompatibles, les valeurs sont différentes). On obtient 
\[x_1 = x_2  \hspace{5mm} x_2 = x_3 \hspace{5mm} ... \hspace{5mm} x_{N-1}=x_N \hspace{5mm} x_N \neq x_1 \]

La nature paradoxale du cycle menteur devient ici flagrante, du fait de l'incohérence de l'équation. On peut voir chaque equation $x_i = x_j$ comme fibré sur l'ensemble des variables dans lequel elles apparaissent.
\[ \{x_1,x_2\}:x_1 = x_2 \hspace{5mm} ... \hspace{5mm} \{x_{N-1},x_N\}:x_{N-1} = x_N \hspace{5mm} \{x_N,x_1\}:x_N \neq x_1\]

Chacune de ces équations, prise indépendemment, est cohérente, et tout sous-ensemble de $n-1$ de ces équations l'est également, alors que l'ensemble global est incohérent, et conduit à un paradoxe. On retrouve ici l'obstruction caractéristique de la contextualité. Une boite PR est par exemple equivalente à un cycle menteur de longueur $N=4$. Réciproquement, la contextualité offre une nouvelle perspective sur ces paradoxes, ces cycles contradictoires donannt lieu exactement à la formation de la cohérence locale et de l'incohérence globale caractéristiques de la contextualité.\\ 

Le paradoxe du menteur peut être transcrit sous la forme d'un cycle menteur de longueur $N=3$, constitué des trois propositions suivantes :
$S_1 := $ "Je suis crétois." ; $S_2 := $ "Je dis la vérité." ; $S_3 := $  "Tous les crétois sont des menteurs." Ces trois phrases sont "localement cohérentes" :\\
"Je suis crétois et je dis la vérité" : $\{x_1,x_2\}:x_1 = x_2 $ \\
"Je dis la vérité et tous les crétois sont des menteurs." $\{x_2,x_3\}:x_2 = x_3 $ \\
"Tous les crétois sont des menteurs et je ne suis pas crétois."$\{x_3,x_1\}:x_3 \neq x_1 $\\
En revanche, lorsqu'on tente de les interpréter globalement, on se retrouve piéger dans un paradoxe tendue par l'auto-référence. On remarque que l'hypothèse selon laquelle Epiménide dit la vérité semble être nécessaire ici. En effet, en ne considérant que les phrases $S_1$ et $S_3$, on peut postuler que le chaman ment lorsqu'il dit qu'il est crétois, et sortir ainsi du paradoxe. \\

L'énoncé le plus simple du paradoxe du menteur reste le suivant :
\begin{center}
Je suis en train de mentir.
\end{center}
celui-ci est équivalent aux deux propositions localement cohérentes :\\
" Je dis la vérité et je dis que je mens."  $\{x_1,x_2\}:x_1 = x_2 $ \\
"Je dis que je mens et je ne dis pas la vérité."  $\{x_2,x_1\}:x_2 \neq x_1 $ \\
Une fois de plus, le paradoxe émerge d'une obstruction globale : "si je dis la vérité quand je dis que je mens, alors je suis un menteur, et je ne dis pas la vérité."\\

\subsection{Contextualité et paradoxes d'infuturabilité.}

Le rapport structurel entre certains paradoxes logiques classiques et la contextualité n'est pas surprenant. Rappelons que le théorème de Kochen-Specker est basé sur une étude logique de la mécanique quantique, inspirée des travaux de Birkhoff et von Neumann, et de l'article de 1960 de Specker. Rappelons également qu'il en résulte que la \textit{\textbf{logique quantique n'est pas booléenne}}. Cela admis, il n'y a rien d'étonnant à constater qu'un résultat de logique quantique tel que la "contextualité" permet de "comprendre" les paradoxes, ou encore que ces paradoxes nous permettent de nous "représenter" l'étrangeté de la mécanique quantique. \\

Dans l'article de 1960, Specker établis un parallèle entre l'étude des "propositions non décidables simultanément" et celle des "infuturabilités" théologiques. Je pense que la définition logique de la contextualité d'Abramsky apporterait une nouvelle perspective à cette analogie.\\

Une infuturabilité, ou futur contingent, est un énoncé décrivant un évènement futur que l'on ne peut pas prédire, i.e. on ne peut pas associer une valeur de vérité "vrai" ou "faux", tant que l'évènement n'a pas eu lieu. Par définition, de tels énoncés ne peuvent donc être décrits par la logique booléenne, et peuvent donc mener dans certains cas à des paradoxes, que nous appelerons \textit{paradoxes d'infuturabilité}.\\

Le paradoxe d'Aristote exposé au chapitre précédent, entre dans cette catégorie. Reprenons le. On suppose qu'une bataille navale n'aura pas lieue demain. Si l'énoncé est vrai aujourd'hui, alors il était aussi vrai hier, le mois dernier, l'année passée... Mais ce qui est vrai hier est nécessairement vrai au présent,ce qui invite à conclure que la bataille n'aura effectivement pas lieue demain. Nous avons exposé deux solutions proposées pour résoudre ce paradoxe : la nature contingente du futur (solution aristotélicienne), et la solution de Diodore, via l'argument dominateur. Par ailleurs, il est, selon moi, possible de reproduire le paradoxe par le modèle suivant :\\

\begin{figure}[ht!]
\centering
\includegraphics[width=4cm]{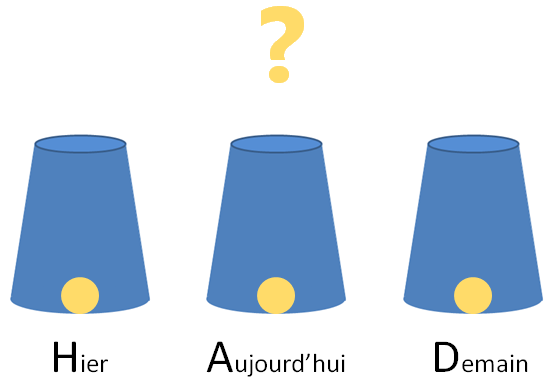}
\caption{Modèlisation du paradoxe de la bataille navale.}
\end{figure}

on considère trois boîtes noires distinctes : $H$ pour "hier", $A$ pour "aujourd'hui", $D$ pour "demain". Chaque boîte peut contenir une "balle" (valeur 1), qui symbolise l'évènement "la bataille navale a lieu". La valeur 0 correspond donc à une absence de balle et ainsi au fait que la bataille n'a pas lieu. L'énoncé "ce qui est vrai hier est nécessairement vrai au présent" se traduit par le fait que deux boites adjacentes ont le même contenu, i.e. $HA \in \{00,11\}$ et $AD \in \{00,11\}$. $H$ et $A$ sont simultanément décidables, de même que $A$ et $D$. En revanche, ce n'est pas nécessairement le cas de $D$ et $H$. Le postulat "tout énoncé portant sur le futur est imprévisible" correspond au fait que les contenus de $D$ et $H$ ne sont pas corrélés, i.e. $DH \in \{00,01,10,11\}$. On obtient donc les tableau de possibilité et faisceau de contextualité suivants : 

\begin{center}
\begin{tabular}{c|cccc}
  &  00 & 01 & 10 & 11 \\
  \hline
  H A & 1 & 0 & 0 & 1 \\
  A D & 1 & 0 & 0 & 1 \\
  D H & 1 & 1 & 1 & 1 \\
\end{tabular}
\end{center}

\begin{figure}[ht!]
\centering
\includegraphics[width=3cm]{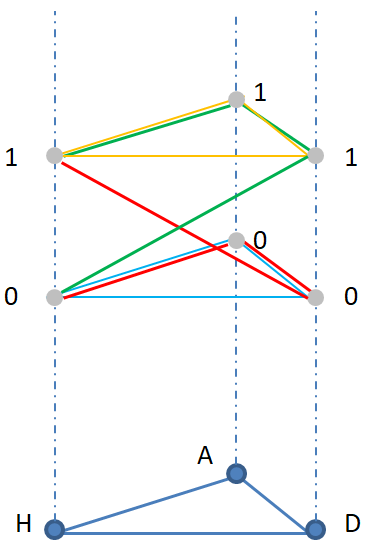}
\caption{Faisceau de contextualité du paradoxe d'Aristote}
\end{figure}

On constate la présence de deux chemins fermés univoques triviaux :\\ $HA,AD,DH = 00$ et $HA,AD,DH = 11$,\\ mais également l'existence de deux sections locales globalement icohérentes :\\ $\{H \rightarrow 1, A \rightarrow 1, D \rightarrow 0 \}$ et $\{H \rightarrow 0, A \rightarrow 0, D \rightarrow 1 \}$. Ainsi, si l'on devait classer ce scenario à l'aide de la hiérarchie de l'approche faisceau, nous concluerions que le paradoxe d'Aristote exhiberait une forme de contextualité logique. \\

Un autre paradoxe célèbre, le paradoxe de "la pendaison inattendue", également connu sous le nom de "paradoxe de l'examen surprise", peut être selon moi caractérisé par la contextualité et classer dans une des catégories de l'approche d'Abramsky. \\
\boitesimple{\begin{quotation}
Un juge déclare à un prisonnier condamné qu'il sera pendu la semaine prochaine, entre lundi et vendredi, à midi, mais que le jour de la pendaison sera une surprise pour le prisonnier : il ne connaîtra le jour de sa mort que lorsque le bourreau viendra frapper à la porte de sa cellule.\\
Le prisionnier réfléchit à ces déclarations puis, soudainement, pousse un cri de joie. "Si la pendaison avait lieu vendredi", se dit-il, "alors ce ne serait plus une surprise, puisque, jeudi prochain, une fois midi passé,  je saurais avec certitude que je serai pendu vendredi. Je ne peux donc pas être pendu vendredi."\\
"Mais," poursuit-il,"cela veut aussi dire que la pendaison ne peut pas avoir lieu jeudi ! Si c'était le cas, ça ne serait pas une surprise, puisque, mercredi prochain, midi passé,  je saurais avec certitude que je serai pendu jeudi. Et un raisonnement semblable démontre que la pendaison ne peut pas non plus avoir lieu mercredi, mardi, ou même lundi !"\\
Le prisonnier retourne à sa cellule, confiant et rassuré. La semaine suivante, le bourreau frappa à sa porte le mercredi midi - à sa grande surprise.\\
\end{quotation}}

Divers solutions ont été proposées pour tenter d'expliquer et de résoudre ce paradoxe. Par exemple, celui-ci pourrait émerger de la nature vague du terme "surprise", rendant l'énoncé du juge imprécis, voir auto-contradictoire. On pourrait également invoquer la contingence du futur - quoiqu'il arrive, le prisonnier ne pourra pas prédire le jour de sa pendaison future ; son raisonnement est donc incorrect - ou encore l'argument dominateur - le possible est nécessairement vrai ; ainsi, le jour de la pendaison est déjà fixé et déterminé, et le paradoxe est du aux tentatives vaines du prisonnier de le prévoir.\\

Une solution épistémologique peut également être défendue ; en réduisant le scénario à deux jours, lundi et mardi, et en décomposant l'assertion du juge en trois conditions :\\

\hspace{-6mm}1- La pendaison aura lieu lundi ou mardi.\\
2- Si la pendaison a lieu lundi, alors le prisonnier ne le saura pas la veille (dimanche).\\
3- Si la pendaison a lieu mardi, alors le prisonnier ne le saura pas la veille (lundi).\\

Le paradoxe émergerait alors d'un raisonnement éronné du prisonnier, qui considère à tord qu'il est impossible qu'il soit pendu mardi alors que, ce qui est impossible, c'est une situation dans laquelle la pendaison a lieu mardi malgré le fait que le prisonnier sait, le lundi soir, que les trois assertions (1), (2) et (3) sont toutes vraies. L'énoncé que le prisonnier est censé reconnaître comme vrai portant sur son incapacité de connaître certains faits, la sentence du juge pourrait donc être reformulée en : "\textit{Tu seras pendu demain, mais tu ne le sais pas.}"\\

Enfin, si le raisonnement du prisonnier peut paraître étonnant, il est néanmoins tout à fait valable d'un point de vue logique. Dans "\textit{A Goedelized Formulation of the Prediction Paradox} (1964)", Frederic Fitch démontre ainsi que le paradoxe ne serait pas dû au raisonnement du prisonnier, mais à la prédiction du juge, qui est une auto-référence.\\

Je propose de reformuler le paradoxe de façon similaire à la modélisation précédemment exposée (pour le paradoxe d'Aristote) :
considérons cinq gobelets, représentant chacun un jour de la semaine (de lundi à vendredi) : $\{V, J, M', M, L \}$. Une balle est cachée sous l'un des gobelets, et correspond à la pendaison (valeur 1). La valeur 0, ou l'absence de balle, signifie donc l'absence de pendaison. Par analogie avec le raisonnement du prisonnier, les gobelets sont considérés par paires (simultanément décidables) : $\{VJ, JM', M'M, ML, LV\}$. La pendaison ne pouvant pas avoir lieu deux fois, aucune des paires ne peut être égale à 11. Si l'on réduit le scenario à deux jours, $\{ VJ, JV\}$, on remarque alors que puisque la pendaison a nécessairement lieue, aucune des paires ne peut être égale à 00. On extrapôle alors ce raisonnement aux cinq couples de jour et on obtient le tableau de possibilité suivant :
\begin{center}
\begin{tabular}{c|ccccc}
  &  00 & 01 & 10 & 11 \\
  \hline
  $V J$ & 0 & 1 & 1 & 0 \\
  $J M'$ & 0 & 1 & 1 & 0 \\
  $M' M$ & 0 & 1 & 1 & 0 \\
  $M L$ & 0 & 1 & 1 & 0 \\
  $L V$ & 0 & 1 & 1 & 0 \\
\end{tabular}
\end{center}
On constate alors qu'il s'agit du même tableau que celui associé à l'inégalité KCBS. La nature paradoxal du scenario émegerait ainsi du fait que la "surprise" annoncée n'est pas envisagée sur l'ensemble global de la semaine, mais sur des paires de jours locales. D'après ce modèle, le paradoxe de la pendaison inattendue partagerait ainsi la même structure topologique qu'un scénario de contextualité forte.\\

Je pense que d'autres paradoxes, qui semblent à la fois mêlés auto-référence et infuturabilités, pourrait également être traité de façon similaire (cf. par exemple paradoxe du pont de Buridan et paradoxe de Pinnochio en annexe).\\

La contextualité ne résout pas ces paradoxes, au sens où elle n'apporte pas de clé permettant de dénouer leurs intrigues. Seulement, elle pourrait permettre d'identifier ou de caractériser leur nature : l'obstruction globale de données localement cohérente constituerait une source de leur étrangeté. Cela ne signifie pas non plus que tous ces paradoxes peuvent être transcrits et traités en version quantique. L'approche logique de la contextualité donnée par Abramsky et Brandenburger est très générale, et inclue des scénarii non quantiques, comme la boite PR ou la parabole de Specker. Néanmoins, au-delà de l'interêt qu'elle offre pour le domaine de la logique et de l'informatique, cette approche a le mérite d'offrir des \textit{mythes}, des histoires permettant de "raconter" la contextualité.

\section{Raconter la contextualité ?}
\label{sraconter}
\begin{quotation}
\textit{Le réel ne peut s’exprimer que par l’absurde.} Paul Valéry
\end{quotation}

\textit{Cette section constitue un essai sur les outils picturaux qu'offre l'approche faisceau de la contextualité. Il s'agit d'une réflexion originale, qui n'a pas la prétention de restituer les pensées des auteurs de cette approche, ou celles des auteurs cités.\\}

La mécanique quantique est souvent dite paradoxale, au sens où elle semble faire appel à des concepts qui, lorsqu'on tente de les représenter, vont à l'encontre de la \textit{doxa}, i.e. bouleverse notre intuition. Comme le remarque Alexei Grinbaum dans \cite{alexei3}, le recours aux métaphores est inévitable lorsqu'il est question d'expliquer la théorie quantique à un public non intitié. La physique semble difficilement pouvoir se passer d'images, et je pense que l'on attend généralement d'elle qu'elle nous offre un discours sur le monde. Cependant, le passage des équations mathématiques aux mots du langage courant ne se fait pas toujours sans dommages. L'atome, d'abord pudding, s'est métamorphosé en système planétaire, avant de devenir une entité vague, "quantique", "à la fois onde et particule".\\

Si les images évoluent avec les théories, l'avènement de la physique quantique a profondément bousculé notre façon de "parler" du monde. Le langage classique ne semble plus suffir. Il ne parvient plus à contenir des concepts qui échappent à sa logique. L'expression "dualité onde-corpuscule", encore utilisée dans la plupart des cours d'introduction à la mécanique quantique, a toujours été obsolète : une particule quantique n'est ni une onde, ni un corpuscule, et certainement pas "les deux à la fois". Face à ces deux représentations contradictoires, plusieurs solutions s'offrent aux "conteurs" : on pourrait considérer que notre langage, nos images, notre "façon de raconter le monde" devraient, peut-être, évoluer avec nos concepts. Jean Marc Levy Leblond et Françoise Balibar proposent ainsi \cite{leblond} d'inventer un nouveau mot, "\textit{quanton}", pour désigner les particules quantiques. Niels Bohr semblerait quant à lui avoir une approche différente, qui consisterait à changer non pas notre langage, mais notre façon d'appréhender la physique.

\begin{quotation}
\textit{Il n'y a pas de monde quantique. Il y a seulement une description physique quantique abstraite. Il est faux de penser que la tâche de la physique soit de découvrir comment la nature se comporte. La physique concerne ce que l’on peut dire de la nature. [...] De quoi dépendons-nous, nous humains ? Nous dépendons de nos mots. Notre tâche est de communiquer expérience et idées à autrui. Nous sommes suspendus dans le langage. ‘We are suspended in language’.} (cité par \cite{petersen}). 
\end{quotation}

 \begin{quotation}
\textit{Dans la mesure où les phénomènes transcendent la portée de l'explication physique classique, l'énoncé de toutes preuves doit être exprimé en termes classiques. (...) L'argument est simplement que, par le mot expérience }[experiment]\textit{, nous nous référons à une situation où nous pouvons dire aux autres ce que nous avons fait et ce que nous avons appris et que, par conséquent, le compte rendu des arrangements expérimentaux et des résultats des observations doivent être exprimées dans un langage sans ambiguïté avec une application appropriée de la terminologie de la physique classique.} \cite{bohr4}
 \end{quotation}

Bohr semble ainsi considèrer que "la tâche" de la physique serait de permettre aux hommes de transmettre, entre eux, « ce qu’ils ont fait et ce qu’ils ont appris ». \\
Cette pensée m'invite à envisager que les équations de la théorie quantique n'autorisent plus à raconter le monde sans prendre en considération notre condition de "conteur", d'observateur. Il ne serait plus question de prétendre à un discours sur le monde, où le physicien accèderait à une forme de transcendance, en attendant d'être omniscient,  mais plutôt d'un récit de notre expérimentation du monde.\\

Aspirer à raconter le monde comme un conteur raconterait des contes de fées au coin du feu, peut se révéler être à double tranchant : ou l'auditoire accepte les images, et il repart avec le sentiment d'avoir pu appréhender une once de réalité, ou bien il se rend compte de l'artifice, et la magie ne peut opérer.  C'est notamment le cas  d'Albert Camus, qui désespère de parvenir à connaître un monde qu'il aime mais qui reste silencieux.

\begin{quotation}
\textit{Voici encore des arbres et je connais leur rugueux, de l’eau et j’éprouve sa saveur. Ces parfums d’herbe et d’étoiles, la nuit, certains soirs où le coeur se détend, comment nierais-je ce monde dont j’éprouve la puissance et les forces ? Pourtant toute la science de cette terre me donnera rien qui puisse m’assurer que ce monde est à moi. Vous me le décrivez et vous m’apprenez à le classer. Vous énumérez ses lois et dans ma soif de mécanisme et mon espoir s’accroît. Au terme dernier, vous m’apprenez que cet univers prestigieux et bariolé lui-même se réduit à l’atome et que l’atome lui-même se réduit à l’électron. Tout ceci est bon et j’attends que vous continuiez. Mais vous me parlez d’un invisible système planétaire où des électrons gravitent autour d’un noyau. \textbf{Vous m’expliquez ce monde avec une image. Je reconnais alors que vous en êtes venus à la poésie} : je ne connaîtrai jamais. Ai-je me temps de m’en indigner ? Vous avez déjà changé de théorie. Ainsi cette science qui devait m’apprendre finit dans l’hypothèse, cette lucidité sombre dans la métaphore, cette incertitude se résout en oeuvre d’art. Qu’avais-je besoin de tant d’efforts ? Les lignes douces de ces collines et la main du soir sur ce coeur agité m’en apprennent bien plus.Je suis revenu à mon commencement. Je comprends que, si je puis par la science saisir les phénomènes et les énumérer, je ne puis pour autant appréhender le monde. Quand j’aurai suivi du doigt son relief tout entier, je n’en saurais pas plus. Et vous me donnez à choisir entre une description qui est certaine mais qui ne m’apprend rien, et des hypothèses qui prétendent m’enseigner, mais qui ne sont point certaines.} \cite{camus} [\textit{notre emphase}] 
\end{quotation}

Le silence du monde auquel fait face Camus semble faire écho au silence des équations de la mécanique quantique. Qu'il s'agisse de métaphores à visée éducative, ou d'une interprétation philosophique, une image reste une image et si elle permet de raconter le réel, elle ne permet pas de l'appréhender. Le réalisme extrait des équations de la physique a pris une saveur de plus en plus âcre. Il a fini par perdre ses couleurs. N'est-il pas au fond une hypothèse ? Un rêve, une promesse projetée sur des objets mathématiques qui restent muets, silencieux.\\

La prise de conscience d'une telle "trahison des images", qui rappelle les travaux picturaux de Magritte, invite selon moi à deux prises de positions : 

la première est d'ordre philosophique.\\

L'entêtement à vouloir sauver une forme de réalité \textit{naïve} du monde me semble absurde, puisque cette réalité ontologique, faite d'entités indépendantes de toute observation, serait non seulement profondément contre-intuitive, mais aussi en grande partie inaccessible. Pourquoi, alors, ne pas simplement accepter l'invitation proposée par les résultats du principe d'indétermination d'Heisenberg et du théorème de Kochen-Specker : abandonner l'idée que les systèmes quantiques (électrons, photons...) seraient réels parce qu'ils sont des objets expérimentaux. Je serais même tenté d'ajouter que ces résultats nous incitent à penser le concept d'\textit{entité} en tant qu'\textit{obstacle épistémologique}, "un concept  qui \textit{arrête la pensée} et dont l'esprit doit se déprendre pour devenir scientifique." (\cite{bontems2}) L'\textit{entité} (du latin \textit{ens}, \textit{entis}, "chose qui existe") désigne un objet qui existe par lui-même, ou l'ensemble des propriétés constitutives d'une chose. Chez Aristote, elle repose sur trois piliers logiques : l'existence, la non-contradiction et l'identité. On voit bien qu'aucun de ces trois piliers ne semble compatible avec les inégalités d'Heisenberg ou le théorème de Kochen-Specker. La présupposition que la réalité est peuplée d'entités "\textit{restreint et conditionne les possibilités de penser la structure formelle de la mécanique quantique en exigeant sa traduction en une ontologie substantialiste, dépassée et inadéquate}" \cite{bontems}.\\

Supposons que l'on puisse représenter une théorie quelconque par une boucle, la séparation entre l'objet et les présupposés de la théorie étant symbolisée par une ouverture en un certain point. Du fait de la nécessité de cette séparation, il est impossible de donner une description théorique de la boucle comme un tout. Une fois la position de cette césure fixée, certains éléments de la boucle sont traités comme des objets de la théorie, tandis que les autres tombent dans un domaine méta-théorique, i.e. sont chargés d'expliquer le fonctionnement des dispositifs de mesure de la théorie considérée, et ne peuvent donc pas être décrits par celle-ci. Pour une césure différente, ces instruments de mesure feront l'objet d'une autre théorie, qui expliquera leur fonctionnement. (source : \cite{alexei4})\\

Du fait de la nécessité de la césure entre observateur et système observé, une théorie ne peut pas décrire l'ensemble de la boucle. Le réalisme entitique (ou intrinsèque) me paraît relever d'une forme de complexe icarien latent, où le physicien pourrait survoler et s'extérioriser du monde. Un observateur ne devrait pas être relégué au rang "d'épiphénomène inessentielle dans l'univers." \cite{fuchs} La physique est une activité humaine, et on ne peut pas prétendre observer la nature et ses lois en adoptant un point de vue qui leur est extérieur et nier la participation active et non négligeable de l'agent observateur. Si cette position peut sembler arrogante (certains y voient une forme d'anti-révolution copernicienne), il me semble plus inconvenant de penser que l'on puisse s'extraire de la nature et viser (inconsciemment ?) une forme d'omniscience transcendantale.\\

Que reste-t-il alors ? La structure mathématique des théories physiques, qui n'implique aucunement la notion d'entité. Les affirmations métaphysiques telles que "La véritable ontologie de la Nature est ceci et cela..." ne peuvent être affiliées à la science physique. Si celle-ci ne précise pas si les états (ou toute autre variable) d'un système sont ontologiques, alors ne les considérons pas comme ontologique : adoptons plutôt une attitude de "\textbf{modestie épistémologique}".\cite{alexei4}\\

Cette position me semble néanmoins difficile à tenir face à une audience "grand public", qui réclamerait qu'on lui apprenne quelque chose sur le réel. Une telle audience risquerait par ailleurs de se perdre dans les méandres de l'abstraction mathématique. \\

Une solution serait d'admettre et d'embrasser l'illusionisme pictural. Le grand public n'est pas dupe. Il sait qu'il ne repartira pas avec un accès direct au monde des vérités mathématiques\footnote{Il y a quelque chose de très platonicien (anamnèse) dans le verbe "représenter", qui sous entend que l'objet concerné a déjà été présenté auparavant, qu'il a déjà été donné à voir, il a déjà été présent, actuel.}. Ce que le non-initié souhaite, c'est d'être initié. Il aimerait qu'on lui raconte une histoire qui lui permetterait d'actualiser son regard sur lui-même et ce qui l'entoure. Il désire repartir avec des images qui lui donneront, peut-être, le sentiment de pouvoir se représenter, imaginer, le monde sous une facette inédite. \footnote{Ayant moi-même été initié à travers différents outils de vulgarisations (vidéos, conférences, oeuvres), je me permet de parler au nom d'un "grand public", qui, il me semble, partage cette envie "d'imaginer le monde différemment".}\\

Alexei Grinbaum propose, par exemple, de faire usage des mythes \cite{alexei3}, afin de concilier l'élégance apolinienne des mathématiques et la puissance suggestive des métaphores.

\begin{quotation}
\textit{Il est inévitable de recourir à des métaphores dans les explications vulgarisées de la théorie quantique. Quelle que soit la métaphore utilisée dans un langage commun, elle conduit nécessairement à la perte du sentiment d'élégance mathématique que le physicien ressent lorsqu'il travaille mathématiquement. Si la beauté ne peut pas être véhiculée par une métaphore, elle peut tout de même être intégrée dans le récit qui englobe cette métaphore. Cela ne se produit pas avec des récits qui sont des histoires rationnelles, comme la parabole de chat de Schrödinger. Mais si un récit peut s'inspirer de sa propre source de beauté, inhérente au genre littéraire de ce récit, il peut provoquer alors un sentiment d'élégance. Ce sentiment de beauté n'est pas basé sur les mathématiques, mais est toujours une notion esthétique à part entière. Russell a comparé la beauté mathématique à la beauté de la poésie: «Le véritable esprit de la joie, de l'exaltation, du sentiment d'être plus que l'homme, qui est la pierre de touche de la plus haute excellence, se trouve dans les mathématiques aussi sûrement que dans la poésie. Le meilleur des mathématiques ne mérite pas simplement d'être appris en tant que tâche, mais d'être assimilé comme une partie de la pensée quotidienne. . . "[59, p. 49]. Je propose de choisir le mythe comme un autre type de récit capable de provoquer un sentiment esthétique et de transmettre l'idée que la beauté fait partie de la pensée quotidienne du savant.}

\end{quotation}

Du fait de la révolution conceptuelle propre à la mécanique quantique, re-présenter au public ces concepts re-visités en les re-contextualisant, sous un angle historique, me semble également pertinent.\\

Dans son ouvrage "Mécanique des étreintes" \cite{alexei6}, Grinbaum propose ainsi une introduction à l'histoire du concept de "composition", qui semble être au coeur du théorème de Bell. Il évoque notamment une similarité assez saisissante entre les corrélations quantiques et la notion théologique de  \textit{périchorèse}, l'interpénétration "sans destruction, fusion, ni confusion" des natures divines et humaine du Christ. Ce terme est également utilisé pour désigner précisément la relation non-contradictoire entre chaque figure de la Trinité : Le Père n'est pas le Fils, qui n'est pas le Saint-Esprit, qui n'est lui-même pas le Père. En revanche, chacun est une figure de Dieu. \\

\begin{figure}[ht!]
\centering
\includegraphics[width=4cm]{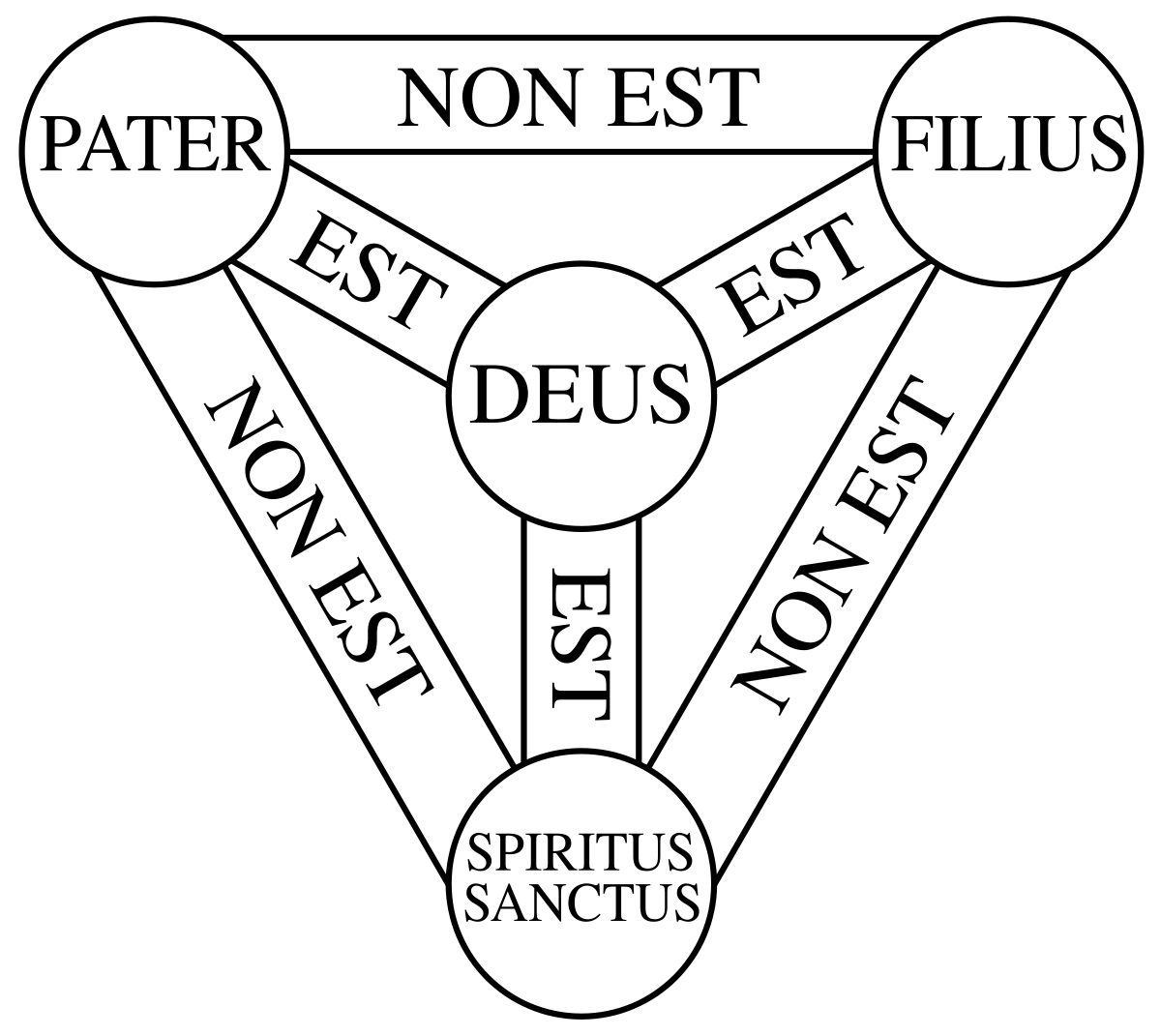}
\caption{Représentation graphique de la périchorèse.}
\end{figure}

Les symboles graphiques représentant la Trinité véhiculent cette notion d'intrication, d'enchevêtrement, comme la triskèle ou la triquetra, d'origine celtique et reprises dans les représentations religieuses médiévales. Les \textit{trois lièvres} est un autre symbole parfois utilisé pour représenter la Trinité. Si son origine est assez mystérieuse, la plus vieille représentation connue à ce jour a été trouvée en Chine, où il symboliserait l'harmonie.  La représentation est parfois accompagnée de la maxime :
\begin{center}
Trois lièvres partagent trois oreilles,\\
Pourtant chacun d'entre eux en a deux.
\end{center}
\begin{figure}[ht!]
\centering
\includegraphics[width=5cm]{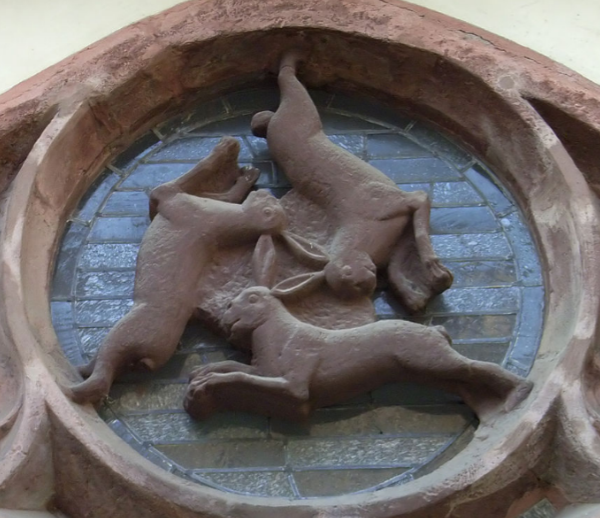}
\caption{Les trois lièvres \cite{lievres}}
\end{figure}
Chaque lièvre peut être individuellement vu comme cohérent, complet. Cependant, lorsque l’on essaye de percevoir les trois lièvres en même temps, globalement, on ne parvient pas à définir la distribution des paires d'oreilles.\\

L'énigme des trois lièvres, et les figures de noeud de trèfle rappellent une célèbre \textbf{figure impossible}, souvent attribuée à Roger Penrose\footnote{Le triangle impossible avait néanmoins déjà été construit indépendamment par l'artiste suédois Oscar Reutesvard en 1934. Si l'on peut retrouver divers paradoxes géométriques dans certaines représentations iconographiques médiévales, celles-ci sont la plupart du temps involontaires, et issues d'une erreur de perspectives. Oscar Reutervärd fut le premier à étudier ces étranges figures, et en dessina pas moins de 2500 tout au long de sa vie. \cite{figimp}}, qui la construisit pour la première fois en 1958 :
\begin{figure}[ht!]
\centering
\includegraphics[width=4cm]{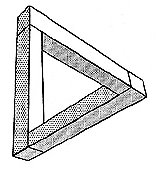}
\caption{Le triangle de Penrose, ou tribar \cite{figimp}}
\end{figure}

\begin{quotation}
\textit{Chaque partie est acceptable en tant que représentation d'un objet tridimensionnel : pourtant du fait de l'assemblage pervers de ces parties, la figure dans son ensemble conduit à produire l'effet d'une figure impossible.}\\   Roger Penrose
\end{quotation}

Les figures impossibles sont généralement construites de la façon suivante : chaque partie de la figure est interprétée immédiatement par notre cerveau comme la représentation d'un objet dans l'espace, i.e. localement cohérent. En tentant de concilier l'interprétation des différentes parties, i.e. en cherchant à donner une cohérence globale à l'objet, le caractère paradoxal de la figure se manifeste. On retrouve à nouveau la contextualité de l'approche faisceau, i.e. l'impossibilité de considérer globalement des données localement cohérentes.\footnote{Ce parallèle est établi par Samson Abramsky, Rui Soares Barbosa, Kohei Kishida, Raymond Lal et Shane Mansfield dans \cite{abramsky3}} \\

Dans les années 50, Roger Penrose, encore étudiant, assiste à un Congrès de Mathématiques à Amsterdam. ll y découvre le travail de M.C. Escher, à l'occasion d'une exposition de certains travaux de l'artiste. Penrose, fasciné, essaye alors de produire lui-même des images paradoxales : des routes, rivières et trains impossibles. Il simplifie alors ses esquisses et découvre l'essence de leur étrangeté : le \textit{tribar} aussi appelé \textit{triangle de Penrose}. Il montre alors le fruit de son travail à son père, Lionel Penrose, qui se passionne tout de suite pour ces curieuses figures géométriques. Il commence lui aussi à dessiner d'impossibles constructions, et finalement, créé l'\textit{escalier impossible}. Père et fils écrivent alors un article, qui sera publié en 1954 dans le magazine anglais Psychology, et envoient une copie à Escher, qui créera à partir de leurs travaux trois lithographies :  
\textit{Waterfall,} inspiré du triangle de Penrose, \textit{Ascending and Descending}, inspiré de l’escalier sans fin, et \textit{Belvedere}. \cite{penrose} \\

\begin{figure}[ht!]
\centering
\includegraphics[width=6cm]{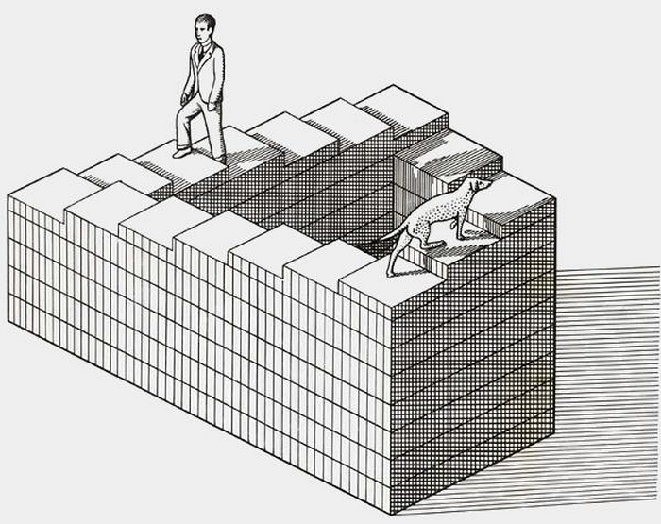}
\caption{L'escalier impossible de Lionel Penrose \cite{figimp}}
\end{figure}
\begin{figure}[ht!]
\centering
\includegraphics[width=10cm]{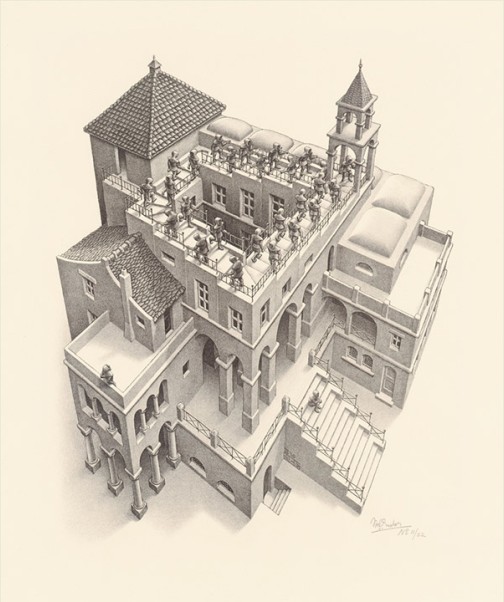}
\caption{Ascending and Descending, de M.C Escher, 1960, All M.C. Escher works © 2017 The M.C. Escher Company - the Netherlands. All rights reserved. Used by permission. www.mcescher.com \textit{Cette figure impossible composé de quatre éléments locaux rappelle le faisceau de contextualité de la boite PR.}}
\end{figure}

La lithographie \textit{Waterfall} d’Escher est particulièrement intéressante, à plusieurs égards. Le tableau représente une machine à mouvement perpétuel, un cycle sans fin où l’eau de la base d'une cascade semble descendre le long d'un chemin avant d'atteindre le sommet de la cascade. Le paradoxe n’est pas simplement issu d’un jeu de perspective, mais il vient du fait que le cours d’eau a la structure de deux triangles de Penrose.\\
\begin{figure}[ht!]
\centering
\includegraphics[width=8cm]{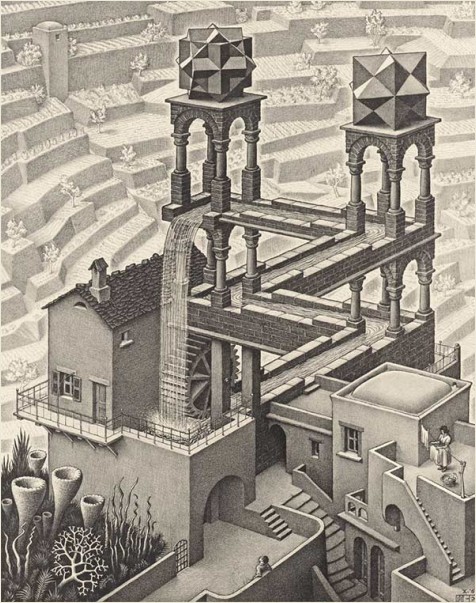}
\caption{Waterfall, de M.C. Escher, 1960, All M.C. Escher works © 2017 The M.C. Escher Company - the Netherlands. All rights reserved. Used by permission. www.mcescher.com}
\end{figure}

Sur le sommet des tours qui soutiennent l’aqueduc, on trouve deux polyèdres. \`A droite, un composé de trois octahèdres non réguliers, aussi connu sous le nom de \textit{solide d'Escher}. Sur la tour de gauche, un \textit{composé de trois cubes}.\\ 

\begin{figure}[ht!]
\centering
\includegraphics[width=5cm]{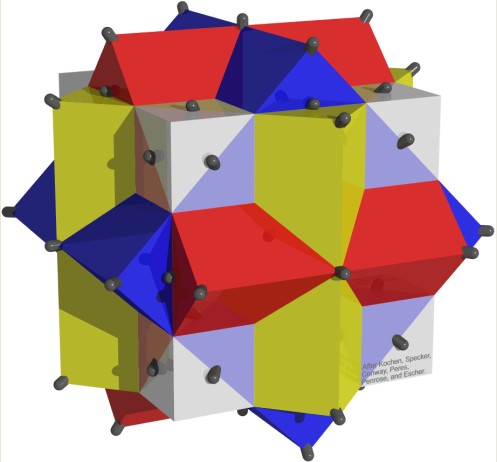}
\caption{Le composé de trois cube, source : \cite{stoica}. Merci à Cristi Stoica pour son autorisation.}
\end{figure}

Fait particulièrement surprenant, Roger Penrose a remarqué que l'ensemble des sommets et des centres des faces et des arêtes de ce tricube correspond aux 33 vecteurs d'une preuve géométrique du théorème de Kochen-Specker, construite par Asher Peres en 1991.  (\cite{peresb} \cite{mermin1}) \\

Il est pour le moins renversant de découvrir deux représentations graphiques (involontaires) du théorème KS (l'approche faisceau et la preuve de Peres) au sein d'un même tableau !\\

Les paradoxes logiques présentés et les figures impossibles ne sont pas quantiques, mais bien classiques. L'objet ou le récit ne représentent pas eux-même la contextualité. En revanche, le point d'ancrage de leur nature paradoxal, leur structure logique, pourrait être comparable à la non-classicité qui semble se manifester à partir des théorèmes de Kochen-Specker et de Bell. Encore une fois, cela n'est pas particulièrement étonnant puisque comme le théorème de Kochen-Specker le démontre, \textbf{la logique quantique ne peut être classique}.\\

Au delà de la simple illustration, ces paradoxes narratifs et graphiques constituent, selon moi, des outils introductifs d'une puissance suggestive difficilement égalable. S'ils rendent l'auditeur perplexe, cette perplexité reste familière et peut être facilement dessinée ou racontée. Par ailleurs, ils ne se contentent pas de laisser l'auditeur songeur, mais lui permettent de repartir avec des images aux saveurs sans doute proches de celles de l'étrangeté quantique.\\

Insistons néanmoins sur un point : ce n'est pas la contextualité en soi qui est illustrée, mais le résultat KS. N'oublions pas que le théorème de Kochen-Specker conduit à une contradiction logique entre trois hypothèses : les prédictions de la mécanique quantique, la non-contextualité des mesures, et le déterminisme des résultats. Parler de contextualité devrait donc supposer que l'on choisit d'abandonner la non-contextualité plutôt que le déterminisme ; une décision qui est loin d'être insignifiante.\\

Pour éviter toute confusion, Ravi Kunjwal, dans sa thèse \textit{Contextuality beyond the Kochen-Specker theorem} \cite{kunjwal2}, propose ainsi d'employer le terme de "noncontextualité-KS" (\textit{KS-noncontextuality}) pour désigner le couple  \{non-contextualité des mesures , déterminisme des résultats\}. Par ailleurs, il faut également noter que ces hypothèses s'appliquent dans le cadre de mesures projectives,  i.e. représentées par des opérateurs hermitiens.\\

Un débat sur la possibilité d'une vérification expérimentale du théorème de Kochen-Specker a longtemps fait rage. L'enjeu est de taille, puisqu'il permettrait de montrer que la notion de non-contextualité n'est pas uniquement en conflit avec le formalisme quantique, mais aussi avec "la nature elle-même, et donc tout successeur de la théorie quantique." \cite{kunjwal2} Certains déclarent ainsi que le théorème limite simplement les possibilités d'interprétations du formalisme quantique, ou encore qu'il s'agit d'un résultat purement logique, et non physique. Dans \cite{cabello3},
David Mermin est par exemple cité pour avoir déclaré : "toute la notion d'un test expérimental [du théorème de Kochen-Specker] passe à côté du problème."\\

L'avènement des "inégalités de contextualité", avec l'inégalité KCBS et l'approche CSW, semblaient avoir tranché en faveur de la faisabilité expérimentale. Cependant, une violation de ces inégalités n'implique pas nécessairement que la nature est contextuelle, puisqu'elles supposent implicitement une réponse déterministe des mesures. La violation pourrait ainsi très bien être due au fait que la nature est "indéterministe et non-contextuelle". Une appelation plus rigoureuse de ces inégalités serait donc "inégalités de contextualité-KS". \\

Ces inégalités sont-elles par ailleurs vraiment comparables aux inégalités de Bell ? Le théorème de Bell et le théorème de Kochen-Specker mènent à un même problème mathématique, qui est de déterminer si une ensemble donné de variables, partagées en plusieurs sous-ensembles distincts, peut admettre une distribution de probabilités jointes, étant données les distributions de probabilités jointes des sous-ensembles. (\cite{kunjwal2}, 1.5)\\

Il existe divers points majeurs de contraste entre les deux théorèmes :\\

\begin{itemize}
\item les corrélations décrites par le théorème de Bell sont nécessairement multipartites, tandis que le théorème de Kochen-Specker peut s'appliquer à un système unique. D'autre part, le théorème de Kochen-Specker ne dépend pas, en général, de l'état du système, alors que c'est le cas pour le théorème de Bell.\\
\item le théorème de Bell est indépendant des théories : une violation expérimentale d'une inégalité de Bell ostracise la causalité locale quel que soit la théorie opératoire qui pourrait modéliser l'expérience. Il n'est par exemple pas nécessaire de supposer que le système est décrit par un espace de Hilbert, comme en théorie quantique. Le théorème de Kochen-Specker est quant à lui un résultat spécifique au formalisme quantique, puisqu'il suppose que les résultats de mesures soient représentés par des projecteurs sur un espace de Hilbert. \\
\item D'autre part, comme le remarquent Kunjwal et Spekkens, ces mesures projectives correspondent à des mesures idéales, parfaites. Cette idéalisation des mesures est issue de la nature fondamentalement logique du théorème, mais, en pratique, toute mesure sur un système physique est sujette à un bruit expérimental, et ne peut pas, par conséquent, être représentée par un ensemble de projecteurs sur l'espace de Hilbert du système. \`A la place, il convient d'utiliser des POVMS (positive operator valued measures), i.e. un ensemble d'opérateurs positifs (appelées "effets") dont la somme est égale à l'opérateur identité. On remarque également que la notion de \textit{contexte de mesure} du théorème de Kochen-Specker est assez réduite, puisqu'elle se restreint à des relations de compatibilité entre mesures projectives. Dans le cas d'une généralisation des mesures aux POVMs, elle devrait également être étendue. \\
\item Le théorème de Bell ne requiert pas d'hypothèse de déterminisme des résulats, alors que c'est le cas pour le théorème de Kochen-Specker.\\
\end{itemize}

Les approches graphique, faisceau et combinatoire semblent apportées une réponse aux deux premières divergences : dans ces approches, les scenarios multipartites de type Bell sont considérés comme des cas particulier de scenarii de contextualité généralisés. Cette généralisation semble également indépendante des théories considérées, puisque les approches graphiques traitent également différents modèles probabilistes, classique, quantique, et super-quantiques. Cependant, dans chacune de ces approches, les mesures sont intrinsèquement projectives, et l'hypothèse du déterminisme des résultats est sous-jacente à la définition de non-contextualité proposée.\footnote{Robert Spekkens a montré dans \cite{spekkens9} que déterminisme des résultats et mesures projectives correspondaient à une seule et même hypothèse. Nous y reviendrons dans le chapitre suivant.} Cela ne remet évidemment pas en cause la généralisation du théorème de Kochen-Specker apportée par ces approches, ou encore le fait que les scenarii de Bell définis sont des cas particuliers de scenarii de contextualité dans l'approche AFLS. Remarquons par ailleurs que si AFLS se limite aux mesures projectives, sa définition de la notion de contexte est déjà généralisée, puisqu'il ne s'agit plus de relations de compatibilité mais d'équivalences opératoires. Néanmoins, les lacunes exposées invitent à rester prudent dans l'emploi des termes que nous faisons, et à éviter toute formule abrégée et sensationelle du type "la non-localité est un cas particulier de contextualité". \\

Il n'est donc pas évident que le résultat de Kochen-Specker soit "robuste" expérimentalement. Celui-ci s'applique parfaitement à la théorie quantique, qui n'est pas, par raisonnement logique, compatible avec la non-contextualité-KS. En revanche, la question de savoir si la nature l'est également ou non est plus complexe, et nécessiterait une généralisation opératoire de la notion de contextualité aux mesures non idéales ("unsharp" dans la littérature), représentées par les POVMS, dépouillée de l'hypothèse de déterminisme.\\

En 2004, Robert Spekkens a proposé une telle généralisation opératoire de la notion de (non-)contextualité, permettant de combler les lacunes du théorème de Kochen-Specker. Cette approche s'applique à toute théorie opératoire, à toutes procédures expérimentales, et à une classe de modèles plus large que les modèles à variables cachées déterministes de la théorie quantique. Ce nouveau cadre de modèles dits \textit{ontologiques} (cf. chapitre suivant), permit également la découverte de nombreux "théorèmes no-go", de défendre une interprétation "épistémique" des états quantiques, et des avancées dans le programme de recherche axiomatique. Elle conduit, enfin, à l'obtention de véritables "inégalités de contextualité", qui peuvent être testées expérimentalement.

\chapter{Critère d'\'Etrangeté : la contextualité de Spekkens}
\label{chapspek}

La mécanique quantique est souvent présentée comme une théorie mystérieuse. Mais comment capturer son étrangeté ? Pour la comprendre, il est primordial d'identifier ce qui la distingue de la physique classique. Cette identification peut être guidée par le critère suivant  :

\boitesimple{\textit{Si un résultat de la physique quantique apparait aussi dans un cadre de physique statistique classique, avec d'éventuelles hypothèses additionnelles mineures qui ne remettent pas violemment en cause nos conceptions quotidiennes, alors il ne devrait pas être considéré comme un résultat intrinsèquement quantique.} \cite{jennings}}

Dans un article paru en janvier 2004 sur \textit{arxiv} \cite{spekkens5}, Robert Spekkens montre qu'une grande partie des résultats de la théorie quantique - du théorème de non-clonage à la téléportation, en passant par la distribution de clés quantiques, l'indiscernabilité des états purs, la non-unicité de décomposition des états mixtes, la monogamie de l'intrication, les interférences dans une expérience de Mach-Zender et bien d'autres - pouvait être obtenus en ajoutant simplement aux théories classiques une restriction sur le types de distributions statistiques autorisées. Ces résultats sont-ils donc vraiment étranges, puisqu'ils ne divergent pas profondément de notre vision classique du monde ? Ils ne sont certes pas classiques, mais ils ne nécessitent pas non plus de bouleversement conceptuel important. Fait remarquable : ce \textit{modèle jouet} de Spekkens ne parvient ni à violer les inégalités de Bell, ni à reproduire le théorème KS. Ces résultats seraient donc plus étranges que leurs congénères, au sens où ils sembleraient profondément non-classiques. Mais comment les comparer entre eux ? \\

Pour ce faire, on pourrait également exiger d'une notion non-classique qu'elle réponde aux conditions suivantes :\\

Premièrement, elle devrait pouvoir être testée par une expérience directe. Il est primordial d'éviter qu'elle ne soit "même pas fausse" \cite{pauli}. Ensuite, on pourrait espérer qu'elle puisse être utilisée sous la forme d'une ressource \cite{coecke1}, qui apporterait de nouveaux avantages pour traiter l'information. Enfin, elle devrait être applicable à un large ensemble de scénarii physiques. \cite{spekkens10}\\

Si la violation des inégalités de Bell respecte les deux premiers critères, elle ne s'applique cependant qu'à un cas particulier de scenarii, mettant en jeu au moins deux parties. Le résultat KS en revanche, n'a pas besoin que d'un système unique pour être vérifié. Mais, bien qu'il ait été identifié comme ressource \cite{comput}, et que divers résultats expérimentaux en sa faveur semblent concluant, il ne pourrait pas, selon certains, constituer une notion non-classique satisfaisante. En effet, pour pouvoir être comparé au théorème de Bell, il serait primordial qu'il soit généralisé à tout type de mesures, et qu'il s'émancipe ainsi des mesures projectives idéales de la théorie quantique.\\ 

Cinq mois après la présentation de son modèle jouet, Spekkens propose une nouvelle définition de la notion de non-contextualité \cite{spekkens1}. Celle-ci s'inscrit dans le cadre des modèles ontologiques de théories opératoires (section \ref{onto}), comme, par exemple, le modèle jouet. Elle s'applique à tout type de procédures expérimentales (section \ref{moncto}), et s'affranchit donc complètement de l'ancrage quantique de la non-contextualité KS, qu'elle parvient néanmoins à reproduire, moyennant une hypothèse supplémentaire (section \ref{ksaspek}). Elle peut alors être rigoureusement comparée au théorème de Bell (section \ref{tbellmo}). En l'associant à un principe philosophique de Leibniz qu'il considère comme fondement de toute théorie physique, Spekkens suggère d'utiliser la \textit{contextualité universelle} comme un critère d'étrangeté (section \ref{leibnizuni}). La violation d'inégalités de non-contextualité universelle par un modèle physique témoignerait ainsi de sa non-classicité (section \ref{inencuni}). Celle-ci étant avérée pour tout modèle quantique, elle nous invite  alors à réfléchir à la construction de nouveaux édifices théoriques (section \ref{sauvleib}).

\section{Modèles ontologiques de théories opératoires}
\label{onto}
\subsection{Théorie opératoire}
Dans une théorie opératoire, un agent, qui n'est pas décrit par la théorie, "sélectionne" une certaine valeur d'entrée $x \in X$, et "obtient" "ensuite" une valeur de sortie $a \in A$, les ensembles $X$ et $A$ correspondant à des alphabets de cardinalité finie. Une théorie opératoire, au sens le plus général du terme, ne nécessite que deux postulats sur la façon dont les entrées sont transformées en sortie :\\

\begin{itemize}
\item ces deux types de données sont parfaitement distinguables;\\
\item le processus de transformation est physique. Cette physique est entièrement contenue dans la distribution de probabilités $p(a|x)$. \cite{alexei2}\\
\end{itemize}

Intuitivement, il nous semble raisonnable de formuler une hypothèse supplémentaire, systémique, selon laquelle les corrélations entre les entrées choisies par l'observateur et les sorties obtenues sont médiées par des porteurs d'information, les systèmes physiques, que l'on dénotera par des lettres capitales $A$,$B$,$C$... D'un certain point de vue, les systèmes correspondent à des "câbles" ou "fils" reliant des "boîtes", qui symbolisent divers opérations effectuées sur l'information dont dispose l'observateur.\\ 

On suppose qu'il est possible que deux systèmes $A$ et $B$ soient considérés conjointement. Le système composé formé est alors noté $A\otimes B$. Cette composition est associative :
\[ \forall A,B,C \hspace{2cm} A\otimes(B\otimes C) = (A\otimes B)\otimes C \]

et on lui attribue un élément neutre $I$, tel que :
\[\forall A \hspace{2cm} A\otimes I=I\otimes A = A \]

$I$ correspond à un système qui ne porte pas d'information pertinente pour l'observateur. Ainsi, la composition du système $A$ avec $I$ revient à considérer uniquement le système $A$. \cite{chiribella1}\\

Nous allons par la suite distinguer deux sortes d'opérations (ou procédures) :

\begin{itemize}
\item Les \textit{procédures de préparation} $P$, de type $I\rightarrow A$, qui rendent le système $A$ disponible.\\

\item Les \textit{procédures de mesure} $M$, de type $A\rightarrow k B = k\cdot I\otimes B$, qui transforment le système d'entrée $A$ en système de sortie $B$ et produit une valeur scalaire k, appelé le résultat de la mesure.
\end{itemize}

\begin{figure}[ht!]
\centering
\includegraphics[width=5cm]{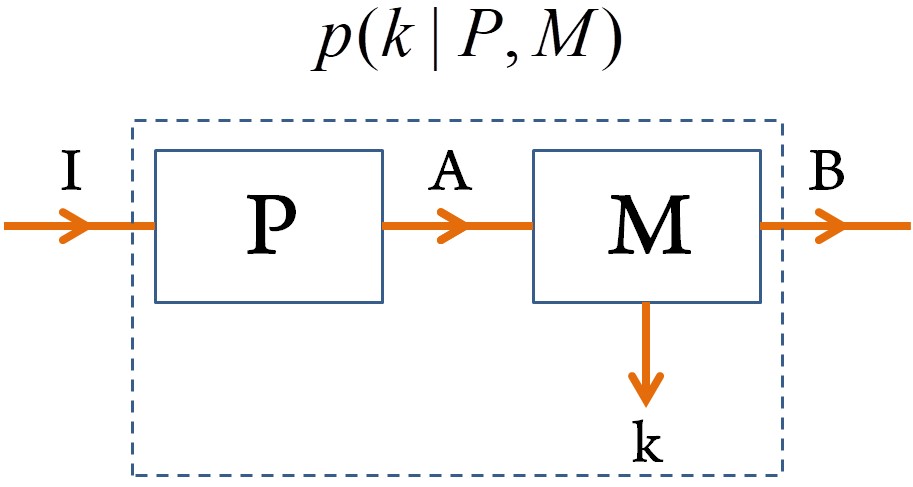}
\caption{Schéma d'une procédure de préparation $P$ suivie d'une mesure $M$}
\end{figure}

Le rôle d’une théorie opératoire est alors simplement de spécifier les probabilités $p(k|P,M)$ qu’une mesure $M$ donne un résultat $k$ pour la procédure de préparation $P$ donnée, où $P$ et $M$ sont des variables que l'on peut contrôler. $k$ correspond a une variable observée. \\

Il est important de souligner que l'opérationnalisation d'une théorie correspond à un mode explicatif de celle-ci, et ne s'apparente en aucun cas à une interprétation philosophique. Une théorie opératoire n'est donc pas l'équivalent d'une position instrumentaliste ou de type "Shut up and calculate."
Si l’approche opératoire a le mérite de permettre de se libérer des préconceptions que l’on peut porter sur le monde, elle ne donne néanmoins pas de sens aux concordances observées entre les prédictions statistiques de la théorie et les résultats expérimentaux. \\

\subsection{Modèle ontologique}

Pour tenter d'expliquer l'éventuel succès d'une théorie opératoire, il est possible de construire un \textit{modèle ontologique}. \cite{spekkens1}. \\

Dans un tel modèle, on suppose qu’il existe des systèmes physiques qui possèdent des caractéristiques indépendantes des tests expérimentaux et de la connaissance que nous en avons. L'ensemble de ces caractéristiques est décrit par l'\textit{état ontologique} du système, du grec ancien \textgreek{ὄντως}, i.e. l'état réel et complet de l'entité physique étudiée. Il est attribué aux quantités physiques qui respectent le critère EPR (1935) : \begin{quotation}
\textit{Si, sans perturber en aucune façon un système physique nous pouvons prévoir avec certitude la valeur d’une quantité physique, alors il existe un élément de la réalité physique qui correspond à cette quantité.} \cite{epr}
\end{quotation}
L'évolution temporelle de ces états est décrite par des théories ontologiques, e.g. la mécanique newtonienne.\\

On peut alors décrire l’ensemble des variables d’un modèle ontologique par $\lambda$, qui représente un état ontologique évoluant dans l’espace des états ontologiques $\Lambda$. Il n'existe pas de restrictions particulières sur la nature de $\lambda$ : il peut correspondre à des trajectoires de particules, des configurations de champs, ou encore des états quantiques.\\

Pour qu'un modèle ontologique puisse reproduire des prédictions opératoires, deux hypothèses sont nécessaires \cite{leifer5}:\\

\begin{itemize}
\item "\textit{Réalisme}" : Pour chaque réalisation expérimentale,  les variables opératoires ($k$,$P$,$M$) et $\lambda$ prennent chacun une valeur définie.\\
\item "\textit{Indépendance}"  : Pour chaque réalisation expérimentale ($k$,$P$,$M$,$\lambda$),  il existe une distribution de probabilité jointe $p(k,\lambda|P,M)$ \end{itemize}
Si ces deux hypothèses sont respectées, on peut donc écrire :
\[ p(k|P,M) = \int_\Lambda d\lambda p(k,\lambda|P,M) = \int_\Lambda d\lambda p(k|\lambda,P,M) p(\lambda|P,M)   \]

Les \textit{procédures de préparation} correspondent à une fixation du système dans un état ontologique particulier. Cet état n'est pas connu de l'observateur, qui ne dispose que des probabilités de fixation ontologique du système. On fait l'hypothèse que ces probabilités sont\textit{ indépendantes de la mesure}\footnote{Cette hypothèse est équivalente à une hypothèse de \textit{libre-arbitre} ou de \textit{liberté du choix de mesure} de l'observateur.}, et donc :
\[ p(\lambda|P,M) = p(\lambda|P) \]
Ainsi, si l'observateur sait qu’un système a été préparé en utilisant la procédure de préparation $P$, il peut le décrire par une densité de probabilité $\mu_P(\lambda) = p(\lambda|P) $ sur l'espace des états ontologiques $\Lambda$, appelée \textit{état épistémique},  telle que pour toute mesure $M$, $\mu_P$ : $\Lambda \rightarrow$ [0, 1],  $\int \mu_P(\lambda) d\lambda = 1$. L'état épistémique, du grec ancien \textgreek{ἐπιστήμη}, correspond donc à l'état de connaissance du système qu'a un agent extérieur à la théorie.\\

De même, des \textit{procédures de mesure} sont des mesures de l’état ontologique du système. Il n'est pas nécessaire que ces procédures dévoilent l’identité de l’état ontologique, ou un ensemble d’états ontologiques auquel appartient l’état ontologique actuel. Elles permettent seulement de déduire les probabilités que le système ait été dans différents états ontologiques. Comme $\lambda$ représente toutes les propriétés physiques du système, on peut supposer qu'il est l'unique médiateur de toutes les corrélations observées entre la préparation et la mesure effectuées. On a donc :
\[ p(k|\lambda,P,M) = p(k|\lambda,M) \]
 Pour chaque valeur de $\lambda$, l'observateur peut associer une probabilité ${\xi_{M,k}(\lambda) = p(k|\lambda,M) }$ appelée \textit{fonction réponse}, qui est la probabilité d’obtenir le résultat $k$ d’une mesure $M$, étant donné que les système est dans l’état ontologique $\lambda$. Pour toute préparation $P$, on a $\xi_{M,k}(\lambda)$ : $\Lambda \rightarrow$ [0, 1],  $\sum \xi_{M,k}(\lambda) = 1$\\
La probabilité d'obtenir le résultat $k$ pour une préparation $P$ et une mesure $M$ correspond donc à la fonction réponse du système, pondérée par l'état épistémique de l'observateur, toutes deux moyennées sur l'ensemble (continu) des états ontologiques :
\[  p(k|P,M) = \int_\Lambda d\lambda \mu_p (\lambda)\xi_M (k|\lambda) \]

La théorie quantique, qui est opératoire, peut être décrite dans ce cadre. On assigne un espace de Hilbert $H$ au système ; un opérateur densité $\rho_P \in H$ tel que pour $\rho$ > 0, Tr($\rho$)=1 à chaque préparation $P$ ;   un POVM\footnote{Un POVM (Positive Operator Valued Measure)  est un ensemble ordonné ${E_k}$ d'opérateurs positifs tels que $\sum_k E_k = I$. On associe au $k^{ième}$ opérateur de l'ensemble, $E_k$, le $k^{ième}$ résultat de mesure. Pour un POVM donné, si $E_k^2 = E_k$ pour tout k, alors les opérateurs positifs $E_k$ sont des projecteurs, et l'ensemble est appelé PVM (Projective Valued Measure). Les mesures associées à ces PVMs sont dites "fines" ou "fortes" (sharp), et correspondent aux mesures standards de la mécanique quantique.
} ${E^M_k}$ à chaque mesure $M$. Les prédictions opératoires de la théorie sont de la forme :
\[ p(k|P,M)  = Tr(\rho_P E^M_k) = \int_\Lambda d\lambda \mu_p (\lambda)\xi_M (k|\lambda)  \]

Tout modèle ontologique associé à la théorie quantique correspond à une interprétation réaliste. On distingue alors trois catégories de \textit{modèles quantiques} \cite{harrigan} \cite{leifer6}:\\

\begin{itemize}
\item les modèles $\psi-ontologiques$, dans lesquels tout état ontologique $\lambda$ est entièrement décrit par un état quantique pur $\ket{\psi}$ (e.g. interprétation de Dirac-von Neumann);\\
\item les modèles $\psi-complétés$, dans lesquels $\lambda = (\ket{\psi},\omega)$ où la variable $\omega$ complète l'information du vecteur d'état $\ket{\psi}$ pour obtenir l'état ontologique (e.g. mécanique bohmienne);\\
\item les modèles $\psi-épistémiques$, dans lesquels deux états quantiques distincts et non orthogonaux correspondent à deux distributions de probabilité qui se recouvrent pour au moins un état ontologique (e.g. interprétation d'Einstein \cite{harrigan}).\\
\end{itemize}

Deux modèles ontologiques seront particulièrement pertinents pour la suite \cite{leifer7} :

\subsubsection{Le modèle de Beltrametti-Bugajski}
Ce modèle $\psi-ontologique$ correspond à l'interprétation orthodoxe de la théorie quantique, adaptée au cadre des modèles ontologiques.
L'espace des états ontologique $\Lambda$ est l'espace de Hilbert des projecteurs. Pour chaque état pur $\ket{\psi}$ préparé, l'état épistémique est :
\[p(\lambda|\psi) = \delta (\ket{\lambda}\bra{\lambda} - \ket{\psi}\bra{\psi})\]
Comme chaque état ontologique est entièrement compris dans un état pur $\ket{\psi}$, connaître l'état quantique du système revient à connaître un élément de réalité. Pour une mesure POVM $M=\{E_k\}$, la fonction réponse du système est :
\[p(k|M,\lambda) = Tr(E_k|\ket{\lambda}\bra{\lambda})\]
Par construction, on obtient :
\[ Tr(E_k|\ket{\lambda}\bra{\lambda}) = \int_\Lambda d\lambda p(k|M,\lambda) p(\lambda|\psi) \]

\subsubsection{Le modèle de Kochen-Specker}
Ce modèle est exposé dans l'article de 1967 de Kochen et Specker, comme contre-exemple de leur théorème dans le cas où l'espace de Hilbert considéré est à deux dimensions.
L'espace des état ontologique $\Lambda$ est une sphère unité, de telle sorte que tout état ontologique peut être écrit comme un vecteur unitaire $\vec{\lambda} \in \Lambda$. On associe à l'état quantique $\psi$ l'état épistémique :
\[p(\vec{\lambda}|\vec{\psi}) = \frac{1}{\pi}H(\vec{\psi}.\vec{\lambda})\vec{\psi}.\vec{\lambda}\]
La fonction $H : x \rightarrow H(x)$ est la fonction de Heaviside : $H(x)=1$ si $x>0$ et $H(x)=0$ sinon.
Pour une mesure $M$ dans la base orthonormée $\{\ket{\psi},\ket{\psi^\perp}\}$ telle que $M=\{\ket{\psi}\bra{\psi},\ket{\psi^\perp}\bra{\psi^\perp} \}$, la fonction réponse est :
\[ p(\vec{\psi}|\vec{\lambda},M) = H(\vec{\psi}.\vec{\lambda}) \]
On a donc 
\[ p(\phi|\psi)  = \int_\Lambda d\lambda p(k|M,\lambda) p(\lambda|\psi)\\ 
				= \int_\Lambda d\lambda \frac{1}{\pi}H(\vec{\psi}.\vec{\lambda})H(\vec{\phi}.\vec{\lambda})\vec{\psi}.\vec{\lambda}\\
                = \frac{1}{2}(1+\vec{\psi}.\vec{\phi})\\
                =|\bra{\psi}\ket{\phi}|^2\]
Le modèle reproduit bien les statistiques quantiques.
D'autre part, 
\[p(\lambda|\psi)p(\lambda|\phi)=\frac{1}{\pi^2}H(\vec{\psi}.\vec{\lambda})H(\vec{\phi}.\vec{\lambda})\vec{\psi}.\vec{\lambda}\vec{\phi}.\vec{\lambda} \]
est non nul pour des états $\psi$ et $\phi$ non-orthogonaux. Ainsi, le modèle de Kochen-Specker est $\psi-épistémique$. Il est aussi déterministe : $p(\vec{\lambda}|\vec{\psi})\in\{0,1\}$; et non-contextuel : $p(\vec{\psi}|\vec{\lambda},M) = p(\vec{\psi}|\vec{\lambda})$, et ne vérifie donc pas le théorème KS.\\

Le cadre des modèles ontologiques et théories opératoires permet  à la fois d'étudier et comparer différents modèles ontologiques pour une théorie donnée, e.g. la théorie quantique, mais aussi de comparer différentes théories opératoires entre elles, via des modèles ontologiques respectifs.\\

Il est par exemple possible d'affilier à toute théorie classique une théorie statistique \cite{spekkens7}, qui décrive les distributions statistiques sur l'espace des états physiques et l'évolution temporelle de ces états. La mécanique de Liouville est ainsi l'équivalent statistique de la mécanique (classique) ; et l'optique statistique l'équivalent statistique de l'optique. Dans une théorie statistique, le système étudié se trouve dans un état physique appartenant à un ensemble de possibilités. Les distributions statistiques associées peuvent être interprétées de deux façons différentes. On peut considérer qu'elles permettent de décrire les fréquences relatives des propriétés physiques au sein de l'ensemble  des états, ou bien qu'elles décrivent la connaissance qu'a un agent de l'état du système lorsqu'il sait que celui-ci est issu de cet ensemble. C'est dans cette dernière approche des probabilités, dite \textit{bayesienne}, que l'on distingue les deux types d'états présentés : l'état ontologique et l'état épistémique. \\

Il est possible d'ajouter à une théorie statistique, à laquelle on aurait associée un modèle ontologique, une contrainte sur les états épistémiques. L'ensemble des théories épistémiquement restreintes sont appelés \textit{théories épistreintes} \cite{spekkens7}.\\

Ces théories épistreintes peuvent être utilisées comme "\textit{foil theories}\footnote{En anglais : foil theories. Un foil de X est quelque chose qui permet de souligner les caractéristiques différente de X par contraste avec lui. Origine : personnage roman}" : elles permettent d'apporter une clarification conceptuelle des principes physiques qui pourraient sous-tendre une théorie, ici la théorie quantique. \\

\subsection{Le modèle jouet de Spekkens}

La première théorie épistreinte, le \textit{modèle jouet} ("toy theory"), a été construite en 2004 par Robert Spekkens (\cite{spekkens5}). Dans cette théorie, la restriction épistémique, appelée "principe d'équilibre de savoir" ("knowledge-balance principle"), impose que dans un état de connaissance maximal, l'agent en sait autant sur le système qu'il en ignore. \cite{leifer5} \\

Prenons l'exemple d'un système pouvant se trouver dans quatre états ontologiques possibles :
\[\Lambda = \{(+,+),(+,-),(-,+),(-,-)\} \]
L'état ontologique peut être par exemple représenté par une balle, qui peut se trouver sous quatre gobelets différents. 
\begin{figure}[ht!]
\centering
\includegraphics[width=6cm]{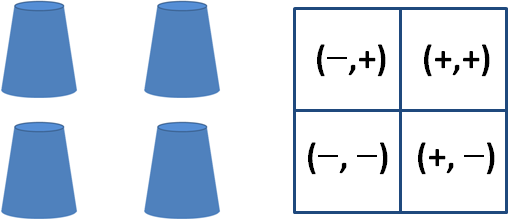}
\caption{Etats ontologiques possibles du système.}
\end{figure}

Pour déterminer l'état ontologique, i.e., sous quel gobelet se trouve la balle, un minimum de deux questions binaires est nécessaire :\\

"la balle est-elle sous un gobelet de la première colonne $(-,.)$ ou de la deuxième colonne $(+,.)$ ?";\\
"la balle est-elle sous un gobelet de la première ligne $(.,+)$ ou la deuxième ligne $(.,-)$ ?".\\

Le principe d'équilibre de savoir rend ces questions incompatibles : une réponse à la première question nous permet de connaître avec certitude la colonne dans laquelle la balle est située, mais ne nous donne aucune information sur la ligne, et vice-versa. \\

On distingue alors 6 préparations possibles, et donc 6 états épistémiques $p(\lambda|P)$ : 

\begin{figure}[ht!]
\centering
\includegraphics[width=10cm]{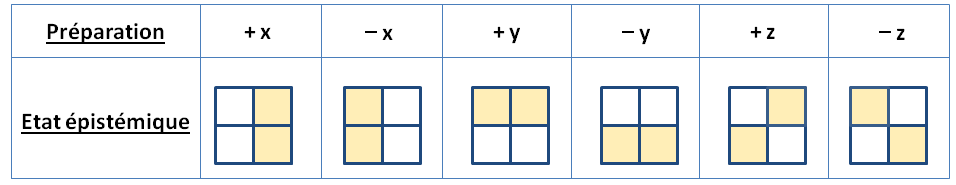}
\caption{Etats épistémiques (maximaux) possibles du système.}
\end{figure}

Ces six états sont les états de connaissance maximale autorisés par le principe d'équilibre de savoir. Il existe également un état de connaissance non maximale : 

\begin{figure}[ht!]
\centering
\includegraphics[width=3cm]{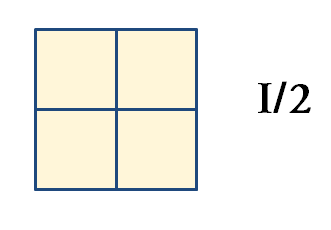}
\caption{Etat épistémique non maximal du système.}
\end{figure}

Pour une préparation $+x$, l'agent sait que la balle se trouve sous un gobelet de la deuxième colonne. Il ignore en revanche dans quelle ligne se trouve la balle : si c'était le cas, il connaitrait avec certitude l'état ontologique du système, ce qui violerait la restriction épistémique. Il en va de même pour toutes les mesures, qui en plus d'être répétables (elles doivent rendre le même résultat si elles sont effectuées deux fois de suite), doivent être compatibles avec le principe d'équilibre de savoir, et donc laisser le système dans un état épistémique valable. Les mesures peuvent uniquement révéler de l'information "coarse-grained", et perturbent l'état ontologique.

\begin{figure}[ht!]
\centering
\includegraphics[width=12cm]{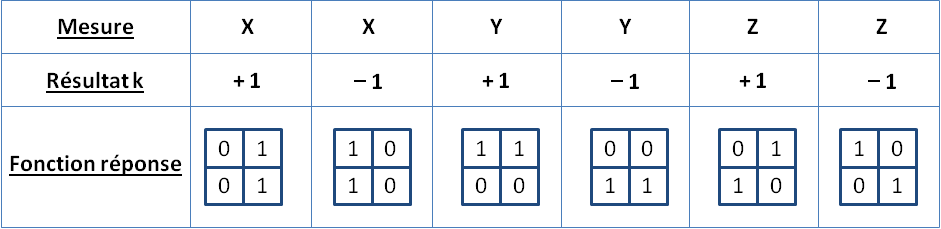}
\caption{Fonctions réponses du système.}
\end{figure}

Il n'existe donc qu'un ensemble fini de mesures adéquates, répertoriées dans le tableau 4.5. 
Prenons par exemple la mesure $M=X$, qui donne une sortie $k=\pm 1$ :\\
\begin{figure}[ht!]
\centering
\includegraphics[width=2cm]{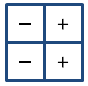}
\caption{Répartitions des résultats pour une mesure X.}
\end{figure}

Si l'on appliquait la mesure X à l'état épistémique $+y$, et que l'on obtenait le résultat $k=+1$, alors on saurait que l'état ontologique est $(+,+)$ avant la mesure.
Pour préserver le principe d'équilibre de savoir, et maintenir la répétabilité, les états $(+,+)$ et $(+,-)$ doivent être échangés avec une probabilité $1/2$ pendant la mesure. Ainsi, l'état épistémique, mis à jour,  reste compatible après mesure. Dans cet exemple, l'état épistémique final est $+x$.\\

\begin{figure}[ht!]
\centering
\includegraphics[width=10cm]{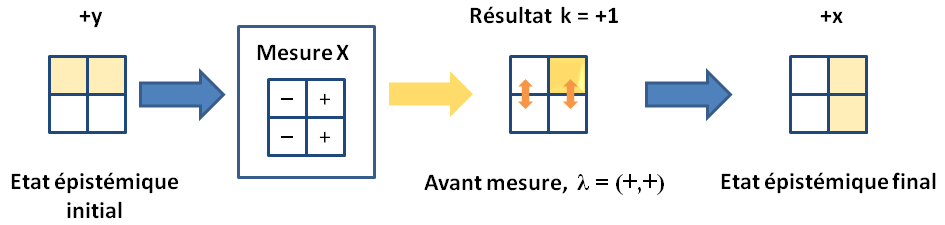}
\caption{Exemple d'évolution de l'état épistémique après mesure.}
\end{figure}

\begin{figure}[ht!]
\centering
\includegraphics[width=12cm]{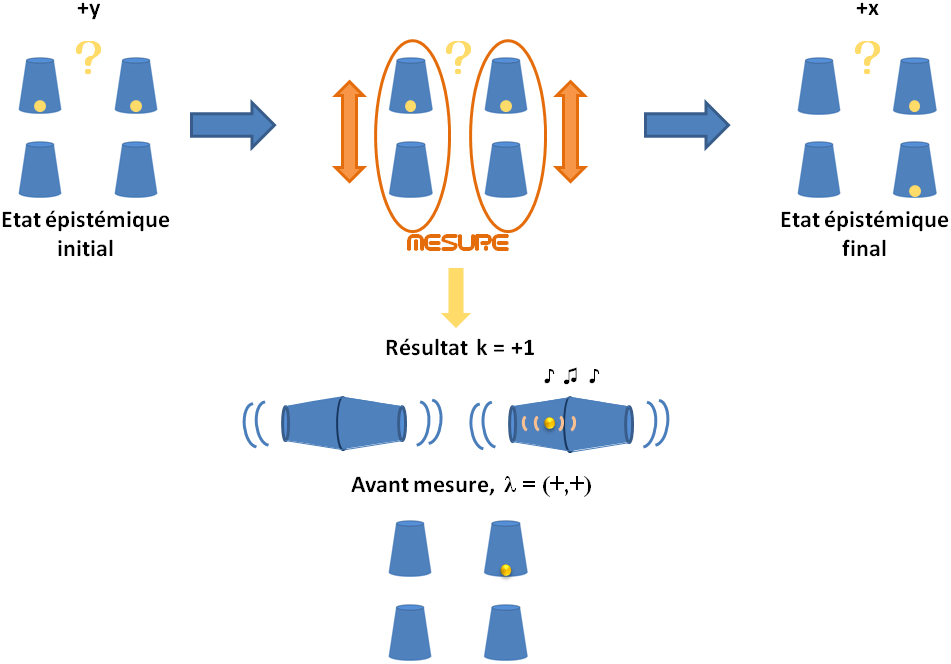}
\caption{Exemple d'évolution de l'état épistémique après mesure (image des gobelets).}
\label{fgob}
\end{figure}

Ramenée à l'image des gobelets et de la balle cachée, la mesure $X$ revient à mélanger les gobelets de chaque colonne entre eux. Dans le cas où l'état épistémique initial est $+y$ et le résultat de la mesure est $k=+1$, on remarque pendant le mélange des gobelets de la deuxième colonne que la balle est présente. On peut donc en conclure que la balle était dans cette colonne et à la première ligne avant la mesure, i.e. dans l'état ontologique $(+,+)$. Cependant, la mesure perturbe la position de la balle : l'état épistémique finale est $+x$. Du fait du mélange, on sait que la balle se trouve dans la deuxième colonne, mais on ne sait pas à quelle ligne (cf. figure \ref{fgob}).\\

Le modèle jouet permet de reproduire un grand nombre de résultats de la mécanique quantique : la non-commutativité des mesures, le phénomène d'interférence, la superposition des états, la multiplicité des décompositions convexes d'un état mixte, l'impossibilité de discriminer des états non-orthogonaux, la distinction entre intrication bi-partite et tri-partite, le non-clonage, la téléportation, et bien d'autres \cite{spekkens5} \cite{leifer5}. Ceux-ci ne sont donc pas particulièrement \textit{étranges}, puisqu'il suffit d'ajouter une \textit{restriction épistémique} à une théorie statisique classique pour les reproduire. Selon Spekkens, cette imitation inviterait à considérer les états quantiques comme états épistémiques.\\

Le modèle jouet ne reproduit néanmoins pas l'ensemble des résultats quantiques : il s'agit d'une "foil theory", et non d'un modèle quantique. Ainsi, il n'est pas causalement local, et ne viole donc pas les inégalités de Bell. Il ne respecte pas non plus le théorème KS.\\

D'une part, le modèle jouet est déterministe : si l'on connait l'état l'ontologique $\lambda$ du système et la mesure effectuée, alors on peut prédire avec certitude le résultat de la mesure. Si la position de la balle pouvait être connue de l'agent, celui-ci serait capable de prédire avec certitude le résultat de toute mesure. La fonction réponse , i.e. la probabilité d'obtenir le résultat $k$ pour une mesure $M$ sur le système dans l'état ontologique $\lambda$, est donc telle que :
\[p(k|M,\lambda) \in\{0,1\}\]
D'autre part, le modèle jouet est également à mesures non-contextuelles, l'état ontologique $\lambda$ étant suffisant pour déterminer la fonction réponse du système :
\[p(k|M,\lambda) =  p(k|\lambda)\]
Le modèle jouet respecte les deux hypothèses sans contradiction. Par construction, ses mesures sont KS-non-contextuelles.\\

Comme le modèle ontologique de Kochen-Specker le montre, le théorème KS ne s'applique pas pour des qubits (système à deux dimensions). Il est par ailleurs restreint à un genre particulier de mesures, projectives : centré sur la mécanique quantique, il s'effondre lorsque des POVMs sont considérées.\\

Ces lacunes ont été comblées par Robert Spekkens \cite{spekkens1}, qui propose, dans son article "\textit{Contextuality for preparations, transformations and unsharp measurements}", une généralisation du concept de non-contextualité s'inscrivant dans le cadre des modèles ontologiques, qui s'applique à tout type de procédures expérimentales (mesures, préparations, transformations) et pour toute théorie opératoire.\\

\section{Modèle ontologique non-contextuel d'une théorie opératoire}
\label{moncto}
La nouvelle approche du concept de (non-)contextualité de Spekkens est la suivante :\\

\boitesimple{\emph{Dans un modèle ontologique non-contextuel d'une théorie opératoire, si deux procédures expérimentales sont opérationnellement équivalentes, alors elles ont des représentations ontologiques équivalentes.} \cite{spekkens1}\\}

Deux procédures de préparations sont dites équivalentes si elles produisent les mêmes données statistiques pour toute procédure de mesure, i.e. 
\[ P \sim Q \hspace{3mm}  si \hspace{3mm}\forall M,  \hspace{3mm} p(k|P,M) = p(k|Q,M)   \]

De même, deux procédures de mesure sont dites équivalentes si elles produisent les mêmes données statistiques pour toute procédure de préparation\footnote{Cette définition a inspiré la notion d'équivalence opératoire à AFLS dans leur approche combinatoire. \cite{sainz3}) }, i.e. 
\[ (M,k) \sim (N,l) \hspace{3mm}  si \hspace{3mm} \forall P, \hspace{3mm} p(k|P,M) = p(l|P,N)   \]

L'ensemble des caractéristiques d'une procédure expérimentale qui ne sont pas désignées en spécifiant la classe d'équivalence opératoire de celle-ci est appelé \textit{contexte} de la procédure expérimentale. On remarque qu'avec cette définition du contexte expérimental, avoir connaissance du contexte ne permet pas de mieux prédire le résultat d'une expérience que dans le cas où l'on a uniquement connaissance de la classe d'équivalence de la procédure expérimentale. Par ailleurs, le contexte ne se limite plus à la commutabilité entre opérateurs hermitiens, comme c'était le cas dans le théorème Kochen-Specker, mais il inclut une notion de compatibilité plus générale, représentée par la mesure jointe.\cite{kunjwal1}\\

Un modèle ontologique est à \textit{préparations non-contextuelles} (ou non-contextualité de préparation) si :
\[  P \sim Q \hspace{3mm} \Rightarrow \hspace{3mm} p(\lambda|P)=p(\lambda|Q)\]
l'équivalence entre deux préparations implique l'identité des états épistémiques.\\

On peut par exemple distinguer différentes façons de préparer l'état mixte maximal. 

\begin{figure}[ht!]
\centering
\includegraphics[width=12cm]{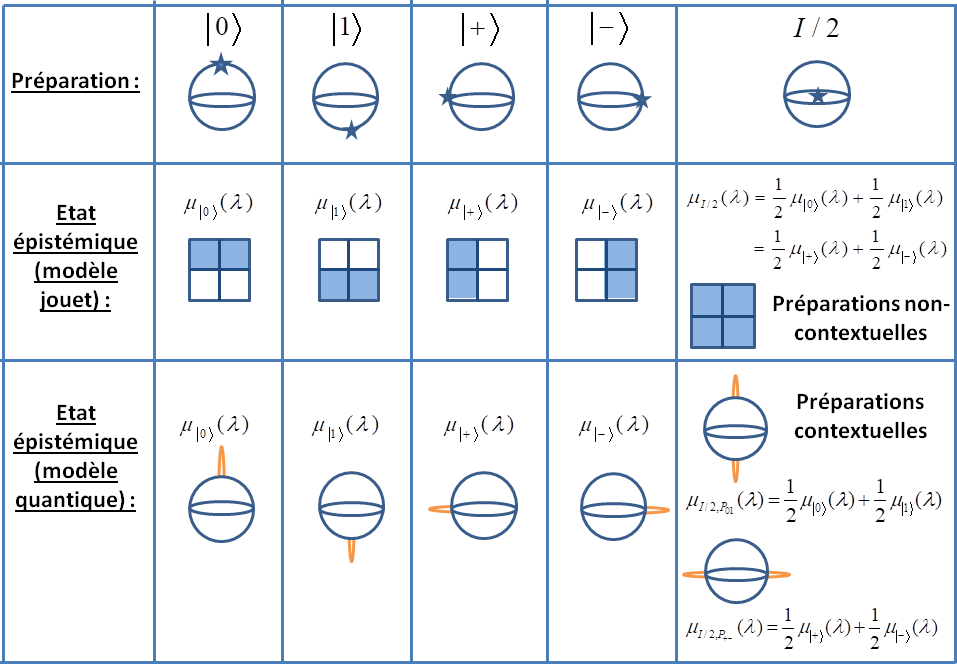}
\caption{Contexutalité et Non-Contextualité de Préparations, d'après \cite{spekkens10}}
\label{prep}
\end{figure}

$P:=\frac{1}{2}I=\frac{1}{2}\ket{0}\bra{0} + \frac{1}{2}\ket{1}\bra{1}$ et $Q:=\frac{1}{2}I=\frac{1}{2}\ket{+}\bra{+} + \frac{1}{2}\ket{-}\bra{-}$. $P$ et $Q$ sont équivalentes. Pour un modèle ontologique à préparations non-contextuelles comme le modèle jouet de Spekkens, les états épistémiques correspondants sont égaux, i.e.\\ $\mu_{I/2}(\lambda)=p(\lambda|I/2)=\mu_{I/2,P_{01}}(\lambda)=\mu_{I/2,P_{+-}}(\lambda)$ est unique (cf. figure \ref{prep}). Pour un modèle ontologique à préparations contextuelles, par exemple un modèle quantique, l'équivalence entre préparations n'implique pas l'identité des états ontologiques \\$\mu_{I/2,P_{01}}(\lambda)\neq\mu_{I/2,P_{+-}}(\lambda)$. \\

Un modèle ontologique est à \textit{mesures non-contextuelles} (ou non-contextualité de mesure) si :
\[  (M,k) \sim (N,l) \hspace{3mm}  \Rightarrow \hspace{3mm} p(k|M,\lambda)=p(l|N,\lambda)\]
Lorsqu'il n'y a pas de distinctions observables entre deux paires de résultats de mesure, elles sont représentées par la même fonction réponse dans le modèle ontologique.\\

\begin{figure}[ht!]
\centering
\includegraphics[width=6cm]{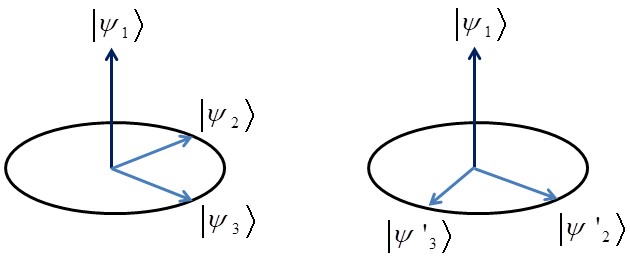}
\caption{Deux contextes de mesures.}
\label{bases}
\end{figure}

Pour la mesure représentée par le PVM $\{\ket{\psi_1}\bra{\psi_1},I-\ket{\psi_1}\bra{\psi_1}\}$, on peut distinguer différents contextes, i.e. différentes bases orthonormées de mesure, par exemple $\{\ket{\psi_1},\ket{\psi_2},\ket{\psi_3}\}$ et  $\{\ket{\psi_1},\ket{\psi_2'},\ket{\psi_3'}\}$ (cf. figure \ref{bases}). On a donc un contexte où\\ $I-\ket{\psi_1}\bra{\psi_1}=\ket{\psi_2}\bra{\psi_2}+\ket{\psi_3}\bra{\psi_3}$, et un autre où $I-\ket{\psi_1}\bra{\psi_1}=\ket{\psi_2'}\bra{\psi_2'}+\ket{\psi_3'}\bra{\psi_3'}$. Si la fonction réponse du modèle ontologique dépend de la base dans laquelle la mesure est effectuée, alors le modèle est à mesures contextuelles.\\

Par souci de clarté, nous ne tenons pas compte ici des procédures de transformation, celles- ci pouvant, dans la plupart des cas, être intégrées aux procédures de préparation. On notera tout de même que l'on peut également définir une non-contextualité de transformation. \cite{spekkens1}

Ces nouvelles notions de non-contextualité sont applicables à toute théorie opératoire, et son indépendantes de la dimension de l'espace des états considérés (cf. sections IV,V,VI de \cite{spekkens1} et l'annexe du mémoire \textit{Preuve de contextualité de préparation pour un qubit}.)\\ 

Ce que nous analysions comme un problème de sémantique ("contextualité-KS" ou "contextualité" ?) ne l'est plus dans l'approche de Spekkens : l'indéterminisme n'implique pas nécessairement la contextualité de mesure, puisqu'il existe des modèles ontologiques réalistes et non-déterministes, qui ne sont pas nécessairement à mesures contextuelles. \\

C'est par exemple le cas du modèle Beltrametti-Bugajski. Ce modèle ne remet pas en cause le théorème de Kochen-Specker. Il stipule simplement que ce n'est pas l'hypothèse de non-contextualité des mesures qui serait la cause de la contradiction logique, mais plutôt l'hypothèse de déterminisme des résultats. Par ailleurs, il n'est pas dépourvu de contextualité. Seulement, celle-ci n'est pas confinée dans les procédures de mesures, mais dans les procédures de préparation. Le modèle est en effet à préparations contextuelles puisque la distribution qui représente une combinaison convexe de procédures de préparations dépend de l'ensemble particulier des états purs, et non uniquement de l'opérateur densité associé au mélange.\\

Il est cependant impossible de convertir un modèle à contextualité de préparations en modèle à contextualité de mesure, autrement dit de confiner la contextualité dans les procédures de mesure, puisque l'hypothèse de non-contextualité de préparation fournit une contradiction par elle-même, sans avoir besoin de faire appel à d'autres hypothèses comme celle du déterminisme.
\begin{quotation}
\textit{En ce sens, la contextualité de préparation est plus fondamentale pour la théorie quantique que la contextualité de mesure.} \cite{spekkens1}
\end{quotation}

Comme le démontre Spekkens dans son article (\cite{spekkens1}, section IV), pour tout modèle ontologique de la théorie quantique, l'hypothèse de non-contextualité des préparations est nécessairement violée, ce qui n'est pas le cas pour les procédures de mesures. Or, la non-contexutalité-KS est composée du couple "non-contextualité de mesure - déterminisme des résultats". Il est donc légitime de se demander comment l'approche de Spekkens intègre-t-elle le résultat de Kochen-Specker au sein de son formalisme.

\section{KS à la Spekkens}
\label{ksaspek}
\textit{Rappel} : 
Le théorème de Kochen-Specker conduit à la contradiction logique suivante :
\[ MQ \wedge NonContextualité-KS \rightarrow contradiction\]
où $MQ$ désigne le formalisme quantique et ses prédictions ;
et où la NonContextualité-KS correspond au couple :
\begin{itemize}
\item non-contexutalité des mesures projectives : $p(\Pi|M,\lambda) = p(\Pi|\lambda)$
\item déterminisme des résultats de mesures projectives : $p(\Pi|M,\lambda)\in\{0,1\}$\\
\end{itemize}

En supposant le formalisme quantique comme théorie opératoire donnée à modéliser, le résultat KS contraint donc à choisir entre abandonner\\
\begin{itemize}
\item la non-contextualité de mesure;
\item le déterminisme des résultats;
\item les deux hypothèses.\\
\end{itemize}

En généralisant la notion de contexte aux \textit{équivalences opératoires entre mesures}, les approches graphiques (AFLS, CSW) ont néanmoins permis d'étendre le résultat KS à l'ensemble des modèles probabilistes, et non exclusivement aux modèles quantiques. \`A l'aune de ces travaux, la contradiction logique présentée ci-dessus peut donc être mise à jour

\[ EOM \wedge NonContextualité-KS \rightarrow contradiction\]

avec $EOM$ correspondant à toutes théories pour lesquelles l'équivalence opératoire pour des ensembles de mesures est définie. \\

L'approche de Spekkens ayant décomposé la non-contextualité-KS en découplant la non-contextualité des mesures et le déterminisme des résultats, comment parvient-elle à reproduire le théorème opérationnellement, autrement dit à justifier le déterminisme des résultats ? 

Il a été montré que pour tout modèles ontologiques de la théorie quantique, la \textit{non-contextualité de préparation} et la \textit{prévisibilité parfaite} des mesures projectives pour les états propres correspondants,  impliquent le déterminisme des résultats. (cf. (\cite{spekkens1}-VIII-A, \cite{kunjwal2}-1.3.2 pour démonstration).\\

Dans \cite{kunjwal3},
cette condition de prévisibilité parfaite est dissociée en deux contraintes :\\

D'une part, l'hypothèse des \textit{corrélations parfaites}, qui stipule que pour $i$ ensembles de $k$ préparations données, la mesure $M_i$ sur la préparation $P_{i,k}$ fournit le $k$-ième résultat avec certitude,

\[\forall i, \forall k, \hspace{2cm} p(k|M_i,P_{i,k})=1. \]

D'autre part, en définissant la préparation effective $P_i^{(ave)}$ comme étant la procédure obtenue en échantillonnant $k$ aléatoirement et uniformément et en implémentant ensuite $P_{i,k}$, on suppose que la condition d'\textit{équivalence opératoire} suivante est vérifiée :
\[ \forall i, \forall i',\hspace{2cm} P_i^{(ave)} \simeq P_i'^{(ave)} \]
Dans le cadre d'une théorie quantique, les $P_i^{(ave)}$ correspondent simplement à différentes façons de préparer l'état mixte maximal. \\

Par exemple, pour $P_{1,2} := \{\ket{0}\bra{0};\ket{1}\bra{1}\}$ et $P_{2,2} := \{\ket{+}\bra{+};\ket{-}\bra{-}\}$ , on obtient $P_1^{(ave)}:=\frac{1}{2}I=\frac{1}{2}\ket{0}\bra{0} + \frac{1}{2}\ket{1}\bra{1}$ et $P_2^{(ave)}:=\frac{1}{2}I=\frac{1}{2}\ket{+}\bra{+} + \frac{1}{2}\ket{-}\bra{-}$\\

\textit{Démonstration} :\\
La non-contextualité de préparation et cette hypothèse d'équivalence opératoire conduisent à l'existence d'une unique distribution $\nu(\lambda)$ sur l'espace des états ontologiques $\Lambda$, telle que
\[ \forall i, \forall i', \hspace{2cm} p(\lambda|P_i^{(ave)})=p(\lambda|P_i'^{(ave)})\equiv\nu(\lambda)\]

Par la définition de $P_i^{(ave)}$, on obtient donc \\

\[ \forall i, \forall \lambda, \hspace{2cm} \nu(\lambda)=\frac{1}{max(k)} \sum_k p(\lambda|P_{i,k}) \]

En se rappelant de la relation liant les prédictions statistiques aux fonctions réponses et états épistémiques,

\[ p(k|P,M) = \sum_\lambda p(k|M,\lambda)p(\lambda|P) \]

et afin de reproduire la condition des corrélations parfaites, on doit avoir

\[ \forall i, \forall k, \hspace{2cm}  \sum_\lambda p(k|M_i,\lambda)p(\lambda|P_{i,k}) = 1   \]

Or, chaque état ontologique $\lambda$ dans le support de $\nu(\lambda)$ apparait dans le support de $p(\lambda|P_{i,k})$ pour un $k$ donné. Il en résulte que si $p(k|M_i,\lambda)$ avait une réponse indéterministe pour un tel $\lambda$, cela entrerait en contradiction avec la condition des corrélations parfaites. Par conséquent, pour tout $i$ et pour tout $k$, le résultat de l'évènement $[k|M_i]$ doit être déterminé pour tout $\lambda$ dans le support de $\nu(\lambda)$ \cite{kunjwal3}.\\

\[ \forall\lambda\in supp(\nu), \hspace{2cm} p(k|M_i,\lambda)\in \{0,1\}  \]

Le déterminisme des résultats est alors justifiée.\\

\begin{figure}[ht!]
\centering
\includegraphics[width=12cm]{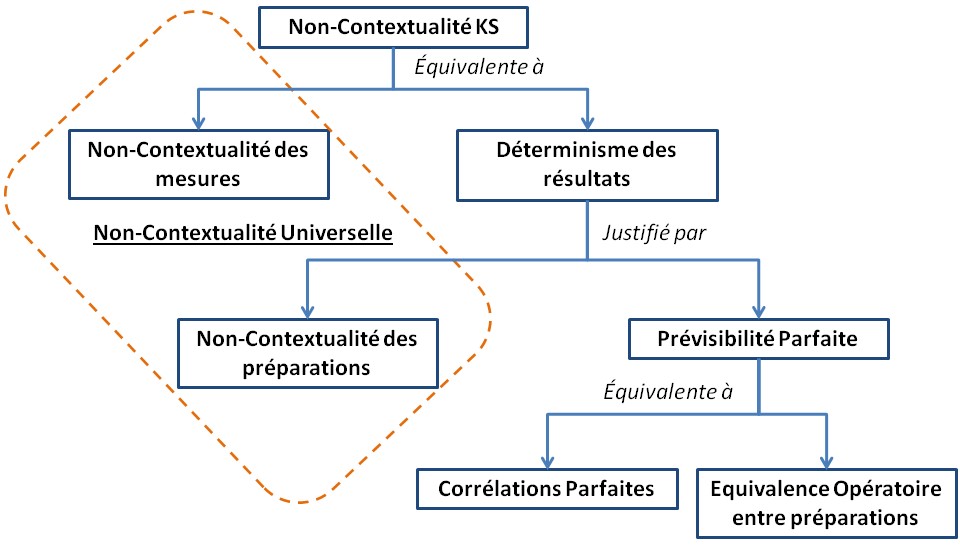}
\caption{Diagramme récapitulatif des relations entre non-contextualité-KS et non-contextualité de Spekkens.}
\label{diag1}
\end{figure}

Ainsi, l'hypothèse de non-contextualité-KS peut être dérivée à partir des deux notions de non-contextualité opératoires de Spekkens. Néanmoins, le résultat KS ne peut être considéré comme un cas particulier de la contextualité opératoire, puisque l'hypothèse du déterminisme de Kochen-Specker n'apparaît pas chez Spekkens, et que la contextualité-KS n'est donc pas une conséquence logique de la contextualité opératoire. De même, le contextualité de Spekkens ne constitue pas une généralisation de la contextualité-KS, au sens où elle ne postule pas le théorème KS comme les approches CSW et AFLS.\\

\`A la différence de la non-contextualité-KS,  l'hypothèse de non-contextualité de mesure de Spekkens est opératoire et indépendante de la théorie considérée. La notion de contexte est devenue opératoire, et n'est plus restreinte à des relations de compatibilité entre observables. C'est également le cas de l'approche AFLS, dont la notion d'équivalence opératoire est inspiré des travaux de Spekkens. Cependant, la contextualité chez AFLS est limitée aux mesures projectives\footnote{Il en va de même pour les approches graphique et faisceau. Celles-ci pouvant être inclues dans l'approche combinatoire, nous ne ferons par la suite uniquement mention des travaux de AFLS, sauf mention contraire.}, tandis que Spekkens la généralise à tout type de mesures, idéales ou non. \\

Par ailleurs, Spekkens a montré qu'un modèle ontologique à préparations et mesures non-contextuelles est déterministe si et seulement si les mesures sont fines, i.e. projectives \cite{spekkens9}. Ainsi, toute hypothèse de déterminisme des résultats implique la réalisation de mesures idéales, non bruitées, et réciproquement, si des mesures projectives sont considérées, cela revient à considérer implicitement que les résultats de mesures sont déterminés. Dans l'approche AFLS, les mesures sont considérées comme projectives : l'hypothèse de non-contextualité est donc couplée à celle du déterminisme des résultats, l'ensemble formant l'hypothèse de non-contextualité-KS.\\

Si une inégalité CSW est violée, l'échappatoire de l'indéterminisme existe toujours. Une telle violation n'est donc pas en soit une preuve de contextualité, mais plutôt une preuve du résultat de Kochen-Specker (ou de contextualité-KS). Il peut alors être argué que la non-localité et la contextualité ne pouvaient pas être rigoureusement comparées puisque c'est la contextualité-KS qui était en fait étudiée.  Quand est-il donc des relations entre non-localité et contextualité au sein de l'approche de Spekkens ? Comment le théorème de Bell s'inscrit-il dans le cadre des modèles ontologiques ?

\section{Théorème de Bell et Modèles Ontologiques}
\label{tbellmo}

Le théorème de Fine (1982) \cite{fine1} \cite{fine2}, stipule que si l'on peut trouver modèle ontologique indéterministe et local pour un certain ensemble de corrélations dans le cadre d'un test de type Bell, alors on peut aussi trouver un modèle ontologique déterministe pour le même ensemble. Il s'en suit que si l'on élimine les modèles déterministes locaux pour un ensemble de corrélations particulier, alors on élimine également les modèles locaux indéterministes. Dans \cite{spekkens9}, 
Spekkens démontre qu'il n'existe pas d'équivalent au théorème de Fine pour les modèles contextuels, autrement dit, éliminer des modèles déterministes pour des mesures POVMS n'implique pas l'élimination des modèles indéterministes pour ces mesures, et inversement. Une preuve intuitive est la suivante \cite{spekkens9} : \\

S'il est vrai que l'on peut toujours représenter une mesure qui dépende de façon indéterminée de l'état ontologique du système par un modèle qui dépend de façon déterminée de l'état ontologique d'un système plus grand (incluant par exemple des variables cachées dans l'appareil de mesure), néanmoins, pour un tel changement de représentation, on découvre que les évènements de mesure qui diffèrent seulement par un choix de contexte sont associés à des évènements de mesure qui se distinguent par leurs statistiques opératoires. Par conséquent, l'application de l'hypothèse de non-contextualité n'est plus justifiée. "\textit{Pour des mesures bruitées, l'état ontologique doit affecter des résultats de façon indéterminée.}" \cite{spekkens10}\\

L'hypothèse du déterminisme des résultats n'a donc pas le même statut dans les théorèmes de Bell et de Kochen-Specker. Ce résultat nous invite à nouveau à rester prudents et à éviter des assertions précipitées du type "la contextualité (de mesure) est une généralisation de la nonlocalité".\footnote{Encore une fois, cela ne remet néanmoins pas en question les résultats de l'approche AFLS, au sein de laquelle, par définition, les scenarii de Bell sont "des cas particuliers" de scenarii de contextualité, au sens où ils sont construits à partir de produits de  Fouli-Randall de ces derniers.} En effet, une telle déclaration impliquerait que toute preuve de nonlocalité serait une preuve de contextualité. Or, puisqu'il existe des preuves de nonlocalité qui ne font pas l'hypothèse du déterminisme, il devrait donc exister des preuves de contextualité de mesure qui ne la font pas non plus. 
\c{C}a ne peut être le cas, puisque l'hypothèse de non-contextualité de mesure n'est pas en soit incompatible avec les prédictions quantiques, comme en atteste l'existence de modèles ontologiques quantiques à mesures non-contextuelles, tel que le modèle de  Beltrametti-Bugajski. La contradiction logique émerge lorsque le déterminisme est également pris en compte. \\

Dans \cite{spekkens1}, Spekkens invite à considérer deux notions distinctes de localité, proposées par Don Howard \cite{howard2} : la séparabilité, ou \textit{indépendance des paramètres}, et la \textit{causalité locale}. C'est seulement dans le cas où cette dernière est violée que l'on pourrait conclure que "la non-localité est un cas particulier de contextualité." \\

Le "\textit{théorème de Bell}" peut en effet faire allusions à deux théorèmes distincts, prouvés par John Bell \cite{wiseman} \cite{wiseman1} : le premier parut en 1964, et correspond à l'incompatibilité des phénomènes quantiques avec deux hypothèses : la "\textit{localité}" (l'indépendance des paramètres) et la "\textit{prédétermination}" (déterminisme des résultats). Le second théorème, daté de 1976, démontre leur incompatibilité avec la \textit{causalité locale}. Bien que ces deux théorèmes soient logiquement équivalents, leurs postulats ne le sont pas. Ils suggèrent ainsi des conclusions distinctes, appréciées en fonction de l'interprétation que l'on donne au formalisme quantique. Ainsi, les "copenhaguiens" semblent préférer la première version du théorème, et conclurent que "la Nature n'est pas déterministe", que "\textit{les mesures non effectuées n'ont pas de résultats.}" L'autre école, "réaliste", utiliserait plutôt la version la plus récente, et interprèterait la violation de la causalité locale comme la preuve du "comportement nonlocale de la Nature". \\

Dans le cadre des modèles ontologiques de théories opératoires, ces hypothèses se traduisent ainsi :\\

Alice et Bob disposent chacun d'une boite noire. Alice choisit l'entrée $x$ (respectivement $y$ pour Bob) de la boite, et enregistre la sortie $a$ (resp. $b$). La préparation $P$ est un évènement passé commun aux évènements $[a|x]$ et $[b|y]$. Nous nous intéressons aux éventuelles corrélations qui peuvent se manifester dans ce type scenario, et qui sont données par les statistiques $p(a,b|x,y,P)$. On suppose qu'il peut exister des paramètres supplémentaires, $\lambda\in\Lambda$, qui pourrait être la cause de ces corrélations. La distribution de probabilités est donc donnée par :
non
\[p(a,b|x,y,P) = \int_\Lambda d\lambda\hspace{2mm} p(a,b|x,y,\lambda,P)\hspace{2mm}p(\lambda|x,y,P) \]

On impose l'hypothèse d'\textit{indépendance de la mesure} : $p(\lambda|x,y,P)=p(\lambda|P)$ ; et du fait que la préparation $P$ est fixée, on s'intéresse donc à :

\[p(a,b|x,y) = \int_\Lambda d\lambda\hspace{2mm}  p(a,b|x,y,\lambda)\hspace{2mm}p(\lambda) \]

Dans le théorème orginal de 1964, l'hypothèse dite de "\textit{localité}" ou "\textit{séparabilité}", aussi appelée "\textit{indépendance des paramètres}" par Abner Shimony \cite{wiseman}, porte sur l'impossibilité de communiquer à une vitesse superluminique. Ainsi, l'obtention d'un résultat $b$ par Bob ne doit pas dépendre du choix $x$ effectué par Alice, et, respectivement, le résultat $a$ obtenu par Alice ne doit pas dépendre de l'entrée $y$ de Bob. 

\[p(b|x,y,\lambda) = p(b|y,\lambda) \hspace{3cm} p(a|x,y,\lambda) = p(a|x,\lambda) \]

Ce postulat n'est cependant pas suffisant pour démontrer le théorème de Bell, qui requiert une hypothèse supplémentaire, dit de \textit{déterminisme des résultats} (ou \textit{prédéterminatio}n), qui stipule que les sorties $a$ et $b$ sont fixées par le paramètre $\lambda$ en fonction des entrées $x$ et $y$, avant que ces entrées soient effectivement choisies par Alice et Bob.

\[p(a,b|x,y,\lambda) =\{0,1\} \]

Dans la seconde version de 1976, Bell réduit son théorème à un seul postulat, la \textit{causalité locale}, i.e. l'hypothèse selon laquelle les distributions de probabilités sur les valeurs d'une variable dans une région de l'espace-temps sont déterminées par les valeurs de toutes les variables présentes dans le cône de lumière passé de cette région.\\

Cette causalité locale comprend en fait deux hypothèses, plus intuitives :\\
\begin{itemize}
\item la \textit{causalité relativiste}, i.e. le postulat que le passé d'un évènement est complètement représenté par son cône de lumière passé ;
\item le \textit{principe de Reichenbach}, qui stipule que si deux évènements $A$ et $B$ sont corrélés, et qu'aucun de ces évènements n'est la cause de l'autre, alors il existe une cause commune $C$ pour $A$ et $B$ tel que $p(A,B|C)=p(A|C)p(B|C)$. \cite{reichenbach} \\
\end{itemize}

Le paramètre $\lambda$ correspond ainsi à une description complète de toute propriété physique dans une région d'espace-temps qui sépare les évènements $[x|a]$ et $[y|b]$.\\%

\begin{figure}[ht!]
\centering
\includegraphics[width=8cm]{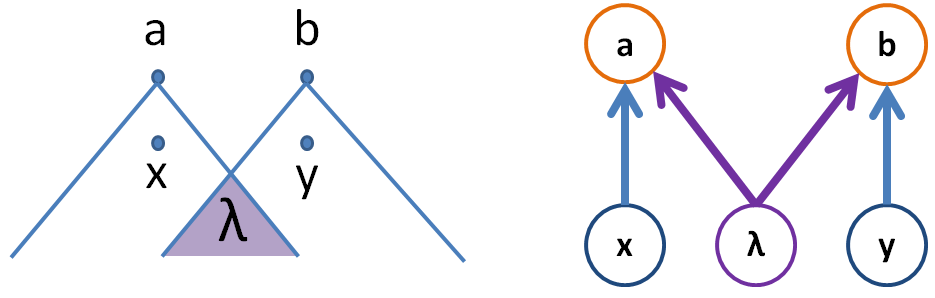}
\caption{Schémas (diagramme d'espace-temps à gauche, structure causale à droite) d'un scenario de Bell bipartite.}
\label{nl}
\end{figure}

Les corrélations obéissent ainsi à la structure causale suivante :

\[p(a,b|x,y,\lambda) = p(a|x,\lambda)\hspace{2mm}p(b|x,\lambda) \]

Selon le théorème de Fine (1982), chaque théorème est un corollaire de l'autre. Autrement dit, il existe un modèle $\theta$ satisfaisant la causalité locale \textit{si et seulement si} il existe un modèle $\theta '$ satisfaisant la prédetermination et l'indépendance des paramètres\footnote{Ce résultat était en fait déjà connu de Bell en 1971 \cite{wiseman1}}. 
Cette équivalence est issue du fait que la causalité locale est semblable au couplage de l'hypothèse "indépendance des paramètres" à l'hypothèse "d'\textit{indépendance des résultats}", qui peut être dérivée du déterminisme des résultats : $p(b|a,x,y,\lambda)=p(b|x,y,\lambda)$ et $p(a|b,x,y,\lambda)=p(a|x,y,\lambda)$.
En effet, si les sorties $a$ et $b$ sont simplement des fonctions de $x$, $y$ et $\lambda$, alors connaître $a$ n'apporte aucune information sur $b$, si $x,y,\lambda$ sont déjà connus.\\

\begin{figure}[ht!]
\centering
\includegraphics[width=12cm]{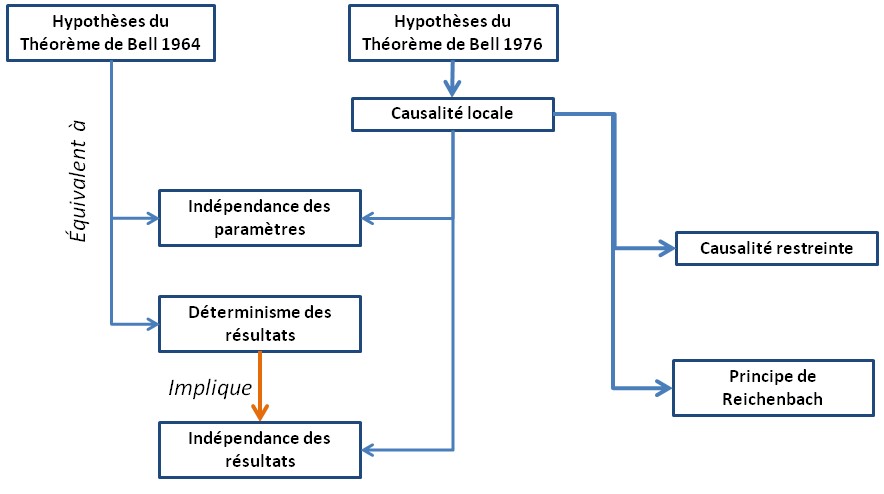}
\caption{Diagramme récapitulatif des relations entre les deux versions du théorème de Bell.}
\label{diag2}
\end{figure}

On retrouve donc l'hypothèse de causalité locale en appliquant ces deux hypothèses aux corrélations :

\begin{align*}
   p(a,b|x,y,\lambda)\hspace{1cm} & = &p(b|a,x,y,\lambda)\hspace{2mm}p(a|b,x,y,\lambda) \hspace{75mm}\\
   & = & p(b|x,y,\lambda)\hspace{2mm}p(a|x,y,\lambda)\hspace{35mm} \hbox{   \textit{indépendance des résultats}}\\
   & = & p(b|y,\lambda)\hspace{2mm}p(a|x,\lambda)\hspace{4cm} \hbox{    \textit{indépendance des paramètres}}
\end{align*}

Or, la sortie $a$ (resp.$b$) correspondant au résultat d'une mesure représentée par l'entrée $x$ (resp. $y$), les paramètres $b$ et $y$ (resp. $a$ et $x$) peuvent ainsi être interprétés comme constituant le contexte de la mesure. La non-contexutalité de mesure peut donc s'écrire :

\[p(a|x,y,b,\lambda) = p(a|x,\lambda) \]

Au sein d'un modèle séparable, i.e. où la condition d'indépendance des paramètres s'applique, la non-contextualité de mesure implique donc l'indépendance des résultats.
Or, une violation de causalité locale au sein d'un modèle séparable implique la dépendance des résultats de mesure, et, par conséquent, la contextualité de mesure. C'est en ce sens que l'on pourrait considérer que la nonlocalité (i.e. l'échec de la causalité locale) est un cas particulier de contextualité de mesure. Il faut néanmoins garder l'idée que l'indépendance des résultats est dérivée du déterminisme des résultats. Autrement dit, une violation de causalité locale implique que le modèle est séparable, à mesures contextuelles, et indéterministe.\\

En revanche, un modèle ontologique peut être nonlocal en violant la séparabilité (l'indépendance des paramètres) et dans ce cas, cela n'implique aucunement que le modèle soit à mesures contextuelles. C'est le cas du modèle  de Beltrametti-Bugajski. Dans ce modèle, les variables pour un système composé ne sont pas simplement le produit cartésien de variables des composés, puisque le produit cartésien de deux espaces de Hilbert projecteurs n'est pas l'espace de Hilbert projecteur du produit tensoriel (il ne parvient pas à inclure les états intriqués). En particulier, on n'associe pas de variables distinctes à des systèmes spatialement séparés. Ainsi, le modèle Beltrametti-Bugajski n'est pas séparable. Comme nous l'avons vu précédemment, ce modèle est contextuel pour les préparations, et non pour les mesures. \\ 

C'est seulement au sein d'une théorie séparable que le théorème de Bell implique la contextualité de mesure.

\section*{Transition}

Dans le cadre d'une étude de la structure mathématique de la théorie quantique, il est possible de concevoir une notion généralisée de contextualité-KS qui engloberait les corrélations multipartites issues de scenarii de type Bell. Néanmoins, ces résultats sont d'ordre logique. Par conséquent, une vérification expérimentale de la violation d'une inégalité graphique pourrait, pour certains, ne pas avoir de sens, puisque l'inégalité est violée \textit{logiquement}. Pour vérifier si en pratique la nature est contextuelle, indépendamment de la théorie choisie pour la modéliser, il est donc nécessaire de se passer de l'hypothèse implicite du déterminisme des résultats, qui sous-tendrait le fait que les mesures considérées sont projectives, idéales parfaites. C'est ce qu'à réaliser Robert Spekkens, en généralisant le concept de contextualité à une notion de mesures plus générale, représentées par des POVMS. Il est alors intéressant de noter que les corrélations du voyant surprotecteur de Specker, irréalisables par des systèmes quantiques dans le cas de mesures projectives, semblent reproductibles quantiquement si l'on utilise certains POVMS \cite{kunjwal2}. \\

La question de l'inclusion de l'approche AFLS (et donc CSW et AB) dans le cadre de l'approche de Spekkens est toujours ouverte à l'heure de la rédaction de ce mémoire. Néanmoins, cette réconciliation semblerait faire écho à la question de l'inclusion de la non-contextualité-KS au sein de l'approche opératoire, étudiée par Kunjwal et Spekkens, qui nécessite le postulat supplémentaire de \textit{perfection des corrélations} \cite{kunjwal2}.\\

La généralisation opératoire a conduit à l'émergence de différents type de non-contextualité, définis en fonction de la procédure expérimentale réalisées (préparation, mesure, ou encore transformation). Néanmoins, Spekkens invite à ne pas les apprécier séparément, mais comme un seul et même postulat, la \textit{non-contextualité universelle}, traduction opératoire d'un principe philosophique de Leibniz ; l'\textit{identité des indiscernables}. \\

\section{L'Identité des Indiscernables de Leibniz}
\label{leibnizuni}

La démarche de généralisation du concept de non-contextualité proposée par Robert Spekkens est basée sur le principe philosophique d'\textit{Identité des Indiscernables} de Gottfried Wilhelm Leibniz (1646 - 1716). Ce principe stipule que deux entités distinctes ne peuvent partager l’ensemble de leurs propriétés. Si deux entités sont distinctes, elles ne se ressemblent jamais exactement, et il existe toujours une propriété mesurable permettant de les différencier.
\begin{quotation}
[...] \textit{Poser deux choses indiscernables est la même chose sous deux noms.} \cite{leibniz}
\end{quotation}
Robert Spekkens remarque que ce principe est très souvent vérifié en physique.

Einstein semble notamment lui être dévoué.\\

Dans sa preuve de l'inexistence d'un référentiel de repos privilégié en électrodynamique, il remarque ainsi que les équations de Maxwell sont incapables de rendre compte de la symétrie dans l'expérience de Faraday, selon qu’on l’observe depuis l’aimant ou le conducteur, en mouvement l’un par rapport à l’autre. Or un même phénomène devrait s’expliquer d’une façon unique.\\

Dans le principe d'équivalence, on ne peut pas distinguer le fait d'accélérer et de se trouver dans un champ gravitationnel uniforme. Il s'agit donc d'une seule et même chose : inertie et gravitation sont "\textit{wesensgleich}", "de même nature".\\

On retrouve également ce principe dans le "hole argument", la motivation de la causalité locale, ou encore l'argument de "no fine-tuning" dans le domaine des inférences causales en machine-learning.\\

Spekkens suggère alors que le principe de Leibniz devrait être considéré comme une "pilier central des fondements conceptuels de la relativité générale", et propose de vérifier si c'est aussi le cas pour les fondements de la mécanique quantique.\\

Transposé au cadre des modèles ontologiques et théories opératoires, le principe de Leibniz établit que si deux scénarios sont empiriquement - c'est-à-dire opératoirement- indiscernables, alors leurs représentations ontologiques sont identiques. Autrement dit, le modèle doit être non-contextuel pour tout type de procédure expérimental, i.e. \textit{universellement non-contextuel}. Cette hypothèse peut être transcrite sous la forme d'une inégalité, testable expérimentalement.

\section{Inégalité de Non-contextualité Universelle}
\label{inencuni}
\textit{D'après la présentation de Robert Spekkens au workshop du 23 juin Solstice of Foundations à Zurich.\\ }

Dans \cite{kunjwal3}, Kunjwal et Spekkens ont développé une méthode permettant de déduire des inégalités de non-contextualité à partir des preuves (indépendantes et dépendantes) du théorème de Kochen-Specker. Cette méthode, applicable aux mesures bruitées et aux états mixtes, peut être testée expérimentalement.\\

Partons de la preuve de Cabello, Estebaranz et Garc\'{\i}a-Alcaine, qui met en jeu 18 projecteurs  dans un espace à 4 dimensions, agencés en 9 mesures PVMs, chacune pouvant donner 4 résultats possibles \cite{cabello12}.

\begin{figure}[ht!]
\centering
\includegraphics[width=6cm]{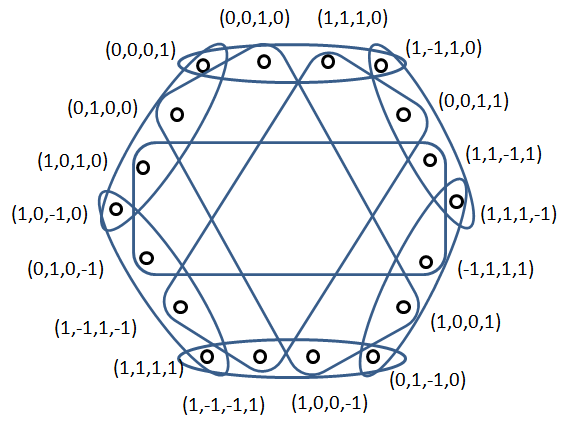}
\caption{Hypergraphe de la preuve de Cabello Estebaranz et Garc\'{\i}a-Alcaine}
\end{figure}

Nous avons déjà vu qu'aucun coloriage de non-contextualité-KS n'était réalisable pour un tel scenario. En effet, dans un modèle KS-non-contextuel de la théorie quantique, la valeur (0 ou 1) assignée à l'évènement $[k|M]$ par $\lambda$ est la même pour l'évènement $[k'|M']$ si ces deux évènements sont représentés par le même projecteur dans l'espace de Hilbert (i.e. le même sommet de l'hypergraphe). D'autre part, lorsque mesure est effectuée, un seul résultat est enregistré. L'affectation doit donc être telle qu'un seul des quatre sommets d'une arrête se voit assigner la valeur 1, ce qui n'est pas possible dans le cas étudié.

\begin{figure}[ht!]
\centering
\includegraphics[width=5cm]{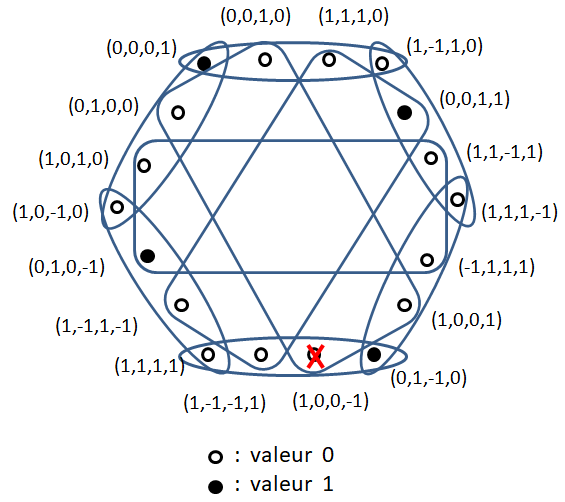}
\caption{Contextualité-KS de l'hypergraphe.}
\end{figure}

Celui-ci constitue ainsi une preuve indépendante de l'état du système du théorème de Kochen-Specker, et donc de la contradiction logique entre la non-contextualité de mesure, le déterminisme des résultats et les équivalences opératoires entre mesures. Or, face à une telle contradiction, il est toujours possible d'abandonner le déterminisme des résultats, et donc de sauver la non-contextualité de mesure.  \\

Cependant, comme montré précédemment, le déterminisme des résultats peut être justifié à partir de la non-contextualité des préparations, de l'hypothèse de perfections des corrélations, et d'équivalences opératoires entre préparations.\\

On peut considérer ainsi trente-six procédures de préparations organisées en 9 ensembles de 4 chacune : $\{P_{i,k} : i\in \{1,...,9\}, k\in \{1,...,4 \}\}$ telles que 
\begin{itemize}
\item les corrélations entre le choix de la préparation $i$ et le résultat de la mesure $M_i$ soient parfaites : $\forall i, \forall k : p(k|M_i,P_{i,k})=1$ ;
\item pour les préparations effectives $P_i^{(ave)}$ obtenues en échantillonnant $k$ uniformément et aléatoirement et en implémentant ensuite $P_{i,k}$, on observe les relations d'équivalence : $P_i^{(ave)} \simeq P_{i'}^{(ave)} \forall i,i' \in \{1,...,9\}$ 
\end{itemize}

\begin{figure}[ht!]
\centering
\includegraphics[width=7cm]{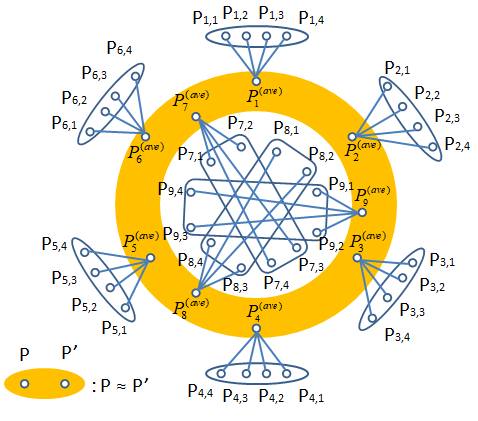}
\caption{36 procédures de préparations organisées en 9 ensembles de 4 chacun. Le point de convergence d'un ensemble de lignes provenant des éléments d'un ensemble représente la préparation effective obtenue par échantillonnage uniforme sur l'ensemble. Une région jaune entourant un ensemble de points implique que ces préparations sont opérationnellement équivalentes. \cite{kunjwal3}}
\end{figure}

La contradiction logique issue de la preuve de Cabello et al. devient donc :
\[\hbox{\textit{non-contexutalité universelle} } \wedge \hbox{ \textit{équivalences opératoires} } \wedge \hbox{ \textit{corrélations parfaites} } \rightarrow \hbox{contradiction}  \]

En pratique, l'hypothèse des corrélations parfaites n'est jamais réalisable. Ainsi, il est impossible de vérifier expérimentalement cette contradiction. Cependant, on peut en déduire la relation suivante :

\[\hbox{\textit{non-contexutalité universelle} } \wedge \hbox{ \textit{équivalences opératoires} } \rightarrow \hbox{corrélations imparfaites}  \]

Ainsi, si les données expérimentales sont universellement non-contextuelle, les corrélations doivent être imparfaites  i.e. la quantité de corrélation, moyennée sur l'ensemble des $i$ et des $k$, devra nécessairement être inférieure à 1. $\forall i, \forall k : p(k|M_i,P_{i,k}) < 1$. La borne maximale correspondante, indépendante de la théorie considérée, correspond à une (véritable) \textit{inégalité de non-contextualité}.\\

Dans le cas étudié, les corrélations sont données par :

\[A \equiv \frac{1}{36} \sum_{i=1}^9 \sum_{k=1}^4 p(k|M_i,P_{i,k}) \]

L'expression de ces corrélations par rapport aux fonctions réponses et aux états épistémiques est : 

\[A = \frac{1}{36} \sum_{i=1}^9 \sum_{k=1}^4 p(k|M_i,\lambda)p(\lambda|P_{i,k}) \]

On définit alors la \textit{prédictibilité maximale} d'une mesure $M$ étant donné l'état ontologique $\lambda$ par 

\[\zeta(M,\lambda) \equiv \max_{k'} p(k'|M,\lambda) \]

et on en déduit 

\[A \leq \sum_\lambda \left(\frac{1}{9}\sum_i \zeta(M_i,\lambda) \left[\frac{1}{4}p(\lambda|P_{i,k})\right] \right) \]
\[A \leq \sum_\lambda \left(\frac{1}{9}\sum_i \zeta(M_i,\lambda) \nu(\lambda) \right) \]
\[A \leq \max_\lambda \left(\frac{1}{9}\sum_i \zeta(M_i,\lambda)\right) \]

\begin{figure}[ht!]
\centering
\includegraphics[width=6cm]{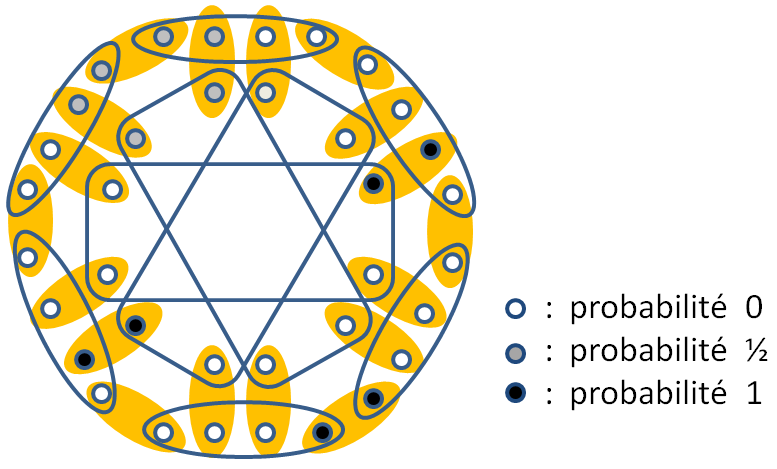}
\caption{Exemple d'une affectation non-contextuelle et indéterministe de résultats de mesure. D'après : \cite{kunjwal3}}
\label{figureravi}
\end{figure}

Les mesures peuvent avoir des réponses indéterminées, i.e. $p(k|M,\cdot) : \Lambda \rightarrow [0,1]$. En revanche, la non-contextualité des mesures implique que $p(k|M_i,\lambda) = p(k'|M_{i'},\lambda)$ pour des paires opérationnellement équivalentes $\{[k|M_i],[k'|M_{i'}]\}$.  Considérons l'assignation de la figure \ref{figureravi}.

Dans cet exemple, six mesures ont une prédictibilité maximale de 1, et trois mesures ont une prédictibilité maximale de $1/2$. Ceci implique
\[ \frac{1}{9}\sum_i \zeta(M_i,\lambda) = \frac{1}{9} \left(6\cdot 1 + 3\cdot \frac{1}{2}\right) = \frac{5}{6}\]
Dans \cite{kunjwal3}, Kunjwal démontre qu'il n'existe pas d'état ontologique possédant une prédictibilité maximale moyenne supérieur à celle de cet exemple. Ainsi, \[\max_\lambda \left(\frac{1}{9}\sum_i \zeta(M_i,\lambda)\right) \leq \frac{5}{6}\]

Il en résulte l'inégalité de non-contextualité universelle :
\[ A \leq  \frac{5}{6}\]

L'exemple présenté d'inégalité de non-contextualité universelle est tiré d'une preuve de non-contextualité-KS indépendante des états.  Il est cependant possible d'appliquer cette méthode de dérivation à tout type de preuve, dépendante \cite{kunjwal3} ou indépendante \cite{krishna}, de non-contextualité. \\

La limite obtenue n'est pas triviale, la valeur logiquement atteignable pour $A$ étant 1. La réalisation quantique des 18 vecteurs viole l'inégalité avec $A_{MQ} = 1$. L'inégalité est par ailleurs robuste expérimentalement, au sens où, contrairement aux inégalités précédemment présentées, elle pourrait ne pas être violée si la réalisation expérimentale ne parvient pas à suffisamment supprimer le bruit.
Inclure le bruit dans les modèles physiques est essentiel si l'on espère utiliser la contextualité comme une ressource pour le traitement de l'information. La violation de l'inégalité présentée a été ainsi pu être vérifiée expérimentalement : $A_{exp} = 0,99709\pm 0,00007$, i.e. une violation de la limite non-contextuelle de $5/6 \approx 0,833$ par $2300 \sigma$. Pour des détails sur divers challenges et échappatoires expérimentaux, se référer à l'article \cite{mazurek} et à la présentation \cite{spekkens11}.\\

Si la théorie quantique n'est pas compatible avec le principe d'identité des indiscernables de Leibniz, c'est semble-t-il aussi le cas de la  Nature, ou si l'on préfère, des données expérimentales que l'on enregistre qui sont manifestement, universellement contextuelles. Ces résultats pourraient donc contrarier la proposition de Spekkens d'ériger le principe de Leibniz en critère physique, que tout modèle théorique devrait respecter \cite{spekkens4} \cite{spekkens11}. Faut-il alors abandonner cette démarche ? Ou bien renier le réalisme intrinsèque ? Robert Spekkens défend une troisième alternative : définir une notion du réalisme qui aille au-delà du cadre des modèles ontologiques, et qui permette de sauver le principe de Leibniz. 

\section{Sauver le principe de Leibniz ?}
\label{sauvleib}
S'adonner au réalisme n'implique pas nécessairement de souscrire aux modèles ontologiques. Il est donc toujours possible d'épargner l'identité des indiscernables, i.e. l'hypothèse de non-contextualité universelle, en s'affranchissant de ce formalisme \cite{spekkens11}.\\

Il peut en effet être reproché aux modèles ontologiques de tenter de donner un sens classique à des résultats qui ne le sont pas. L'incompatibilité de la mécanique quantique avec le principe de Leibniz, l'impossibilité de décrire les prédictions quantiques par un modèle ontologique non-contexutel, témoignerait simplement de la nature non-classique des résultats opératoires obtenus. Essayer de décrire en termes classiques des résultats quantiques semble avoir atteint certaines limites. C'est le résultat que l'on pourrait tirer du théorème de Kochen-Specker et de la violation des inégalités non-contexutelles : \begin{quotation}
La mécanique quantique ne peut être contenue dans une algèbre booléenne.
\end{quotation}

Il semblerait donc nécessaire de développer des conditions de probabilités (et, a fortiori, une logique) analogues à celles utilisées jusqu'alors, classiques, mais qui soient compatibles avec les résultats quantiques et le principe de Leibniz. Une telle démarche a notamment été entreprise par Robert Spekkens et Mattew Leifer \cite{leifer8}.\\

Leur approche part du constat qu'il est possible d'interpréter les modèles ontologiques comme des " réseaux causaux"(aussi appelés "réseaux bayésiens" \footnote{Cette approche est inspiré des travaux de Pearl et de Spirtes, Glymour and Scheines (causal discovery algorithms) \cite{pearl} \cite{spirtes}}), objets étudiés dans le domaine des inférences causale (sous-domaine du machine learning). \cite{wood} On y modélise des expériences par leurs structures causales, i.e. des relations de causes à effets entre variables, qui spécifient les conditions de probabilités nécessaires. \footnote{Pour une introduction à se formalisme, cf. \cite{wood} \cite{cavalcanti} et \cite{cavalcanti1}} Le principe de Leibniz, et donc l'hypothèse de non-contextualité universelle, sont retranscrits sous la forme du concept de \textit{fidélité}, aussi nommé \textit{stabilité}, ou encore de  \textit{non fine-tuning} ("pas de réglages fins").
Basé sur l'idée que l'indiscernabilité empirique implique l'identité des représentations ontologiques, ce postulat stipule qu'un modèle causal doit être fidèle aux évènements expérimentaux décrits, i.e. que chaque indépendance conditionnelle entre variables doit émerger comme conséquence du graphe de causalité (la structure de causalité qui sous-tend les corrélations, cf. figure par exemple) et non d'un réglage fin des paramètres statistiques causaux. Toute explication causale qui n'est pas expérimentalement apparente doit être abandonnée.\\

Dans l'article pionnier \cite{wood}, Christopher J. Wood et Robert Spekkens montrent ainsi que la plupart des mécanismes causaux classiques utilisés pour expliquer la violation des inégalités de Bell - tels que les influences causales superluminiques, le super-déterminisme (i.e. le déni de la liberté de choix des paramètres, parfois appelé "libre-arbitre") ou encore les influences rétrocausales (qui vont du futur vers le passé) - requièrent nécessairement un "\textit{fine-tuning}" (ajustement, réglage fin). Toute tentative de fournir une explication causale de corrélations non-communicantes qui violent une inégalité de Bell entre en conflit avec le postulat de non fine-tuning. Autrement dit, l'idée selon laquelle on ne doit pas postuler l'existence de relations causales supraluminiques si l'on ne peut pas communiquer supraluminiquement est violée par tout modèle violant une inégalité de Bell. \\

Contextualité et nonlocalité requerraient donc, au sein d'un modèle ontologique ou causal classique, l'invocation d'une forme de "conspiration", décrite par Spekkens dans son article original, alors qu'il s'interroge sur l'éventuelle étrangeté de l'hypothèse de non-contextualité, i.e. si  une violation de l'hypothèse de non-contextualité universelle bouleverse notre représentation de la réalité.\cite{spekkens1}

\begin{quotation}
\textit{Sans tenir compte des préjugés classiques, la nonlocalité n'est pas une hypothèse déraisonnable. Cependant, si l'univers est fondamentalement non-séparable ou est tel que des influences causales peuvent se propager plus vite que la vitesse de la lumière, alors pourquoi devrait-il aussi être tel que l'on ne puisse pas utiliser ces effets pour réaliser des communications supraluminiques ? Étant donné la présence de la nonlocalité au niveau ontologique, il semble presque conspiratoire que l'on ne puisse pas faire usage de cette nonlocalité pour communiquer. De même, il n'est certainement pas déraisonnable de considérer que pour un état ontologique donné, les statistiques des résultats expérimentaux dépendent de détails de la procédure expérimentale. Mais en supposant que ce soit le cas, il est très surprenant que lorsque l'on considère une distribution de probabilité quelconque sur l'ensemble des états ontologiques (c'es-à-dire, toute distribution qui caractérise ce que quelqu'un qui ne connait que la procédure de préparation sait sur l'état ontologique), la moyenne pondérée sur les statistiques des résultats ne dépende pas des détails de la procédure expérimentale. Une fois encore, cela semble presque conspiratoire.}\\
\end{quotation}

Si nos modèles physiques décrivent une différence ontologique, par exemple deux objets $A$ et $B$ possédant une propriété physique qui les distingue, alors on s'attendrait à pouvoir l'observer expérimentalement, i.e. on s'attend à ce qu'il existe une mesure permettant de les différencier. Si ce n'est pas le cas, si le principe de Leibniz, et donc de non fine-tuning, est violé, alors nos mesures doivent seulement révéler de l'information "coarse-grained" qui est "finement réglée" afin que cette différence entre $A$ et $B$ ne puisse être révélée. Cette "aspect conspiratoire" me rappelle l'image des figures impossibles : chaque partie localement cohérente du triangle de Penrose correspond à l'information "coarse-grained" que l'on peut obtenir, tandis que l'incohérence globale à l'origine du paradoxe visuel peut être identifiée à cette "différence qui ne peut être perçue."
Mais, à nouveau, il suffit de remarquer que le paradoxe et le bouleversement apparent semblent uniquement émerger du fait que l'on cherche à exprimer un résultat qui n'obéit pas à la logique classique en termes classiques. Un autre langage, quantique, évaporerait l'étrangeté en permettant de conserver principe de Leibniz et cohérence logique, dissipant ainsi les paradoxes. \\

Spekkens et Leifer ont ainsi entrepris la construction d'un tel langage, où \textit{états conditionnels} et structures causales quantiques se substituent aux conditions et structures causales classiques. La non-classicité, l'étrangeté n'est donc pas l'échec de la non-contextualité, mais l'échec du "classicisme", de la projection des concepts et notions classiques (logique booléenne, probabilités classiques non bayésiennes ou encore réalisme "naïf"...) sur des systèmes quantiques \cite{leifer8} \cite{spekkens4}. \\

Un autre programme de reconstruction est également exploré par Chris Fuchs et l'école QBiste (prononcé "cubiste") \cite{fuchs2} \cite{fuchs5}. Fondée sur les principes de John Wheeler de "loi sans loi" ou encore "d'observation participative", et la philosophie pragmatique de William James, l'interprétation QBiste conclue des résultats de Bell et de Kochen-Specker que "toute mesure non-effectuée n'a pas de résultat." Autrement dit, il n'est pas sain de spéculer sur des expériences non réalisées, d'assigner des valeurs à des paramètres non-observés dans des dispositions expérimentales qui ne peuvent pas être (simultanément) réalisées en pratique. Si l'état futur n'existe pas, i.e. le pari est subjectif, l'argument de non fine-tuning pourrait, dans une telle approche, être vu comme un pari sur notre futur état de connaissance, et perd tout sens au moment de la mesure. Les états quantiques n'ont pas de réalité ontologique, mais correspondent à des degrés de croyances personnels, utilisés pour effectuer ces "paris subjectifs sur le monde". \\

C'est peut-être ainsi que devrait être reçu tout "théorème no-go" : non pas comme des résultats qui imposent des contraintes sur le monde, qui nous contraindraient alors à nous confronter à des "phénomènes" étranges et paradoxaux, mais comme des invitations à abandonner une vision réaliste naïve (entitique), au profit, par exemple d'un réalisme participatif, où l'agent observateur ne se contenterait pas de révéler passivement des propriétés préexistantes mais participerait, à chaque mesure, à la création de la réalité. La magie quantique s'affranchirait alors de son inquiétante étrangeté, pour devenir une source de pouvoir de l'agent.\\

Indépendamment de l'interprétation qui en est faite, la violation d'une inégalité de non-contextualité pourrait ainsi être utilisée comme une ressource\footnote{Pour une étude sur cette notion de ressource, cf. par exemple l'introduction de l'article \cite{coecke1}} dans le cadre du domaine de l'information quantique. Divers applications ont ainsi été identifiées\footnote{Les paragraphes suivants sont inspirés de la vidéo \cite{pusey3}} : 

Une tâche particulière de traitement de l'information, nommée "parity-oblivious multiplexing", consiste à encoder deux bits dans un unique qubit de telle sorte que si l'on choisit aléatoirement d'avoir accès au premier ou au second bit, il est impossible d'y accéder parfaitement, i.e. la probabilité d'obtenir exactement le bit choisi est limitée à 85\%. Or, dans le cas d'un modèle non-contextuel, la borne supérieure de l'inégalité est de 75\%. La nature contextuelle des préparations quantiques permet donc de surpasser les capacités de traitement de l'information classiques pour une telle tâche, et fournit également des exemples de "codes d'accès aléatoires quantiques". \cite{spekkens3}\\

La contextualité a également été identifiée comme avantage pour un cas particulier de discrimination d'états. Cette tâche consiste à deviner quel état quantique décrit un système quantique donné lorsque l'état du système est tiré d'un ensemble connu de possibilités avec une distribution antérieure connue, et l'estimation est basée sur le résultat d'une des  mesures choisies. Dans la variante à "erreur minimale", l'objectif est de minimiser la probabilité d'une erreur d'estimation. Or, il a été démontré que les contraintes statistiques correspondantes constituent des inégalités de non-contextualité violées par la théorie quantique. Cette violation implique donc un avantage quantique vis à vis des modèles non-contextuels pour la discrimination d'états. \cite{schmid}\\ 

La création d'un ordinateur quantique, dont la vitesse de calcul serait supérieure à celle d'un ordinateur classique, est l'un des grands enjeux du domaine de l'information quantique. Il est souvent défendu, dans la presse de vulgarisation scientifique, que cette rapidité serait due au fait que "toutes les réponses possibles sont explorées en même temps". Il a cependant été démontré que ce "parallélisme" \cite{para} ne constituait pas une source d'accélération satisfaisante \cite{nop1} \cite{nop2}. Afin d'identifier cette source, divers modèles de calculs sont explorés. Un de ces modèles est basé sur la mesure : avant de décider du calcul, l'expérimentateur prépare d'abord un certain état intriqué, et encode ensuite le calcul à effectuer sur la façon dont l'état est mesuré. L'aspect non-classique de cette démarche est assez intuitif : il semble absurde d'essayer d'encoder une décision importante sur la façon dont on mesure un système classique. La non-contextualité se trouve à nouveau être un excellent critère de classicité puisqu'elle est violée par ce modèle de calcul basé sur la mesure quantique. \cite{raussendorf}\\

Un autre exemple de mode de calcul exploré consiste à réaliser une correction d'erreur de telle sorte qu'il existe certaines portes logiques facilement implémentables. Celles-ci ne sont cependant pas suffisantes pour construire un ordinateur quantique complet, et nécessitent la préparation d'états particuliers, appelés états magiques, qui permettent d'une certaine façon de "promouvoir l'ordinateur en ordinateur quantique universel complet". Or, il se trouve que ces états magiques, couplés aux portes logiques facilement exécutables, permettent de construire une preuve de contextualité, dans un espace de dimension impaire et premier. \cite{comput} La question d'une extension aux qubits (systèmes de dimension 2) est toujours ouverte, mais de premiers résultats encourageants ont été obtenus \cite{bermejo-vega}. La contextualité pourrait donc être la ressource nécessaire à la construction du Graal de l'information quantique.\\

L'échec de non-contextualité, en plus de constituer un critère de détection d'étrangeté, se trouve être une ressource offrant une variété d'avantages quantiques dans les domaines de la cryptographie et de l'informatique. Il se cache également derrière certains paradoxes à post-sélection et à mesures faibles stupéfiants, développés par Yakir Aharonov.

\chapter{Contextualité et Paradoxes Quantiques }
\label{chappara}
Dans les années 60, l’indéterminisme quantique, jusque-là considéré comme "l'insatisfaisante" caractéristique d’une théorie "très impressionnante" \cite{pais1}, devient subitement le point d’ancrage de nouvelles possibilités inhérentes à la mécanique quantique, avec la découverte des corrélations nonlocales et contextuelles. Les théorèmes de Bell et de Kochen-Specker semblent avoir mis en évidence des caractéristiques fondamentalement non-classiques, i.e. étranges.\\

Un autre résultat, publié en 1964, va conduire à alimenter le débat sur la nature de cette étrangeté quantique. Yakir Aharonov, Peter Bergmann et Joel Lebowitz (ABL) démontrent dans l’article intitulé "Time Symmetry in the Quantum Process of Measurement"\cite{aharonov2}, qu'il est possible de sélectionner de façon indépendante l’état initial et l’état final d’un système quantique. \`A partir de ces pré- et post- sélections, Yakir Aharonov (et al.) va construire un nouveau formalisme, dit "à deux vecteurs" \cite{aharonov3} \cite{aharonov4}, et un motto :\begin{quotation}
"Progress through paradox." \cite{aharonov8}
\end{quotation}

Il propose ainsi d'étudier la théorie quantique à travers divers paradoxes qui émergent de cette double sélection, tels que le paradoxe des trois boîtes, le chat de Cheshire ou encore le pigeonhole effect. Or le caractère quantique de chacun de ces scenarii a été sujet à débat. Afin d'en identifier la nature, il serait intéressant de les soumettre au critère de non-classicité de Spekkens, i.e. la noncontextualité universelle. Ce travail a notamment été réalisé par Matthew Leifer, Robert Spekkens et Matthew Pusey.\\

Ces nouvelles notions de \textit{pré-} et \textit{post-sélection} (section \ref{spps} ) ont permis de découvrir de nouveaux paradoxes logiques, à l'instar du \textit{paradoxe des trois boîtes}, qui constituent des preuves de contextualité quantique (section \ref{scpps}). Associés à une \textit{mesure faible}, des paradoxes comme celui du \textit{chat de Cheshire} permettent d'observer directement la contextualité en la restreignant à un contexte particulier (section \ref{schess}).

\section{Pré- et Post-Sélections}
\label{spps}
\textit{Dans cette section, nous présenterons succinctement les concepts de pré- et post-sélection introduits par Aharonov, Bergmann et Lebowitz.  Pour une introduction plus complète, voir \cite{aharonov3} \cite{aharonov4} \cite{tamir}.}

Considérons que l'on prépare à un temps $t_0$ un système quantique en notre possession dans l'état $\ket{\psi}$ (pré-sélection). \`A un temps intermédiaire $t$, on soumet notre système à une mesure projective $M=\{P_j\}$. Enfin, à un temps $t_1$, une autre mesure projective est effectuée. \`A l'issue de cette projection, le système est sélectionné s'il se trouve dans l'état propre $\ket{\phi}$.  (voir figure)\\

\begin{figure}[ht!]
\centering
\includegraphics[width=6cm]{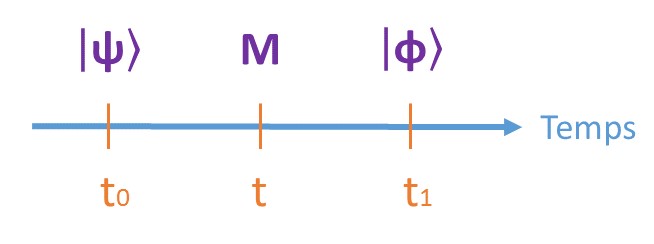}
\caption{Schéma des opérations effectuées lors des procédures de pré- et post-sélection.}
\end{figure}

En supposant qu'aucune autre évolution n'ait eu lieu, la probabilité jointe d'obtenir le résultat $P_j$ pour la mesure intermédiaire et que la projection sur $\ket{\phi}$ soit réalisée à $t_1$, i.e. que le système "passe la post-sélection", est : 
\[p(P_j,\phi(t_1)|\psi(t_0),M) = |\bra{\psi(t_0)} P_j\ket{\phi(t_1)}|^2  \]

La probabilité de passer la post-sélection est donc :

\[ p(\phi(t_1)|\psi(t_0),M) = \sum_j p(P_j,\phi(t_1)|\psi(t_0),M) = \sum_j |\bra{\psi(t_0)} P_j\ket{\phi(t_1)}|^2 \]

On peut alors calculer les probabilités d'obtenir le résultat $P_k$ de la mesure intermédiaire $M$ au temps $t$ pour un système passant les pré- et post-sélections :

\[ p(P_k|\psi(t_0),M,\phi(t_1)) = \frac{p(P_k,\phi(t_1)|\psi(t_0),M)}{p(\phi(t_1)|\psi(t_0),M)} = \frac{|\bra{\psi(t_0)} P_k\ket{\phi(t_1)}|^2}{\sum_j |\bra{\psi(t_0)} P_j\ket{\phi(t_1)}|^2 } \]

L'état du système au temps $t$ est ainsi inféré par les états pré- et post-sélectionnés $\ket{\psi}$ et $\ket{\phi}$. Les probabilités associées à cet état sont données par la formule ci-dessus, appelée "règle ABL". \cite{aharonov2}   \\

\section{Contextualité et Paradoxes "PPS" logiques.}
\label{scpps}
De nouveaux paradoxes étonnants, dits « paradoxes PPS », émergent de la possibilité de pré- et post- sélectionner un état quantique. \`A travers l'un de ces paradoxes, dit "paradoxe des trois boîtes" (\ref{boites}), la question de la nature contextuelle de ces paradoxes peut être étudiée (\ref{contextuel?}). Celui-ci appartient à une catégorie particulière de paradoxes PPS, dits "logiques", qui peuvent être convertis en preuves de contextualité-KS (\ref{deapreuve}). \`A partir de ces résultats, Matthew Pusey et Matthew Leifer sont parvenus à établir un théorème selon lequel tout paradoxe PPS logique constitue intrinsèquement une preuve de contextualité universelle (\ref{theoremeLP}). Enfin, à partir de nouvelles notions de mesures et valeurs "faibles", développées par Aharnov et al, (\ref{faible}), ce théorème peut être généralisé à l'aide d'un résultat obtenu par Matthew Pusey, selon lequel un certain type de valeur faible constitue, lui aussi, une preuve de contextualité universelle (\ref{vf}). \\

\subsection{Le paradoxe des trois boîtes}
\label{boites}
Imaginons que nous disposions d'une particule quantique qui puisse se trouver dans trois boîtes distinctes \cite{3boites}. L'état initial de la particule (la pré-sélection) est :
\[ \ket{\psi} = \frac{1}{\sqrt{3}} (\ket{1} + \ket{2} + \ket{3}) \]
où $\ket{i}$ signifie "la particule est dans la boîte $i$".\\

L'état final (la post-sélection) est, par supposition :
\[ \ket{\phi} = \frac{1}{\sqrt{3}} (\ket{1} + \ket{2} - \ket{3}) \]
La question "La particule est-elle dans la boîte 1 ?" est une mesure "idéale" ("fine", sharp), associée à la mesure projective\footnote{Mesure projective ou PVM (Projector Valued Measures), i.e. un ensemble de projecteurs ${P_j}$ dont la somme est égale à l'opérateur identité} : $M_1 =\{P_{1} , P_{1}^{\perp}\}$ avec 
\[ P_1 = \ket{1} \bra{1} \hspace{1cm} P_1^{\perp} = \ket{2} \bra{2} + \ket{3} \bra{3}\]
En appliquant, la règle ABL, on obtient la probabilité de trouver la particule dans la boîte 1 à un temps intermédiaire entre la pré-sélection et la post-sélection :
\[ p(P_1| \psi,M_1,\phi) = \frac{|\bra{\phi}\ket{1}\bra{1}\ket{\psi}|^2}{|\bra{\phi}\ket{1}\bra{1}\ket{\psi}|^2 + |\bra{\phi}\ket{2}\bra{2}\ket{\psi} + \bra{\phi}\ket{3}\bra{3}\ket{\psi}|^2} = \frac{(1/3)^2}{(1/3)^2 + 0} = 1 \]

On a donc 100\% de chances de trouver la particule dans la boîte 1.\\

Cependant, en effectuant un raisonnement et un calcul analogue pour la boîte 2, avec, $ P_2 = \ket{2} \bra{2}$ , $ P_2^{\perp} = \ket{1} \bra{1} + \ket{3} \bra{3}$ et $M_2 = \{P_2,P_2^{\perp}\}$, on remarque que \[ p(P_2 | \psi,M_2,\phi) = 1 \]
On a donc également 100\% de chances de trouver la particule dans la boîte 2 !\\

Dans le cadre d'une interprétation où les états quantiques $\psi$ et $\phi$ seraient ontologiques, il faudrait donc conclure que l'’observateur trouverait la particule avec certitude dans le cas où il ouvrirait la première boîte, mais aussi s'il ouvrait la seconde boîte.
\subsection{Un paradoxe contextuel ?}
\label{contextuel?}
On peut remarquer que le PVM utilisé établi une distinction entre la première boîte, et le système formé par les boîtes 2 et 3. On pourrait cependant mesurer $P_1$ dans le contexte : $M_1' = \{P_{1} , P_t\}$, avec $P_t = \ket{1} \bra{1} + \ket{2} \bra{2} + \ket{3} \bra{3}$, qui n’introduit pas une telle discrimination. On obtient alors le résultat suivant :
\[ p(P_1| \psi,M_1',\phi) = \frac{|\bra{\phi}\ket{1}\bra{1}\ket{\psi}|^2}{|\bra{\phi}\ket{1}\bra{1}\ket{\psi}|^2 + |\bra{\phi}\ket{2}\bra{2}\ket{\psi}|^2 + |\bra{\phi}\ket{3}\bra{3}\ket{\psi}|^2} = \frac{1}{3} \]De même, pour $M_2'=\{P_{2} , P_t\}$, on a : \[ p(P_2| \psi,M_2',\phi) = \frac{1}{3}\]
Dans ce contexte, le paradoxe précédent disparait !
Il semble donc légitime de supposer qu’une forme de contextualité-KS puisse être à l'origine du paradoxe.\\

Mais l'est-elle vraiment ? Le paradoxe des trois boîtes peut-il exister dans un cadre KS-non-contextuel ?\\

Considérons un modèle ontologique contextuel semblable au modèle de jouet de Spekkens (cf. chapitre 3). On rappelle que le système peut se trouver dans quatre états ontologiques différents : $\{(+,+) ; (+,-) ; (-,+) ; (-,-) \}$. On prépare l'état ontologique de telle sorte que l'état épistémique pré-sélectionné soit $+y = \{(-,-);(+,-)\}$. On choisit alors l'état post-sélectionné $-y = \{(-,+);(+,+)\}$, et l'on réalise intermédiaire projective $X$, qui transforme l'état ontologique $(-,-)$ en $(-,+)$, et $(+,-)$ en $(+,+)$.\\

Interprété en termes de gobelets et de balle cachée, le modèle correspond au scenario suivant. On dispose de quatre gobelets. Une balle est dissimulée sous l'un des deux gobelets "avant", et l'expérimentateur ignore lequel. La postsélection indique seulement qu'après mesure, la balle se trouvera sous l'un des deux gobelets "arrière". Or la mesure effectuée consiste à mélanger les deux gobelets d'une colonne entre eux. \\

Si l'on mélangeait les deux gobelets de "gauche", alors il serait nécessaire que la balle s'y trouve. Autrement, elle n'aurait pas pu passer du compartiment "avant" au compartiment "arrière". De même, avec un argument identique, si l'on secouait le compartiment "droit", on y trouverait également la balle.\\

En ce sens, pour l'observateur la balle est à la fois dans le compartiment "gauche" et dans le compartiment "droit" ! Le paradoxe des trois boîtes peut donc être simulé par un modèle KS-noncontextuel. Une assignation KS-non-contextuelle de valeurs à la mesure intermédiaire peut ainsi apparaître comme KS-contextuelle sous la post-sélection, et ce du fait de la \textit{perturbation} de l'état ontologique par la mesure : en secouant les gobelets "gauche (droit)", du fait de la post-sélection, le résultat "trouver la balle dans un gobelet droit (gauche)" devient impossible. \\

Malgré la similarité apparente entre contextualité et paradoxes PPS, de tels paradoxes peuvent donc avoir lieu dans des modèles ontologiques KS-non-contextuels si l'on considère que les mesures intermédiaires peuvent \textit{perturber} les valeurs des états ontologiques. Le paradoxe a la saveur de la contextualité-KS, mais est différent : dans ce scenario, les prédictions quantiques en jeu s'appliquent à différentes mesures actuelles, non pas à des valeurs assignées à un système à un temps fixé. \\

\subsection{Du paradoxe à une preuve de contextualité}
\label{deapreuve}
En 2005, Matthew Leifer et Robert Spekkens ont cependant démontré \cite{leifer4} que certains paradoxes PPS, nommé "paradoxes PPS logiques", dont fait partie le paradoxe des trois boîtes, peuvent être convertis en preuve de contextualité-KS. Il suffit pour cela de considérer une expérience opératoire (préparation et mesure) sans post-sélection, dans laquelle les mesures intermédiaires, ainsi que deux mesures correspondantes à la pré- et à la post-sélection, sont considérées comme des alternatives contrefactuelles, i.e. ne correspondent plus à des mesures successives dans le temps, mais à des mesures pouvant toutes se réaliser à un même temps t.\\

\boitesimple{Un paradoxe PPS logique est un paradoxe où toutes les probabilités ABL sont égales à 0 ou à 1, et où les états de pré- et post-sélection sont non orthogonaux. Les probabilités correspondantes\\ $f(P) =p(P|\psi,\phi)$ issues de la règle ABL  violent nécessairement une des \textit{conditions algébriques} suivantes :\\
$(\alpha) \hspace{5mm}\forall P, 0 \leq f(P) \leq 1 $;\\
$(\beta) \hspace{5mm} f(I) = 1$ ; $f(0) = 0$;\\
$(\gamma) \hspace{5mm} \forall P,Q$ tels que $PQ = QP$,\\ $f(P + Q - PQ) = f(P) + f(Q) - f(PQ)$  }
Reprenons l'exemple du paradoxe des trois boîtes.\\

On remarque que $P_1^{\perp}$ peut être décomposé en une somme de projecteurs sur les vecteurs $\ket{2} + \ket{3}$ et $\ket{2} - \ket{3}$. Par ailleurs, on constate que $\ket{2} + \ket{3}$  est orthogonal à l'état de post-sélection ; et que $\ket{2} - \ket{3}$ est orthogonal à l'état de pré-sélection. Ainsi, la probabilité d'obtenir le résultat associé à  $P_1^{\perp}$, étant donné $\ket{\psi}$ et $\ket{\phi}$, est nulle. Par conséquent, la mesure $M_1$ donnera le résultat associé à $P_1$.  De même, en considérant $P_2^{\perp}$ et les vecteurs $\ket{1} + \ket{3}$ et $\ket{1} - \ket{3}$, on peut conclure que la mesure $M_2$ donnera le résultat associé à $P_2$. On retrouve alors le paradoxe : si l'on effectue la mesure $M_1$ afin de voir si la particule est dans la première boîte, on l'y trouvera avec certitude ; si l'on effectue la mesure $M_2$ afin de voir si la particule est dans la seconde boîte, on l'y trouvera aussi avec certitude.\\
Considérons à présent les 8 vecteurs mentionnés dans ce paradoxe :\\
$\{\ket{1};\ket{2}; \ket{2} + \ket{3} ; \ket{2} - \ket{3} ; \ket{1} + \ket{3} ; \ket{1} - \ket{3} ; \ket{\psi} ; \ket{\phi}\}$\\

Imaginons que chacun de ces vecteurs décrit une mesure alternative possible à un temps t donné. Dans le cadre d'une théorie à variables cachées
, i.e. d'un modèle ontologique déterministe et à mesures non-contextuelles, on peut assigner la valeur 1 ou 0 à chaque vecteur, selon la réussite ou l’échec du test associé.\\

L'assignation suit les règles suivantes :\begin{itemize}
\item chaque élément d'une paire orthogonale de vecteurs ne peut se voir assigner la même valeur (exclusivité);
\item un seul et unique élément d'un triplet orthogonal de vecteurs doit se voir assigner la valeur 1 (normalisation)
\end{itemize} 

\begin{figure}[ht!]
\centering
\includegraphics[width=10cm]{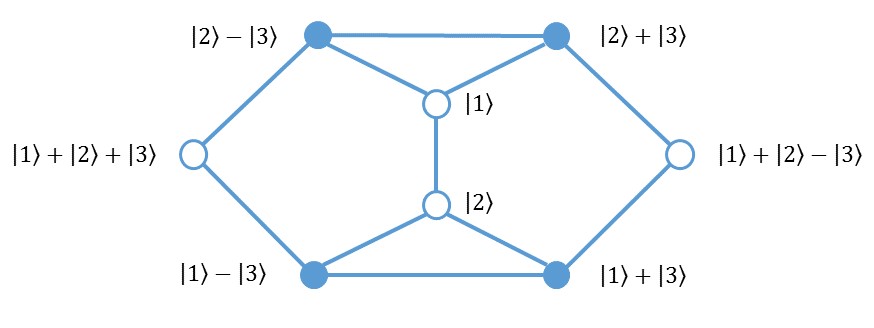}
\caption{Graphique associé au paradoxe des trois boîtes}
\label{graphtroisboites}
\end{figure}

On peut ainsi dessiner un graphique\footnote{Ce graphe est une preuve géométrique de Clifton \cite{clifton}, qui est un sous-graphe de la preuve originale du théorème de Kochen-Specker.} où chaque vecteur est représenté par un point ; la relation d'orthogonalité entre deux vecteurs correspond à une ligne reliant deux points ; l'assignation de la valeur 0 à un vecteur est associée à la coloration d'un point en bleue, la couleur blanche étant associée à la valeur 1. (cf. figure \ref{graphtroisboites}) \\

Comme les états $\ket{\psi}$ et $\ket{\phi}$ sont préparés, le test qui leur est respectivement associé est toujours satisfait. Ainsi, l'état ontologique devrait assigner la valeur 1 à chacun d'entre eux. $\ket{1} - \ket{3}$ et $\ket{2} - \ket{3}$ étant orthogonaux à  $\ket{\psi}$, on leur assigne la valeur 0. $\ket{1} + \ket{3}$ et $\ket{2} + \ket{3}$ étant orthogonaux à  $\ket{\phi}$, on leur assigne également la valeur 0. Puisque $\ket{2} - \ket{3}$, $\ket{2} + \ket{3}$ et $\ket{1}$ forment un triplet orthogonal, on doit assigner à ce dernier la valeur 1. D'autre part, $\ket{1} - \ket{3}$, $\ket{1} + \ket{3}$ et $\ket{2}$ formant aussi un triplet orthogonal, on doit assigner à $\ket{2}$ la valeur 1, ce qui conduit à une contradiction,  $\ket{1}$  et  $\ket{2}$ étant orthogonaux. \\

Dans \cite{leifer3}\cite{leifer4}, Leifer et Spekkens ont démontré que l'on pouvait construire une preuve de contextualité-KS semblable à celle présentée ci-dessus pour tout paradoxe PPS logique. \\

Par ailleurs, il semblerait que la contextualité-KS de ces paradoxes puisse être restreinte à un contexte particulier(\cite{waegell1}), en reconstruisant le scenario paradoxal en carré de Mermin-Peres. Le "quantum pigeonhole effect" est l'un de ces paradoxes.

\subsection{Exemple de l'effet des casiers à pigeons ("Pigeonhole effect")}
Imaginons que nous possédions trois pigeons, et que nous souhaitions les ranger dans deux casiers. Il semble tout à fait raisonnable d'affirmer que l'on trouverait au moins deux pigeons dans le même casier.
\begin{figure}[ht!]
\centering
\includegraphics[width=10cm]{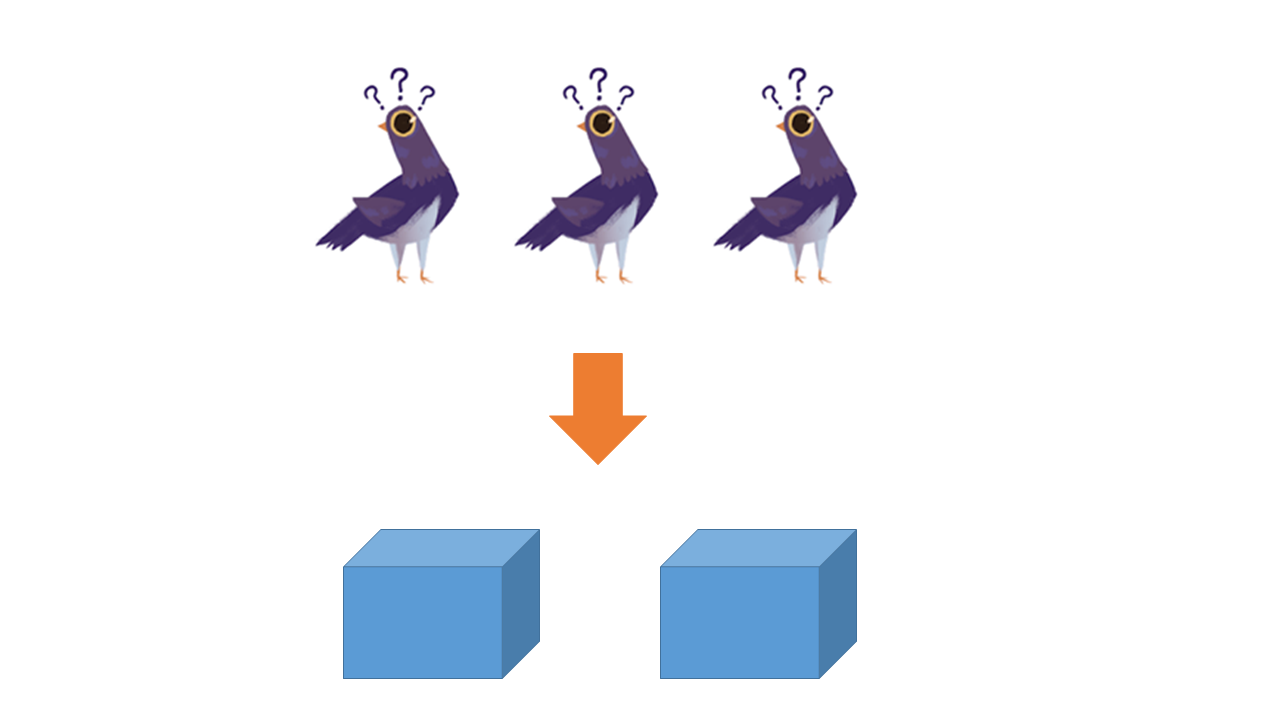}
\caption{Trois pigeons et deux boîtes}
\end{figure}
En mécanique quantique, cependant, ce n'est pas toujours le cas. (source : \cite{aharonov6}) \\

Considérons deux casiers : le casier gauche "L" et le casier droit "R". Remplaçons les pigeons par des particules quantiques, étant chacune dans l'état :
\[ \ket{LR} = \frac{1}{\sqrt{2}}(\ket{L} + \ket{R}) \]
L'état initial de notre système, i.e. notre pré-sélections, est donc le suivant :
\[ \ket{\psi} = \ket{LR}_1 \ket{LR}_2 \ket{LR}_3 \]
On post-sélectionne l'état suivant :
\[ \ket{\phi} = \ket{+}_1 \ket{+}_2 \ket{+}_3 \]
avec \[\ket{+} = \frac{1}{\sqrt{2}}(\ket{L} + i \ket{R}) \]
Considérons une paire de particule, par exemple la paire formée par les particules 1 et 2.\\

Si les particules sont dans le même casier, alors l'état de la paire est $\ket{L}_1 \ket{L}_2$ si les particules sont dans le casier de gauche ; et $\ket{R}_1 \ket{R}_2$ si elles sont dans le casier de droite.\\

De même, si les elles se trouvent dans des casiers différents, l'état de la paire est décrit par $\ket{L}_1 \ket{R}_2$ et $\ket{R}_1 \ket{L}_2$.\\

Les projecteurs associés aux questions : les particules sont-elles dans le même casier ? et les particules sont-elles dans des casiers différents ? sont respectivement :

\[\Pi_{1,2}^{=} = \Pi_{1,2}^{LL} + \Pi_{1,2}^{RR}  \]
\[\Pi_{1,2}^{\neq} = \Pi_{1,2}^{LR} + \Pi_{1,2}^{RL}  \]
avec $\Pi_{1,2}^{LL} = \ket{L}_1 \ket{L}_2 \bra{L}_1 \bra{L}_2$

\`A l'état initial, la probabilité de trouver les particules 1 et 2 dans le même casier et la probabilité de les trouver dans des casiers différents sont de 50\%.\\

Cependant, étant donné l'état final, on trouve toujours les particules 1 et 2 dans des casiers différents !\\

Supposons en effet qu'à un temps intermédiaire, on trouve les particules dans un même casier.\\
L'état intermédiaire est donc : 
\[ \ket{\psi'} = \Pi_{1,2}^{=} \ket{\psi} = \frac{1}{2} ( \ket{L}_1 \ket{L}_2 + \ket{R}_1 \ket{R}_2)\ket{LR}_3 \]
et donc \[ \bra{\phi}\ket{\psi'} = \bra{\phi}\Pi_{1,2}^{=} \ket{\psi} = 0  \]
Ainsi, dans le cas où les particules sont dans un même casier au temps intermédiaire, on ne peut pas les trouver dans l'état final $\ket{\phi}$. Dans le cas de la post-sélection effectuée, les particules 1 et 2 ne sont jamais dans le même casier. Par ailleurs, les états considérés étant symétriques sous permutation, ce qui est vrai pour la paire de particules (1,2) est vrai pour toute autre paire. \\

En conclusion, étant données les pré- et post-sélections considérées, malgré le fait que les trois particules soient disposées dans deux casiers, on découvre qu'aucune paire de particules ne peut être trouvée dans le même casier. Ce paradoxe est appelé "quantum pigeonhole effect", littéralement "l'effet des casiers à pigeons quantique"\footnote{Une réalisation expérimentale de ce scenario a été effectuée par interférométrie neutronique, sur un ensemble de 17 qubits. \cite{waegell2}}. Il est possible de le représenter sous la forme d'un carré de Mermin-Peres. \\

Ici, trois qubits, correspondant aux trois pigeons, sont mis en jeu. On définit la base : $\{\ket{L};\ket{R}\} \equiv \{\ket{Z=+1}\equiv\ket{+Z};\ket{Z=-1}\equiv\ket{-Z}\}$
On obtient le carré suivant :\\
\begin{figure}[ht!]
\centering
\includegraphics[width=5cm]{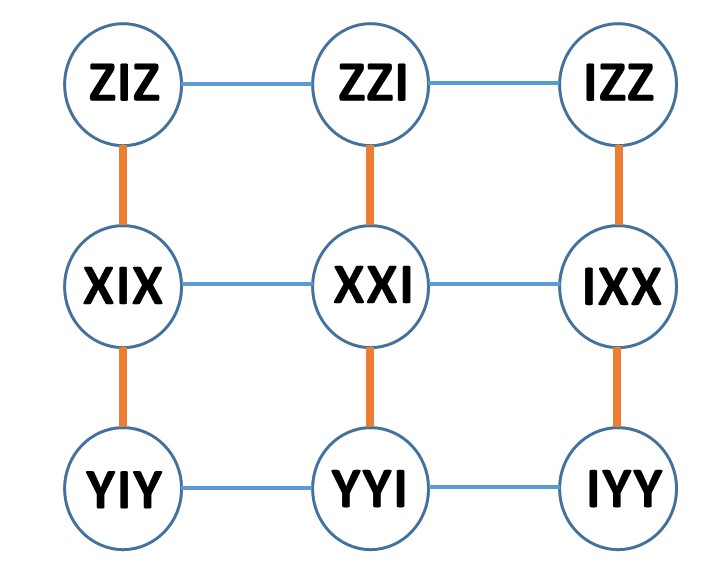}
\caption{Carré de Mermin Peres (3 qubit)}
\end{figure}

Rappelons que chaque ligne et chaque colonne correspond à un contexte d'observables qui commutent, où I, X, Y, Z sont les matrices de Pauli, avec $XIX := X_1 \otimes I_2 \otimes X_3$ .\\ Trois observables reliées par un trait bleu signifient que le produit de ces observables est égal à $+ I^{\otimes 3}$, tandis que les traits oranges correspondent à un produit d'observables égal à  $- I^{\otimes 3}$.\\

Dans l'ensemble considéré, trois contextes donnent un produit $+ I^{\otimes 3}$ ; un contexte donne un produit égal à $- I^{\otimes 3}$. La mécanique quantique prédit ainsi que le produit de la totalité des observables est égal à -1. \\

Supposons qu'un modèle ontologique KS-non-contextuel puisse décrire le système étudié. On peut donc assigner à chacune des neufs observables une valeur $\pm 1$. Comme chaque observable appartient à deux contextes (un contexte "ligne" et un contexte "colonne"), et que l'assignation est supposée KS-non-contextuelle, chacune des valeurs associées apparaît deux fois dans le calcul du produit total. Ce dernier devrait donc être égal à +1, ce qui entre en contradiction avec le résultat précédemment établi. \\

Il est possible de réduire à un contexte particulier cette violation de noncontextualité-KS par l'intermédiaire de pré- et post-sélections bien choisies.\\

Dans le cas de l'effet "pigeonhole", 
la pré-sélection et la post-sélection peuvent être réécrites sous les formes $\ket{\psi}=\ket{+X}\ket{+X}\ket{+X}$ et $\ket{\phi}=\ket{+Y}\ket{+Y}\ket{+Y}$. \\
Elles fixent une valeur $+ 1$ aux observables des deux lignes inférieures du carré.\\
Enfin, il a été montré par Leifer et Pusey (\cite{pusey1}) que si la probabilité ABL d'un projecteur est égale à 1 pour une pré-post-sélection particulière, alors on doit assigner la valeur 1 à ce projecteur dans le cadre d'un modèle ontologique non-contextuel (cf. chapitre). Il en résulte les valeurs "forcées" (\cite{waegell1}) par la règle ABL suivantes :\\

\hspace{-6mm}$ZIZ = - 1$, "le premier et le troisième pigeons sont dans des boîtes différentes";\\
$ZZI = -1$, "le premier et le deuxième pigeons sont dans des boîtes différentes";\\
$IZZ = -1$, "le deuxième et le troisième pigeons sont dans des boîtes différentes".\\

Ce triplet constitue une preuve de contextualité-KS, restreinte au contexte $\{ZIZ, ZZI, IZZ\}$.
\begin{figure}[!h]
\centering
\subfigure[Carré 3 qubit]{\includegraphics[height =4cm]{3qubit.jpg}}
\quad
\subfigure[PPS]{\includegraphics[height =4cm]{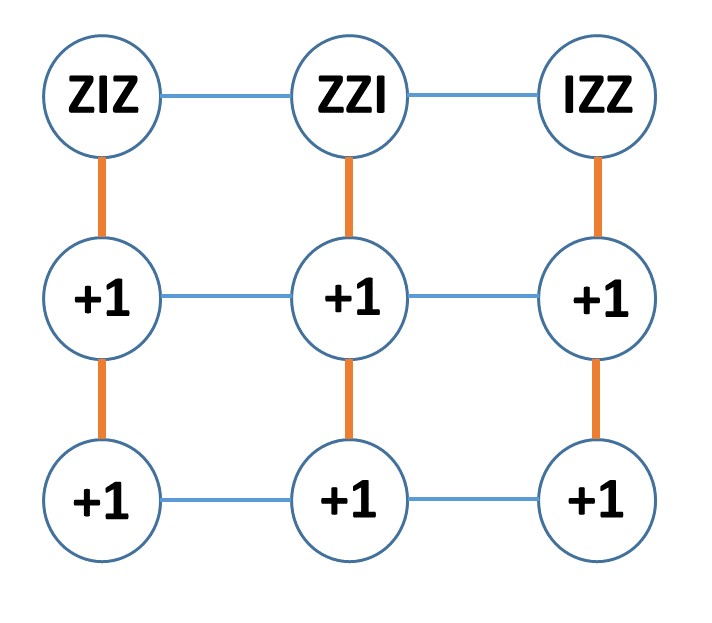}}
\quad
\subfigure[ABL]{\includegraphics[height =4cm]{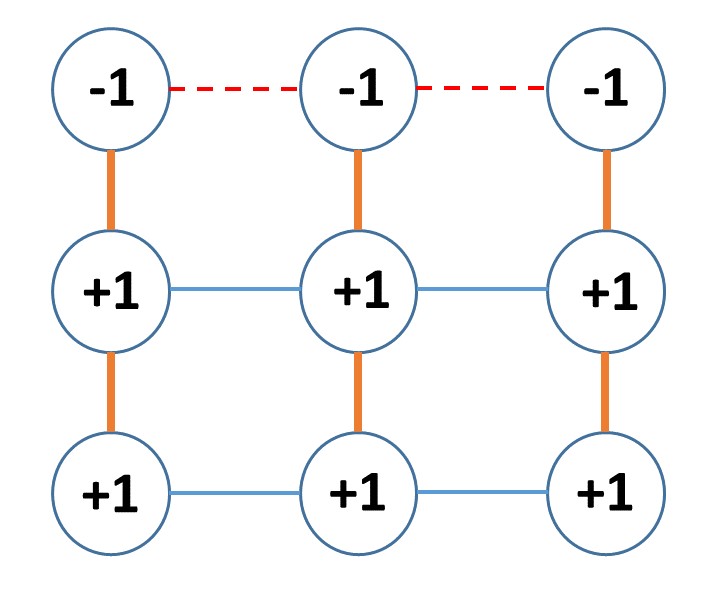}}
\caption{3 qubit}
\end{figure}

Cette contextualité-KS restreinte a l'avantage d'identifier le contexte de mesure dans lequel la contradiction logique a lieue.\\

Du fait des résultats précédemment exposés, la relation entre contextualité et paradoxe PPS logique semble assez floue. D'une part, les paradoxes PPS logiques peuvent être reproduits par un modèle KS-noncontextuel ; mais, d'autre part, ils peuvent chacun être convertis en preuve de contextualité-KS. La notion opératoire de noncontextualité universelle va permettre de surmonter cette impasse, celle-ci étant violée via les procédures de transformation au sein du modèle KS-noncontextuel. \\

\subsection{Le Théorème de Leifer-Pusey}
\label{theoremeLP}

Dix ans après l'article \cite{leifer4}, Matthew Leifer et Matthew Pusey \cite{pusey1} sont parvenus à démontrer le théorème suivant :

\boitesimple{Tout paradoxe PPS logique est une preuve de contextualité universelle.}
 
En analysant les perturbations potentielles provoquées par une mesure intermédiaire dans le cadre d'un modèle non-contextuel de Spekkens\footnote{Spekkens, on le rappelle, généralise la non-contextualité de Kochen-Specker en incluant les notions de préparation et de POVMs}, ils ont découvert que la perturbation d'un état ontologique par une mesure projective ne peut pas rendre impossible le résultat d'une autre mesure. Or, comme nous l’avons vu dans le cas du paradoxe des trois boîtes, il s’agit précisément du type de perturbation utilisé pour démontrer que la contextualité (et a fortiori la nature quantique) d’un paradoxe PPS logique n’est qu’apparente.  \\

Dans le cas d'un espace fini des états ontologiques $\Lambda$, pour un modèle ontologique universellement contextuel, la preuve du théorème de Leifer-Pusey peut s'expliquer ainsi (source :\cite{pusey1}) : \\

Du fait de la non-orthogonalité des états de pré- et post-sélections dans un paradoxe PPS logique, il résulte que l'état post-sélectionné peut toujours exister dans le cas où aucune mesure intermédiaire n'est effectuée. Ainsi, il existe un état ontologique $\lambda$ compatible avec la préparation P qui permet à la post-sélection de se produire. On considère la probabilité ABL $p(P|\psi, \phi) = 1$. Si $p(P|\lambda) = 0$, alors la mesure intermédiaire doit toujours empêcher la post-sélection de se produire, ce qui est impossible. Or, d’après le déterminisme des résultats de mesures fines, $p(P|\lambda) \in \{0,1\}$. Il en résulte que $p(P|\lambda) = 1$.  Ainsi, pour toute préparation P telle que $p(P|\psi, \phi) = 1$, on obtient : $p(P|\lambda) = p(P|\psi, \phi)$. Cependant, le modèle ontologique étant universellement non-contextuel, $p(P| \lambda)$ doit vérifier les conditions algébriques $(\alpha) , (\beta), (\gamma)$, ce qui n'est pas le cas de $p(P|\psi, \phi)$\footnote{cf. définition d'un paradoxe PPS logique}. Cette contradiction implique donc que tout paradoxe PPS logique est une preuve de contextualité universelle.\footnote{Pour une preuve formelle appliquée à un espace d'états ontologiques arbitraire, voir \cite{pusey1}. Pour une preuve impliquant la noncontextualité des procédures de transformations, voir \cite{leifer9}} Autrement dit, la nature paradoxale de ces paradoxes est bien d'origine quantique, non-classique. Ces scenarii sont donc étranges, car ils violent le principe fondamental de Leibniz lorsqu'ils sont modélisés ontologiquement.\\

D'autres types de paradoxes PPS sont également universellement contextuels. Dans ces nouveaux scenarii, les mesures intermédiaires entre pré- et post- sélections ne sont pas projectives, mais "faibles".\\

\subsection{Mesure Faible et Valeur Faible}
\label{faible}

Une mesure faible est un genre particulier de POVM qui donne en moyenne très peu d'information sur le système, et dont la perturbation sur l'état du système est négligeable. \\

On remarque qu'il semble y avoir une certaine dissonance entre l'approche originale des mesures faibles, controversée, qui est basée sur la notion de dynamique d'interaction de mesure, et l'approche plus générale des mesures basées sur les opérateurs de mesure (POVM) largement utilisés dans l'information quantique. \footnote{Pour une discussion à ce sujet, voir \cite{vaidman} } Les deux approches s'accordent cependant à dire que la mesure faible peut être considérée comme une généralisation de la mesure "forte" projective. \cite{tamir}\cite{landsman}\\

Par soucis de clarté et de cohérence vis-à-vis des travaux sur lesquels nous nous appuyons dans ce mémoire, nous présenterons la notion de mesure faible introduite par Aharonov et Vaidman \cite{aharonov5} dans cette section. Une présentation de la mesure faible d'un point de vue opérationnel peut être trouvée en annexe.\\

Le modèle considéré par Aharonov s'appuie sur le concept de "mesure idéale de von Neumann" : la mesure y est définie par une interaction entre le système quantique étudié et le dispositif de mesure, représenté par un pointeur, qui est également un système quantique. On obtient la valeur d'une observable $\widehat{A}$ en effectuant une mesure projective sur le pointeur. Dans le cas d’un processus idéal, la position $\widehat{x}$ du pointeur est bien déterminée. Or comme $[\widehat{x},\widehat{p}]=i\hbar$ d'après le principe d'Heisenberg, le moment conjugué $\widehat{p}$ du pointeur est indéterminé. A l'issu du processus, le pointeur s'est alors déplacé de $dx = gT \langle \widehat{A} \rangle$ ; avec $T$ la durée de la mesure et $g$ est une "constante de couplage" : dans le cas d'une mesure idéale, où le système étudié et le dispositif de mesure sont parfaitement intriqués, $g=1$. La nouvelle position du pointeur correspond alors à une valeur propre de $\widehat{A}$. \\

Cette interaction peut être décrite par l'hamiltonien suivant :
\[ H_{meas} = gT \widehat{A}\otimes\widehat{p} \]

D'après l'équation de Schrödinger, on peut lui associer la transformation unitaire :
\[\widehat{U} = \exp(-igT\widehat{A}\otimes\widehat{p})\]
\`A l'issu d'une "mesure" idéale, on connait donc parfaitement la valeur de l'observable considérée. Cependant, l'intrication avec le dispositif de mesure induit un effondrement de la fonction d'onde du système.\\

Dans le cas d'une "mesure" faible, on sacrifie notre connaissance "parfaite" de la valeur de l'observable au profit d'une perturbation négligeable de l'état du système. Cela se traduit par une indétermination de la position du pointeur, au profit d'un moment bien défini. L'intensité du couplage entre le système et le dispositif de mesure est alors tel que $gt \ll \Delta x$, avec $\Delta x$ l'incertitude sur la position du pointeur.\footnote{On remarque qu'après un nombre suffisamment grand de mesures sur un ensemble de systèmes identiques, la moyenne de la position du pointeur peut être déterminée à n’importe quel degré de précision.} \\

Dans le cadre d'une pré-sélection $\ket{\psi}$ au temps $t_0$ et d'une post-sélection $\ket{\phi}$ au temps $t_1$ , on peut ainsi réaliser une mesure intermédiaire faible d'une observable $A$. Ceci permet "d'observer" le comportement du système entre les deux sélections, sans perturber de façon significative son état.\\

On peut décrire évolution du système par :
\[ \bra{\phi(t_1)} \widehat{U} \ket{\psi(t_0)} = \bra{\phi(t_1)} e^{-igT\widehat{A}\otimes\widehat{p}} \ket{\psi(t_0)} \simeq \bra{\phi(t_1)} \ket{\psi(t_0)} \left( 1 - i g T w(A|\psi,\phi)\widehat{p}\right) \]

Au premier ordre en $gT$, le pointeur s'est donc déplacé de $gTw(A|\psi,\phi) $, où $w(A|\psi,\phi)$ correspond au résultat de la mesure faible, appelé \textit{la valeur faible } de $A$\footnote{De nombreux débats toujours d'actualité portent sur l'interprétation physique de ces valeurs. Voir \cite{vaidman}}. On a finalement :

\[w(A|\psi,\phi) = \hbox{Re}\left(\frac{\bra{\phi}\widehat{A}\ket{\psi}}{\bra{\phi}\ket{\psi}}\right) \]

Une valeur faible peut être située en dehors de l'intervalle de valeurs propres de l'observateur : elle est alors appelée \textit{valeur faible anomale}. On remarque que dans le cas où la pré- et la post-sélection sont égales, la valeur faible correspond simplement à la valeur moyenne de l'observable A.\\

\subsection{Valeurs Faibles Anomales et Contextualité}
\label{vf}

En 2014, Matthew Pusey a également démontré \cite{pusey2} le théorème suivant :
\boitesimple{Toute \textit{valeur faible anomale} $w(P|\psi,\phi)$ d'un opérateur P, i.e. située en dehors de l'intervalle de valeurs propres de l'opérateur, est une preuve de contextualité universelle.}
Le dénominateur de $w(P|\psi,\phi)$ étant indépendant du projecteur mesuré, la valeur faible d'un projecteur P vérifie toujours les conditions algébriques $(\beta)$ et $(\gamma)$. Cependant, une valeur faible anomale viole par définition la condition $(\alpha)$, i.e. $w(P|\psi,\phi)$ peut être négative ou supérieure à 1.
Il est également facile de vérifier la propriété suivante : si $ P(P|\psi,\phi) \in \{0,1\}$ alors $w(P|\psi,\phi) = P(P|\psi,\phi)$. Or, par définition, aucune probabilité ABL ne respecte les conditions algébriques pour un paradoxe PPS logique. La condition $(\alpha)$ étant la seule qui puisse être violée par les valeurs faibles, il en résulte que \textit{pour tout paradoxe PPS logique, il existe un projecteur P qui possède une valeur faible anomale. }\\

Dans l’exemple du paradoxe des trois boîtes, on peut ainsi obtenir les valeurs faibles anomales suivantes :
\[ w(\ket{1}\bra{1} + \ket{2}\bra{2}| \psi, \phi) = 2  \hspace{2cm} w(\ket{3}\bra{3}| \psi, \phi) = -1 \]
Ces valeurs faibles suffisent à démontrer l'incompatibilité de la théorie quantique avec tout modèle ontologique universellement non-contextuel. Combiné au théorème de Leifer-Pusey, le résultat de Pusey indique donc que tout paradoxe PPS logique est une preuve de contextualité, indépendamment de la nature « forte » ou « faible » de la mesure intermédiaire.\footnote{Une preuve expérimentale du théorème de Pusey a été réalisée : \cite{expwv} }\\

Il est intéressant de remarquer que certains paradoxes PPS (non logiques) peuvent être représentés par un modèle universellement non-contextuel si les mesures intermédiaires effectuées sont fortes, et peuvent néanmoins être universellement contextuels dans le cas de mesures intermédiaires faibles. C'est le cas du paradoxe du "chat de Cheshire", présenté dans la section suivante.\\

\section{Le chat de Cheshire}
\label{schess}
\begin{figure}[ht!]
\centering
\includegraphics[width=4cm]{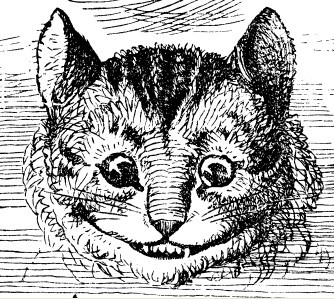}
\caption{Le chat de Cheshire}
\end{figure}
\begin{quotation}
"Très bien", dit le Chat, et cette fois il disparut lentement, en commençant par la fin de la queue, et en finissant par son large sourire, qui demeura quelque temps après que le reste soit parti.  "Et bien ! J’ai souvent vu un chat sans sourire, pensait Alice, mais un sourire sans chat ! C’est la chose la plus curieuse que j’ai vu de ma vie !" \cite{alice}\\
\end{quotation}

\textit{La présentation suivante est issue de \cite{chess}.\\}

Considérons un photon (notre chat de Cheshire) pouvant prendre deux états de polarisation : une polarisation circulaire droite, $\ket{+}$, et une polarisation circulaire gauche $\ket{-}$. Dans la base orthonormée $\{\ket{H};\ket{V}\}$ des polarisations linéaire horizontale et verticale, les deux états de polarisations du photon s'écrivent :
\[ \ket{+} = \frac{\ket{H} + i \ket{V}}{\sqrt{2}}    \hspace{2cm}   \ket{-} = \frac{\ket{H} - i \ket{V}}{\sqrt{2}} \]
Le photon, après être passé à travers une lame séparatrice, peut prendre deux positions : il peut passer par la gauche, $\ket{L}$, ou par la droite $\ket{R}$.
Nous pré-sélectionnons l'état d'un photon de polarisation linéaire horizontale après son passage à travers la séparatrice :
\[\ket{\psi} = \frac{1}{\sqrt{2}} \left( \ket{L} + \ket{R} \right) \ket{H}\]
Nous post-sélectionnons l'état intriqué du photon suivant :
\[\ket{\phi} = \frac{1}{\sqrt{2}} \left( \ket{L}\ket{H} + \ket{R}\ket{V} \right)\]
On se pose dans un premier temps la question suivante : "considérant nos pré- et post-sélections, peut-on trouver le photon à droite ?". On effectue alors la mesure intermédiaire (supposée non-destructive) : \[\Pi_R = \ket{R}\bra{R}\] On remarque que si le photon était à droite, l'état intermédiaire $\ket{\psi '} = \ket{R}\ket{H}$ serait orthogonal à $\ket{\phi}$.
Il en résulte que le photon passe toujours par la gauche.\\
Si notre "chat" est à gauche, peut-on trouver son "sourire" ailleurs ? Pour répondre à cette question, on effectue une mesure intermédiaire (non-destructive) de spin :  \[\sigma_z^{(R)} = \Pi_R \left(\ket{+}\bra{+} - \ket{-}\bra{-} \right) \] à laquelle trois réponses (valeurs propres) sont associées : \\
$-1$ si le photon est dans l'état de spin $\ket{R}\ket{-}$\\
$+1$ si le photon est dans l'état de spin $\ket{R}\ket{+}$\\
$0$ si le photon est dans un état de spin appartenant au sous-espace dégénéré couvert par $\ket{L}\ket{+}$ et $\ket{L}\ket{-}$\\

En considérant que $p(\Pi|\psi,\phi)= p(\Pi|\psi,\{\Pi,I-\Pi\},\phi)$ , on obtient les probabilités ABL suivantes :

\[p(\Pi_R\otimes I | \psi, \phi ) = 0  \]
\[p(\Pi_R\otimes \ket{+}\bra{+} | \psi, \phi) = p(\Pi_R\otimes \ket{-}\bra{-} | \psi, \phi) = \frac{1}{4} \]
On a donc :
\[p(\Pi_R\otimes I | \psi, \phi ) = p(\Pi_R\otimes (\ket{+}\bra{+} + \ket{-}\bra{-}) | \psi, \phi) \]
\[p(\Pi_R\otimes I | \psi, \phi ) \neq p(\Pi_R\otimes \ket{+}\bra{+} | \psi, \phi) + p(\Pi_R\otimes \ket{-}\bra{-} | \psi, \phi) \]

La condition algébrique $(\gamma)$ est violée. D'autre part, la probabilité de trouver le spin à droite, i.e. que le résultat de la mesure donne $+1$ ou $-1$, est non nulle. \\

Le photon semble ainsi séparé de son état de polarisation ! On peut néanmoins constater que dans l'exemple ci-dessus, position et spin ne sont jamais mesurés simultanément. Or $[\Pi_R,\sigma_z^{(R)}]=0$. Les deux mesures sont donc compatibles. Le paradoxe apparent est en fait contrafactuel : une mesure simultanée de la position et du spin fait disparaître la violation de $(\gamma)$ et, donc, la contradiction. Par ailleurs, bien qu'elles violent une condition algébrique, les probabilités ABL n'appartiennent pas à $\{0,1\}$. Le paradoxe du chat de Cheshire n'est donc pas un paradoxe PPS logique. Le paradoxe peut ainsi être reproduit à l'aide du modèle jouet noncontextuel de Spekkens.\\

Pour un système composite dans le modèle jouet, chaque partie possède son propre état ontologique. Il existe donc 16 états ontologiques possibles, et 4 questions binaires sont nécessaires pour spécifier exactement cet état. Du fait de la restriction épistémique, on ne peut  obtenir une réponse qu'à deux de ces questions. Par ailleurs, le principe d'équilibre des connaissances  ne s'applique pas uniquement au système composé, mais aussi à chaque partie. Dynamique et existence des états ontologiques sont donc restreintes.\\

Dans le scenario du chat de Cheshire, le premier qubit transporte l'information sur le chemin choisi, tandis que le second qubit indique l'état de polarisation. L'état $\ket{0}$ correspond à l'état $\ket{-x}$, et l'état $\ket{1}$ à l'état $\ket{+x}$ (cf. chapitre 3). On définit les notations :
\[\ket{0}_A \equiv \ket{L} \hspace{3cm} \ket{1}_A \equiv \ket{R} \]
\[\ket{0}_B \equiv \ket{H} \hspace{3cm} \ket{1}_B \equiv \ket{V} \]

Ainsi, les états de pré- et post- sélections s'écrivent :
\[\ket{\psi} = \frac{1}{\sqrt{2}} \left( \ket{0}_A + \ket{1}_A \right) \ket{0}_B\]
\[\ket{\phi} = \frac{1}{\sqrt{2}} \left( \ket{0}_A\ket{0}_B + \ket{1}_A\ket{1}_B \right)\]

\begin{figure}[ht!]
\centering
\includegraphics[width=6cm]{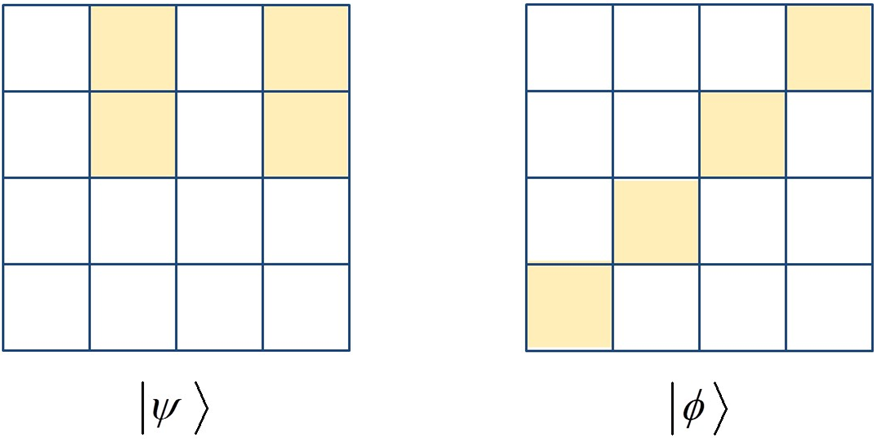}
\caption{Etats de pré- et post sélections dans le modèle jouet}
\end{figure}

Les trois mesures intermédiaires projectives sont les suivantes :

\begin{figure}[ht!]
\centering
\includegraphics[width=8cm]{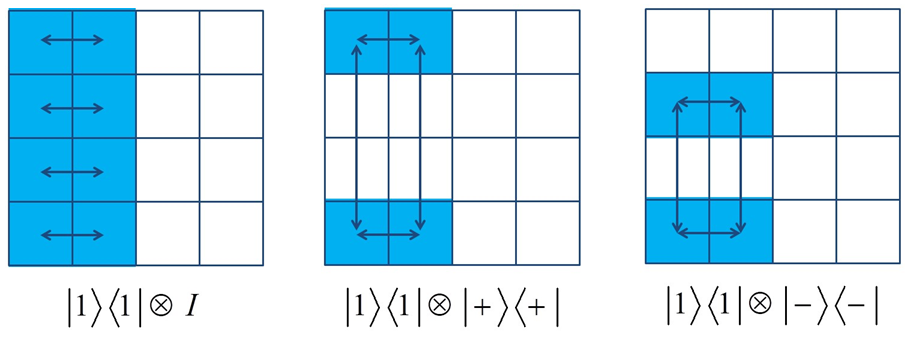}
\caption{Représentation des trois mesures projectives dans le modèle jouet}
\end{figure}

Appliquées à la pré-sélection, elles donnent les états épistémiques intermédiaires suivants :

\begin{figure}[ht!]
\centering
\includegraphics[width=8cm]{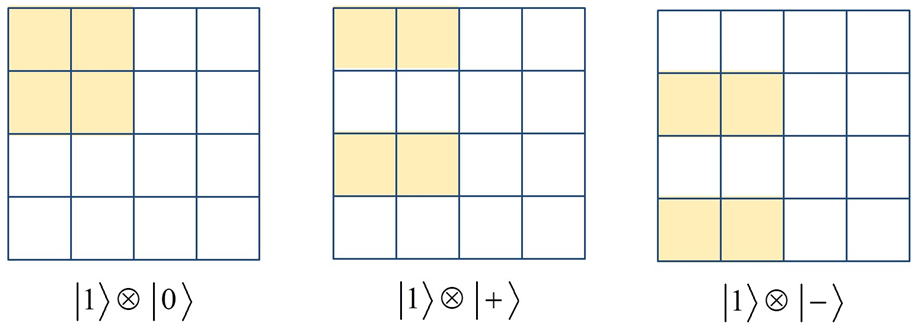}
\caption{Etats post-mesures dans le modèle jouet}
\end{figure}

On constate qu'aucun des états ontologiques de l'état $\ket{1}_A\otimes\ket{0}_B$ ne coïncide avec ceux de la post-sélection. La probabilité que l'état de spin du photon se trouve dans la branche gauche de l'interféromètre est donc nulle. En revanche, pour les deux autres états, $\ket{1}_A\otimes\ket{+}_B$ et $\ket{1}_A\otimes\ket{-}_B$, dans chaque cas, on constate qu'un état ontologique sur quatre est identique à l'un de ceux de la post-sélection. Ainsi, il y a une chance sur 4 de trouver, dans la branche droite, l'état de spin $\ket{+}$, et une chance sur quatre de trouver l'état de spin $\ket{-}$. Le modèle jouet reproduit donc exactement les prédictions quantiques.\\ 

Mais que se passerait-il si, plutôt que des mesures projectives, nous effectuions des mesures faibles ? \\

Les valeurs faibles obtenues sont les suivantes :
\[<\Pi_L>_w = \frac{\bra{\phi}\Pi_L\ket{\psi}}{\bra{\phi}\ket{\psi}} = 1 \hspace{1cm} <\Pi_R>_w = 0  \hspace{1cm} <\sigma_z^{(L)}>_w = 0  \hspace{1cm} <\sigma_z^{(R)}> = 1\]
On constate que l’on retrouve, dans le cas de mesures faibles, que le photon se situe à gauche, et son spin à droite ! \\

Nous avons donc bien obtenu un "chat de Cheshire".\footnote{L'expérience de pensée décrite par Aharonov \cite{chess} a été réalisée en laboratoire \cite{denkmayr}.}
D'autre part,
\[ <\Pi_R>_w = w(\Pi_R\otimes I|\psi,\phi) = 0 \]
\[ w(\Pi_R\otimes\ket{+}\bra{+} |\psi,\phi) = \frac{1}{2} \]
\[ w(\Pi_R\otimes\ket{-}\bra{-} |\psi,\phi) = - \frac{1}{2} \]

Les conditions $(\beta)$ et $(\gamma)$ sont satisfaites. En revanche, la valeur faible $-1/2$ n'appartient pas à l'intervalle des valeurs propres de $\Pi_R\otimes\ket{-}\bra{-}$ : elle est donc anomale, et viole la condition $(\alpha)$. Le paradoxe du chat de Cheshire, ainsi formulé, est incompatible avec l'hypothèse de non-contextualité universelle. Si la version forte du théorème peut être reproduite "classiquement", sa version faible est bien étrange, non-classique.\\

\section{Conclusion}

Les paradoxes PPS logiques sont étranges, au sens où ils violent l'hypothèse de non-contextualité universelle. Néanmoins, cette violation n'est pas triviale : bien que chacun de ces paradoxes permette de construire une preuve de contextualité-KS, l'aspect intuitivement contextuel de ces scenarii est trompeur. Il peut être expliqué par une perturbation ( ou transformation), lors de la mesure, de l'état ontologique du système, et ainsi être reproduit par un modèle ontologique à mesures non-contextuelles. Ce sont, en revanche, ces transformations qui sont, elles, contextuelles, et violent ainsi le principe de Leibniz. \\

Enfin, si certains paradoxes PPS non-logiques, tels que le chat de Cheshire quantique, peuvent être reproduits par des modèles ontologiques universellement non-contextuels comme le modèle jouet, la version faible de ces paradoxes est également étrange, puisqu'elle implique l'existence de valeurs faibles anomales, non-classiques selon le critère de Spekkens.\\

\clearemptydoublepage
\chapter{Conclusion}

Août 2017. Près de huits mois se sont écoulés depuis le début de cette quête épistémique, et il est à présent temps de conclure, ou du moins de faire le point.\\

J'espère que cette introduction a pu convaincre le lecteur de l'importance fondamentale du théorème de Kochen-Specker et de ses implications philosophiques et théoriques, et lui a également offert divers outils lui permettant d'\textit{imaginer} et de \textit{raconter} la contextualité. \\

J'espère également avoir pu attirer son attention et son intérêt sur un nouveau champ d'investigation, "au-delà du théorème de Kochen-Specker" (\cite{kunjwal2}), ancré dans le contexte de la seconde révolution quantique : la contextualité est aussi bien étudiée comme pilier théorique dans le domaine des fondements quantiques qu'en tant que ressource dans celui de l'information quantique.\\

Je tiens à rappeler que les présentations des différentes approches de généralisation sont sommaires et non exhaustives. J'aimerai qu'elles soient considérées comme des apéritifs, qui vous inviteraient vous, lecteur, à vous pencher sur ces travaux. \\

Le mémoire peut être vu comme étant plutôt orienté "en faveur" de l'approche de Robert Spekkens, qui y prend une plus grande place. Je dois dire que sa notion de contextualité comme témoin de non-classicité, que j'associe au terme "étrangeté", m'a particulièrement interpellé. Bien sûr, les autres approches n'en sont pas moins intéressantes. Vous avez pu constater que les différents chemins explorés n'étaient pas de simples lignes droites, tracées à partir du point d'origine qu'est le théorème Kochen-Specker, mais qu'ils s'entrecroisaient, s'inspiraient et s'influencaient mutuellement. L'étude des relations qu'entretiennent ces démarches entre elles est par ailleurs un sujet actif de recherche. \\

Enfin, je souhaiterai ouvrir cette conclusion sur de récents résultats obtenus dans les différentes pistes de recherche mentionnées.\\

L'approche CSW a ainsi permis d'identifier que la monogamie - principe fondamental qui stipule que si deux qubits $A$ et $B$ sont maximalement intriqués, il ne peut être corrélés à un troisième qubit $C$ - pouvais être comprise comme une conséquence du principe de Specker. \cite{jia}\\ 

Une étude récente des relations de compositions entre scenarii de contextualité dans l'approche AFLS est récemment parue sur arxiv \cite{sainz4} ; 
ainsi qu'une méthode permettant de mesurer le degré de contextualité d'un scenario dans l'approche faisceau. \cite{abramsky4}\\ 

Enfin, Ravi Kunjwal et Robert Spekkens ont étendu leur technique permettant de dériver des inégalités de non-contextualité universelle à toute preuve arbitraire du théorème de Kochen-Specker \cite{kunjwal4}, et Kunjwal semble avoir clarifié le lien entre l'approche de Spekkens et celle de CSW \cite{kunjwal5}.\\ 

Un lecteur curieux pourrait également se diriger vers les enregistrements du séminaire "Contextuality : Conceptual issues, operational signatures, and applications" qui a eu lieu en juillet dernier. \cite{pirsa}

\chapter{Annexe : La contextualité de Bohr}
\begin{quotation}
\vspace{-6mm}
\hspace{-6mm}\textit{-Qu’est-ce que tu étudie ?\\
-          La contextualité quantique !\\
-          Ha ! Bohr, Copenhague, tout ça !\\}
\end{quotation}
\small
Niels Bohr est souvent vu comme le Maître d'une "école de Copenhague" \footnote{“L’école de Copenhague” ou “l’interprétation de Copenhague” est un terme-valise qui semble faire une synthèse des interprétations de Bohr, Heisenberg, Pauli, Born et al. mais qui reste très ambigu. \cite{leifer1} \cite{howard1} }, une "interprétation orthodoxe", à laquelle serait ralliée une grande partie des physiciens.\cite{schlo} Cette interprétation, pragmatique selon certains\footnote{\textit{J'aime parlé de deux Bohrs : le premier est quelqu'un de très pragmatique qui insiste sur le fait que l'appareil de mesure est classique, et l'autre est très arrogant, un homme pontifiant qui fait d'énormes déclarations sur ce qu'il a accompli.}\cite{bell2}}, anti-réaliste selon d'autres,\footnote{\textit{Si j'étais contraint de résumer en une phrase ce que dit l'interprétation de Copenhague d'après moi, ce serait 'Shut up and calculate!'} David Mermin} est profondément marquée par la césure entre observateur et système observé. Les positions philosophiques de Bohr étant complexes et ayant évoluées au fil des ans, il me semble difficile de présenter succinctement sa pensée.\footnote{D'autant plus que : "\textit{Tout commentaire sur ce que dit Bohr est une interprétation de la pensée de Bohr.}" \cite{alexei5}}\\

Je retiendrai de mes lectures que l’on peut mettre en évidence un avant et un après EPR dans la pensée de Bohr. (\cite{held}, \cite{beller}, \cite{howard1}, \cite{saunders}).\\
 
Avant EPR, Bohr considérait que le système et l'observateur étaient deux ingrédients nécessaires à une description complète de la théorie quantique. \cite{bohr0}
Sa réponse à EPR marque quant à elle une nouvelle insistance sur le concept de \textit{phénomène}, défini par son indivisibilité entre objet et observateur, qui n’ont plus d’identités propres. L’usage du terme “phénomène” en mécanique quantique doit être ainsi exclusivement utilisé “pour faire référence aux observations obtenues sous des circonstances spécifiques, incluant un compte rendu de l’arrangement expérimental dans sa totalité.” \cite{bohr2}. 
Cette pensée le mène à la conclusion suivante : 
\begin{quotation}
Il n’y a pas de monde quantique. Il y a seulement une description physique quantique abstraite. Il est faux de penser que la tâche de la physique soit de découvrir comment la nature se comporte. La physique concerne ce que l’on peut dire de la nature. [...] De quoi dépendons-nous, nous humains ? Nous dépendons de nos mots. Notre tâche est de communiquer expérience et idées à autrui. Nous sommes suspendus dans le langage. ‘We are suspended in language’. (cité par \cite{petersen}). 
\end{quotation}

La référence au langage, cette doctrine des concepts classiques, établit une continuité dans la pensée de Bohr.
 \begin{quotation}\textit{Dans la mesure où les phénomènes transcendent la portée de l'explication physique classique, l'énoncé de toutes les preuves doit être exprimé en termes classiques. (...) L'argument est simplement que, par le mot expérience [experiment], nous nous référons à une situation où nous pouvons dire aux autres ce que nous avons fait et ce que nous avons appris et que, par conséquent, le compte rendu des arrangements expérimentaux et des résultats des observations doive être exprimé dans un langage sans ambiguïté avec une application appropriée de la terminologie de la physique classique.} \cite{bohr4}
 \end{quotation}

Bohr considère que "la tâche" de la physique est de permettre aux hommes de se transmettre « ce qu’ils ont fait et ce qu’ils ont appris ». \footnote{Cette position est bien loin du réducteur « Shut up and calculate ! ».}. Ces communications classiques sont non ambigües : elles peuvent être étudiées indépendamment de l’observateur. \footnote{La nature de la médiation de ces communications était un sujet de discorde entre Bohr et Heisenberg, qui la voyaient respectivement comme l'ensemble des mots et des images, et le formalisme mathématique abstrait de la théorie quantique.}
 
 \begin{quotation}
\textit{Toute description d'expériences a jusqu'à présent été basée sur l'hypothèse, déjà inhérente aux conventions ordinaires du langage, qu'il est possible de distinguer nettement le comportement des objets et les moyens d'observation. Cette hypothèse n'est pas seulement pleinement justifiée par toute l'expérience quotidienne, mais constitue même toute la base de la physique classique...}

 \end{quotation}
 
La physique devant établir des communications non ambiguës avec une description objective, Bohr conclut que cela doit également s’appliquer à la physique quantique, malgré le fait qu’elle nécessite la présence d’un observateur (ou appareil de mesure, Bohr ne fait pas de distinction entre les deux termes). On peut ainsi, en se rapprochant d’une interprétation kantienne (\cite{hooker} ; \cite{murdoch}, \cite{bitbol}]), donner une origine épistémologique à sa doctrine des concepts classiques, qui émergerait d’une analyse des conditions du savoir humain. (cf. titre de \cite{bohr2} ) \\
\begin{quotation}
\textit{La physique ne doit pas être vue comme l’étude de quelque chose a priori donné, mais plutôt comme le développement de méthodes permettant d’ordonner et d’examiner l’expérience humaine. À cet égard, notre tâche doit être de rendre compte de cette expérience d'une manière indépendante du jugement subjectif individuel et donc de manière objective dans la mesure où elle peut être communiquée sans ambiguïté dans le langage humain ordinaire.} \cite{bohr2} 
\end{quotation}
 
Quoiqu’il en soit, la notion de réalité est bien présente chez Bohr.  Il invite cependant à être prudent lorsqu’elle est utilisée : elle ne doit pas être érigée en concept absolu. Bohr n’indique jamais dans ses écrits qu’il serait impossible d’attribuer une réalité aux phénomènes quantiques. En revanche, il est impossible selon lui d’attribuer une réalité "indépendante" des observateurs, qu'ils soient humains ou appareils de mesure. Puisque, lors d’une observation, l’objet et l’appareil de mesure sont corrélés, on ne peut pas, selon lui, leur accorder des "éléments de réalité" séparés tels que les définissent Einstein, Podolsky et Rosen. (\cite{epr}, \cite{bohr}, \cite{howard2}) \\
 
\section*{La complémentarité}
De la corrélation entre objet et appareil de mesure découle le principe de \textit{complémentarité} : la mécanique quantique est caractérisée par une "discrimination entre différents arrangements expérimentaux qui permet d'utiliser de façon non-ambigüe des concepts classiques complémentaires".
Bohr présenta pour la première le concept de complémentarité en 1927, à la conférence de Como, lors d’un cours intitulé “The Quantum Postulate and the Recent Development of Atomic Theory”. La même année, Heisenberg introduisit son fameux principe d’indétermination (cf. encadré), dans lequel Bohr voyait l’expression symbolique de la complémentarité. \footnote{Ce n’était pas l’avis de Heisenberg, qui interprétait ses inégalités comme les conséquences de changements discontinues ayant lieu pendant un processus de mesure. \cite{philocop}}
 \small
\boitesimple{Soient   et \^B deux opérateurs hermitiens, et $\ket{\psi}$ un état quantique. Le principe d'indétermination d'Heisenberg s'écrit :
\[\Delta(A)\Delta(B) \ge \frac{\abs{\bra{\psi}[A,B]\ket{\psi}}}{2} \]
Dans le cas où les opérateurs ne commutent pas, les déviations standards des observables mesurées sont alors toutes les deux strictement supérieures à $0$. D'après Bohr, il n'existe donc pas de description "classique" commune liée à ces deux observables. Prenons un exemple avec l'opérateur position \^X et l'opérateur quantité de mouvement \^P. Le commutateur de ces deux opérateur est \\$[X,P]$ = i$\hbar$ . Il en résulte que
\[\Delta(X)\Delta(P) \ge \frac{\hbar}{2} \] En appliquant l'interprétation de Bohr, il en résulte que la description "corpusculaire" (position parfaitement définie) et la description "ondulatoire" (quantité de mouvement parfaitement définie) ne sont possibles que dans des arrangements expérimentaux distincts.}
 
\normalsize
Bohr ne donna pas de définition précise de cette notion philosophique\footnote{Il pensait peut-être qu'une définition ne permettait pas d’appréhender le concept dans son intégralité. Le couple définition/observation était en effet un de ses exemples favoris de paires complémentaires.}. Il se contenta d’analyser de nombreux exemples, qui n'étaient pas nécessairement issus de la physique (vérité/clarté, science/religion, pensées/émotions, objectivité/introspection... \cite{pais2}) ; l'exemple le plus connu restant certainement celui du couple particule/onde, qui permit d'interpréter les expériences des fentes d'Young à travers l'union de deux descriptions classiques.\\
 
Notons que le concept de complémentarité s'adapta également à l'évolution de la pensée de Bohr, se concentrant de plus en plus sur l'idée que la spécification des conditions expérimentales est cruciale pour l’usage non ambigu de concepts classiques nécessaires en théorie quantique. \cite{held} Einstein lui-même concédait qu'à travers son débat avec Bohr, il n'était jamais parvenu à comprendre la notion de complémentarité.\\
 
Les deux plus proches collaborateurs du physicien danois avaient chacun leur propre interprétation du concept de complémentarité : si Heisenberg identifiait les images de complémentarité à un système quantique avec des représentations mathématiques équivalentes (Complémentarité de x et p comme représentation de Schrödinger), Pauli considérait que deux observables sont complémentaires si les opérateurs correspondant ne commutent pas. Il déduisait ainsi des inégalités d'Heisenberg que les observables complémentaires ne peuvent pas être simultanément observées avec une précision arbitraire. La complémentarité suggérerait alors qu'elles devraient être mesurées indépendamment, en utilisant des arrangements expérimentaux mutuellement exclusifs. \footnote{Dans ces premiers écrits post-Como, il semble que Bohr ne fût pas entièrement en accord avec l'approche de Pauli \cite{held} : dans le cas de l'expérience des fentes d’Young, Bohr considérait que la complémentarité intervenait entre les concepts de particule et d’onde ; tandis que Pauli l’associait aux observables position et moment. Dans ses derniers écrits cependant, Bohr reprend l'interprétation de Pauli.}

 Cette discrimination des agencements expérimentaux vis à vis de la mesure d’observables non commutatives est à l'origine de l'association du terme "contextualité" aux idées de Bohr.
Néanmoins, à ma connaissance, Bohr n'a jamais utilisé le terme de "contexte expérimental", ou de "contextualité". Il employait plutôt les termes "d'arrangement expérimental" ou "d'appareil de mesure" pour identifier "l'agent observateur". Si "contexte expérimental" est synonyme "d'appareil de mesure", alors on pourrait considérer que la complémentarité est une forme de "contextualité".\\

\chapter{Annexe : Futurs contingents, d'Aristote à Leibniz}
\normalsize
La contingence aristotélicienne entre en conflit avec le dogme de la nécessité absolue, exposé un siècle auparavant par Leucippe, le "père de l'atomisme" :
\begin{quotation}
\textit{Rien ne se produit par hasard, mais tout se produit pour une raison et par nécessité.} 
 (Leucippus, 440 BCE) \cite{leucippe}

\end{quotation}

Près de deux millénaires après Leucippe, Spinoza va s'opposer à la vision aristotélicienne du monde, et ravive cette position "super-déterministe" :
\begin{quotation}
\textit{Il n'y a rien de contingent dans la nature des êtres ; toutes choses au contraire sont déterminées par la nécessité de la nature divine à exister et à agir d'une manière donnée.} \cite{spinoza}
\end{quotation}

Pour Spinoza, Dieu, ou la Nature, est nécessaire, et ne peut être contingent. Or, "Tout ce qui est, est en Dieu". Le philosophe reprend alors l'argument "dominateur" de Diodore dans sa démonstration :
\begin{quotation}
 \textit{Ainsi donc, Dieu est la cause de tous ces modes, non seulement en tant qu'ils existent purement et simplement (par le corollaire de la proposition 24), mais aussi en tant qu'on les connaît comme déterminés à telle ou telle action (par la proposition 26). Que si Dieu ne les détermine en aucune façon, toute détermination qu'on leur attribuera, sera, non pas une chose contingente, mais une chose impossible (par la même proposition), et au contraire si Dieu les détermine de quelque façon, supposer qu'ils se rendent eux-mêmes indéterminés, ce ne sera pas supposer une chose contingente, mais une chose impossible.} \cite{spinoza}

\end{quotation}

Trois ans après la publication de l'\textit{Ethique}, Leibniz imagine un scientifique devenu prophète, étant capable de prédire le futur à partir d'une connaissance complète du présent : \begin{quotation}
\textit{On voit [...] que tout est mathématique, c'est-à-dire, que tout arrive infailliblement dans le vaste monde tout entier, de telle sorte que, si quelqu'un pouvait avoir une vue suffisante des parties intérieures des choses et en même temps suffisamment de mémoire et de compréhension, il serait un prophète et verrait le futur dans le présent comme dans un miroir.} \cite{bouveresse1}

 \end{quotation}
 Le prophète de  Leibniz sera plus tard repris par Pierre-Simon Laplace dans son \textit{Essai philosophique sur les probabilités}, et entrera dans la postérité sous l'appelation de "démon de Laplace" : 
 \begin{quotation}
\textit{Les événemens actuels ont, avec les précédens, une liaison fondée sur le principe évident, qu’une chose ne peut pas commencer d’être, sans une cause qui la produise. Cet axiome, connu sous le nom de principe de la raison suffisante, s’étend aux actions mêmes que l’on juge indifférentes [...].
Nous devons donc envisager l’état présent de l’univers, comme l’effet de son état antérieur, et comme la cause de celui qui va suivre. Une intelligence qui, pour un instant donné, connaîtrait toutes les forces dont la nature est animée, et la situation respective des êtres qui la composent, si d’ailleurs elle était assez vaste pour soumettre ces données à l’analyse, embrasserait dans la même formule les mouvemens des plus grands corps de l’univers et ceux du plus léger atome : rien ne serait incertain pour elle, et l’avenir comme le passé, serait présent à ses yeux.} \cite{laplace}\\
\end{quotation}

Comme Aristote, qui dans sa \textit{Métaphysique $\Lambda$} écrivait que : " \textit{rien n'est mû par le hasard, mais il faut toujours que soit quelque cause} ", Leibniz affirme que :

\begin{quotation} \textit{jamais rien n'arrive sans qu'il y ait une cause ou du moins une raison déterminante, c'est-à-dire qui puisse servir à rendre raison a priori pourquoi cela est existant plutôt que non existant et pourquoi cela est ainsi plutôt que de toute autre façon.} \cite{leibniz1} \end{quotation}

Ce \textit{principe de raison suffisante}, i.e. qu'il y ait une raison suffisante à tout évènement, n'implique pas que cette raison soit toujours accessible à l'entendement humain \cite{leibniz1}.  Leibniz distingue ainsi deux types de nécessités : les nécessités nécessaires (ou universelles), qui portent sur des vérités universelles, comme par exemple le fait que trois fois trois font neuf ; et les nécessités contingentes (ou singulières), qui correspondent à des vérités "certaines" mais qui ne sont pas "nécessaires".  Les raisons suffisantes de ces vérités qui auraient pu ne pas être dans un autre monde, à un autre temps, ne sont connues que de Dieu. La bataille navale potentielle d'Aristote est un exemple de nécessité contingente. Le paradoxe n'en est un que pour l'entendement humain, car il n'a pas accès aux raisons suffisantes de cet évènement avant qu'il ait lieu, contrairement à Dieu.\\
 
D'après Leibniz, "tout est possible". Cependant, d'après la proposition 2 de Diodore, l'impossible ne peut pas émerger du possible. Leibniz imagine alors une ensemble infini de mondes possibles, mais qui ne sont pas \textit{compossibles} entre eux. Ainsi, "tout est possible, mais tous les possibles ne sont pas compossibles les uns avec les autres." \cite{deleuze} "César a franchi le Rubicon" est un énoncé qui est "compris" dans notre monde. Il existe un monde possible où "César n'a pas franchi le Rubicon", mais celui-ci n'est pas compossible avec le notre.\footnote{Deleuze remarque que le concept de compossibilité ne peut se réduire au principe de contradiction. "Ce n'est pas possible puisque Adam non pécheur n'est pas contradictoire en soi et que la relation de compossibilité est absolument irréductible à la simple relation de possibilité logique. " } Enfin, à la question : "Pourquoi  sommes nous dans ce monde et pas un autre ?", Leibniz répond que "Dieu choisit le meilleur des mondes possibles".\\

Leibniz semble ainsi réconcilier la notion de contingence d'Aristote et l'argument de Diodore, en invoquant une pluralité de mondes possibles, liés entre eux par des relations de compossibilité ou d'incompossiblité. \\

Le problème des futurs contingents et l'étude des contrafactualités reste encore aujourd'hui, des sujets ouverts. Ainsi, au début du XXe siècle, le logicien polonais Jan Lukasiewicz proposa d'abandonner la logique classique bivalente pour une logique ternaire : le vrai, le faux, et l'\textit{encore indéterminé}. Le philosophe américain David Lewis s'est quant à lui particulièrement intéressé à la question des contrafactuelles  et défend l'idée d'une réalité modale faite d'une infinité de mondes, une vérité étant nécessaire si elle est vrai dans tous les mondes possibles. \cite{lewis}\\

\chapter{Annexe : Le paradoxe de Hardy}

Le paradoxe de Hardy est une expérience de pensée dans laquelle une particule et son antiparticule peuvent interagir sans s'annihiler, \cite{hardy3} et qui a conduit à la découverte d'une preuve de non-localité quantique sans inégalités. \cite{hardy4} Cette preuve peut être très facilement exposée, à l'aide d'une simple analogie. (Source : \textit{The mystery of the quantum cakes}, par Kwiat et Hardy \cite{kwiat})\\

Considérons une fabrique artisanale à partir de laquelle sortent de deux côtés opposés, des fours fermés, contenant chacun un gâteau "quantique". Les fours se déplacent  sur un tapis roulant, et deux testeurs, Alice et Bob, placés chacun d'un côté, sont chargés d'effectuer des mesures en ouvrant les fours. Ils ont le choix entre deux types de tests :\\

\begin{itemize}
\item un \textit{test de goût} : l'expérimentateur ouvre le four en fin de parcours, et goûte le gâteau, qui peut être bon ou mauvais ;
\item un \textit{test de levage} : l'expérimentateur ouvre le four à mi-parcours, et observe si le gâteau est prématurément gonflé ou non.\\
\end{itemize}

Ces deux "observables" sont incompatibles entre elles : en effet, on peut facilement imaginer qu'après avoir effectué un test de levage, refermer le four conduirait à l'affaissement du gâteau et cela impacterait alors certainement son goût.\\

Chaque expérimentateur effectue aléatoirement et librement les tests qu'il souhaite effectués et enregistre les résultats. Alice et Bob se réuniront plus tard pour les comparer : ils découvriront alors d'étranges corrélations entre les gâteaux, qui émergent de leur nature quantique.\\

En effet, pour des gâteaux décrits par un état quantique intriqué particulier, si Alice et Bob ont tous les deux ouvert leur four à mi-chemin, ils trouvent alors, dans 9\% du temps, que les deux gâteaux ont gonflés prématurément. Le reste du temps, seulement un des gâteaux (ou aucun d'entre eux) est déjà levé (1).\\

Dans le cas où l'un des expérimentateurs regarde à mi-chemin, et l'autre à la fin : si le gâteau d'Alice est prématurément gonflé, le gâteau de Bob a bon goût (2) ; si le le gâteau de Bob est prématurément gonflé, le gâteau d'Alice a bon goût (2').\\

Enfin, considérons le cas où ils ont tous les deux effectué un test gustatif. Raisonnons dans un premier temps contrafactuellement, et considérons les 9\% de cas dans lesquels deux gâteaux auraient été trouvés prématurément gonflés, si Alice et Bob avaient tous les deux effectué un test de levage plutôt qu'un test gustatif. Puisque dans ces 9\% de cas, le gâteau d'Alice aurait été gonflé, (2) implique que le gâteau de Bob aurait eu bon goût. De même, puisque dans ces 9\% de cas, le gâteau de Bob aurait été gonflé, (2') implique que le gâteau d'Alice a bon goût. Ainsi, sur la base de ce raisonnement contrafactuel, on peut inférer que deux gâteaux seront bons dans au moins 9\% des cas. \'Etrangement, le résultat donné par les corrélations quantiques est tout autre : \\
les gâteaux d'Alice et de Bob ne sont jamais tous les deux bons, i.e. au moins un des gâteaux est toujours mauvais. (3)\\

Si (2), (2') et (3) sont prises comme conditions de départ, la contradiction entre les prédictions quantiques et le réalisme locale émerge dès qu'un évènement de type (1) est observé. Il s'agit donc d'une violation quantique "non-statistique" de l'hypothèse de réalisme local.

\chapter{Annexe : Autres paradoxes}
\vspace{-1.5cm}
\small
Le \textbf{Pont de Buridan} (aussi connu sous le nom de Sophisme 17) fut énoncé par Jean Buridan, prêtre et philosophe français du XIVe siècle, dans son ouvrage \textit{Sophismata}. \\

\begin{figure}[ht!]
\centering
\includegraphics[width=4cm]{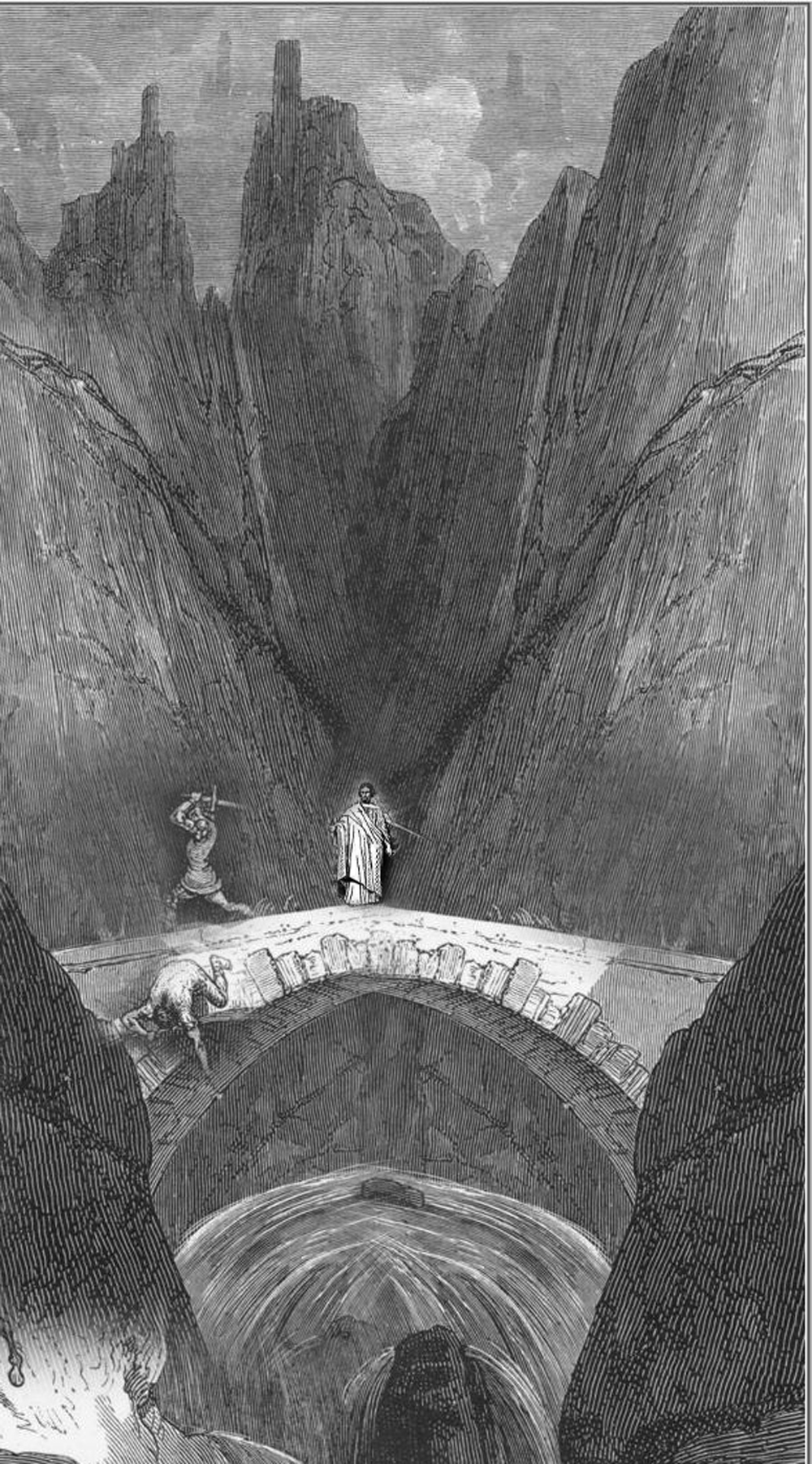}
\caption{Le pont de Buridan. \cite{buridan} }
\end{figure}

Socrate souhaite traverser une rivière et arrive devant un pont, gardé par Platon, qui l'apostrophe : "Socrate, si la première chose que tu me dis est vrai, je t'autoriserai à passer. Mais si tu mens, je te jetterai dans l'eau." [Socrate est embêté : il ne sait pas nager, et veut absoluement traverser ce pont. Soudain, son démon apparait, et lui murmure quelque chose à l'oreille.] Socrate fait alors face à Platon et lui répond :"Tu me jetteras dans l'eau."\\

La réponse de Socrate met Platon dans l'embarras : il ne peut pas jeter Socrate à l'eau car s'il le faisait, il rendrait la déclaration de Socrate vraie, violant ainsi sa promesse de ne pas le jeter s'il dit la vérité. D'autre part,  Platon ne peut pas laisser Socrate traverser le pont, car s'il le faisait, la réponse de Socrate deviendrait fausse et il devrait le jeter dans l'eau. \\

Buridan propose la solution suivante : il est impossible de déterminer si l'énoncé de Socrate est vrai ou faux, puisqu'il s'agit d'un futur contingent. Il propose donc de considérer que la promesse de Platon devrait être fausse, puisque celui-ci à donner sa parole négligemment, et qu'il n'est pas obligé de tenir cette promesse. Par aileurs, Platon aurait du être plus prudent en formulant sa condition, afin d'éviter toute contradiction. \\

\vspace{-5mm}
Walter Burley, un philosophe scolastique anglais contemporain de Buridan, appliqua le principe suivant afin de résoudre le sophisme : "\textit{nihil est verum nisi in hoc instanti}", "Rien n'est vrai si ce n'est à l'instant." L'énoncé de Socrate n'est donc pas vrai, puisqu'il porte sur le futur : Platon doit le jeter dans l'eau. Ce sophisme fut repris par Miguel de Cervantes dans son \textit{Don Quixote}. Sancho doit résoudre un dilemme similaire, et propose la solution suivante : [Platon] devrait jeter à l'eau la partie [de Socrate] qui ment, et laisser passer la partie qui dit vrai. \cite{buridan}\\

Ici, l'aspect paradoxal de la réponse de Socrate, vient à la fois fait qu'elle est une infuturabilité et une auto-référence.
"Socrate ment quand il dit que Platon va le jeter à l'eau."
"Platon le jette à l'eau, donc Socrate disait la vérité quand il déclarait que Platon le jetterait à l'eau."\\

Un paradoxe similaire, dit \textbf{paradoxe de Pinocchio}, est devenu un "mème" sur Internet.\footnote{Un "mème" est  élément d'une culture ou d'un ensemble de comportements qui se transmet d'un individu à l'autre par imitation ou par un quelconque autre moyen non-génétique.}\\

\begin{figure}[ht!]
\centering
\includegraphics[width=3cm]{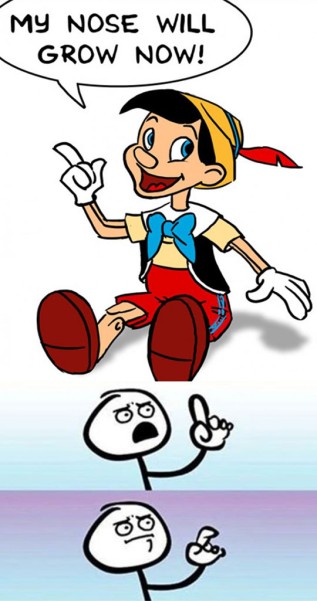}
\caption{Le paradoxe de Pinocchio }
\end{figure}

 Lorsque Pinocchio ment, son nez s'allonge. Que se passerait-il si Pinocchio déclarait : "Mon nez va grandir !" ? Si Pinocchio déclare que son nez va grandir, mais que ce n'est pas le cas, alors il est en train de mentir. Mais son nez grandit lorsqu'il ment, alors il devrait être en train de dire la vérité. Du coup, son nez continue à grandir, alors qu'il dit la vérité ! Une nouvelle fois, la contextualité semble se cacher derrière le paradoxe : on distingue deux sections locales cohérentes, qui deviennent incohérentes globalement :\\
"Pinocchio ment quand il dit que son nez va grandir et donc son nez grandit." \\
"Le nez de Pinocchio grandit, donc il disait la vérité quand il déclarait que son nez allait grandir."
\normalsize
\clearemptydoublepage

\chapter{Annexe : Preuve de contextualité de préparation pour un qubit}
\small
\vspace{-1.5cm}
Considérons les 6 états suivants sur la sphère de Bloch. (\cite{spekkens1})

\begin{figure}[ht!]
\centering
\includegraphics[width=4cm]{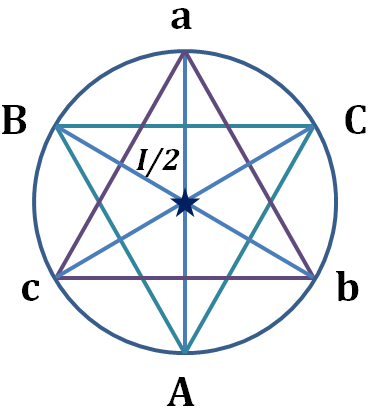}
\caption{Représentation des 6 états purs sur la sphère de Bloch }
\end{figure}

Ces états sont tels que :
\[ \ket{a}\bra{A} =\ket{b}\bra{B} = \ket{c}\bra{C} = 0\]
ce qui implique
\[ \Lambda_a\cap\Lambda_A = \Lambda_b\cap\Lambda_B = \Lambda_c\cap\Lambda_C = \emptyset \]
où \[\Lambda_\psi = \{\lambda|p(\lambda|\psi) > 0 \}\]

Lemme :\\
$P$ est une préparation de $\ket{\psi}$
$M$ est une mesure dans la base orthonormée qui inclue $\ket{\psi}$
si
\[ \Lambda^P_\psi = \{ \lambda \in \Lambda|p(\lambda|P)>0 \} \]
\[ \Gamma^M_\psi = \{ \lambda \in \Lambda|p(\psi|M,\lambda)=1 \} \]

alors
\[ \Lambda^P_\psi \subseteq \Gamma^M_\psi  \]

Démonstration du lemme :\\

\[ |\bra{\psi}\ket{\psi}|^2 = \int_\Lambda d\lambda p(\psi|M,\lambda)p(\lambda|P)\\ 
							= \int_{\Lambda^P_\psi} d\lambda p(\psi|M,\lambda)p(\lambda|P)\\
                            = 1\]
Cependant, puisque \[ \int_{\Lambda^P_\psi} d\lambda p(\lambda|P) = 1\]
et $p(\lambda|P)>0$ sur $\Lambda^P_\psi$, la fonction réponse $p(\psi|M,\lambda)$ doit être égale à 1 sur l'ensemble $\Lambda^P_\psi$.\\

On a donc $\Lambda^P_\psi \subseteq \Gamma^M_\psi$\\

Cependant, $p(a|M,\lambda) + p(A|M,\lambda) = 1$ d'où $\forall \lambda \in \Lambda$
 $p(a|M,\lambda)=1 \Rightarrow p(A|M,\lambda)=0$ et vice-versa.
Ainsi, 
\[\Gamma^M_a \cap \Gamma^M_A = \emptyset \Rightarrow \Lambda^P_a \cap \Lambda^P_A = \emptyset  \]

L'état mixte maximal $I/2$ peut être obtenu avec différentes préparations.
\[ \frac{I}{2} = \frac{1}{2}(\ket{a}\bra{a}+\ket{A}\bra{A})
			   = \frac{1}{2}(\ket{b}\bra{b}+\ket{B}\bra{B})
               = \frac{1}{2}(\ket{c}\bra{c}+\ket{C}\bra{C})\]
               \[ \frac{I}{2}= \frac{1}{3}(\ket{a}\bra{a}+\ket{b}\bra{b}+\ket{c}\bra{c})
               = \frac{1}{3}(\ket{A}\bra{A}+\ket{B}\bra{B}+\ket{C}\bra{C})\]
Préparer l'état $\ket{a}\bra{a}$ avec une probabilité $1/2$ et l'état $\ket{A}\bra{A}$ avec une probabilité $1/2$ est opérationnellement équivalent à préparer l'état $\ket{b}\bra{b}$ avec une probabilité $1/2$ et l'état $\ket{B}\bra{B}$ avec une probabilité $1/2$ etc.\\

Supposons par l'absurde que les préparations ne sont pas contextuelles. On peut donc écrire :

\[ p(\lambda|\frac{I}{2}) = \frac{1}{2}(p(\lambda|a)+p(\lambda|A))
			   = \frac{1}{2}(p(\lambda|b)+p(\lambda|B))
               = \frac{1}{2}(p(\lambda|c)+p(\lambda|C))\]
              \[ p(\lambda|\frac{I}{2}) = \frac{1}{3}(p(\lambda|a)+p(\lambda|b)+p(\lambda|c))\\
               = \frac{1}{3}(p(\lambda|A)+p(\lambda|B)+p(\lambda|C))\]

Du fait de l'orthogonalité des couples ($\ket{a}$,$\ket{A}$),($\ket{b}$,$\ket{B}$), ($\ket{c}$,$\ket{C}$), tout état ontologique $\lambda$ donné peut seulemet être dans au plus $\Lambda_a$ ou $\Lambda_A$, $\Lambda_b$ ou $\Lambda_B$, $\Lambda_c$ ou $\Lambda_C$.
Choisissons un $\lambda$ qui n'est ni dans $\Lambda_a$, ni dans $\Lambda_b$, ni dans $\Lambda_C$. On a donc $p(\lambda|a)=p(\lambda|b)=p(\lambda|C)=0$
d'où 
\[p(\lambda|\frac{I}{2}) = \frac{1}{2} p(\lambda|c) = \frac{1}{3}p(\lambda|c) \]
et donc, pour cet état ontologique $\lambda$ particulier :
\[p(\lambda|\frac{I}{2}) = 0 \]
On trouve un résultat similaire pour chaque choix d'état $\lambda \notin
 \Lambda_{a,A},\Lambda_{b,B},\Lambda_{c,C}$
On épuise insi l'espae des état ontologiques $\Lambda$ : $p(\lambda|\frac{I}{2}) = 0$ sur tout $\Lambda$, ce qui ne peut être vrai pour une distribution de probabilité.

\chapter{Annexe : La notion de mesure faible comme POVM}
\small
\vspace{-3mm}
Considérons\footnote{Nous reprenons ici le paragraphe de Quantum Error Correction publié par Daniel A. Lidar,Todd A. Brun , \cite{lidar} } un système dans l'état $\ket{\psi} = \alpha\ket{0}+\beta\ket{1}$. On définit les opérateurs positifs suivants :

\[ E_0 \equiv \ket{0}\bra{0} + \left(1-\epsilon\right) \ket{1}\bra{1} = A_0^2 \]
\[ E_1 \equiv \epsilon \ket{1}\bra{1} = A_1^2 \]
\[ A_0 \equiv \ket{0}\bra{0} + \sqrt{1-\epsilon} \ket{1}\bra{1} \]
\[ A_1 \equiv \sqrt{\epsilon} \ket{1}\bra{1} \]

avec $\epsilon \ll 1$. Ces opérateurs constituent un POVM ($E_0 + E_1 = I$). La probabilité d'obtenir le résultat 0 ;  $p_0 = \bra{\psi}E_0\ket{\psi}=1-\epsilon|\beta|^2$ ; est proche de 1, tandis que la probabilité d'obtenir le résultat 1 ; $p_1 = \bra{\psi}E_1\ket{\psi}=\epsilon|\beta|^2$ ;  est peu plausible. Ainsi, la plupart des mesures de ce type donneront le résultat 0, et peut d'information sur le système sera alors obtenue. On peut en effet quantifier l'information disponible à l'aide de l'entropie de Shannon. On obtient : $H_{meas} \leq h(\epsilon) = -\epsilon\log_2\epsilon - (1-\epsilon)\log_2(1-\epsilon)$ avec $h(\epsilon)\rightarrow 0$ et $\epsilon \rightarrow 0$.\\

L'état change très peu après l'obtention du résultat de mesure 0, le nouvel état étant :

\[\ket{\psi_0} = A_0\ket{\psi}/\sqrt{p_0} = (\alpha\ket{0} + \beta\sqrt{1-\epsilon}\ket{1})/\sqrt{p_0} \]

Cependant, après un résultat de mesure 1, l'état du système change drastiquement, le laissant dans l'état $\ket{1}$. Cette mesure faible donne ainsi généralement peu d'information sur l'état du système et à un effet négligeable sur lui ; en revanche, il arrive rarement qu'elle fournisse une grande quantité d'information et qu'elle perturbe fortement l'état.

On peut également donner un autre exemple de mesure faible avec les opérateurs positifs suivants :
\vspace{-3mm}
\[ E_0 \equiv \left(\frac{1+\epsilon}{2}\right) \ket{0}\bra{0} + \left(\frac{1-\epsilon}{2}\right) \ket{1}\bra{1} = A_0^2 \]

\[ E_1 \equiv \left(\frac{1-\epsilon}{2}\right) \ket{0}\bra{0} + \left(\frac{1+\epsilon}{2}\right) \ket{1}\bra{1} = A_1^2 \]

\[ A_0 \equiv \sqrt{\frac{1+\epsilon}{2}} \ket{0}\bra{0} + \sqrt{\frac{1-\epsilon}{2}} \ket{1}\bra{1} \]

\[ A_1 \equiv \sqrt{\frac{1-\epsilon}{2}} \ket{0}\bra{0} + \sqrt{\frac{1+\epsilon}{2}} \ket{1}\bra{1} \]

Ces opérateurs constituent un POVM ($E_0 + E_1 = I$). Avec $\epsilon \ll 1$ $E_0$ et $E_1$ sont très proches de $I/2$, et on a $p_0 \simeq p_1 \simeq 1/2$. L'information obtenue à la suite de cette mesure est donnée par l'entropie de Shannon (voir encadré), et est environ égale à un bit : $1-\epsilon^2/2 \ln 2 \inf H_{meas} \leq 1$.\\
\vspace{-5mm}
\boitesimple{L'information moyenne obtenue à la suite d'une mesure projective est donnée par l'entropie de Shannon : \[ H_{meas} = -p_0\log_2p_0 - p_1\log_2p_1  \]}

L'évolution de l'état du système après mesure est négligeable. En effet, le nouvel état est donné par $\ket{\psi_k} = A_k\ket{\psi}/\sqrt{p_k}$. On a donc :

\[ \ket{\psi_0} = \frac{1}{\sqrt{1+\epsilon (|\alpha|^2 -|\beta|^2}}(\alpha\sqrt{1+\epsilon}\ket{0}+\beta\sqrt{1-\epsilon}\ket{1}) \simeq \alpha(1+\epsilon|\beta|^2)\ket{0} + \beta(1-\epsilon|\alpha|^2)\ket{1}\]

\[ \ket{\psi_1} = \frac{1}{\sqrt{1-\epsilon (|\alpha|^2 -|\beta|^2}}(\alpha\sqrt{1-\epsilon}\ket{0}+\beta\sqrt{1+\epsilon}\ket{1}) \simeq \alpha(1-\epsilon|\beta|^2)\ket{0} + \beta(1+\epsilon|\alpha|^2)\ket{1}\]

Pour cette mesure faible, le résultat de mesure est presque aléatoire, mais comporte une petite quantité d'information sur l'état. Dans ce cas, on ne peut pas quantifier l'information sur l'état en utilisant l'entropie de Shannon, qui est toujours très proche de 1. En revanche, on peut regarder comment l'entropie de von Neumann peut décroître : si l'on commence avec un mélange maximal $\rho = I/2$; dont l'entropie de von Neumann est égale à 1, et que l'on effectue une mesure faible, l'entropie de von Neumann décroit de $\epsilon^2/2\ln2$. \\

On remarque enfin que dans le cas où $\epsilon =1$, $\{E_0,E_1\}$ constituent un PVM. Une mesure forte est donc un cas particulier de mesure faible. (source : \cite{lidar} )\\
\normalsize
\backmatter

\clearemptydoublepage

\bibliography{biblio}

\begin{thebibliography}{100}

\bibitem{pirsa}
Contextuality:conceptual issues, operational signatures, and applications, july
  2017.

\bibitem{deleuze}
Gilles deleuze - cinéma cours 48 du 06/12/83 - 2 transcription : Sabine mazé.

\bibitem{figimp}
Impossible figures.

\bibitem{lievres}
Trois lièvres, wikipedia.

\bibitem{buridan}
Wikipedia : Buridan's bridge.

\bibitem{eco1}
Quantum leaps, 2017.
\newblock
  http://www.economist.com/news/leaders/21718503-strangeness-quantum-realm-opens-up-exciting-new-technological-possibilities-quantum?fsrc=scn

\bibitem{eco2}
Technology quaterly, here, there and everywhere, 2017.
\newblock
  http://www.economist.com/technology-quarterly/2017-03-09/quantum-devices.

\bibitem{abramsky3}
S.~Abramsky, R.~S. Barbosa, K.~Kishida, R.~Lal, and S.~Mansfield.
\newblock {Contextuality, Cohomology and Paradox}.
\newblock 41:211--228, 2015.

\bibitem{abramsky4}
S.~Abramsky, R.~S. Barbosa, and S.~Mansfield.
\newblock Contextual fraction as a measure of contextuality.
\newblock {\em Phys. Rev. Lett.}, 119:050504, Aug 2017.

\bibitem{abramsky1}
S.~Abramsky and A.~Brandenburger.
\newblock The sheaf-theoretic structure of non-locality and contextuality.
\newblock {\em New Journal of Physics}, 13(11):113036, 2011.

\bibitem{abramsky2}
S.~Abramsky and L.~Hardy.
\newblock Logical bell inequalities.
\newblock {\em Phys. Rev. A}, 85:062114, Jun 2012.

\bibitem{acin1}
A.~Ac{\'i}n, T.~Fritz, A.~Leverrier, and A.~B. Sainz.
\newblock A combinatorial approach to nonlocality and contextuality.
\newblock {\em Communications in Mathematical Physics}, 334(2):533--628, Mar
  2015.

\bibitem{aharonov5}
Y.~Aharonov, D.~Z. Albert, and L.~Vaidman.
\newblock How the result of a measurement of a component of the spin of a
  spin-1/2 particle can turn out to be 100.
\newblock {\em Physical Review Letters. 60, 1351–1354}, 1988.
\newblock doi:10.1103/PhysRevLett.60.1351.

\bibitem{aharonov2}
Y.~Aharonov, P.~G. Bergmann, and J.~L. Lebowitz.
\newblock Time symmetry in the quantum process of measurement.
\newblock {\em Phys. Rev. 134, B1410}, jun 1964.

\bibitem{aharonov6}
Y.~Aharonov, F.~Colombo, S.~Popescu, I.~Sabadini, D.C.Struppa, and
  J.~Tollaksen.
\newblock Quantum violation of the pigeonhole principle and the nature of
  quantum correlations.
\newblock {\em PNAS 113, 532}, 2016.
\newblock doi: 10.1073/pnas.1522411112 ; arxiv version : arXiv:1407.3194.

\bibitem{aharonov3}
Y.~Aharonov, S.~Popescu, and J.~Tollaksen.
\newblock A time-symmetric formulation of quantum mechanics.
\newblock {\em Phys. Today 63, 27–32}, 2010.

\bibitem{aharonov8}
Y.~Aharonov and D.~Rohrlich.
\newblock {\em Quantum Paradoxes: Quantum Theory for the Perplexed}.
\newblock Wiley-VCH, 2003.

\bibitem{chess}
Y.~Aharonov, D.~Rohrlich, S.~Popescu, and P.~Skrzypczyk.
\newblock Quantum cheshire cats.
\newblock {\em New J. Phys. 15, 113015}, 2013.
\newblock arxiv version : arXiv:1202.0631.

\bibitem{aharonov4}
Y.~Aharonov and L.~Vaidman.
\newblock The two-state vector formalism of qauntum mechanics: an updated
  review.
\newblock 2001.

\bibitem{appleby2}
D.~M. Appleby.
\newblock Existential contextuality and the models of meyer, kent, and clifton.
\newblock {\em Phys. Rev. A 65, 022105}, 2002.
\newblock doi.org/10.1103/PhysRevA.65.022105.

\bibitem{appleby}
D.~M. Appleby.
\newblock The bell-kochen-specker theorem.
\newblock {\em Stud.Hist.Philos.Mod.Phys. 36}, 2005.

\bibitem{aristote}
Aristote.

\bibitem{aspect3}
A.~Aspect, J.~Dalibard, and G.~Roger.
\newblock Experimental test of bell's inequalities using time-varying
  analyzers.
\newblock {\em Phys. Rev. Lett., 49 (25): 1804–7}, 1982.

\bibitem{aspect2}
A.~Aspect, P.~Grangier, and G.~Roger.
\newblock Experimental tests of realistic local theories via bell's theorem.
\newblock {\em Phys. Rev. Lett., 47 (7): 460–3}, 1981.

\bibitem{primas}
H.~Atmanspacher and H.~Primas.
\newblock {\em Recasting Reality: Wolfgang Pauli's Philosophical Ideas and
  Contemporary Science}.
\newblock Springer Science and Business Media, 2008.

\bibitem{csm1}
A.~Auffèves and P.~Grangier.
\newblock Contexts, systems and modalities: A new ontology for quantum
  mechanics.
\newblock {\em P. Found Phys 46}, page 121, 2016.

\bibitem{barrett1}
J.~Barrett and A.~Kent.
\newblock Noncontextuality, finite precision measurement and the kochen-specker
  theorem.
\newblock {\em Stud. Hist. Philos. Mod. Phys. 35, 151}, 2004.
\newblock arxiv version :arXiv:quant-ph/0309017.

\bibitem{bell1}
J.~Bell.
\newblock On the einstein-podolsky-rosen paradox.
\newblock {\em Physics 1}, page 195–200, 1964.

\bibitem{bell}
J.~Bell.
\newblock On the problem of hidden variables in quantum. mechanics.
\newblock {\em Rev. Mod. Phys. 38}, page 447, 1966.

\bibitem{bell3}
J.~Bell.
\newblock The theory of local beables.
\newblock 1975.

\bibitem{davies}
J.~Bell and P.~Davies.
\newblock Adaptation from the edited transcript of a bbc radio interview. see
  the ghost in the atom: A discussion of the mysteries of quantum physics, by
  paul c. w. davies and julian r. brown, 1986/1993, pp. 45-46, 1985.

\bibitem{bell2}
J.~S. Bell.
\newblock {\em Speakable and Unspeakable in Quantum Mechanics.}
\newblock Cambridge University Press, 1987.

\bibitem{beller}
M.~Beller.
\newblock {\em Quantum Dialogue.}
\newblock University of Chicago Press, 1999.

\bibitem{bermejo-vega}
J.~Bermejo-Vega, N.~Delfosse, D.~E. Browne, C.~Okay, and R.~Raussendorf.
\newblock Contextuality as a resource for qubit quantum computation.
\newblock 2016.

\bibitem{birkhoff}
G.~Birkhoff and J.~von Neumann.
\newblock The logic of quantum mechanics.
\newblock {\em Annals of Mathematics 37}, page 823–843, 1936.

\bibitem{bitbol}
M.~Bitbol and S.~Osnaghi.
\newblock Bohr’s complementarity and kant’s epistemology in niels bohr,
  1913-2013 volume 68 of the series progress in mathematical physics pp
  199-221, 1972.

\bibitem{tamir}
E.~C. Boaz~Tamir.
\newblock Introduction to weak measurements and weak values.
\newblock {\em Quanta 2: 7–17}, 2013.

\bibitem{bohr0}
N.~Bohr.
\newblock Bohr's como lecture, originally published in a supplement to nature,
  april 14, 1928, p.580., sep 1927.

\bibitem{bohr}
N.~Bohr.
\newblock Can quantum-mechanical description of physical reality be considered
  complete?
\newblock {\em Phys. Rev. 48}, page 696, 1935.

\bibitem{bohr4}
N.~Bohr.
\newblock Discussion with einstein on epistemological problems in atomic
  physics. in p.a. schlipp (ed.), albert einstein: Philosopher-scientist, pages
  201–241. la salle: Open court, 1949.

\bibitem{bohr2}
N.~Bohr.
\newblock {\em Atomic Physics and Human Knowlegde.}
\newblock New York: Wiley, 1958.

\bibitem{bontems2}
V.~Bontems.
\newblock {\em Bachelard}.
\newblock Les Belles Lettres, 2010.

\bibitem{bontems}
V.~Bontems and C.~de~Ronde.
\newblock La notion d’entité en tant qu’obstacle épistémologique.
  bachelard, la mécanique quantique et la logique.
\newblock {\em Bulletin des Amis de Gaston Bachelard}, 2011.

\bibitem{bouveresse1}
J.~Bouveresse.
\newblock {\em Dans le labyrinthe : nécessité, contingence et liberté chez
  Leibniz : Cours 2009-2010, p.63}.
\newblock Collège de France, Paris.

\bibitem{bouveresse}
J.~Bouveresse.
\newblock {\em Qu'est-ce qu'un système philosophique ?}
\newblock Collège de France, Paris, 2012.

\bibitem{brassard}
G.~Brassard, H.~Buhrman, N.~Linden, A.~A. M\'ethot, A.~Tapp, and F.~Unger.
\newblock Limit on nonlocality in any world in which communication complexity
  is not trivial.
\newblock {\em Phys. Rev. Lett.}, 96:250401, Jun 2006.

\bibitem{magic}
S.~Bravyi and A.~Kitaev.
\newblock Universal quantum computation with ideal clifford gates and noisy
  ancillas.
\newblock {\em Phys. Rev. A 71}, 2005.

\bibitem{breuer}
T.~Breuer.
\newblock Kochen-specker theorem for finite precision spin-one measurements.
\newblock {\em Phys. Rev. Lett. 88, 240402}, 2002.
\newblock doi.org/10.1103/PhysRevLett.88.240402.

\bibitem{nop3}
A.~Brodutch.
\newblock Discord and quantum computational resources.
\newblock {\em Phys. Rev. A 88}, 2003.

\bibitem{brunner}
N.~Brunner and N.~Gisin.
\newblock Partial list of bipartite bell inequalities with four binary
  settings.
\newblock {\em Physics Letters A}, 372(18):3162 -- 3167, 2008.

\bibitem{bub3}
J.~Bub.
\newblock Von neumann's 'no hidden variables' proof: A re-appraisal.
\newblock {\em Found Phys 40: 1333–1340}, 2010.
\newblock doi:10.1007/s10701-010-9480-9 ; arxiv version : arXiv:1006.0499
  [quant-ph].

\bibitem{bub4}
J.~Bub and A.~Stairs.
\newblock Contextuality in quantum mechanics: Testing the klyachko inequality.
\newblock 2010.

\bibitem{cabello5}
A.~Cabello.
\newblock Finite-precision measurement does not nullify the kochen-specker
  theorem.
\newblock {\em Phys. Rev. A 65, 052101}, 2002.
\newblock doi.org/10.1103/PhysRevA.65.052101.

\bibitem{cabello14}
A.~Cabello.
\newblock Experimentally testable state-independent quantum contextuality.
\newblock {\em Phys. Rev. Lett.}, 101:210401, Nov 2008.

\bibitem{cabello8}
A.~Cabello.
\newblock Specker's fundamental principle of quantum mechanics.
\newblock 2011.
\newblock arXiv:1212.1756 [quant-ph].

\bibitem{cabello15}
A.~Cabello.
\newblock Simple explanation of the quantum violation of a fundamental
  inequality.
\newblock {\em Phys. Rev. Lett.}, 110:060402, Feb 2013.

\bibitem{cabello13}
A.~Cabello.
\newblock Interpretations of quantum theory: A map of madness.
\newblock 2015.
\newblock arXiv:1509.04711 [quant-ph].

\bibitem{cabello6}
A.~Cabello and M.~T. Cunha.
\newblock Proposal of a two-qutrit contextuality test free of the finite
  precision and compatibility loopholes.
\newblock {\em Phys. Rev. Lett. 106, 190401}, 2011.
\newblock doi.org/10.1103/PhysRevLett.106.190401.

\bibitem{cabello12}
A.~Cabello, J.~M. Estebaranz, and G.~G. Alcaine.
\newblock Bell-kochen-specker theorem: A proof with 18 vectors.
\newblock {\em Phys.Lett. A212 (1996) 183}, 1996.
\newblock arxiv version :arXiv:quant-ph/9706009.

\bibitem{cabello3}
A.~Cabello and G.~Garcia-Alcaine.
\newblock Proposed experimental tests of the bell-kochen-specker theorem.
\newblock {\em Phys. Rev. Lett.}, 80:1797--1799, Mar 1998.

\bibitem{cabello9}
A.~Cabello, S.~Severini, and A.~Winter.
\newblock (non-)contextuality of physical theories as an axiom.
\newblock 2010.
\newblock arXiv:1010.2163 [quant-ph].

\bibitem{cabello10}
A.~Cabello, S.~Severini, and A.~Winter.
\newblock Graph-theoretic approach to quantum correlations.
\newblock {\em Phys. Rev. Lett.}, 112:040401, Jan 2014.

\bibitem{camus}
A.~Camus.
\newblock {\em Le Mythe de Sisyphe}.
\newblock Gallimard.

\bibitem{alice}
L.~Carroll.
\newblock {\em Alice's Adventures in Wonderland}.

\bibitem{nat3}
D.~Castelvecchi.
\newblock Ibm's quantum cloud computer goes commercial.
\newblock {\em Nature 543}, mar 2017.

\bibitem{nat1}
D.~Castelvecchi.
\newblock Quantum computers ready to leap out of the lab in 2017.
\newblock {\em Nature 541}, jan 2017.

\bibitem{cavalcanti1}
E.~G. Cavalcanti.
\newblock Nonlocality and contextuality as fine-tuning ; pirsa 17070039.

\bibitem{cavalcanti}
E.~G. Cavalcanti.
\newblock Quantum nonlocality and contextuality as fine-tuning.

\bibitem{chiribella1}
G.~Chiribella, G.~M. D'Ariano, and P.~Perinotti.
\newblock {\em Quantum from principles, book chapter in "Quantum Theory:
  Informational Foundations and Foils", G. Chiribella AND R. Spekkens}.
\newblock eds. Springer, 2016.

\bibitem{clifton}
R.~Clifton.
\newblock Getting contextual and nonlocal elements-of-reality the easy way.
\newblock {\em American Journal of Physics}, 61(5):443--447, 1993.

\bibitem{clifton2}
R.~Clifton and A.~Kent.
\newblock Simulating quantum mechanics by non-contextual hidden variables.
\newblock {\em Proc. Roy. Soc. Lond. A 456, 2101}, 2000.
\newblock doi:10.1098/rspa.2000.0604.

\bibitem{coecke1}
B.~Coecke, T.~Fritz, and R.~W. Spekkens.
\newblock A mathematical theory of resources.
\newblock {\em Information and Computation}, 250:59 -- 86, 2016.
\newblock Quantum Physics and Logic.

\bibitem{coecke}
B.~Coecke, D.~Moore, and A.~Wilce.
\newblock {\em Operational Quantum Logic: An Overview}, pages 1--36.
\newblock Springer Netherlands, Dordrecht, 2000.

\bibitem{gilles}
G.~Cohen-Tannoudji.

\bibitem{gilles1}
G.~Cohen-Tannoudji.
\newblock Actualit{\'e} de la philosophie de ferdinand gonseth.
\newblock {\em Revue de Synth{\`e}se}, 126(2):417--429, Jun 2005.

\bibitem{dis}
A.~Datta, A.~Shaji, and C.~M. Caves.
\newblock Quantum discord and the power of one qubit.
\newblock {\em Phys. Rev. Lett. 100, 050502}, page 117, 2008.

\bibitem{desilva}
N.~de~Silva.
\newblock Graph-theoretic strengths of contextuality.
\newblock {\em Phys. Rev. A}, 95:032108, Mar 2017.

\bibitem{nop4}
M.~den Nest.
\newblock Universal quantum computation with little entanglement.
\newblock {\em Phys. Rev. Lett. 110, 060504}, 2013.

\bibitem{denkmayr}
T.~Denkmayr, H.~Geppert, S.~Sponar, H.~Lemmel, A.~Matzkin, J.~Tollaksen, and
  Y.~Hasegawa.
\newblock Observation of a quantum cheshire cat in a matter-wave interferometer
  experiment.
\newblock {\em Nature Communications 5, 4492}, 2014.
\newblock doi:10.1038/ncomms5492.

\bibitem{para}
D.~Deutsch.
\newblock Quantum theory, the church-turing principle and the universal quantum
  computer.
\newblock {\em Proc. R. Soc. A. 400 1818 97}, page 117, 1985.

\bibitem{dieks}
D.~Dieks.
\newblock Von neumann's impossibility proof: Mathematics in the service of
  rhetorics.
\newblock {\em Studies in History and Philosophy of Modern Physics}, 2016.
\newblock https://doi.org/10.1016/j.shpsb.2017.01.008.

\bibitem{epr}
A.~Einstein, B.~Podolsky, and N.~Rosen.
\newblock Can quantum-mechanical description of physical reality be considered
  complete?
\newblock {\em Phys. Rev. 47}, page 777, 1935.

\bibitem{emery}
E.~Emery.
\newblock {\em Ferdinand Gonseth: pour une philosophie dialectique ouverte à
  l'expérience}.
\newblock L'AGE D'HOMME, 1985.

\bibitem{fine1}
A.~Fine.
\newblock Hidden variables, joint probability, and the bell inequalities.
\newblock {\em Phys. Rev. Lett.}, 48:291--295, Feb 1982.

\bibitem{fine2}
A.~Fine.
\newblock Joint distributions, quantum correlations, and commuting observables.
\newblock {\em Journal of Mathematical Physics}, 23(7):1306--1310, 1982.

\bibitem{foulis}
D.~J. Foulis and C.~H. Randall.
\newblock {\em Empirical logic and tensor products}, pages 9 -- 20.
\newblock Wissenschaftsverlag, Bibliographisches Institut 5, Zurich, 1981.

\bibitem{fritz2}
T.~Fritz, A.~B. Sainz, R.~Augusiak, J.~B. Brask, A.~L. R.~Chaves, and A.~Acín.
\newblock Local orthogonality as a multipartite principle for quantum
  correlations.
\newblock {\em Nature Communications 4, 2263}, 2013.
\newblock arxiv version : arXiv:1210.3018 [quant-ph].

\bibitem{fuchs3}
C.~Fuchs.
\newblock Quantum mechanics as quantum information (and only a little more).
\newblock 2002.

\bibitem{fuchs2}
C.~Fuchs.
\newblock Qbism, the perimeter of quantum bayesianism.
\newblock 2010.

\bibitem{fuchs}
C.~Fuchs.
\newblock On participatory realism.
\newblock 2016.

\bibitem{fuchs4}
C.~Fuchs, M.~Schlosshauer, and B.~C. Stacey.
\newblock My struggles with the block universe.
\newblock 2014.

\bibitem{fuchs5}
C.~Fuchs and B.~Stacey.
\newblock Qbism: Quantum theory as a hero's handbook.
\newblock 2016.

\bibitem{giustina}
M.~Giustina, M.~A.~M. Versteegh, S.~Wengerowsky, J.~Handsteiner, A.~Hochrainer,
  K.~Phelan, F.~Steinlechner, J.~Kofler, J.-A. Larsson, C.~Abellan, W.~Amaya,
  V.~Pruneri, M.~W. Mitchell, J.~Beyer, T.~Gerrits, A.~E. Lita, L.~K. Shalm,
  S.~W. Nam, T.~Scheidl, R.~Ursin, B.~Wittmann, and A.~Zeilinger.
\newblock Significant-loophole-free test of bell's theorem with entangled
  photons.
\newblock {\em Phys. Rev. Lett. 115, 250401}, 2015.
\newblock doi:10.1103/PhysRevLett.115.250401 ; arxiv version :
  arXiv:1511.03190.

\bibitem{gleason}
A.~M. Gleason.
\newblock Measures on the closed subspaces of a hilbert space.
\newblock {\em J. Math. Mech. 6}, page 885, 1957.

\bibitem{ghz}
D.~M. Greenberger, M.~A. Horne, A.~Shimony, and A.~Zeilinger.
\newblock Bell's theorem without inequalities.
\newblock 58:1131--43, 1990.

\bibitem{ghz2}
D.~M. Greenberger, M.~A. Horne, and A.~Zeilinger.
\newblock Going beyond bell's theorem.
\newblock {\em in 'Bell's Theorem, Quantum Theory, and Conceptions of the
  Universe', M. Kafatos (Ed.), Kluwer, Dordrecht, 69-72 (1989)}.

\bibitem{alexei5}
A.~Grinbaum.
\newblock communication privée.

\bibitem{alexei7}
A.~Grinbaum.
\newblock On the notion of reconstruction of quantum theory.
\newblock 2005.
\newblock arXiv:quant-ph/0509104.

\bibitem{alexei4}
A.~Grinbaum.
\newblock On epistemological modesty.
\newblock {\em Philosophica 83 pp. 139-150}, 2008.

\bibitem{alexei6}
A.~Grinbaum.
\newblock {\em Mécanique des étreintes}.
\newblock Les Belles Lettres, 2014.

\bibitem{alexei2}
A.~Grinbaum.
\newblock How device-independent approaches change the meaning of physical
  theory.
\newblock {\em Studies in History and Philosophy of Modern Physics 58 (2017)
  22-30}, 2015.

\bibitem{alexei1}
A.~Grinbaum.
\newblock Quantum theory as a critical regime of language dynamics.
\newblock {\em Found Phys 45}, 2015.

\bibitem{alexei3}
A.~Grinbaum.
\newblock Narratives of quantum theory in the age of quantum technologies.
\newblock 2017.

\bibitem{hardy3}
L.~Hardy.
\newblock Quantum mechanics, local realistic theories, and lorentz-invariant
  realistic theories.
\newblock {\em Phys. Rev. Lett.}, 68:2981--2984, May 1992.

\bibitem{hardy4}
L.~Hardy.
\newblock Nonlocality for two particles without inequalities for almost all
  entangled states.
\newblock {\em Phys. Rev. Lett.}, 71:1665--1668, Sep 1993.

\bibitem{hardy2}
L.~Hardy.
\newblock Quantum theory from five reasonable axioms.
\newblock 2001.
\newblock arXiv:quant-ph/0101012.

\bibitem{harrigan}
N.~Harrigan and R.~Spekkens.
\newblock Einstein, incompleteness, and the epistemic view of quantum states.
\newblock {\em Foundations of Physics}, 40(2):125--157, Feb 2010.

\bibitem{havlicek}
H.~Havlicek, G.~Krenn, J.~Summhammer, and K.~Svozil.
\newblock Coloring the rational quantum sphere and the kochen-specker theorem.
\newblock {\em J. Phys. A 34, 3071}, 2001.

\bibitem{held}
C.~Held.
\newblock The meaning of complementarity.
\newblock {\em Studies in History and Philosophy of Science 25, 871-893}, 1994.

\bibitem{hensen}
B.~Hensen, H.~Bernien, A.~Dréau, A.~Reiserer, N.~Kalb, M.~Blok, J.~Ruitenberg,
  R.~Vermeulen, R.~Schouten, C.~Abellán, W.~Amaya, V.~Pruneri, M.~W. Mitchell,
  M.~Markham, D.~Twitchen, D.~Elkouss, S.~Wehner, T.~Taminiau, and R.~Hanson.
\newblock Loophole-free bell inequality violation using electron spins
  separated by 1.3 kilometres.
\newblock {\em Nature. 526: 682–686}, 2015.
\newblock doi:10.1038/nature15759 ; arxiv version : arXiv:1508.05949.

\bibitem{henson}
J.~Henson.
\newblock Quantum contextuality from a simple principle?
\newblock Oct 2012.

\bibitem{hooker}
C.~A. Hooker.
\newblock The nature of quantum mechanical reality: Einstein versus bohr. in j.
  colodny (ed.). paradigms and paradoxes: The philosophical challenges of the
  quantum domain, pages 67–302. pittsburgh: University of pittsburgh press,
  1972.

\bibitem{houzel}
C.~Houzel.
\newblock Histoire de la théorie des faisceaux.
\newblock {\em AMS 1991 Mathematics Subject Classification: 01A65, 55-03,
  14-03, 35A27}.

\bibitem{howard2}
D.~Howard.
\newblock Einstein on locality and separability.
\newblock {\em Studies in History and Philosophy of Science 16: 171–201},
  1985.

\bibitem{howard1}
D.~Howard.
\newblock Who invented the copenhagen interpretation?
\newblock {\em Philos. Sci. 71}, page 669–682, 2004.

\bibitem{comput}
M.~Howard, J.~Wallman, V.~Veitch, and J.~Emerson.
\newblock Contextuality supplies the magic for quantum computation.
\newblock {\em Nature}, 510:351--355, May 2014.

\bibitem{jammer1}
M.~Jammer.
\newblock {\em The Philosophy of Quantum Mechanics}.
\newblock Wiley, New York, 1974.

\bibitem{jammer2}
M.~Jammer.
\newblock John stewart bell and his work - on the occasion of his sixtieth
  birthday.
\newblock {\em Found. Phys. 20, 1139}, 1990.

\bibitem{jennings}
D.~Jennings and M.~Leifer.
\newblock No return to classical reality.
\newblock {\em Contemporary Physics}, 57(1):60--82, 2016.

\bibitem{jia}
Z.-A. Jia, G.-D. Cai, Y.-C. Wu, G.-C. Guo, and A.~Cabello.
\newblock The exclusivity principle determines the correlation monogamy.
\newblock 2017.

\bibitem{kent}
A.~Kent.
\newblock Noncontextual hidden variables and physical measurements.
\newblock {\em Phys. Rev. Lett. 83, 3755}, 1999.
\newblock doi.org/10.1103/PhysRevLett.83.3755.

\bibitem{klein}
E.~Klein.
\newblock {\em Petit voyage dans le monde des quanta}.
\newblock Flammarion, 2004.

\bibitem{kleinmann}
M.~Kleinmann, O.~Gühne, J.~R. Portillo, J.~Åke Larsson, and A.~Cabello.
\newblock Memory cost of quantum contextuality.
\newblock {\em New Journal of Physics}, 13(11):113011, 2011.

\bibitem{kcbs}
A.~A. Klyachko, M.~A. Can, S.~Binicio\ifmmode~\breve{g}\else \u{g}\fi{}lu, and
  A.~S. Shumovsky.
\newblock Simple test for hidden variables in spin-1 systems.
\newblock {\em Phys. Rev. Lett.}, 101:020403, Jul 2008.

\bibitem{ks}
S.~Kochen and E.~Specker.
\newblock The problem of hidden variables in quantum mechanics.
\newblock {\em J. Math. Mech. 17}, page~59, 1967.

\bibitem{krishna}
A.~Krishna, R.~W. Spekkens, and E.~Wolfe.
\newblock Deriving robust noncontextuality inequalities from algebraic proofs
  of the kochen-specker theorem: the peres-mermin square.
\newblock 2017.

\bibitem{kunjwal5}
R.~Kunjwal.
\newblock Beyond the cabello-severini-winter framework: making sense of
  contextuality without sharpness of measurements.

\bibitem{kunjwal1}
R.~Kunjwal.
\newblock A note on the joint measurability of povms and its implications for
  contextuality.
\newblock 2014.

\bibitem{kunjwal2}
R.~Kunjwal.
\newblock {\em Contextuality beyond the Kochen-Specker theorem}.
\newblock PhD thesis, Institute of Mathematical Sciences, Chennai, 2016.

\bibitem{kunjwal4}
R.~Kunjwal and R.~W. Spekkens.
\newblock From statistical proofs of the kochen-specker theorem to noise-robust
  noncontextuality inequalities.

\bibitem{kunjwal3}
R.~Kunjwal and R.~W. Spekkens.
\newblock From the kochen-specker theorem to noncontextuality inequalities
  without assuming determinism.
\newblock {\em Phys. Rev. Lett.}, 115:110403, Sep 2015.

\bibitem{kwiat}
P.~G. Kwiat and L.~Hardy.
\newblock The mystery of the quantum cakes.
\newblock {\em American Journal of Physics}, 68(1):33--36, 2000.

\bibitem{landsman}
N.~Landsman.
\newblock Between classical and quantum.
\newblock 2005.
\newblock arXiv:quant-ph/0506082.

\bibitem{laplace}
J.-S. Laplace.
\newblock {\em Essai philosophiques sur les probabilités}.
\newblock Courcier, 1814.

\bibitem{larsson1}
J.-. Larsson.
\newblock A kochen-specker inequality.
\newblock {\em EPL (Europhysics Letters)}, 58(6):799, 2002.

\bibitem{leibniz}
G.~Leibniz.
\newblock {\em Quatrième écrit de Leibniz, Correspondance Leibniz-Clarke,
  lettre du 2 juin 1716}.

\bibitem{leibniz1}
G.~Leibniz.
\newblock {\em Théodicée, I, 44}.

\bibitem{leifer1}
M.~Leifer.
\newblock The dirty secrets of quantum foundations, 2016.

\bibitem{leifer9}
M.~S. Leifer.
\newblock Aharonov vs. spekkens round ii : Contextuality in pre- and
  post-selection paradoxes, pirsa 17010041.

\bibitem{leifer5}
M.~S. Leifer.
\newblock Foundations of quantum mechanics (review) lecture 6, pirsa 17010038.

\bibitem{leifer7}
M.~S. Leifer.
\newblock Foundations of quantum mechanics (review) lecture 7, pirsa 17010039.

\bibitem{leifer6}
M.~S. Leifer.
\newblock Is the quantum state real? an extended review of psi-ontology
  theorems.
\newblock {\em Quanta}, 3(1):67--155, 2014.

\bibitem{leifer3}
M.~S. Leifer and R.~W. Spekkens.
\newblock Logical pre- and post-selection paradoxes, measurement-disturbance
  and contextuality.
\newblock {\em Int. J. Theor. Phys. 44, 1977-1987}, 2005.

\bibitem{leifer4}
M.~S. Leifer and R.~W. Spekkens.
\newblock Pre- and post-selection paradoxes and contextuality in quantum
  mechanics.
\newblock {\em Phys. Rev. Lett. 95, 200405}, 2005.

\bibitem{leifer8}
M.~S. Leifer and R.~W. Spekkens.
\newblock Towards a formulation of quantum theory as a causally neutral theory
  of bayesian inference.
\newblock {\em Phys. Rev. A}, 88:052130, Nov 2013.

\bibitem{leucippe}
Leucippus.
\newblock Fragment 569 - from fr. 2 actius i, 25, 4.

\bibitem{lewis}
D.~Lewis.
\newblock {\em Counterfactuals}.
\newblock Oxford, Blackwell Press, 21973.

\bibitem{liang}
Y.-C. Liang, R.~W. Spekkens, and H.~M. Wiseman.
\newblock Specker’s parable of the overprotective seer: A road to
  contextuality, nonlocality and complementarity.
\newblock {\em Physics Reports}, 506(1):1 -- 39, 2011.

\bibitem{lidar}
A.~Lidar and T.~A. Brun.
\newblock {\em Quantum Error Correction.}
\newblock Cambridge University Press, 2013.

\bibitem{leblond}
J.-M. Lévy-Leblond and F.~Balibar.
\newblock {\em Quantique}.
\newblock Inter-Editions, 1984.

\bibitem{mackey}
G.~W. Mackey.
\newblock {\em Mathematical Foundations of Quantum Mechanics.}
\newblock Benjamin, New York, 1963.

\bibitem{mazurek}
M.~D. Mazurek, M.~F. Pusey, R.~Kunjwal, K.~J. Resch, and R.~W. Spekkens.
\newblock An experimental test of noncontextuality without unwarranted
  idealizations.
\newblock {\em Nat. Commun}, 7, June 2016.

\bibitem{specker3}
J.~Meon and E.~Specker.
\newblock {\em A Story of a Friend, introduction of the book 'Ernst Specker
  Selecta'}.
\newblock Birkhäuser Verlag, Basel, Switzerland, 1990.

\bibitem{mermin1}
D.~Mermin.
\newblock Hidden variables and the two theorems of john bell.
\newblock {\em Rev. Mod. Phys.}, 65:803--815, Jul 1993.

\bibitem{meyer}
D.~A. Meyer.
\newblock Finite precision measurement nullifies the kochen-specker theorem.
\newblock {\em Phys. Rev. Lett. 83, 3751}, 1999.

\bibitem{nat2}
M.~Mohseni, P.~Read, H.~Neven, S.~Boixo, V.~Denchev, R.~Babbush, A.~Fowler,
  V.~Smelyanskiy, and J.~Martinis.
\newblock Commercialize quantum technologies in five years.
\newblock {\em Nature 543}, mar 2017.

\bibitem{murdoch}
D.~Murdoch.
\newblock {\em Niels Bohrs Philosophy of Physics.}
\newblock Cambridge University Press, 1987.

\bibitem{navascues}
M.~Navascu{\'e}s and H.~Wunderlich.
\newblock A glance beyond the quantum model.
\newblock {\em Proceedings of the Royal Society of London A: Mathematical,
  Physical and Engineering Sciences}, 466(2115):881--890, 2010.

\bibitem{connorrob}
J.~J. O'Connor and E.~F. Robertson.

\bibitem{philocop}
S.~E. of~Philosophy.
\newblock Copenhagen interpretation of quantum mechanics, 2002-2014.
\newblock https://plato.stanford.edu/entries/qm-copenhagen/.

\bibitem{pais1}
A.~Pais.
\newblock {\em Subtle is the Lord… The Science and Life of Albert Einstein.}
\newblock Toronto: Oxford University Press, 1982.

\bibitem{pais2}
A.~Pais.
\newblock {\em A Tale of Two Continents : A Physicist's Life in a Turbulent
  World.}
\newblock Princeton University Press, 1997.

\bibitem{pal}
K.~F. P\'al and T.~V\'ertesi.
\newblock Maximal violation of a bipartite three-setting, two-outcome bell
  inequality using infinite-dimensional quantum systems.
\newblock {\em Phys. Rev. A}, 82:022116, Aug 2010.

\bibitem{pauli}
W.~Pauli.

\bibitem{pawlowski}
M.~Pawowski, T.~Paterek, D.~Kaszlikowski, V.~Scarani, A.~Winter, and
  M.~Zukowski.
\newblock Information causality as a physical principle.
\newblock {\em Nature}, 461:1101--1104, 2009.

\bibitem{pearl}
J.~Pearl.
\newblock {\em Causality: models, reasoning, and inference.}
\newblock Cambridge University Press, 2009.

\bibitem{penrose}
R.~Penrose and J.~al~Khalili.
\newblock Roger penrose on black holes.

\bibitem{peres2}
A.~Peres.
\newblock Unperformed experiments have no results.
\newblock {\em American Journal of Physics, Volume 46, Issue 7, pp. 745-747},
  1978.
\newblock doi:10.1119/1.11393.

\bibitem{peres3}
A.~Peres.
\newblock Incompatible results of quantum measurements.
\newblock {\em Physics Letters A}, 151(3):107 -- 108, 1990.

\bibitem{peres}
A.~Peres.
\newblock {\em Quantum Theory, Concepts and Methods, chap. 7}.
\newblock Kluwer Academic, Dordrecht, 1993.

\bibitem{peresb}
A.~Peres.
\newblock {\em Quantum Theory, Concepts and Methods, p.212}.
\newblock Kluwer Academic, Dordrecht, 1993.

\bibitem{petersen}
A.~Petersen.
\newblock The philosophy of niels bohr.
\newblock {\em Bulletin of the Atomic Scientists, 19: 8–14}, 1963.

\bibitem{expwv}
F.~Piacentini, A.~Avella, M.~P. Levi, R.~Lussana, F.~Villa, A.~Tosi, F.~Zappa,
  M.~Gramegna, G.~Brida, I.~P. Degiovanni, and M.~Genovese.
\newblock An experiment investigating the connection between weak values and
  contextuality.
\newblock {\em Phys. Rev. Lett. 116, 180401}, 2016.
\newblock arxiv version : arXiv:1602.02075.

\bibitem{popescu}
S.~Popescu.
\newblock Nonlocality beyond quantum mechanics.
\newblock {\em Nat Phys}, 10:264--270, Apr 2014.

\bibitem{pr}
S.~Popescu and D.~Rohrlich.
\newblock Quantum nonlocality as an axiom.
\newblock {\em Foundations of Physics}, 24(3):379--385, Mar 1994.

\bibitem{pusey3}
M.~F. Pusey.
\newblock Contextuality as a litmus test for quantum weirdness ; pirsa
  114100106.

\bibitem{pusey2}
M.~F. Pusey.
\newblock Anomalous weak values are proofs of contextuality.
\newblock {\em Phys. Rev. Lett.113 , 200401}, 2014.

\bibitem{pusey1}
M.~F. Pusey and M.~S. Leifer.
\newblock Logical pre- and post-selection paradoxes are proofs of
  contextuality.
\newblock {\em EPTCS 195}, pages 295--306, 2015.

\bibitem{raussendorf}
R.~Raussendorf.
\newblock Contextuality in measurement-based quantum computation.
\newblock {\em Phys. Rev. A}, 88:022322, Aug 2013.

\bibitem{reichenbach}
H.~Reichenbach.
\newblock {\em The Direction of Time}.
\newblock University of California Press, Berkeley, 1956.

\bibitem{rovelli}
C.~Rovelli.
\newblock Relational quantum mechanics.
\newblock {\em Int. J. of Theor. Phys., 35: 1637}, 1996.
\newblock doi: 10.1007/BF02302261 ; arxiv version : arXiv:quant-ph/9609002.

\bibitem{sainz3}
A.~B. Sainz.

\bibitem{sainz}
A.~B. Sainz, T.~Fritz, R.~Augusiak, J.~B. Brask, R.~Chaves, A.~Leverrier, , and
  A.~Acín.
\newblock Exploring the local orthogonality principle.
\newblock {\em Phys. Rev. A 89, 032117}, 2013.
\newblock doi:10.1103/PhysRevA.89.032117.

\bibitem{sainz2}
A.~B. Sainz, T.~Fritz, R.~Augusiak, J.~B. Brask, R.~Chaves, A.~Leverrier, and
  A.~Acín.
\newblock Exploring the local orthogonality principle.
\newblock {\em Phys. Rev. A 89, 032117}, 2014.
\newblock doi: 10.1103/PhysRevA.89.032117 ; arxiv version : arXiv:1311.6699
  [quant-ph].

\bibitem{sainz4}
A.~B. Sainz and E.~Wolfe.
\newblock Multipartite composition of contextuality scenarios.
\newblock 2017.

\bibitem{saunders}
S.~Saunders.
\newblock Complementarity and scientific rationality.
\newblock {\em Foundations of Physics 35, Issue 3, pp 417–447}, 2005.
\newblock doi:10.1007/s10701-004-1982-x.

\bibitem{schlo}
M.~Schlosshauer, J.~Kofler, and A.~Zeilinger.
\newblock A snapshot of foundational attitudes toward quantum mechanics.
\newblock 2013.

\bibitem{schmid}
D.~Schmid and R.~W. Spekkens.
\newblock Contextual advantage for state discrimination.
\newblock 2017.

\bibitem{shalm}
L.~K. Shalm, E.~Meyer-Scott, B.~G. Christensen, P.~Bierhorst, M.~A. Wayne,
  M.~J. Stevens, T.~Gerrits, S.~Glancy, D.~R. Hamel, M.~S. Allman, K.~J.
  Coakley, S.~D. Dyer, C.~Hodge, A.~E. Lita, V.~B. Verma, C.~Lambrocco,
  E.~Tortorici, A.~L. Migdall, Y.~Zhang, D.~R. Kumor, W.~H. Farr, F.~Marsili,
  M.~D. Shaw, J.~A. Stern, C.~Abellán, W.~Amaya, V.~Pruneri, T.~Jennewein,
  M.~W. Mitchell, P.~G. Kwiat, J.~C. Bienfang, R.~P. Mirin, E.~Knill, and S.~W.
  Nam.
\newblock A strong loophole-free test of local realism.
\newblock {\em Phys. Rev. Lett. 115, 250402}, 2015.
\newblock doi:10.1103/PhysRevLett.115.250402 ; arxiv version :
  arXiv:1511.03189.

\bibitem{shimony}
A.~Shimony.
\newblock {\em The Search for a naturalistic World View.}
\newblock Cambridge University Press, 1993.

\bibitem{simon1}
C.~Simon, C.~Brukner, and A.~Zeilinger.
\newblock Hidden-variable theorems for real experiments.
\newblock {\em Phys. Rev. Lett. 86, 4427- 4430}, 2001.
\newblock doi:10.1103/PhysRevLett.86.4427 ; arxiv version :
  arXiv:quant-ph/0006043.

\bibitem{simon2}
C.~Simon, i.~c.~v. Brukner, and A.~Zeilinger.
\newblock Hidden-variable theorems for real experiments.
\newblock {\em Phys. Rev. Lett.}, 86:4427--4430, May 2001.

\bibitem{solana}
M.~Solana.
\newblock {\em Historia de la filosofia española}.
\newblock Madrid : Real Academia de Ciencias Exectas, Fisicas y Naturales,
  1941.

\bibitem{specker4}
E.~Specker.
\newblock {\em Wie ein Dieb in der Nacht: 35 Kurzpredigten}.
\newblock TVZ, Theol. Verlag, 2008.

\bibitem{specker}
E.~P. Specker.
\newblock Die logik nicht gleichzeitig entscheidbarer aussagen.
\newblock {\em Dialectica 14}, pages 239--246, 1960.

\bibitem{specker2}
E.~P. Specker and A.~Cabello.
\newblock https://vimeo.com/52923835.

\bibitem{spekkens8}
R.~Spekkens.
\newblock Ernst paul specker (1920–2011).
\newblock {\em Mind and Matter Vol. 9(2), pp. 121–128}, 2011.

\bibitem{spekkens10}
R.~W. Spekkens.
\newblock Experimental quantum foundations.

\bibitem{spekkens11}
R.~W. Spekkens.
\newblock Noncontextuality: how we should define it, why it is natural, and
  what to do about its failure ; pirsa 17070035.

\bibitem{spekkens1}
R.~W. Spekkens.
\newblock Contextuality for preparations, transformations, and unsharp
  measurements.
\newblock {\em Phys. Rev. A}, 71:052108, May 2005.

\bibitem{spekkens5}
R.~W. Spekkens.
\newblock Evidence for the epistemic view of quantum states: A toy theory.
\newblock {\em Phys. Rev. A}, 75:032110, Mar 2007.

\bibitem{spekkens9}
R.~W. Spekkens.
\newblock The status of determinism in proofs of the impossibility of a
  noncontextual model of quantum theory.
\newblock {\em Foundations of Physics}, 44(11):1125--1155, Nov 2014.

\bibitem{spekkens4}
R.~W. Spekkens.
\newblock Leibniz's principle of the identity of indiscernibles as a
  foundational principle for quantum theory, conference on
  information-theoretic interpretations of quantum mechanics, university of
  western ontario, london, canada, jun 2016.
\newblock https://www.youtube.com/watch?v=HWOkjisIxc4.

\bibitem{spekkens7}
R.~W. Spekkens.
\newblock {\em Quasi-Quantization: Classical Statistical Theories with an
  Epistemic Restriction}, pages 83--135.
\newblock Springer Netherlands, Dordrecht, 2016.

\bibitem{spekkens3}
R.~W. Spekkens, D.~H. Buzacott, A.~J. Keehn, B.~Toner, and G.~J. Pryde.
\newblock Preparation contextuality powers parity-oblivious multiplexing.
\newblock {\em Phys. Rev. Lett.}, 102:010401, Jan 2009.

\bibitem{spinoza}
B.~Spinoza.
\newblock {\em Ethics, Proposition 29}.
\newblock 1677.

\bibitem{spirtes}
P.~Spirtes, C.~N. Glymour, and R.~Scheines.
\newblock {\em Causation, prediction, and search.}
\newblock The MIT Press, 2001.

\bibitem{nop1}
A.~M. Steane.
\newblock A quantum computer only needs one universe.
\newblock {\em Preprint available at http://arxiv.org/abs/quantph/ 0003084},
  2000.

\bibitem{stoica}
C.~Stoica.

\bibitem{suarez}
A.~Suarez.
\newblock communication privée de suzanne specker à a. suarez.

\bibitem{svozil1}
K.~Svozil.
\newblock On counterfactuals and contextuality.
\newblock {\em AIP Conference Proceedings 750. Foundations of Probability and
  Physics-3, ed. by Andrei Khrennikov (American Institute of Physics,Melville,
  NY, 2005) pp. 351-360}, 2005.
\newblock doi : 10.1063/1.1874586 ; arxiv version : arXiv:quant-ph/0406014.

\bibitem{svozil2}
K.~Svozil.
\newblock Quantum scholasticism: On quantum contexts, counterfactuals, and the
  absurdities of quantum omniscience.
\newblock {\em Information Sciences 179, 535-541}, 2009.
\newblock doi : 10.1016/j.ins.2008.06.012; arxiv version : arXiv:0711.1473
  [quant-ph].

\bibitem{tsirelson}
B.~Tsirelson.
\newblock Quantum generalizations of bell's inequality.
\newblock {\em Lett. Math. Phys.}, 4:93--100, 1980.

\bibitem{3boites}
L.~Vaidman.
\newblock Weak-measurement elements of reality.
\newblock {\em Foundations of Physics 26, 895-906}, 1996.

\bibitem{vaidman2}
L.~Vaidman.
\newblock Counterfactuals in quantum mechanics.
\newblock 2007.
\newblock arXiv:0709.0340 [quant-ph].

\bibitem{vaidman3}
L.~Vaidman.
\newblock Quantum theory and determinism.
\newblock {\em Quantum Studies: Mathematics and Foundations}, September 2014.
\newblock doi : 10.1007/s40509-014-0008-4 ; arxiv version : arXiv:1405.4222
  [quant-ph].

\bibitem{vaidman}
L.~Vaidman.
\newblock Weak value controversy.
\newblock 2017.
\newblock arXiv:1703.08870 [quant-ph].

\bibitem{quint}
G.~Vidal.
\newblock Efficient classical simulation of slightly entangled quantum
  computations.
\newblock {\em Phys. Rev. Lett. 91, 147902}, 2003.

\bibitem{nop2}
V.~Vidal.
\newblock The elusive source of quantum speedup.
\newblock {\em Found. Physics, 40 022307}, 2010.

\bibitem{vonneum}
J.~von Neumann.
\newblock {\em Mathematische Grundlagen der Quantenmechanik}.
\newblock Springer, Berlin, 1932.

\bibitem{waegell2}
M.~Waegell, T.~Denkmayr, H.~Geppert, D.~Ebner, T.~Jenke, Y.~Hasegawa,
  S.~Sponar, J.~Dressel, and J.~Tollaksen.
\newblock Confined contextuality in neutron interferometry: Observing the
  quantum pigeonhole effect.
\newblock 2016.
\newblock arXiv:1609.06046.

\bibitem{waegell1}
M.~Waegell and J.~Tollaksen.
\newblock Contextuality, pigeonholes, cheshire cats, mean kings, and weak
  values.
\newblock 2015.
\newblock arXiv:1505.00098.

\bibitem{wilce}
A.~Wilce.
\newblock {\em Test Spaces}, pages 443 -- 549.
\newblock Elsevier, Amsterdam, 2009.

\bibitem{wiseman}
H.~Wiseman and E.~G. Cavalcanti.
\newblock Causarum investigatio and the two bell's theorems of john bell.

\bibitem{wiseman1}
H.~M. Wiseman.
\newblock The two bell's theorems of john bell.
\newblock {\em Journal of Physics A: Mathematical and Theoretical},
  47(42):424001, 2014.

\bibitem{wood}
C.~J. Wood and R.~W. Spekkens.
\newblock The lesson of causal discovery algorithms for quantum correlations:
  causal explanations of bell-inequality violations require fine-tuning.
\newblock {\em New Journal of Physics}, 17(3):033002, 2015.

\bibitem{zeilinger}
A.~Zeilinger.
\newblock {\em Dance of the Photons}.
\newblock Farrar, Straus and Giroux, New York, 2010.

\end{thebibliography}
\bibliographystyle{abbrv}

\end{document}